\documentclass[11pt]{article}
\def\IRIScombinedmode{}
\usepackage{graphicx}
\usepackage{pdflscape}
\usepackage{afterpage}
\usepackage[letterpaper,margin=1in]{geometry}

\usepackage[T1]{fontenc}
\usepackage{mathptmx}
\usepackage{amsmath,amssymb,bm}
\usepackage{booktabs,tabularx,array,longtable}
\usepackage{enumitem}
\usepackage{xcolor}
\usepackage[numbers,sort&compress]{natbib}
\usepackage{xr-hyper}
\usepackage[hidelinks]{hyperref}
\usepackage{siunitx}
\usepackage[font=small,labelfont=bf]{caption}
\usepackage{etoc}

\newif\ifIRIScombined
\ifdefined\IRIScombinedmode \IRIScombinedtrue \fi

\ifIRIScombined
  \newcommand{\SIref}[1]{\ref{#1}}
  \newcommand{\Mref}[1]{\ref{#1}}
\else
  \newcommand{\SIref}[1]{\ref{SI-#1}}
  \newcommand{\Mref}[1]{\ref{M-#1}}
\fi
\newcommand{\Meqref}[1]{(\Mref{#1})}
\newcommand{\SFig}[1]{Supplementary Fig.~\SIref{#1}}
\newcommand{\STab}[1]{Supplementary Table~\SIref{#1}}
\newcommand{\SSec}[1]{Supplementary Section~\SIref{#1}}

\renewcommand{\arraystretch}{1.15}
\newcolumntype{L}[1]{>{\raggedright\arraybackslash}p{#1}}

\newcommand{\e}{\mathbf{e}}
\newcommand{\eh}{\hat{\mathbf{e}}}
\newcommand{\et}{\tilde{\mathbf{e}}}
\newcommand{\eb}{\bar{\mathbf{e}}}
\newcommand{\rv}{\mathbf{r}}
\newcommand{\Q}{\mathbf{Q}}
\newcommand{\Aset}{\mathcal{A}}
\newcommand{\SG}{\mathrm{SG}}

\newcommand{\IRISauthors}{Sukriti Manna\textsuperscript{1,2,*}, Sraddha Agrawal\textsuperscript{1}, Ramamoorthy Ramesh\textsuperscript{3},\\ Panchapakesan ~Ganesh\textsuperscript{4,*}, Pierre Darancet\textsuperscript{1,*}, S.~K.~R.~S.~Sankaranarayanan\textsuperscript{1,2,*}}
\newcommand{\IRISaffils}{%
\textsuperscript{1}Center for Nanoscale Materials, Argonne National Laboratory, Lemont, IL 60439, USA\\
\textsuperscript{2}Department of Mechanical and Industrial Engineering, University of Illinois Chicago, Chicago, IL 60607, USA\\
\textsuperscript{3}Department of Materials Science and Engineering, University of California, Berkeley, CA 94720, USA\\
\textsuperscript{4}Center for Nanophase Materials Sciences, Oak Ridge National Laboratory, Oak Ridge, TN 37831, USA\\[3pt]
\textsuperscript{*}Corresponding authors: \href{mailto:smanna@anl.gov}{smanna@anl.gov} (S.~M.), \href{mailto:ganeshp@ornl.gov}{ganeshp@ornl.gov} (P.~G.), \href{mailto:pdarancet@anl.gov}{pdarancet@anl.gov} (P.~D.), \href{mailto:skrssank@uic.edu}{skrssank@uic.edu} (S.~K.~R.~S.~S.)}

\newcommand{\IRIStitle}{IRIS: an agentic eye on symmetry-derived paths and energy landscapes of structural phase transitions}

\ifdefined\IRIScombinedmode
  \newcommand{\IRISsibibliography}{\begin{thebibliography}{54}
\providecommand{\natexlab}[1]{#1}
\providecommand{\url}[1]{\texttt{#1}}
\expandafter\ifx\csname urlstyle\endcsname\relax
  \providecommand{\doi}[1]{doi: #1}\else
  \providecommand{\doi}{doi: \begingroup \urlstyle{rm}\Url}\fi

\bibitem[{LangChain, Inc.}(2026)]{SI-LangGraph}
{LangChain, Inc.}
\newblock {LangGraph}: a library for building stateful, multi-actor
  applications with language models.
\newblock \url{https://github.com/langchain-ai/langgraph}, 2026.

\bibitem[Jain et~al.(2013)Jain, Ong, Hautier, Chen, Richards, Dacek, Cholia,
  Gunter, Skinner, Ceder, and Persson]{SI-Jain2013}
Anubhav Jain, Shyue~Ping Ong, Geoffroy Hautier, Wei Chen, William~Davidson
  Richards, Stephen Dacek, Shreyas Cholia, Dan Gunter, David Skinner, Gerbrand
  Ceder, and Kristin~A. Persson.
\newblock Commentary: The {Materials Project}: A materials genome approach to
  accelerating materials innovation.
\newblock \emph{APL Materials}, 1\penalty0 (1):\penalty0 011002, 2013.

\bibitem[Kresse and Furthm{\"u}ller(1996)]{SI-Kresse1996a}
G.~Kresse and J.~Furthm{\"u}ller.
\newblock Efficient iterative schemes for ab initio total-energy calculations
  using a plane-wave basis set.
\newblock \emph{Physical Review B}, 54:\penalty0 11169--11186, 1996.

\bibitem[Yoo et~al.(2003)Yoo, Jette, and Grondona]{SI-Yoo2003}
Andy~B. Yoo, Morris~A. Jette, and Mark Grondona.
\newblock {SLURM}: Simple {Linux} utility for resource management.
\newblock In \emph{Job Scheduling Strategies for Parallel Processing}, volume
  2862 of \emph{Lecture Notes in Computer Science}, pages 44--60. Springer,
  2003.

\bibitem[Batatia et~al.(2022)Batatia, Kov{\'a}cs, Simm, Ortner, and
  Cs{\'a}nyi]{SI-Batatia2022}
Ilyes Batatia, D{\'a}vid~P. Kov{\'a}cs, Gregor N.~C. Simm, Christoph Ortner,
  and G{\'a}bor Cs{\'a}nyi.
\newblock {MACE}: Higher order equivariant message passing neural networks for
  fast and accurate force fields.
\newblock In \emph{Advances in Neural Information Processing Systems},
  volume~35, pages 11423--11436, 2022.
\newblock \doi{10.52202/068431-0830}.

\bibitem[Togo et~al.(2024)Togo, Shinohara, and Tanaka]{SI-Togo2024}
Atsushi Togo, Kohei Shinohara, and Isao Tanaka.
\newblock Spglib: a software library for crystal symmetry search.
\newblock \emph{Science and Technology of Advanced Materials: Methods},
  4:\penalty0 2384822, 2024.
\newblock \doi{10.1080/27660400.2024.2384822}.

\bibitem[Stokes and Hatch(2005)]{SI-Stokes2005}
Harold~T. Stokes and Dorian~M. Hatch.
\newblock {FINDSYM}: program for identifying the space-group symmetry of a
  crystal.
\newblock \emph{Journal of Applied Crystallography}, 38:\penalty0 237--238,
  2005.

\bibitem[Aroyo(2016)]{SI-ITA2016}
Mois~I. Aroyo, editor.
\newblock \emph{International Tables for Crystallography, Volume A: Space-group
  symmetry}.
\newblock Wiley, Chichester, 6th edition, 2016.

\bibitem[Bl{\"o}chl(1994)]{SI-Blochl1994}
P.~E. Bl{\"o}chl.
\newblock Projector augmented-wave method.
\newblock \emph{Physical Review B}, 50:\penalty0 17953--17979, 1994.

\bibitem[Kresse and Joubert(1999)]{SI-Kresse1999}
G.~Kresse and D.~Joubert.
\newblock From ultrasoft pseudopotentials to the projector augmented-wave
  method.
\newblock \emph{Physical Review B}, 59:\penalty0 1758--1775, 1999.

\bibitem[Perdew et~al.(1996)Perdew, Burke, and Ernzerhof]{SI-Perdew1996}
John~P. Perdew, Kieron Burke, and Matthias Ernzerhof.
\newblock Generalized gradient approximation made simple.
\newblock \emph{Physical Review Letters}, 77:\penalty0 3865--3868, 1996.

\bibitem[Ceperley and Alder(1980)]{SI-Ceperley1980}
D.~M. Ceperley and B.~J. Alder.
\newblock Ground state of the electron gas by a stochastic method.
\newblock \emph{Physical Review Letters}, 45\penalty0 (7):\penalty0 566--569,
  1980.

\bibitem[Perdew and Zunger(1981)]{SI-Perdew1981}
J.~P. Perdew and Alex Zunger.
\newblock Self-interaction correction to density-functional approximations for
  many-electron systems.
\newblock \emph{Physical Review B}, 23\penalty0 (10):\penalty0 5048--5079,
  1981.

\bibitem[Dudarev et~al.(1998)Dudarev, Botton, Savrasov, Humphreys, and
  Sutton]{SI-Dudarev1998}
S.~L. Dudarev, G.~A. Botton, S.~Y. Savrasov, C.~J. Humphreys, and A.~P. Sutton.
\newblock Electron-energy-loss spectra and the structural stability of nickel
  oxide: an {LSDA+U} study.
\newblock \emph{Physical Review B}, 57:\penalty0 1505--1509, 1998.

\bibitem[Liechtenstein et~al.(1995)Liechtenstein, Anisimov, and
  Zaanen]{SI-Liechtenstein1995}
A.~I. Liechtenstein, V.~I. Anisimov, and J.~Zaanen.
\newblock Density-functional theory and strong interactions: orbital ordering
  in {Mott-Hubbard} insulators.
\newblock \emph{Physical Review B}, 52:\penalty0 R5467--R5470, 1995.

\bibitem[Perdew et~al.(2008)Perdew, Ruzsinszky, Csonka, Vydrov, Scuseria,
  Constantin, Zhou, and Burke]{SI-Perdew2008}
John~P. Perdew, Adrienn Ruzsinszky, G{\'a}bor~I. Csonka, Oleg~A. Vydrov,
  Gustavo~E. Scuseria, Lucian~A. Constantin, Xiaolan Zhou, and Kieron Burke.
\newblock Restoring the density-gradient expansion for exchange in solids and
  surfaces.
\newblock \emph{Physical Review Letters}, 100:\penalty0 136406, 2008.

\bibitem[Glazer(1972)]{SI-Glazer1972}
A.~M. Glazer.
\newblock The classification of tilted octahedra in perovskites.
\newblock \emph{Acta Crystallographica Section B}, 28\penalty0 (11):\penalty0
  3384--3392, 1972.

\bibitem[Benedek and Fennie(2011)]{SI-Benedek2011}
Nicole~A. Benedek and Craig~J. Fennie.
\newblock Hybrid improper ferroelectricity: a mechanism for controllable
  polarization-magnetization coupling.
\newblock \emph{Physical Review Letters}, 106:\penalty0 107204, 2011.

\bibitem[Nowadnick and Fennie(2016)]{SI-Nowadnick2016}
Elizabeth~A. Nowadnick and Craig~J. Fennie.
\newblock Domains and ferroelectric switching pathways in
  {Ca}$_3${Ti}$_2${O}$_7$ from first principles.
\newblock \emph{Physical Review B}, 94:\penalty0 104105, 2016.

\bibitem[Grote and Berger(2015)]{SI-Grote2015}
Christopher Grote and Robert~F. Berger.
\newblock Strain tuning of tin--halide and lead--halide perovskites: a
  first-principles atomic and electronic structure study.
\newblock \emph{The Journal of Physical Chemistry C}, 119\penalty0
  (40):\penalty0 22832--22837, 2015.
\newblock \doi{10.1021/acs.jpcc.5b07446}.

\bibitem[Meier et~al.(2020)Meier, Stucky, Teyssier, Griffin, van~der Marel, and
  Spaldin]{SI-Meier2020}
Quintin~N. Meier, Adrien Stucky, Jeremie Teyssier, Sin{\'e}ad~M. Griffin, Dirk
  van~der Marel, and Nicola~A. Spaldin.
\newblock Manifestation of structural {Higgs} and {Goldstone} modes in the
  hexagonal manganites.
\newblock \emph{Physical Review B}, 102:\penalty0 014102, 2020.

\bibitem[Fennie and Rabe(2005)]{SI-Fennie2005}
Craig~J. Fennie and Karin~M. Rabe.
\newblock Ferroelectric transition in {YMnO}$_3$ from first principles.
\newblock \emph{Physical Review B}, 72:\penalty0 100103(R), 2005.

\bibitem[Artyukhin et~al.(2014)Artyukhin, Delaney, Spaldin, and
  Mostovoy]{SI-Artyukhin2014}
Sergey Artyukhin, Kris~T. Delaney, Nicola~A. Spaldin, and Maxim Mostovoy.
\newblock Landau theory of topological defects in multiferroic hexagonal
  manganites.
\newblock \emph{Nature Materials}, 13:\penalty0 42--49, 2014.

\bibitem[Sm{\aa}br{\aa}ten et~al.(2018)Sm{\aa}br{\aa}ten, Meier,
  Skj{\ae}rv{\o}, Inzani, Meier, and Selbach]{SI-Smabraten2018}
Didrik~R. Sm{\aa}br{\aa}ten, Quintin~N. Meier, Sandra~H. Skj{\ae}rv{\o},
  Katherine Inzani, Dennis Meier, and Sverre~M. Selbach.
\newblock Charged domain walls in improper ferroelectric hexagonal manganites
  and gallates.
\newblock \emph{Physical Review Materials}, 2:\penalty0 114405, 2018.

\bibitem[Gonze et~al.(2016)Gonze, Jollet, Abreu~Araujo, Adams, Amadon,
  Applencourt, Audouze, Beuken, Bieder, Bokhanchuk, et~al.]{SI-Gonze2016}
X.~Gonze, F.~Jollet, F.~Abreu~Araujo, D.~Adams, B.~Amadon, T.~Applencourt,
  C.~Audouze, J.-M. Beuken, J.~Bieder, A.~Bokhanchuk, et~al.
\newblock Recent developments in the {ABINIT} software package.
\newblock \emph{Computer Physics Communications}, 205:\penalty0 106--131, 2016.

\bibitem[Xu et~al.(2017)Xu, Garcia, Fusil, Bibes, and Bellaiche]{SI-Xu2017}
Bin Xu, Vincent Garcia, St{\'e}phane Fusil, Manuel Bibes, and L.~Bellaiche.
\newblock Intrinsic polarization switching mechanisms in {BiFeO}$_3$.
\newblock \emph{Physical Review B}, 95:\penalty0 104104, 2017.

\bibitem[Ederer and Spaldin(2005)]{SI-Ederer2005}
Claude Ederer and Nicola~A. Spaldin.
\newblock Weak ferromagnetism and magnetoelectric coupling in bismuth ferrite.
\newblock \emph{Physical Review B}, 71:\penalty0 060401(R), 2005.

\bibitem[Neaton et~al.(2005)Neaton, Ederer, Waghmare, Spaldin, and
  Rabe]{SI-Neaton2005}
J.~B. Neaton, C.~Ederer, U.~V. Waghmare, N.~A. Spaldin, and K.~M. Rabe.
\newblock First-principles study of spontaneous polarization in multiferroic
  {BiFeO}$_3$.
\newblock \emph{Physical Review B}, 71:\penalty0 014113, 2005.

\bibitem[Lebeugle et~al.(2007)Lebeugle, Colson, Forget, and
  Viret]{SI-Lebeugle2007}
D.~Lebeugle, D.~Colson, A.~Forget, and M.~Viret.
\newblock Very large spontaneous electric polarization in {BiFeO$_3$} single
  crystals at room temperature and its evolution under cycling fields.
\newblock \emph{Appl. Phys. Lett.}, 91:\penalty0 022907, 2007.
\newblock \doi{10.1063/1.2753390}.

\bibitem[Di{\'e}guez et~al.(2011)Di{\'e}guez, Gonz{\'a}lez-V{\'a}zquez,
  Wojde{\l}, and {\'I}{\~n}iguez]{SI-Dieguez2011}
Oswaldo Di{\'e}guez, O.~E. Gonz{\'a}lez-V{\'a}zquez, Jacek~C. Wojde{\l}, and
  Jorge {\'I}{\~n}iguez.
\newblock First-principles predictions of low-energy phases of multiferroic
  {BiFeO}$_3$.
\newblock \emph{Physical Review B}, 83:\penalty0 094105, 2011.

\bibitem[Fedorova et~al.(2022)Fedorova, Nikonov, Li, Young, and
  {\'I}{\~n}iguez]{SI-Fedorova2022}
Natalya~S. Fedorova, Dmitri~E. Nikonov, Hai Li, Ian~A. Young, and Jorge
  {\'I}{\~n}iguez.
\newblock First-principles {Landau}-like potential for {BiFeO}$_3$ and related
  materials.
\newblock \emph{Phys. Rev. B}, 106:\penalty0 165122, 2022.
\newblock \doi{10.1103/PhysRevB.106.165122}.

\bibitem[Kubel and Schmid(1990)]{SI-Kubel1990}
F.~Kubel and H.~Schmid.
\newblock Structure of a ferroelectric and ferroelastic monodomain crystal of
  the perovskite {BiFeO$_3$}.
\newblock \emph{Acta Crystallogr. B}, 46:\penalty0 698--702, 1990.

\bibitem[Kornev et~al.(2006)Kornev, Bellaiche, Janolin, Dkhil, and
  Suard]{SI-Kornev2006}
Igor~A. Kornev, L.~Bellaiche, P.-E. Janolin, B.~Dkhil, and E.~Suard.
\newblock Phase diagram of {Pb(Zr,Ti)O}$_3$ solid solutions from first
  principles.
\newblock \emph{Physical Review Letters}, 97:\penalty0 157601, 2006.

\bibitem[Duchon(1977)]{SI-Duchon1977}
Jean Duchon.
\newblock Splines minimizing rotation-invariant semi-norms in {Sobolev} spaces.
\newblock In \emph{Constructive Theory of Functions of Several Variables},
  volume 571 of \emph{Lecture Notes in Mathematics}, pages 85--100. Springer,
  Berlin, 1977.

\bibitem[Virtanen et~al.(2020)Virtanen, Gommers, Oliphant, Haberland, Reddy,
  Cournapeau, Burovski, Peterson, Weckesser, Bright, et~al.]{SI-Virtanen2020}
Pauli Virtanen, Ralf Gommers, Travis~E. Oliphant, Matt Haberland, Tyler Reddy,
  David Cournapeau, Evgeni Burovski, Pearu Peterson, Warren Weckesser, Jonathan
  Bright, et~al.
\newblock {SciPy} 1.0: fundamental algorithms for scientific computing in
  {Python}.
\newblock \emph{Nature Methods}, 17:\penalty0 261--272, 2020.

\bibitem[E et~al.(2002)E, Ren, and Vanden-Eijnden]{SI-E2002String}
Weinan E, Weiqing Ren, and Eric Vanden-Eijnden.
\newblock String method for the study of rare events.
\newblock \emph{Physical Review B}, 66\penalty0 (5):\penalty0 052301, 2002.

\bibitem[King-Smith and Vanderbilt(1993)]{SI-KingSmith1993}
R.~D. King-Smith and David Vanderbilt.
\newblock Theory of polarization of crystalline solids.
\newblock \emph{Physical Review B}, 47\penalty0 (3):\penalty0 1651--1654, 1993.

\bibitem[Manna et~al.(2026)]{SI-Manna2026hafnia}
Sukriti Manna et~al.
\newblock Equivariant neural network potential for hafnia.
\newblock submitted, 2026.

\bibitem[Wang et~al.(2023)Wang, Wei, Schuurmans, Le, Chi, Narang, Chowdhery,
  and Zhou]{SI-Wang2023SC}
Xuezhi Wang, Jason Wei, Dale Schuurmans, Quoc Le, Ed~Chi, Sharan Narang,
  Aakanksha Chowdhery, and Denny Zhou.
\newblock Self-consistency improves chain of thought reasoning in language
  models.
\newblock In \emph{International Conference on Learning Representations
  (ICLR)}, 2023.

\bibitem[Geifman and El-Yaniv(2017)]{SI-Geifman2017}
Yonatan Geifman and Ran El-Yaniv.
\newblock Selective classification for deep neural networks.
\newblock In \emph{Advances in Neural Information Processing Systems},
  volume~30, pages 4878--4887, 2017.

\bibitem[Quach et~al.(2024)Quach, Fisch, Schuster, Yala, Sohn, Jaakkola, and
  Barzilay]{SI-Quach2024}
Victor Quach, Adam Fisch, Tal Schuster, Adam Yala, Jae~Ho Sohn, Tommi~S.
  Jaakkola, and Regina Barzilay.
\newblock Conformal language modeling.
\newblock In \emph{International Conference on Learning Representations
  (ICLR)}, 2024.

\bibitem[Zheng et~al.(2024)Zheng, Zhou, Meng, Zhou, and Huang]{SI-Zheng2024MCQ}
Chujie Zheng, Hao Zhou, Fandong Meng, Jie Zhou, and Minlie Huang.
\newblock Large language models are not robust multiple choice selectors.
\newblock In \emph{International Conference on Learning Representations
  (ICLR)}, 2024.

\bibitem[{Anthropic}(2026{\natexlab{a}})]{SI-ModelSonnet55}
{Anthropic}.
\newblock {Claude Sonnet 5.5} ({API} model \texttt{claude-sonnet-5-5}).
\newblock Models overview, Claude Platform Docs,
  \url{https://platform.claude.com/docs/en/models/overview},
  2026{\natexlab{a}}.
\newblock Documentation accessed 19 Oct 2026; model used 17--29 Oct 2026.

\bibitem[{Anthropic}(2026{\natexlab{b}})]{SI-ModelOpus55}
{Anthropic}.
\newblock {Claude Opus 5.5} ({API} model \texttt{claude-opus-5-5}).
\newblock Models overview, Claude Platform Docs,
  \url{https://platform.claude.com/docs/en/models/overview},
  2026{\natexlab{b}}.
\newblock Documentation accessed 19 Oct 2026; model used 17--29 Oct 2026.

\bibitem[{OpenAI}(2026)]{SI-ModelGPT61Sol}
{OpenAI}.
\newblock {GPT-6.1 Sol} ({API} model \texttt{gpt-6.1-sol}).
\newblock OpenAI API documentation, model page,
  \url{https://developers.openai.com/api/docs/models/gpt-6.1-sol}, 2026.
\newblock Documentation accessed 19 Oct 2026; model used 17--29 Oct 2026.

\bibitem[{Google}(2026)]{SI-ModelGemini4Argon}
{Google}.
\newblock {Gemini 4 Argon} ({API} model \texttt{gemini-4-argon}).
\newblock Gemini API, Google AI for Developers,
  \url{https://ai.google.dev/gemini-api/docs/models/gemini-4-argon}, 2026.
\newblock Documentation accessed 19 Oct 2026; model used 17--29 Oct 2026.

\bibitem[Yao et~al.(2023)Yao, Zhao, Yu, Du, Shafran, Narasimhan, and
  Cao]{SI-Yao2023}
Shunyu Yao, Jeffrey Zhao, Dian Yu, Nan Du, Izhak Shafran, Karthik~R.
  Narasimhan, and Yuan Cao.
\newblock {ReAct}: Synergizing reasoning and acting in language models.
\newblock In \emph{International Conference on Learning Representations
  (ICLR)}, 2023.

\bibitem[Zou et~al.(2025)Zou, Cheng, Aldossary, Bai, Leong,
  Campos-Gonzalez-Angulo, Choi, Ser, Tom, Wang, et~al.]{SI-Zou2025}
Yunheng Zou, Austin~H. Cheng, Abdulrahman Aldossary, Jiaru Bai, Shi~Xuan Leong,
  Jorge~Arturo Campos-Gonzalez-Angulo, Changhyeok Choi, Cher~Tian Ser, Gary
  Tom, Andrew Wang, et~al.
\newblock {El Agente}: An autonomous agent for quantum chemistry.
\newblock \emph{Matter}, 8\penalty0 (7):\penalty0 102263, 2025.
\newblock \doi{10.1016/j.matt.2025.102263}.

\bibitem[Pham et~al.(2026)Pham, Tanikanti, and Ke{\c{c}}eli]{SI-Pham2025}
Thang~D. Pham, Aditya Tanikanti, and Murat Ke{\c{c}}eli.
\newblock {ChemGraph} as an agentic framework for computational chemistry
  workflows.
\newblock \emph{Communications Chemistry}, 9:\penalty0 33, 2026.
\newblock \doi{10.1038/s42004-025-01776-9}.

\bibitem[Chen et~al.(2025)Chen, Chen, Ning, Zhang, Wang, Yu, Li, Liao, Wei, Lu,
  et~al.]{SI-Chen2025}
Ziru Chen, Shijie Chen, Yuting Ning, Qianheng Zhang, Boshi Wang, Botao Yu,
  Yifei Li, Zeyi Liao, Chen Wei, Zitong Lu, et~al.
\newblock {ScienceAgentBench}: Toward rigorous assessment of language agents
  for data-driven scientific discovery.
\newblock In \emph{International Conference on Learning Representations
  (ICLR)}, 2025.

\bibitem[Tang et~al.(2025)Tang, Jin, Zhu, Yuan, Zhang, Zhou, Qu, Zhao, Tang,
  Zhang, et~al.]{SI-Tang2025}
Xiangru Tang, Qiao Jin, Kunlun Zhu, Tongxin Yuan, Yichi Zhang, Wangchunshu
  Zhou, Meng Qu, Yilun Zhao, Jian Tang, Zhuosheng Zhang, et~al.
\newblock Risks of {AI} scientists: prioritizing safeguarding over autonomy.
\newblock \emph{Nature Communications}, 16:\penalty0 8317, 2025.
\newblock \doi{10.1038/s41467-025-63913-1}.

\bibitem[Cohen(1992)]{SI-cohen1992origin}
Ronald~E. Cohen.
\newblock Origin of ferroelectricity in perovskite oxides.
\newblock \emph{Nature}, 358\penalty0 (6382):\penalty0 136--138, 1992.

\bibitem[King-Smith and Vanderbilt(1994)]{SI-kingsmith1994first}
R.~D. King-Smith and David Vanderbilt.
\newblock First-principles investigation of ferroelectricity in perovskite
  compounds.
\newblock \emph{Physical Review B}, 49\penalty0 (9):\penalty0 5828--5844, 1994.

\bibitem[Campbell et~al.(2006)Campbell, Stokes, Tanner, and
  Hatch]{SI-Campbell2006}
Branton~J. Campbell, Harold~T. Stokes, David~E. Tanner, and Dorian~M. Hatch.
\newblock {ISODISPLACE}: a web-based tool for exploring structural distortions.
\newblock \emph{Journal of Applied Crystallography}, 39:\penalty0 607--614,
  2006.

\end{thebibliography}
}
\else
  \newcommand{\IRISsibibliography}{\bibliographystyle{unsrtnat}\bibliography{refs}}
\fi

\newcommand{\IRISsinumbering}{%
  \setcounter{section}{0}\setcounter{figure}{0}\setcounter{table}{0}\setcounter{equation}{0}%
  \renewcommand{\thesection}{S\arabic{section}}%
  \renewcommand{\thefigure}{S\arabic{figure}}%
  \renewcommand{\thetable}{S\arabic{table}}%
  \renewcommand{\theequation}{S\arabic{equation}}%
  \renewcommand{\theHsection}{S\arabic{section}}%
  \renewcommand{\theHfigure}{S\arabic{figure}}%
  \renewcommand{\theHtable}{S\arabic{table}}%
  \renewcommand{\theHequation}{S\arabic{equation}}%
  \renewcommand{\figurename}{Figure}\renewcommand{\tablename}{Table}%
  \renewcommand{\refname}{Supplementary References}}

\makeatletter
\renewcommand*\l@section[2]{%
  \ifnum \c@tocdepth >\z@
    \addpenalty\@secpenalty
    \addvspace{1.0em \@plus\p@}%
    \setlength\@tempdima{2.2em}%
    \begingroup
      \parindent \z@ \rightskip \@pnumwidth
      \parfillskip -\@pnumwidth
      \leavevmode \bfseries
      \advance\leftskip\@tempdima
      \hskip -\leftskip
      #1\nobreak\hfil
      \nobreak\hb@xt@\@pnumwidth{\hss #2\kern-\p@\kern\p@}\par
    \endgroup
  \fi}
\makeatother

\ifx\iPrompt\undefined
\usepackage{listings}
\usepackage[most]{tcolorbox}
\definecolor{iSys}{HTML}{2A5DB0}\definecolor{iUser}{HTML}{6B6B66}\definecolor{iGuide}{HTML}{D9822B}
\definecolor{iOut}{HTML}{1E8A5A}\definecolor{iGate}{HTML}{6250D6}
\lstdefinestyle{iris}{basicstyle=\ttfamily\scriptsize,breaklines=true,columns=fullflexible,keepspaces=true,showstringspaces=false,breakindent=0pt,upquote=true,postbreak=\mbox{\textcolor{iUser}{$\hookrightarrow$}\space},literate={Å}{{\AA}}1 {—}{{---}}1 {→}{{$\to$}}1}
\newtcblisting{iPrompt}[2]{listing only,breakable,enhanced,colback=#1!4,colframe=#1,colbacktitle=#1,coltitle=white,fonttitle=\sffamily\bfseries\footnotesize,title={#2},boxrule=0.6pt,arc=1.5mm,left=2mm,right=2mm,top=1mm,bottom=1mm,before skip=6pt,after skip=6pt,listing engine=listings,listing options={style=iris}}
\newtcolorbox{iBox}[2]{breakable,enhanced,colback=#1!5,colframe=#1,colbacktitle=#1,coltitle=white,fonttitle=\sffamily\bfseries\footnotesize,title={#2},boxrule=0.6pt,arc=1.5mm,left=2mm,right=2mm,top=1mm,bottom=1mm,before skip=6pt,after skip=6pt}
\newtcolorbox{iCall}[1]{enhanced,breakable,colback=white,colframe=iUser!60,colbacktitle=iUser!12,coltitle=black,fonttitle=\sffamily\bfseries\footnotesize,title={#1},boxrule=0.5pt,arc=1.5mm,fontupper=\small,left=2mm,right=2mm,top=1mm,bottom=1mm,before skip=4pt,after skip=4pt}
\newcommand{\iTag}[2]{\makebox[1.75cm][l]{\colorbox{#1}{\textcolor{white}{\sffamily\bfseries\scriptsize #2}}}}
\newcommand{\iLine}[3]{\par\noindent\hangindent=1.75cm\hangafter=1\iTag{#1}{#2}#3\par}
\newcommand{\iReq}[1]{\iLine{iUser}{REQUEST}{\texttt{#1}}}
\newcommand{\iAns}[1]{\iLine{iOut}{MODEL}{#1}}
\newcommand{\iVer}[1]{\iLine{iGate}{GATE $\boldsymbol{G_0}$}{#1}}

\fi

\makeatletter\renewcommand*\l@subsection{\@dottedtocline{2}{1.5em}{3.4em}}\renewcommand*\l@subsubsection{\@dottedtocline{3}{4.9em}{4.2em}}\makeatother

\title{\textbf{\IRIStitle}}
\author{\parbox{\dimexpr\textwidth-2\tabcolsep\relax}{\centering\normalsize \IRISauthors\\[6pt]
{\small\IRISaffils}}}

\begin{document}
\maketitle
\begin{abstract}

Structural phase transitions underpin ferroelectric and multiferroic functionality, but constructing interpretable energy landscapes remains dependent on specialist expertise. Predicting this behaviour rests on Landau free energies expanded in symmetry-adapted order parameters: their coefficients classify a transition, expose the couplings between distortions, set switching paths and barriers, and parametrize phase-field and effective-Hamiltonian models at larger scales. Obtaining these coefficients from first principles means sampling the Born--Oppenheimer energy surface along symmetry-derived structural paths, which requires crystallographic analysis, reliable structure generation and extensive computation, especially for coupled multidimensional order parameters. Here we introduce IRIS, an agentic workflow that converts plain-language requests and crystal structures into symmetry-adapted energy landscapes and fitted Landau energy models. Deterministic gates govern acceptance of structures, calculations and fits, while language models interpret requests and select constrained recovery actions. Applications across six compounds include a six-dimensional BiFeO$_3$ landscape that separates competition between polar and tilt amplitudes from directional locking and ranks homogeneous switching pathways. Controlled replays detected all 53 injected faults; disabling gates allowed 14 silent errors. Four prospective BaTiO$_3$ runs produced fitted landscapes from 96 density-functional calculations in 34 minutes total without intervening human decisions. IRIS makes physical model construction a reusable computational capability and provides a framework for scientific automation grounded in explicit verification.

\vspace{4pt}\noindent\textbf{Keywords:} agentic AI; structural phase transitions; symmetry-mode analysis; Landau theory; first-principles calculations
\end{abstract}

\section{Introduction}
Structural phase transitions drive functional behaviour in ferroelectrics, multiferroics, shape-memory alloys, correlated oxides and topological systems, because the collective atomic distortions that change the crystal structure couple to electronic, magnetic and mechanical degrees of freedom~\cite{Carpenter2009,PerezMato2010,cheong2007,imada1998}. The energy landscape of a transition, with its minima, barriers and couplings, constrains which phases are thermodynamically stable and sets the energetics of homogeneous switching pathways, and it supplies the coefficients of the Landau and phase-field models that describe domains and switching on larger scales, where kinetics also depend on nucleation, defects, interfaces and dynamics; without first-principles landscapes, these models rest on assumed or fitted parameters. Symmetry-adapted mode analysis decomposes the distortion between two phases into displacement modes that transform as irreducible representations (irreps) of the higher-symmetry parent~\cite{howard1998,Campbell2006,PerezMato2010}. When the lower-symmetry phase is a subgroup of the parent, the distortion components along the symmetry-breaking irreps are the order parameters: vectors in irrep space, each with an amplitude, a direction and as many components as its irrep has dimensions. Primary order parameters produce the symmetry lowering, and secondary ones are induced through their couplings. A transition is then a path on the energy landscape spanned by the order parameters, and phases without a group--subgroup relation can be connected through a common higher-symmetry phase.

Computing energy landscapes in the space of these order parameters becomes harder as the order parameters gain components, grow in number or couple: the two-dimensional order parameter in hexagonal manganites needs a basis that a single decomposition does not provide; the coupled polar and antiferrodistortive three-dimensional order parameters of BiFeO$_3$ span a space too large to sample naively; and trilinear (hybrid-improper) couplings in layered perovskites tie a polar mode to non-polar distortions~\cite{Benedek2011,Nowadnick2016}. 

The tools for each step, from structure standardization and symmetry-mode analysis to first-principles codes and model fitting, are mature~\cite{Stokes2005,Campbell2006,Kresse1996a}; the bottleneck, and the risk, lie in the workflow that chains them. Workflow engines such as AiiDA and atomate automate execution, provenance and error handling~\cite{Huber2020,Mathew2017}, but not the domain knowledge: which branch the crystallography calls for (the number and dimension of the active irreps decide whether a line, a plane or a high-dimensional grid is needed) and how to verify that each generated structure is the intended one. Because each tool is internally consistent, a mismatch of conventions, such as origin, atom ordering or amplitude normalization, can pass every check and yield a landscape that is smooth but wrong.

Automating such workflows requires adapting each step to what the previous one found while keeping every adaptation crystallographically sound. Large language model (LLM) agents, which interleave reasoning with calls to external tools~\cite{Yao2023,Schick2023} and can work as coordinated teams~\cite{Hong2024MetaGPT}, can interpret intermediate results and recover from failures~\cite{Shinn2023}; they now plan chemical syntheses and autonomous experiments~\cite{Boiko2023,Bran2024,Szymanski2023}, run atomistic~\cite{Zou2025,Pham2025,Campbell2025} and continuum~\cite{Manna2026AutoMOOSE,Ni2023MechAgents,Pandey2025OpenFOAMGPT} simulation workflows from plain-language requests, design alloys in multi-agent teams~\cite{Ghafarollahi2025}, retrieve materials data~\cite{Chiang2025} and draft research studies~\cite{Lu2024}. The responses of these agents, however, can be fluent and internally consistent yet physically wrong~\cite{Ji2023}, the underlying language models do not reliably detect such errors in their own reasoning~\cite{Huang2024}, and benchmarks of chemical knowledge, of scientific agents and of general agents measure how often a language model answers correctly or completes a task~\cite{Mirza2025,Chen2025,Liu2024AgentBench,Jimenez2024} rather than whether an error is caught before it propagates; safeguards on what an agent may do have been argued to matter more than its autonomy~\cite{Tang2025}, and emulated tool use shows how often agents take risky actions~\cite{Ruan2024ToolEmu}. Hence our design principle: a language model may propose what to do next, but deterministic checks of the crystallography, the calculations and the fit decide whether a result is accepted, as in planning frameworks in which a model proposes and external verifiers decide~\cite{Kambhampati2024} and in verifier-guided reasoning~\cite{Lightman2024}.

Here we present IRIS (Irrep Route and Interpolation System), an agentic workflow built on this principle. Seven agents carry a plain-language request and two crystal structures to an energy landscape and a fitted Landau model. Branching decisions, such as which modes become axes and how many dimensions the grid needs, are explicit policies, and known failures are repaired by stored rules. A deterministic gate on every hand-off checks crystallographic consistency, mode fidelity and convergence, and every structure against the group--subgroup tree, so that no result moves on unless its gate passes. A language model acts only where no policy or stored rule decides, interpreting the request and choosing among the fixes that the code lists, and it never routes the workflow or accepts a result. We demonstrate IRIS on six compounds of increasing order-parameter complexity. The density functional theory (DFT) campaigns were run once, block by block, with the same inputs, outputs and checks as in IRIS, and failures were fixed by hand; those fixes became the stored rules. The complete gated graph was then rerun end to end on the stored energies and broken deliberately, to measure what the gates and the language model each contribute. The language model makes the workflow usable from a plain-language request and answers judgment calls without stopping, whereas failures already seen are better handled by stored rules; in these replays, the gates contained wrong results whichever of the two chose the fix. Replay measures the parts; to show the whole, we also ran IRIS prospectively on BaTiO$_3$, from a typed request to the energy landscape: the structures were fetched from the Materials Project, the symmetry analysis was run live, and the three polar paths and the $(P_z,P_x)$ plane were computed with the agent submitting and collecting its own density-functional calculations on a high-performance computing (HPC) cluster, every gate passing without a human decision.

\section{Results}
\subsection{IRIS: a gated agentic workflow from two structures to a Landau model}\label{sec:results1}
\begin{figure}[!htbp]
  \centering
  \includegraphics[width=\textwidth,height=0.645\textheight,keepaspectratio]{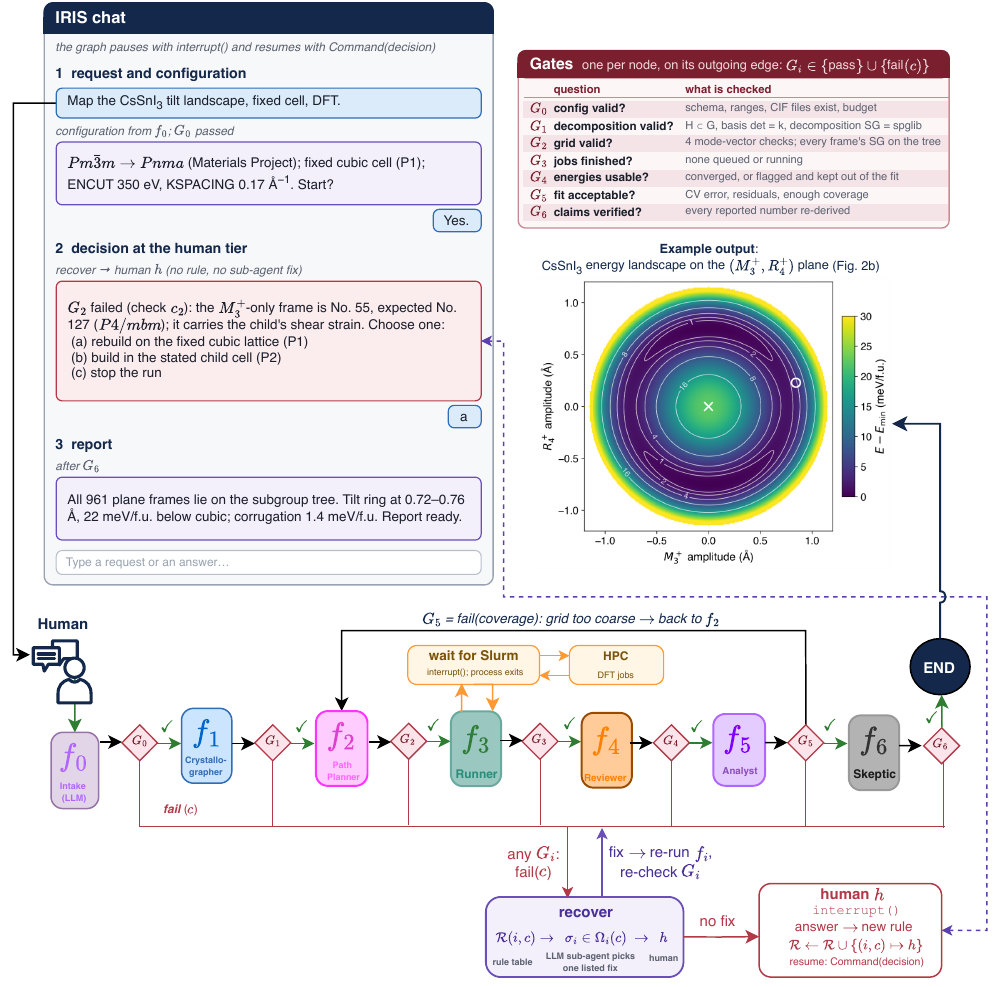}
  \caption{\textbf{Overview of IRIS as a gated state graph.}
IRIS runs from the human (left) to END through seven agent nodes, $f_0$--$f_6$: $f_0$ Intake, the only node that calls a language model directly, turns a plain-language request and two structures into a validated configuration; $f_1$ Crystallographer decomposes the distortion into irrep modes; $f_2$ Path Planner selects and certifies the modes and builds the grid; $f_3$ Runner submits the DFT jobs, and the graph pauses at a checkpoint until Slurm reports them finished (orange); $f_4$ Reviewer checks every energy; $f_5$ Analyst fits the Landau model; and $f_6$ Skeptic re-derives every reported number.
Each node is followed by a gate on its outgoing edge, $G_i\in\{\mathrm{pass}\}\cup\{\mathrm{fail}(c)\}$: deterministic code (top right) that returns pass ($\checkmark$, green arrows) or a failure with an error code $c$: $G_0$, configuration valid; $G_1$, decomposition valid ($H\subset G$, basis $\det=k\,z_H/z_G$); $G_2$, grid valid (certified mode vectors, every frame on the tree); $G_3$, jobs finished; $G_4$, energies usable; $G_5$, fit acceptable (cross-validation, coverage); $G_6$, claims verified; checks and codes in \STab{tab:gates}.
Any $\mathrm{fail}(c)$ travels along the fail bus (red) to the recovery node (purple), which applies a stored rule $\mathcal{R}(i,c)$ if one exists, otherwise asks the gate's sub-agent $\sigma_i$, a language-model call that must choose one fix from the listed set $\Omega_i(c)$, and otherwise the human $h$, reached through \texttt{interrupt()}; the answer resumes the graph through \texttt{Command(decision)} and is stored as a new rule, $\mathcal{R}\leftarrow\mathcal{R}\cup\{(i,c)\mapsto h\}$. The fix is applied, $f_i$ is re-run and $G_i$ re-checks it; $\mathrm{fail}(\text{coverage})$ at $G_5$ returns to $f_2$.
Top left: the chat interface, illustrated with a CsSnI$_3$ exchange (request, a human decision at a gate, report) composed for the figure; the interface itself is the one used in the prospective BaTiO$_3$ run of Section~\ref{sec:live}, in which no decision went to the human; the intake was evaluated on text requests (Section~\ref{sec:autonomy}), the front end was not. Right: the output (Fig.~\ref{fig:perovskite}b).
Full description: Methods (Section~\ref{sec:agent}) and \SSec{sec:si_leg_fig1}.}
  \label{fig:pipeline}
\end{figure}

IRIS takes a plain-language request and two crystal structures, a high-symmetry parent and a low-symmetry child, and returns a symmetry-adapted energy landscape, a fitted Landau model and an audit trail, a hashed log of every structure, calculation, gate result and decision behind each reported number (Fig.~\ref{fig:pipeline}; Methods). It is a state graph of seven agent nodes, each a component with its own inputs, outputs, decisions and gate rather than an independently reasoning language model: $f_0$ Intake, $f_1$ Crystallographer, $f_2$ Path Planner, $f_3$ Runner, $f_4$ Reviewer, $f_5$ Analyst and $f_6$ Skeptic, which validate the request, decompose the distortion into irrep modes, build a certified grid of structures, run and check the calculations, fit the Landau model and re-derive every reported number (Extended Data Table~\ref{tab:blocks}). Each hand-off passes a gate, deterministic code that returns a pass or an error code. A failure goes to a recovery node that tries a stored rule, then the gate's sub-agent, which must choose from the fixes listed for that code, then a human, whose answer becomes a new rule; the same gate re-checks the result, so a language model never decides whether a step succeeded (Methods, Section~\ref{sec:agent}; Supplementary Tables~\SIref{tab:gates}--\SIref{tab:subagents}). Each node $f_i$ has a gate $G_i$ (defined in Fig.~\ref{fig:pipeline}). Choices that shape a landscape are explicit policies, not model decisions: R1--R5 select the active modes (R5 refers a borderline selection to the sub-agent or the human), D1--D5 set the dimension and sampling of the grid (D4 treats hard secondary modes; D3 is not yet implemented) and P1--P5 fix the lattice (P1--P3 in the workflow; the anchors of P4 and the bias statement of P5 are still produced outside it; Methods, Section~\ref{sec:planner}; \STab{tab:si_policies}).

Unlike a scripted interpolation between two structures, every coordinate and structure is fixed by symmetry (the two constructions are compared in \SSec{sec:si_interp}). Each axis is one component of an active order parameter, the amplitude of an irrep along a fixed direction of its irrep space, so a two- or three-component order parameter spans two or three axes; each frame is the parent plus certified mode vectors, so its space group is known before any calculation and is checked against the group--subgroup tree. Certification removes errors that symmetry checks cannot see (mismatched atoms, the rigid translation carried by polar modes, mis-scaled amplitudes), and the grid size follows the number of free order-parameter components (Extended Data Figs~\ref{fig:workflow} and~\ref{fig:planner}; \SFig{fig:s_iso}). 

\subsection{Validation on perovskite tilt transitions}\label{sec:res_perov}
\begin{figure}[!htbp]
  \centering
  \includegraphics[width=\textwidth]{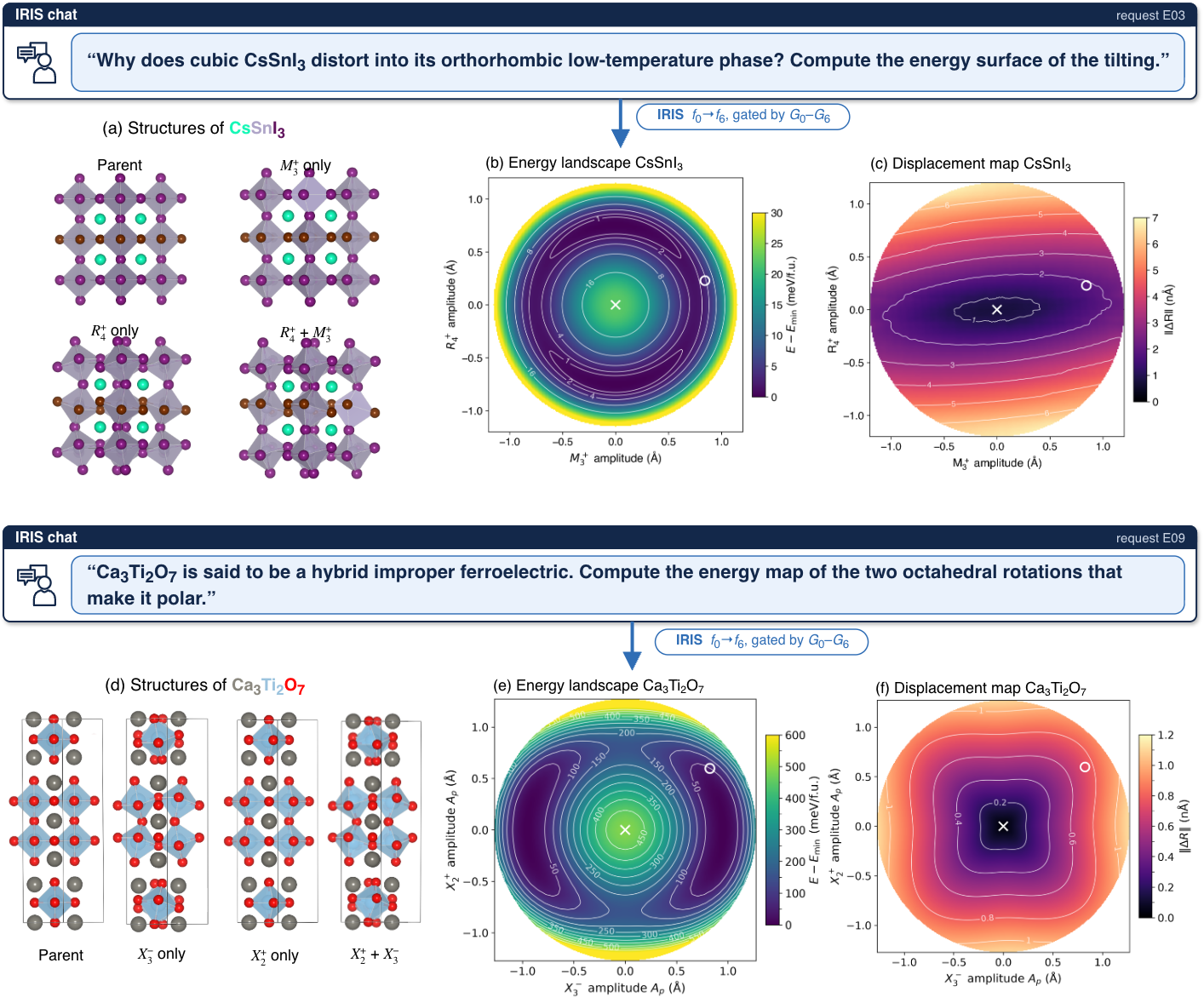}
  \caption{\textbf{IRIS energy landscapes on irrep planes and validation of the structure generation.}
\textbf{a},~CsSnI$_3$: the cubic parent ($Pm\bar{3}m$), $M_3^+$ alone ($P4/mbm$), $R_4^+$ alone ($Imma$) and both ($Pnma$), drawn at 1.2~\AA\ per irrep; Cs, teal; Sn, lilac; I, purple.
\textbf{b},~Energy on the ($M_3^+$, $R_4^+$) plane at the fixed parent lattice, $E-E_{\min}$ per formula unit; contours at 1, 2, 4, 8 and 16~meV/f.u.
\textbf{c},~Difference $\lVert\Delta R\rVert$ ($10^{-9}$\AA) between each IRIS frame and the same frame rebuilt by linear interpolation from the parent to $P4/mbm$ and $Pnma$, on a $101\times101$ grid over the plane in \textbf{b}; the maximum, $8.7\times10^{-9}$~\AA, is the rounding of the stored vectors.
\textbf{d},~Ca$_3$Ti$_2$O$_7$: the parent ($I4/mmm$, constructed from the child), $X_3^-$ alone ($Cmcm$), $X_2^+$ alone ($Cmce$) and both ($A2_1am$); Ca, grey; Ti, blue; O, red.
\textbf{e},~Energy on the ($X_3^-$, $X_2^+$) plane at the fixed parent lattice with $\Gamma_5^-$ and strain at zero; contours every 50~meV/f.u.
\textbf{f},~As \textbf{c}, interpolating to $Cmcm$ and $A2_1am$ (maximum $1.1\times10^{-9}$~\AA\ per atom over the grid). The agreement in \textbf{c} and \textbf{f} is an algebraic closure test of the frame construction, not a test of the energies.
Amplitudes are $A_p$ in \AA; white $\times$, parent; white $\circ$, child (CsSnI$_3$: (0.843, 0.229)~\AA; Ca$_3$Ti$_2$O$_7$: (0.822, 0.599)~\AA).
Energies are DFT (PBE; CsSnI$_3$ on a $31\times31$ grid over $\pm1.2$~\AA, Ca$_3$Ti$_2$O$_7$ on a $37\times37$ grid over $\pm1.27$~\AA; \STab{tab:dft}), shown within an inscribed disk; colour scales are capped at 30 (\textbf{b}) and 600~meV/f.u.\ (\textbf{e}).
Bubbles: human requests E03 and E09 (\STab{tab:si_requests}), whose configurations reproduce \textbf{b} and \textbf{e} in replay. Details: \SSec{sec:si_leg_fig2}.}
  \label{fig:perovskite}
\end{figure}

The six compounds, CsSnI$_3$, Ca$_3$Ti$_2$O$_7$, InMnO$_3$, ErMnO$_3$, BiFeO$_3$ and HfO$_2$, are ordered by order-parameter complexity, so that each exercises a part of the workflow that the previous one does not (Extended Data Table~\ref{tab:systems}). We first establish that the structures IRIS generates are the intended ones, using two perovskite tilt transitions in which two primary irreps span a plane (Fig.~\ref{fig:perovskite}; \STab{tab:perov}; decompositions in \STab{tab:dec_perov} and \STab{tab:dec_ca3ti2o7}; DFT settings for all systems in \STab{tab:dft}). The two transitions test the construction in complementary ways. In cubic CsSnI$_3$ ($Pm\bar{3}m\to Pnma$), which adopts the most common perovskite tilt pattern, the two irreps move disjoint displacement components; in the layered hybrid improper ferroelectric Ca$_3$Ti$_2$O$_7$ ($I4/mmm\to A2_1am$) they share 16 of the 48 atoms, and only the child is available, so IRIS must first construct the parent by symmetrizing it (\STab{tab:ca3ti2o7_parent}). Symmetry also fixes what each plane contains. In CsSnI$_3$ the rotation $M_3^+$ and tilt $R_4^+$ have isotropy subgroups that intersect in $Pnma$, so every interior point has the ground-state symmetry; in Ca$_3$Ti$_2$O$_7$, by contrast, the tilt $X_3^-$ and rotation $X_2^+$ condense on one arm, allowing their trilinear coupling to the polar $\Gamma_5^-$~\cite{Benedek2011,Nowadnick2016}. All 961 frames of the CsSnI$_3$ plane and all 1{,}372 of the Ca$_3$Ti$_2$O$_7$ plane (and all 10{,}201 nodes of its finer check grid) have the expected space groups, yet certification caught errors that leave the space group unchanged: four in Ca$_3$Ti$_2$O$_7$ (permuted atoms, mis-scaled vectors, norms short by $\sqrt2$, caught by the norm check of $G_1$, and a rigid translation making up 75\% of the squared norm of the raw $\Gamma_5^-$ vector; \SFig{fig:s_certify}b) and mis-scaled vectors in CsSnI$_3$, besides a single-mode frame on the child's sheared lattice that the three-way symmetry check caught (\SFig{fig:s_three}c; \SSec{sec:si_perov_ext}).

Because a correct space group does not by itself prove that a frame is the intended structure, we rebuilt every frame independently, by linear interpolation from the parent to two phases. In both perovskites the displacement fields to these end phases contain only the two primary irreps, so by symmetry any interpolation along them stays on the irrep plane, carries the correct irreps and must reproduce the IRIS frame (\SSec{sec:si_interp}). It does so to $3.1\times 10^{-9}$ and $1.1\times 10^{-9}$\AA\ per atom ($8.7\times 10^{-9}$\AA\ for the whole CsSnI$_3$ cell) over 10{,}201 grid nodes (Fig.~\ref{fig:perovskite}c,f; \SFig{fig:s_census}, \SFig{fig:s_dgrid}), and to $4$--$8\times10^{-15}$~\AA\ with sixteen stored decimals (\STab{tab:cssni3_audit}, \STab{tab:ca3ti2o7_audit}, \SFig{fig:s_ca3ti2o7_coords}, \SFig{fig:s_ca3ti2o7_energy}). This particular algebraic test is possible due to symmetry and confirms that IRIS builds the frames it claims, not that the landscape is physically right. We note that interpolating instead between the published CsSnI$_3$ structures as found~\cite{Jain2013} involves secondaries that shift the Sn--I--Sn angles by up to 2.1$^\circ$ (\STab{tab:cssni3_angles}, \SFig{fig:s_angles}).

Beyond the structure generation and validation, the computed potential energy landscape of CsSnI$_3$ is shown in Fig.~\ref{fig:perovskite}b. As expected, it shows a nearly isotropic ring of tilt minima, 22~meV per formula unit (f.u.) deep and corrugated by 1.4~meV/f.u. (Landau fit in \STab{tab:cssni3_landau} and \SFig{fig:s_landau}). This plane validates the structure generation, but we note that the structural ground state lies outside of this ring: the $Pnma$ structure is stabilized by the $X_5^+$ secondary and by strain, which alone contributes about 16 of the 19.2~meV/f.u.\ separating relaxed $Pnma$ from $P4/mbm$ (\SFig{fig:s_x5}); the relaxed anchors lie within 2~meV/f.u.\ of published values~\cite{Grote2015} (\STab{tab:cssni3_anchors}). The Ca$_3$Ti$_2$O$_7$ plane (Fig.~\ref{fig:perovskite}e) is about twenty times deeper, with four minima 489~meV/f.u.\ below the parent and the primaries-only child 51~meV/f.u.\ above them (\SFig{fig:s_ct_mirror}): here too the two primaries alone do not select the child.

\subsection{Computing the interplay of the near-U(1) potential and improper ferroelectric order in hexagonal manganites}\label{sec:res_mn}
The construction of the potential energy surface of hexagonal manganites $R$MnO$_3$ and subsequent extraction of the associated Landau components (see ref.~\cite{Meier2020} for InMnO$_3$ and ErMnO$_3$) are an interesting test for our approach based on symmetry-derived pathways: the improper ferroelectric ground state ($P6_3/mmc\to P6_3cm$) in these compounds results from the interaction between a two-dimensional primary order parameter (the trimerizing $K_3$ distortion, $(K_{3a},K_{3b})=(Q\cos\Phi,\,Q\sin\Phi)$, with amplitude $Q$ and phase $\Phi$, the angle in its two-dimensional irrep space) and the polar $\Gamma_2^-$  distortion of amplitude $P$. Our goal in this section is to use our method to fit the Landau expression of refs~\cite{Artyukhin2014,Meier2020}: $F=\tfrac{a}{2}Q^2+\tfrac{b}{4}Q^4+\tfrac{Q^6}{6}(c+c'\cos6\Phi)-gQ^3P\cos3\Phi+\tfrac{g'}{2}Q^2P^2+\tfrac{a_P}{2}P^2$ (eq.~\eqref{eq:meier}), where $a$, $b$ and $c$ are the radial coefficients, $c'$ the six-fold angular term, $g$ the trilinear and $g'$ the biquadratic coupling to the polarization, and $a_P$ the polar stiffness. Due to the improper ferroelectric nature of these materials, the polar mode enters through the trilinear invariant $Q^3P\cos3\Phi$~\cite{Fennie2005,Artyukhin2014}. 

Our IRIS decomposition returns the child's direction (\STab{tab:dec_mn}), correctly generates the second component with a parent translation lost in the child, and verifies the $K_3$ representation matrices for all 72 operations (\SSec{sec:si_k3}, \STab{tab:k3}). We note that in InMnO$_3$ the polar mode passes the amplitude threshold but leads the largest inactive irrep by less than the 0.1~Å margin of rule R5 and is flagged as borderline (\SFig{fig:s_select}f; Section~\ref{sec:planner}), suggesting that users might need to weaken this rule when studying improper ferroelectrics. 

\begin{figure}[!htbp]
  \centering
  \includegraphics[width=0.93\textwidth,height=0.68\textheight,keepaspectratio]{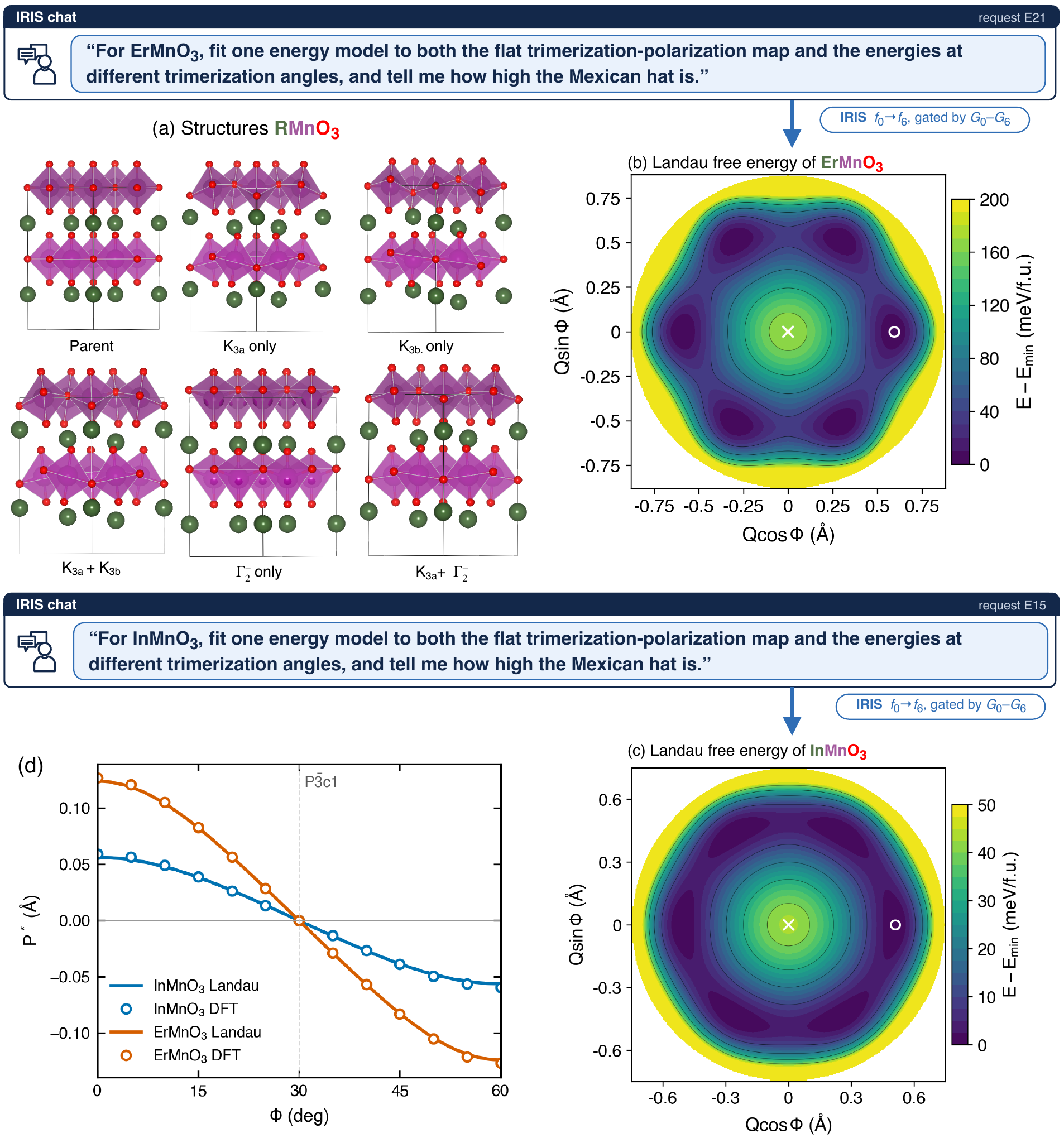}
  \caption{\textbf{Improper ferroelectric behaviour from the coupling with a two-dimensional order parameter: the U(1)-like landscape of hexagonal manganites.}
The two-dimensional $K_3$ order parameter, $(K_{3a},K_{3b})=(Q\cos\Phi,\,Q\sin\Phi)$, couples to the polar $\Gamma_2^-$ (amplitude $P$) through $Q^3P\cos3\Phi$; energies on a ($Q$, $\Phi$, $P$) grid are fitted to eq.~\eqref{eq:meier}.
\textbf{a},~$R$MnO$_3$ structures: the parent ($P6_3/mmc$), $K_{3a}$ alone ($P6_3cm$), $K_{3b}$ alone ($P\bar{3}c1$), $K_{3a}+K_{3b}$ ($P3c1$), $\Gamma_2^-$ alone ($P6_3mc$) and $K_{3a}+\Gamma_2^-$ ($P6_3cm$). $R$, green; Mn, purple; O, red.
\textbf{b},~Landau free energy of ErMnO$_3$ on the $K_3$ plane, $E-E_{\min}$ per formula unit, with $P$ relaxed to $P^*$ everywhere: a ring of radius $Q^*$ with six minima at $\Phi=n\cdot60^\circ$ (six $P6_3cm$ domains) and saddles at $\Phi=30^\circ+n\cdot60^\circ$ ($P\bar{3}c1$).
\textbf{c},~As \textbf{b} for InMnO$_3$.
\textbf{d},~Optimal polar amplitude $P^*$ against $\Phi$ for InMnO$_3$ (blue) and ErMnO$_3$ (orange) from the Landau model (lines) and DFT (circles); the dashed line marks the centrosymmetric $P\bar{3}c1$ saddle ($\Phi=30^\circ$).
In \textbf{b} and \textbf{c}, amplitudes are $A_p$ in Å; white $\times$, parent; white $\circ$, relaxed child (InMnO$_3$ 0.528~Å, ErMnO$_3$ 0.604~Å); colour scales are capped at 200 (\textbf{b}) and 50~meV/f.u.\ (\textbf{c}).
DFT: LDA$+U$ ($U=8$, $J=0.88$~eV), A-type antiferromagnetic order, fixed child lattice (\STab{tab:dft}); per compound, a $15\times15$ ($K_3$, $\Gamma_2^-$) plane at $\Phi=0$ (225 structures) and a ($Q$, $\Phi$, $P$) grid of five radii, seven angles ($0$--$30^\circ$) and five polar amplitudes (175), plus 11 points on the $\Gamma_2^-$ axis for ErMnO$_3$, fitted jointly (RMS 3.5 for InMnO$_3$ and 6.5~meV/f.u.\ for ErMnO$_3$; coefficients in \STab{tab:mn_fig4}).
In \textbf{d}, DFT values for $30^\circ<\Phi\le60^\circ$ follow by symmetry, $P^*(60^\circ-\Phi)=-P^*(\Phi)$; $P$ is translation-free. Bubbles: human requests E21 and E15 (\STab{tab:si_requests}), whose configurations reproduce \textbf{b} and \textbf{c} in replay.}
  \label{fig:manganite}
\end{figure}

The fitted Landau landscapes on the IRIS grids are shown in Fig.~\ref{fig:manganite}b,c (the DFT points in \SFig{fig:s_mn_validation}). They display a U(1)-like profile with a gapped, pseudo-Goldstone phase mode and six $P6_3cm$ minima and $P\bar{3}c1$ saddles~\cite{Artyukhin2014} (energy differences of 43  and 170~meV/f.u. for InMnO$_3$ and ErMnO$_3$, respectively). The domain barrier along the brim is ten times larger in ErMnO$_3$ (3.3 and 34.3~meV/f.u.\ in DFT; \SFig{fig:s_mn_validation}) and comes from the polar coupling, which lowers the energy by $(gQ^3\cos3\Phi)^2/[2(a_P+g'Q^2)]$, not from the sixth-order angular term, which alone would favour the saddles. The analytical form for the polarization~\cite{Fennie2005,Artyukhin2014}, $P^*=gQ^3\cos3\Phi/(a_P+g'Q^2)$, follows the DFT results at every $\Phi$ (Fig.~\ref{fig:manganite}d; \SFig{fig:s_mn_pstar}).

A fit to the full analytical expression is used to derive Landau coefficients based on the DFT data (as detailed in \STab{tab:mn}, \STab{tab:mn_grids}, \STab{tab:mn_fig4}), with an associated root-mean-square (RMS) error of 3.5 (InMnO$_3$) and 6.5~meV/f.u.\ (ErMnO$_3$; \SFig{fig:s_mn_plane}). Against the coefficients of Meier \textit{et al.}~\cite{Meier2020,Smabraten2018} (Extended Data Table~\ref{tab:meier}), the shape of the hat agrees, with the reduced brim parameter $\tilde b=bQ_0^2/|a|$ ($Q_0$, the brim radius at $P=0$) at 0.69 against 0.72 (InMnO$_3$) and 0.91 in both (ErMnO$_3$) in the joint fits (\STab{tab:mn_fig4}; 0.70--0.72 against 0.73 for the InMnO$_3$ plane fits with $c'$ omitted, \STab{tab:mn_red}), but its scale differs in  the harmonic term $a$, found 1.47--1.65 times larger in InMnO$_3$, and the anharmonic and coupling terms, 0.40--0.64 times the reference, in ErMnO$_3$ (this difference is not resolved by changing the $+U$ scheme, magnetic order or polar convention). Possible origins, which differ between the two calculations, are discussed in \SSec{sec:si_mn_ext}.

\subsection{Application to coupled three-dimensional order parameters in multiferroic BiFeO$_3$}\label{sec:res_bfo}
We now turn to the coupled polar--tilt landscape of the multiferroic BiFeO$_3$~\cite{Neaton2005,Ederer2005}. The transition $Pm\bar{3}m\to R3c$ couples two three-dimensional order parameters, the polar $\Gamma_4^-$ ($\mathbf P$) and the tilt $R_5^-$ ($\mathbf R$), with a resulting six-dimensional landscape. This space therefore tests the dimensionality policy. 

\begin{figure}[!htbp]
  \centering
  \includegraphics[width=\textwidth]{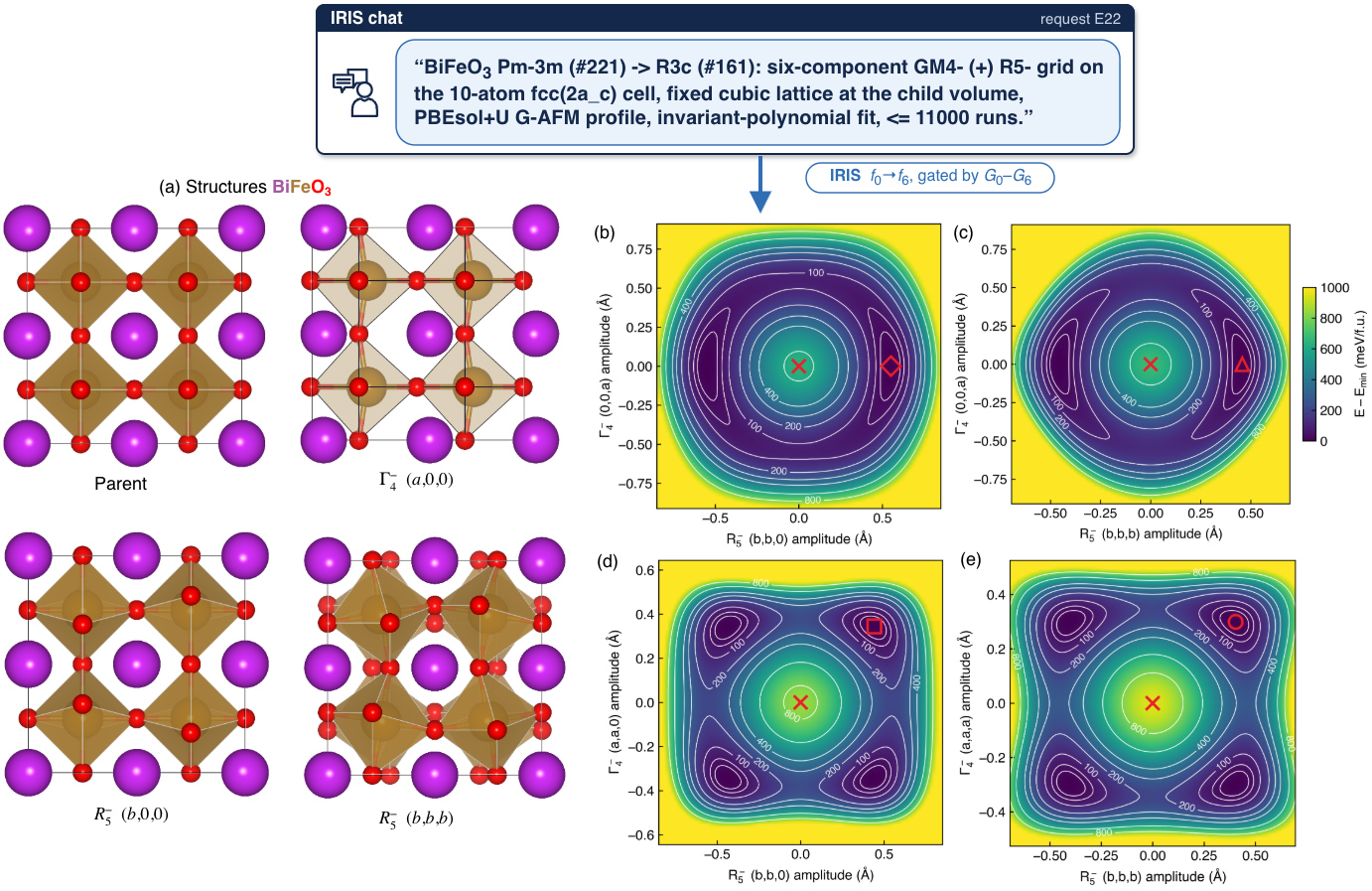}
  \caption{\textbf{Three-dimensional order parameters and polar--tilt coupling in multiferroic BiFeO$_3$.}
In $Pm\bar{3}m\to R3c$ both primary irreps are three-dimensional: the ferroelectric $\Gamma_4^-$, $\mathbf{P}=(P_x,P_y,P_z)$, and the antiferrodistortive tilt $R_5^-$, $\mathbf{R}=(R_x,R_y,R_z)$.
Landscapes are planes through the six-dimensional ($\mathbf{P}$, $\mathbf{R}$) space, from 10{,}054 DFT calculations at two volumes (PBEsol$+U$, $U_{\mathrm{eff}}=4$~eV on Fe $d$; \STab{tab:dft}), shown at the fixed cubic lattice of the $R3c$ volume ($a=3.9407$~\AA) with G-type antiferromagnetic order.
\textbf{a},~Structures: the parent ($Pm\bar{3}m$), $\Gamma_4^-$ along $(a,0,0)$ ($P4mm$), $R_5^-$ along $(b,0,0)$ ($I4/mcm$) and $R_5^-$ along $(b,b,b)$ ($R\bar{3}c$). Bi, purple; Fe, brown; O, red.
\textbf{b}--\textbf{e},~Energy landscapes, $E-E_{\min}$ per formula unit ($E_{\min}$, each plane's minimum: 313, 275, 101 and 0~meV/f.u.\ above $R3c$ for \textbf{b}--\textbf{e}; \STab{tab:bfo_planes}); axes give component $a$ or $b$ along each direction; contours at 100, 200, 400 and 800~meV/f.u.
\textbf{b},~$\mathbf{P}\parallel[001]$, $\mathbf{R}\parallel[110]$ and \textbf{c},~$\mathbf{P}\parallel[001]$, $\mathbf{R}\parallel[111]$: axes misaligned; minima at pure tilt phases $Imma$ ($\diamond$) and $R\bar{3}c$ ($\triangle$).
\textbf{d},~$\mathbf{P}\parallel\mathbf{R}\parallel[110]$ and \textbf{e},~$\mathbf{P}\parallel\mathbf{R}\parallel[111]$: axes aligned; four equivalent minima at $Ima2$ ($\square$) and the $R3c$ ground state ($\circ$).
Amplitudes are translation-free $A_p$ in \AA; colour scale capped at 1{,}000~meV/f.u.; red $\times$, parent; all 17 inequivalent high-symmetry planes in \SFig{fig:s_bfo_17planes}. Bubble: human request E22 (\STab{tab:si_requests}), whose configuration reproduces these planes in replay.}
  \label{fig:bfo}
\end{figure}

For BiFeO$_3$, certification first removes a rigid translation between child and parent that made up 76\% of the squared norm of the raw $\Gamma_4^-$ vector (\SSec{sec:origin}), both obtained through the Materials Project~\cite{Jain2013}. The certified amplitudes are 0.526 and 0.694~\AA\ (Fig.~\ref{fig:bfo}a; \STab{tab:bfo}; \STab{tab:dec_bfo}, \STab{tab:bfo_iso}). The dimensionality policy then computes one point per symmetry orbit: 10{,}054 PBEsol$+U$ calculations on fixed cubic lattices (\SFig{fig:s_strain}) cover 786{,}410 grid nodes at two volumes (9{,}597 symmetry orbits, 6{,}724 at the $R3c$ volume and 2{,}873 at the cubic one, plus refinement points and controls; with the 253 later validation points, 10{,}307 runs in all; \SFig{fig:s_dim}f, \SSec{sec:si_bfo_grid}); the 65 failed runs were handled by fixes that now seed the rule table (\STab{tab:rules}).

A degree-12 invariant polynomial reproduces 361 refinement energies withheld from its fit (274 off the grid lattice, 87 on grid nodes) to 1.15~meV/f.u.\ RMS and 253 later validation points to 0.57~meV/f.u.\ (\STab{tab:bfo_fits}). Its global minimum is $R3c$, 950.0~meV/f.u.\ below the cubic phase; $Ima2$ is an index-1 saddle (\STab{tab:bfo_stat}). Symmetry forbids trilinear couplings, so the lowest are biquadratic (\STab{tab:bfo_coef}): $\lambda_1P^2R^2$ makes the magnitudes compete, each suppressing the other by 12--13\%, whereas $\lambda_2(\mathbf P\cdot\mathbf R)^2$ locks their axes, so that a $10^\circ$ misalignment costs 28~meV/f.u.\ and beyond about $35^\circ$ one order parameter collapses (\STab{tab:bfo_hess}, \SSec{sec:si_bfo_coupling}). Thus with $\mathbf P$ and $\mathbf R$ on different axes the minima are pure tilt phases (Fig.~\ref{fig:bfo}b,c), whereas with aligned axes both survive (Fig.~\ref{fig:bfo}d,e); joint minima exist on only 6 of the 17 inequivalent high-symmetry planes (\SFig{fig:s_bfo_17planes}, \STab{tab:bfo_planes}). An independent interpolant gives the same minimum to 0.5~meV/f.u.\ (\STab{tab:bfo_rbf}, \STab{tab:bfo_extra}), and no low-energy region lies more than 1.13 grid steps from a DFT point (\SSec{sec:si_bfo_cov}).

Because the model is continuous in all six components, it also yields switching paths. In the homogeneous, fixed-cell limit, rotating $\mathbf P$ with $\mathbf R$ following crosses $Ima2$-type saddles at 101.3~meV/f.u., whereas reversing $\mathbf P$ at fixed tilt~\cite{Xu2017} costs 274~meV/f.u.\ (\SFig{fig:s_bfo_paths}; \STab{tab:bfo_paths}, \STab{tab:bfo_vol}). Because the Dzyaloshinskii--Moriya vector follows $\mathbf R$~\cite{Ederer2005}, the joint rotation also reverses the weak magnetization (\STab{tab:bfo_states}, \SSec{sec:si_bfo_magnet}). These are 0~K barriers without field, strain relief or domain walls, with the magnetization inferred from symmetry, so they rank homogeneous routes rather than predict which one a field selects. Polarization, relaxed anchors and structure agree with the literature~\cite{Neaton2005,Dieguez2011,Fedorova2022,Kubel1990} (\SSec{sec:si_berry}, \STab{tab:bfo_berry}, \STab{tab:bfo_lit}, \STab{tab:bfo_structs}; coercive fields with ref.~\cite{Lebeugle2007} in \SSec{sec:si_bfo_ext}). 

\subsection{Delineating language model and gate contributions}\label{sec:autonomy}
To separate what the gates and the language model each contribute, we reran every result end to end on the stored DFT energies and then broke the runs deliberately. In replay, all nodes and gates run as in production except $f_3$, which links each frame to the energy of a matching stored structure instead of running a new DFT calculation ($f_6$'s second route also reads the stored campaign; Methods, Section~\ref{sec:eval}); an unmatched frame stops the run. Replay therefore tests everything except the DFT calculations themselves; the plain-language intake is tested separately below. In all 13 run configurations of the six systems (Extended Data Table~\ref{tab:reruns}), every frame was matched, every gate passed and $f_6$ re-derived every reported number. Human decisions were of only two kinds: the borderline InMnO$_3$ selection (R5), where the human kept $\Gamma_2^-$ as an axis, and the CsSnI$_3$ secondary $X_5^+$ (D4), kept at zero with a caveat. Both answers matter; for the first, in the option replay below, excluding $\Gamma_2^-$ instead builds a one-axis grid on which the $(Q,P)$ model cannot be fitted, and the run stops at $G_5$ (\STab{tab:si_C}).

To test the gates, we injected 53 faults, each a single setting in the run configuration that reproduces a failure met in the campaigns, such as a job part ending at the walltime, a rigid translation left in a mode vector or an inconsistent DFT setting; data and code were never altered (Methods, Section~\ref{sec:eval}; every fault is listed in \STab{tab:si_faultlist}). Of the 53, 30 are errors, which corrupt a result if they pass, and 23 are limit faults, which tighten a threshold on correct data (Fig.~\ref{fig:agentic}d; \STab{tab:si_faults}). Every fault was caught by its expected gate, and none corrupted a result in replay; 21 errors were fixed by stored rules with identical numbers, and 9 went to the human. With the gates off, 14 errors corrupted results silently, 12 errors and 3 limit faults crashed, and 4 errors were harmless. With the catching gate removed, later checks stopped 27 errors only after the work was done (23 by checks that also run in production, 3 by a check that exists only in replay and 1 by a crash that the ablation itself causes), at a cost of 27{,}138 counterfactual DFT submissions; 3 were not caught (\STab{tab:si_ablation}). Runs killed after $f_2$, $f_3$ or $f_5$ and resumed from their checkpoint matched uninterrupted ones (4 of 4: three CsSnI$_3$ kill points and one HfO$_2$ path). Taken together, these experiments ablate the four layers of IRIS (deterministic nodes, gates, graph and language model) separately: the deterministic nodes alone reproduce the published numbers on clean inputs but return 14 silently wrong results on the faulted ones; each gate stops one class of fault that no other gate or the graph stops (54 removals of a non-catching gate change nothing); the graph contributes resumability, the recovery order and the audit, not the numbers; and the model contributes the intake and seven fewer hand-offs (\SSec{sec:si_layers}). The validity of a result rests on the gates, not on the agentic structure around them.

When a gate fails, the recovery node tries a stored rule first, then a language model restricted to a listed menu, then the human (\SSec{sec:si_recovery}). Rules alone were correct on 96\% of 53 benchmark cases, whereas nine models chose the correct fix in 72--91\% of decisions, erring mainly on failures that should be stopped (Fig.~\ref{fig:agentic}c; \STab{tab:si_models}, \STab{tab:si_A}, \SFig{fig:s_agentic_models}, \SFig{fig:s_agentic}, \SFig{fig:s_agentic_robust}). The rules were written from the campaigns the cases come from; scored with every rule seeded by the case's own campaign removed, they are correct on 89\%, and each lost case goes to the human, none to a wrong fix (\STab{tab:si_loso}). The guided prompt carries a one-sentence transcription of the rule table (the guidance line; Methods, Section~\ref{sec:eval}). With that sentence removed from the prompt, the models chose the right fix without being told in 79\% of decisions on the seven cases affected, with 2.5\% harmful (guided 85\% and 0.2\%; bare prompt 74\% and 11\%). The models' harmful choices fall almost entirely on the decoys and follow three patterns, each traceable to an option offered for a failure it does not address; the structural fix is that every recovery option now names in code the error codes it is admissible for, so that the model chooses only among admissible actions and a failure with no admissible fix is stopped by code without a model call. The numbers above and Fig.~\ref{fig:agentic}d,e use the unrestricted menus; Fig.~\ref{fig:agentic}c scores the same 53 cases both ways: with the admissible menus and the same prompts (five repeats), the nine models are 85--97\% correct and 0.17\% harmful, the residue being the R5 judgment call, the decoy harm of the two reference models falls from 3.2 and 1.0 of 11 to 0, and the IRIS order (rules first, Sonnet~5.5 for the rest) rises from 83\% to 92\% correct with no harmful decision, after one correction to the conditions that is disclosed (\SSec{sec:si_eligibility}; a post-hoc prompt repair is reported alongside in \SSec{sec:si_decoys}); every failure mode met in this work, grouped into intake ambiguities, crystallographic rejections and convergence or fit limits with the recovery path of each, is in \STab{tab:si_taxonomy} and \STab{tab:si_taxonomy_c}; and the model layer's cost, cents and seconds per decision against the calculation counts and wall time of a campaign, is in \SSec{sec:si_cost}. With the model as fallback, hand-offs to the human fell from 34 to 27 (Fig.~\ref{fig:agentic}b; \STab{tab:si_interventions}). With the gates on, no wrong result was accepted in replay, against 20 among 124 option branches with them off (Fig.~\ref{fig:agentic}d,e; \STab{tab:si_C}, \STab{tab:si_stack}). All 20 were stopped by checks that also run in production (\STab{tab:si_farr}). This was not so in a first replay: seven of its 22 wrong results, fits on the frames left after a failed job part, had been stopped only by the comparison with the stored campaign that exists in replay alone, and would have been false acceptances in production. A check that refuses a fit on fewer frames than planned unless a human has authorized it (code \texttt{MISSING\_FRAMES}, at $G_4$ and $G_5$) closed that gap, and the replays reported here are the re-run with it (\SSec{sec:si_replay_found}). Containment is still shown under replay only.

To test the effect of how a request is phrased, on 50 requests written and fixed with a hash before any model saw them (Fig.~\ref{fig:agentic}a; \STab{tab:si_E}, \STab{tab:si_requests}; prompts in \SSec{sec:si_prompts}) four models produced configurations with every critical field right in 92--95\% of calls on clear requests, terse and non-specialist ones included, against 64\% for a rule-based parser written from the schema and the tables and frozen before scoring (74\% after one declared bug fix); on expert requests alone the parser (77\%) is within the range of the models (77--86\%), but not on terse and non-specialist ones (parser 62--77\% and 54--69\%, models 94--100\%; \SSec{sec:si_parser}), and 51\% for a keyword lookup (\STab{tab:si_E}); they also handled out-of-scope and new-material requests correctly in every call, and underspecified ones in 23--25 of 25 calls, and every correct configuration reproduced the published numbers in replay. The remaining errors, 9--15 calls per model that passed $G_0$ with a wrong field, were choices that are valid but wrong, which $G_0$ cannot catch because it checks only that a configuration is complete and consistent. Almost all came from two InMnO$_3$ requests for LDA$+U$: the prompt described the stored InMnO$_3$ profiles in identical terms, so the models picked a PBEsol$+U$ profile instead. When the profile descriptions were rewritten to name the functional and the magnetic order, such errors fell to 0--5 per model (\SSec{sec:si_eval_ext}); the guidelines this implies for adapting the intake, every option described by its physics and defaults applied by code, are in \SSec{sec:si_intake_guidelines}.

\begin{figure}[p]
  \centering
  \includegraphics[width=\textwidth,height=0.92\textheight,keepaspectratio]{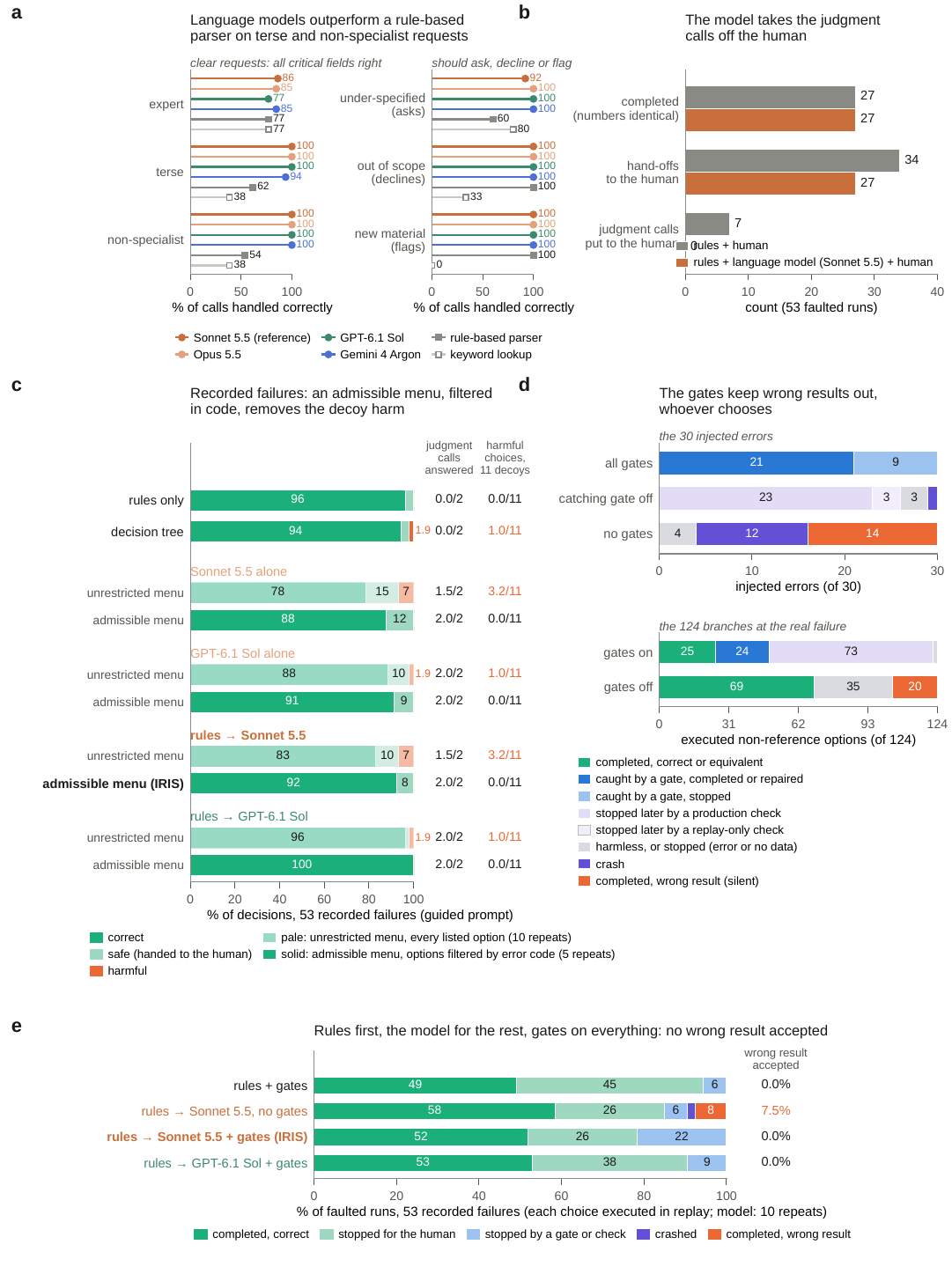}
  \caption{\textbf{What the language model adds to a gated workflow.} Legend continued on the next page.}
  \label{fig:agentic}
\end{figure}
\begin{figure}[!tb]
  \ContinuedFloat
  \caption[]{(continued) \textbf{What the language model adds to a gated workflow.} All runs are replays on the stored DFT energies (Methods, Section~\ref{sec:eval}); definitions in \STab{tab:si_defs}.
\textbf{a},~Plain-language intake on 50 requests written and fixed before any model saw them (39 clear requests, 13 run configurations in three styles; 5 underspecified, 3 out of scope, 3 new materials): share of calls with every critical field right, and share that correctly ask, decline or flag, for four models (five repeats; 1{,}000 calls), a rule-based parser (written from the schema, the tables and the system prompt only, frozen before scoring; v1, \SSec{sec:si_parser}) and a keyword lookup, the simplest deterministic interface.
\textbf{b},~The 53 faulted runs with rules and a human only, and with the reference model (Sonnet~5.5) as sub-agent: completed runs (numbers identical to the fault-free replay), hand-offs to the human, and published judgment calls put to the human (the two judgment calls, R5 and D4, recur in seven of the 53 faulted runs); both the D4 and the R5 answer change the result (excluding the borderline irrep removes an axis and stops the run at $G_5$). Retrieval and the decision tree are compared in \STab{tab:si_A}.
\textbf{c},~Recovery decisions on the 53 recorded failures (guided prompt): the stored rules and a decision tree with the same knowledge; then two models alone and the order IRIS uses (a stored rule first, the model only where no rule applies), each scored twice on the same 53 cases and labels, with the unrestricted menu (every listed option; ten repeats) and with the admissible menu (options filtered by error code in code, a failure with no admissible fix stopped without a model call; five repeats, \SSec{sec:si_eligibility}); pale bars, unrestricted menu; solid bars, admissible menu; bold label, the IRIS configuration. Right, judgment calls answered as published and harmful choices on the 11 decoys, whose correct answer is to stop. All nine models are compared in \SFig{fig:s_agentic_models} (unrestricted) and \SSec{sec:si_eligibility} (admissible).
\textbf{d},~Top, the 30 injected errors with all gates, with the gate that catches each removed (stopped later by a production check, stopped later only by a check that exists in replay alone, not caught and harmless in replay, or the one crash that the ablation itself causes), and with no gates. Bottom, every menu option other than stop and the reference applied at the real failure (124 branches; two asked for a frame that the stored plane does not contain and are counted as stopped), with all gates on and with the gates switched off after the fix.
\textbf{e},~The full recovery stack on the 53 recovery cases of the benchmark (built from the faulted runs): the stored rules alone (stop where no rule applies), and a stored rule first with Sonnet~5.5 or GPT-6.1 Sol for the rest (ten repeats), each choice executed in replay with the gates on; the Sonnet~5.5 order also with the gates switched off after the fix. Stopped for the human includes the limit faults, whose correct answer is to stop; stopped by a gate or check includes fixes refused or rejected after they were applied (by checks that also run in production or by the replay-only comparison with the stored campaign; the split is given in \STab{tab:si_stack}). A reference option that was not re-run without gates is counted with its gated outcome.}
\end{figure}

\providecommand{\livetodo}[1]{\textcolor{red}{[#1]}}
 
\subsection{A prospective end-to-end run}\label{sec:live}
 
All results above were replayed on stored energies, so the Runner never submitted a calculation. To test the workflow prospectively we added a live mode in which $f_0$ fetches the parent and child from the Materials Project, $f_1$ runs FINDSYM and ISODISTORT on the ISOTROPY server, and $f_3$ writes the VASP inputs, submits PBS jobs to an HPC cluster and collects them, with the same gates as in replay; a chat interface shows every node, gate and job as it happens (Methods, Section~\ref{sec:livemode}; \SFig{fig:s_live_page_path}, \SFig{fig:s_live_page_plane}). We chose BaTiO$_3$ because its polar phases are a textbook case: the tetragonal, orthorhombic and rhombohedral wells deepen in that order, most of their depth comes from the strain that accompanies the polar mode, and at a clamped cubic cell the mode alone gives only a shallow well whose anisotropy already favours the rhombohedral and orthorhombic directions~\cite{cohen1992origin,kingsmith1994first}. From requests of the form ``Map the straight-line irrep path of BaTiO3 from cubic Pm-3m (221) to rhombohedral R3m (160), 10 points, PBE'' (verbatim in \SFig{fig:s_live_page_path}), one per child phase, the intake model (Claude Sonnet~5.5) wrote the three path configurations (a fourth request, with \texttt{-{}-plane}, asked for the polar plane described below), and IRIS fetched the lowest-energy entry of each space group (mp-2998, mp-5986, mp-5777 and mp-5020; the $Amm2$ and $R3m$ cells have 10 and 15 atoms), decomposed each child into $\Gamma_4^-$ with its secondary strains, built ten-point paths with the strain interpolated alongside the mode, submitted two jobs of five static calculations per path and fitted each profile (\SFig{fig:s_live}a). The minima lie at $-32.6$, $-40.9$ and $-43.3$~meV per formula unit for $P4mm$, $Amm2$ and $R3m$, all at the lowest-energy sampled image, $\lambda=8/9$ of the Materials Project amplitude: the known order, consistent in magnitude with the Materials Project's own PBE energies of the relaxed phases relative to cubic ($-27.6$, $-37.3$ and $-37.3$~meV per formula unit, from their energies above the hull); the three paths share the cubic end point, where the 5-, 10- and 15-atom settings agree pairwise to 0.04--0.11~meV per formula unit. In a separate run, a two-component $(P_z,P_x)$ plane of $\Gamma_4^-$ at the fixed cubic cell, a $21\times21$ grid computed from its 66 symmetry-unique points in 14 jobs, gives shallow minima of $-4.7$~meV per formula unit along $[001]$ and $-7.5$ along $[101]$, with a fitted $4mm$ Landau anisotropy that favours the diagonal (coefficients in \STab{tab:si_batio3_plane}; \SFig{fig:s_live}b,c). This plane was launched from an earlier configuration template that carried PBEsol, at the PBE cubic cell of the paths (\SSec{sec:si_live_settings}). The two data sets are therefore interpreted on their own, the paths for the ordering of the three wells and the plane for the anisotropy of the polar mode at a fixed cell, and are not subtracted to quantify the contribution of strain to the well depths; that strain supplies most of the depth is the established fixed-cell picture of BaTiO$_3$~\cite{cohen1992origin,kingsmith1994first}, not a result of this comparison. Every gate passed on the first attempt; the gates validated each run but did not then check that two separately configured runs shared a functional, a gap closed afterwards by a metadata check that refuses the joint interpretation (code \texttt{SETTINGS\_MISMATCH}; \SSec{sec:si_live_settings}). $f_6$ re-derived 543 reported numbers (192 for the paths, 351 for the plane), no decision went to the human, and the four runs took 34~min of wall time in total (6--11~min each, dominated by the queue and the 2--7~min jobs). A second, independent live run of the plane reproduced all 66 energies and every coefficient. The recording is Supplementary Video~1 and the complete run directories, with their audit logs and hashes, are deposited (Data availability).
\section{Discussion}
Beyond reproducing known structures, the landscapes give physical insight: the tilts alone do not select the ground state in either perovskite; the manganite domain barrier comes from the trilinear polar coupling; BiFeO$_3$ separates the competition of the polar and tilt magnitudes from the locking of their axes; the polar mechanism of HfO$_2$ depends on the reference phase; and in BaTiO$_3$ the PBE paths recover the order of the three ferroelectric wells while the separate fixed-cell plane resolves the anisotropy of the polar mode; because the two runs used different functionals, their depths are not compared to extract the strain contribution, which also marks a boundary of the verification: the gates validated each campaign but did not yet enforce consistency between independently configured runs used in one interpretation (now a metadata check, \SSec{sec:si_live_settings}). Each check in Extended Data Table~\ref{tab:errors} answers an error met in a campaign or in the option replay, and most of these errors leave the space groups unchanged: checking only symmetry, or only energies, would have produced smooth but wrong landscapes.

Because the model chooses only among listed fixes and human answers become rules, its contribution can be measured directly. The recovery benchmark measures the case most favourable to rules, failures for which a rule was written after the fact; there the model is not needed. It is needed where no rule can act: to turn a plain-language request into a checked configuration and to answer judgment calls, which removed seven of 34 hand-offs to the human with every result identical to the human-answered runs (Section~\ref{sec:autonomy}), while in replay the gates kept all its mistakes, concentrated on failures that should be stopped, from being accepted. The prospective BaTiO$_3$ campaign shows the two roles together in production: the model wrote the four configurations from typed requests (the plane's functional came from its configuration template, Methods), no rule or sub-agent was needed because no gate failed, and the gates, the executor and the re-derivation carried 96 calculations from request to landscape without a human decision (Section~\ref{sec:live}). On failures never seen, for which a rule table can only stop, the listed-option sub-agent is the intended fallback; with the guidance sentence seeded by a case's own campaign removed from their prompt, the models chose the right fix without being told in 79\% of decisions and a harmful one in 2.5\% (\STab{tab:si_loso}), a first measure of that role, though on failure mechanisms of this work. Unlike tool-calling agents that route a workflow themselves~\cite{Yao2023,Zou2025,Pham2025}, listed options make each decision checkable and measurable~\cite{Chen2025} and keep the safeguards in code~\cite{Tang2025}.

\paragraph{Limitation: Application to irrep paths between the polymorphs of HfO$_2$}\label{sec:res_hfo2}

Finally, we apply the framework to a polymorphic material, where part of the task lies beyond its intended use. The cases above each link a single parent structure to a single child structure.  Polymorph families require a stronger form of consistency than the cases considered above: all phases must be represented within the same order-parameter space so that their relative stability, connectivity and competing transformation channels can be compared directly. This requires a common parent-referenced crystallographic basis together with a shared set of symmetry-adapted coordinates. In the current framework, IRIS can construct symmetry-consistent paths for individual pairs of structures, while each path remains defined by the local symmetry relation of that pair. As a result, independently generated pairwise paths do not necessarily combine into a single global energy surface.

\begin{figure}[!htbp]
  \centering
  \includegraphics[width=\textwidth]{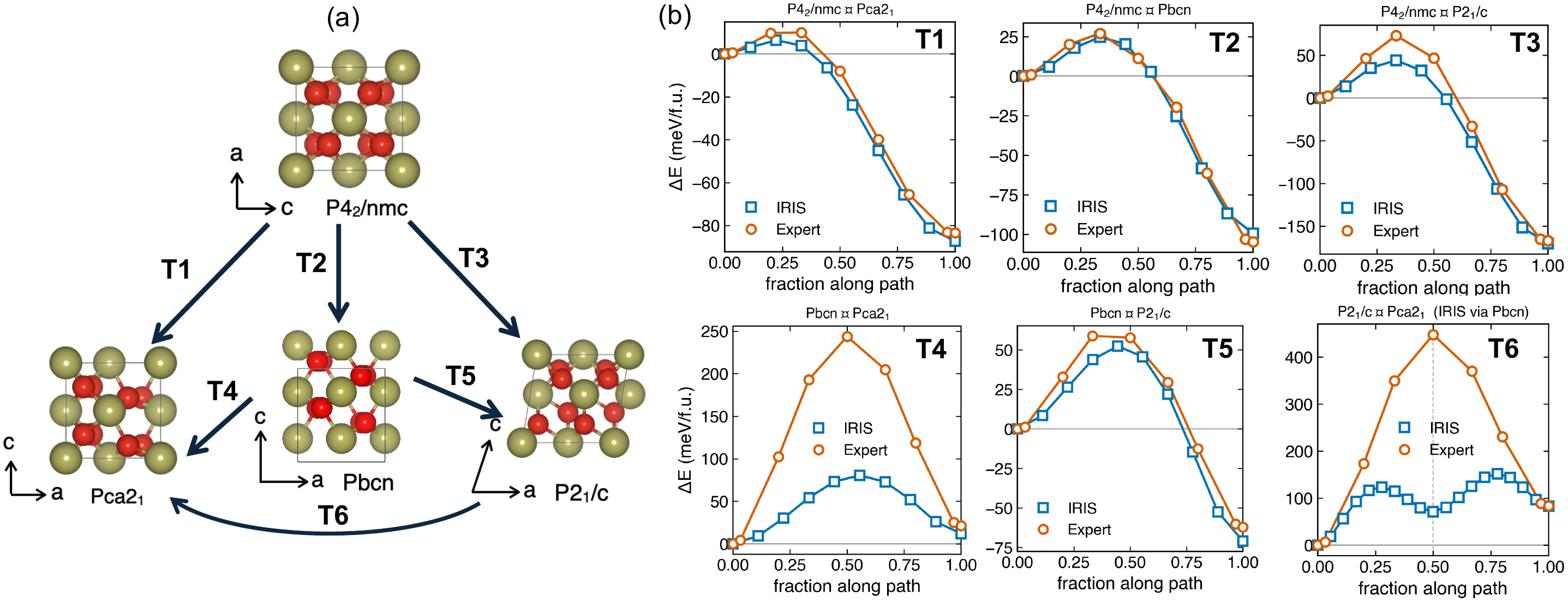}
  \caption{\textbf{Irrep paths between the polymorphs of HfO$_2$.}
\textbf{a},~Transitions between the tetragonal ($P4_2/nmc$), antipolar orthorhombic ($Pbcn$), polar orthorhombic ($Pca2_1$) and monoclinic ($P2_1/c$) phases. T1--T5 connect a parent to one of its subgroups; $Pca2_1$ and $P2_1/c$ are not group--subgroup related, so IRIS routes T6 through $Pbcn$ (T5 reversed, then T4). Hf, olive; O, red.
\textbf{b},~DFT energy along each path (PBE, non-magnetic; \STab{tab:dft}), $\Delta E$ per formula unit relative to the start, against the fraction $\lambda$ along the path; symbols are calculated images. IRIS paths (blue squares) scale every mode amplitude and the strain by $\lambda$ from parent ($\lambda=0$) to child ($\lambda=1$), each in the reference phase of its transition; expert paths (orange circles) express every structure in a common $P4_2/nmc$ basis (Extended Data Table~\ref{tab:expert}). For T6, the IRIS path passes through $Pbcn$ at $\lambda=0.5$ (dashed line). The expert end structures were relaxed separately, so the end-point energies of the two constructions differ by 3--5~meV/f.u. Barriers along straight (frozen) paths are upper bounds; values in \STab{tab:hfo2}.}
  \label{fig:hafnia}
\end{figure}

An illustration of these current limitations is  hafnium dioxide (HfO$_2$), a leading ferroelectric for non-volatile memory, neuromorphic computing and logic-in-memory~\cite{boscke2011,muller2012,park2015,cheema2020}: HfO$_2$'s polar $Pca2_1$ phase arises from the coupled polar and antipolar distortions~\cite{raeliarijaona2023hafnia}, and its polymorphs, which lie within narrow energy windows and are linked by several group--subgroup relations, have so far needed sustained expert intervention to map~\cite{delodovici2021,Manna2026hafnia}, with an example of such an expert-derived map~\cite{delodovici2021} shown in Fig.~\ref{fig:hafnia}a. 

The expert HfO$_2$ construction expresses all phases relative to a tetragonal (and unstable) $P4_2/nmc$ reference, so that the three low-energy polymorphs occupy a single symmetry-defined coordinate system. In this basis, the tetragonal-to-child paths T1–T3 are conventional parent-referenced transitions, and IRIS reproduces the corresponding expert constructions and energy barriers closely (Fig.~\ref{fig:hafnia}b). The child-to-child paths T4–T6 place an additional requirement on the representation. When treated independently, IRIS identifies a symmetry-consistent connection between the selected endpoints and constructs a valid path for each pair. The resulting coordinates, however, are defined by the symmetry relation and structural mapping specific to that transition and may differ from those obtained when all polymorphs are embedded simultaneously in a common parent basis. Within this representation, T5 remains consistent with the pairwise IRIS construction, whereas T4 and T6 follow different intermediate symmetry branches.

Their frozen-path barriers therefore differ despite connecting the same endpoint phases (Fig.~\ref{fig:hafnia}b, \SFig{fig:s_hfo2_sym}, \SSec{sec:si_hfo2_ext}. Extended Data Table~\ref{tab:expert}, Section~\ref{sec:paths}). This comparison shows that a locally valid path between two phases can differ from the path obtained when both phases are constrained to lie within a common polymorph-wide order-parameter manifold.

We note that even when an appropriate common parent can be identified, as for tetragonal HfO$_2$, each polymorph must be mapped into that reference with a consistent cell transformation, origin choice, atomic correspondence, lattice convention, strain definition and mode normalization. The symmetry-adapted modes must then be defined jointly so that the same coordinates retain the same physical meaning across all phases and paths. IRIS already performs many of these operations for individual transitions, including structure standardization, symmetry decomposition, crystallographic transformations, and structural certification. Extending the framework to polymorph families would require an upstream graph-based construction that identifies a common reference, embeds all relevant phases into that basis, determines a shared set of symmetry coordinates, and then constructs transition paths and fits a unified energy landscape.

\section{Methods}
\subsection{Workflow, agents and gates}\label{sec:agent}
IRIS is organized as the gated state graph of Fig.~\ref{fig:pipeline}; the graph, with its checkpoints and interrupts, is implemented with the LangGraph library~\cite{LangGraph}. Seven agent nodes, $f_0$--$f_6$, carry out eight blocks of work (Extended Data Table~\ref{tab:blocks}); each node is followed by a gate $G_i$ on its outgoing edge, and a single recovery node handles every failure. Extended Data Fig.~\ref{fig:workflow} expands the graph into the eight blocks, each with its owner agent, inputs (IN), steps (step $\to$ product), outputs (OUT) and the CsSnI$_3$ example. This section follows that figure block by block, then describes the gates, the recovery node with its rule table and sub-agents, and the role of the language model; execution on HPC systems is part of block~6. The Path Planner, which carries most of the crystallographic logic, is described separately in the next section (Extended Data Fig.~\ref{fig:planner}).

\paragraph{State and hand-offs.} Each node receives the state of the run from the previous node and returns an updated state; formally, $S_{i+1}=f_i(S_i)$, with $S_0$ the request and the two structures. In practice, the state is the run directory of one system: every block writes named files (for example \texttt{config.yaml}, \texttt{setting.json}, \texttt{metadata.json}, \texttt{grid\_manifest.json}, \texttt{energies.csv}) that the next block reads, and every structure and input file is stored with a hash. A gate reads only the files its node has written. The DFT campaigns reported here ran these blocks as staged scripts with the same inputs, outputs and checks; the end-to-end reruns and the evaluation of Section~\ref{sec:autonomy} use the graph implementation, with its checkpoint after every node.

\paragraph{Gating and recovery.} Each hand-off passes through a gate $G_i$, deterministic code that returns either a pass ($\checkmark$) or a failure with an error code $c$ that names what went wrong:
\begin{equation*}
S_{i+1}=f_i(S_i),\qquad G_i(S_{i+1})\in\{\mathrm{pass}\}\cup\{\mathrm{fail}(c)\},\qquad i=0,\dots,6 ,
\end{equation*}
where $i$ indexes the nodes and their gates and $S_i$ is the state on entering $f_i$. The graph advances only on a pass, and a failure travels to a recovery node:
\begin{equation*}
\mathrm{next}(i)=
\begin{cases}
f_{i+1}, & G_i=\mathrm{pass},\\
\mathrm{recover}(i,c), & G_i=\mathrm{fail}(c),
\end{cases}
\end{equation*}
with $f_7\equiv\mathrm{END}$. The recovery node tries three sources of a fix in turn: a stored rule table $\mathcal{R}$; the gate's sub-agent $\sigma_i$, which must choose from the list $\Omega_i(c)$ of fixes the code provides; and a human $h$, whose answer is added to $\mathcal{R}$:
\begin{equation*}
\mathrm{recover}(i,c)=
\begin{cases}
\mathcal{R}(i,c), & (i,c)\in\operatorname{dom}\mathcal{R},\\
\sigma_i(c,S)\in\Omega_i(c), & \text{else, if } \sigma_i \text{ returns a fix},\\
h(c,S),\quad \mathcal{R}\leftarrow\mathcal{R}\cup\{(i,c)\mapsto h\}, & \text{otherwise}.
\end{cases}
\end{equation*}
Here $\operatorname{dom}\mathcal{R}$ is the set of (step, error code) pairs for which a rule is stored and $\mathcal{R}(i,c)$ is that rule's fix; $\sigma_i(c,S)$ is the fix that the sub-agent of $G_i$ selects after reading the error code and the part of the state the gate checked; and $h(c,S)$ is the fix chosen by the human. The human is reached through \texttt{interrupt()}, which pauses the graph, and the answer resumes it through \texttt{Command(decision)}. The update $\mathcal{R}\leftarrow\mathcal{R}\cup\{(i,c)\mapsto h\}$ stores that answer as a rule for the pair $(i,c)$, so the same failure is fixed without asking again. The fix is a configuration patch merged into the overrides of the node it routes to (by default $f_i$); the graph re-runs from that node on the current state, with no rewind of earlier results, and the same $G_i$ re-checks the result; every attempt is logged (the predicates of every gate are set out in \SSec{sec:si_gates_formal}), and a fit that fails $G_5$ for lack of coverage returns the graph to $f_2$ to refine the grid (Fig.~\ref{fig:pipeline}, top). The user works with IRIS through a chat interface (Fig.~\ref{fig:pipeline}; the intake step behind it is evaluated in Section~\ref{sec:autonomy} and the page was used in the prospective run of Section~\ref{sec:live}, but the front end itself was not formally evaluated): the request and the configuration resolved from it are exchanged there, and so is every question that recovery passes to a human, always with the listed fixes to choose from. Long calculations use the same machinery: $f_3$ submits the jobs and the graph pauses at its checkpoint until the scheduler reports them finished (Fig.~\ref{fig:pipeline}, orange loop).

\paragraph{Block 0, Intake ($f_0$).} IN: a plain-language request, the policy file (\texttt{policy.yaml}) and an optional system profile (\texttt{systems/<name>.json}). $f_0$ is the only node that calls a language model directly. It (i) parses the request into a draft intent (system, scientific question, engine, budget); (ii) resolves the two structures from file paths or Materials Project~\cite{Jain2013} identifiers; (iii) fills every remaining configuration field from the policy defaults and records where each value came from (request, language model or default); and (iv) passes the result to $G_0$. OUT: \texttt{config.yaml}, \texttt{intent.json} (the question and the source of every field) and a provenance record. For CsSnI$_3$, the request ``map the CsSnI$_3$ tilt landscape, fixed cell, DFT'' resolves to $Pm\bar{3}m\to Pnma$ from the Materials Project, strain policy P1 (fixed cell) and the electronic-structure settings of the CsSnI$_3$ system profile (\STab{tab:dft}). In the reruns and fault experiments of Section~\ref{sec:autonomy} step (i) was not used: the configuration was given as a file, and $f_0$ validated it, pinned the structure files by their SHA-256 digests and recorded any injected faults. Step (i) was evaluated separately on 50 frozen requests (Section~\ref{sec:eval}). The model returns the draft intent as structured output~\cite{BeurerKellner2024} whose fields can only take values from the stored tables of systems, transitions, analyses, strain policies, system profiles and outputs; the only number it may give is the budget, and it may instead ask one question or decline. $G_0$ gained three codes for this step: \texttt{AMBIGUOUS} (a required field cannot be decided, so the human is asked one listed question), \texttt{OUT\_OF\_SCOPE} and \texttt{NEEDS\_NEW\_DFT} (a valid request for a material with no stored calculations).

\paragraph{Block 1, Input ($f_1$).} IN: two structures (CIF, POSCAR or Materials Project identifier), or the child alone. $f_1$ reads and parses both structures (species, atom counts), matches their compositions (reduced formula and formula units per cell), provisionally assigns the roles parent ($G$) and child ($H$) from their space-group numbers (the assignment is confirmed by the $H\subset G$ test of $G_1$), and constructs a parent when only the child is given, recording the method used. OUT: \texttt{parent.cif}, \texttt{child.cif}, the roles and their provenance (identifier, hash or ``constructed''). For CsSnI$_3$ both structures come from the Materials Project (5 and 20 atoms) and no parent is built.

\paragraph{Block 2, Preprocess ($f_1$).} IN: the two structures and their roles. $f_1$ assigns space groups over a ladder of tolerances with two independent symmetry engines, FINDSYM~\cite{Stokes2005} and spglib~\cite{Togo2024}, and keeps the assignment on which both agree and which is stable across the ladder, together with the standardized structures. The group--subgroup relation $H\subset G$ is then established, with the index $i=t\cdot k$ (translationengleiche index $t$ times klassengleiche index $k$) and the descent tree of intermediate subgroups. Finally the supercell basis (a $3\times3$ integer matrix with $\det=k\,z_H/z_G$, where $z$ is the number of lattice points in the conventional cell, so that $\det=k$ between primitive settings and $\det=6$ for $R3c$ on hexagonal axes from cubic $Pm\bar3m$ with $k=2$) and origin shift that map the parent onto the child are taken from the decomposition step (block 3) or, when it offers none, found by the basis search of $f_1$, which matches atoms between the two structures and accepts right-handed bases with this determinant (\SFig{fig:s_basis}). OUT: the standardized structures, $G$, $H$, the index and tree, and \texttt{setting.json} (basis and origin). For HfO$_2$ $P4_2/nmc\to Pbcn$, the decomposition listed no basis because of 11.3\% strain along the parent $c$ axis, and the basis search supplied one. The tolerance ladder runs from $10^{-4}$ to 0.3~Å (the gate requires the $10^{-3}$ and $10^{-2}$~Å rungs to agree), and generated frames are audited at $10^{-2}$~Å ($10^{-3}$~Å for the six-component grid; \SSec{sec:si_gates_formal}).

\paragraph{Block 3, Decompose ($f_1$).} IN: the standardized structures and \texttt{setting.json}. $f_1$ performs the symmetry-mode decomposition itself, using the ISODISTORT engine of the ISOTROPY suite~\cite{Campbell2006,StokesISOTROPY} as its computational back end: it uploads the parent once per session (and caches it), submits the child with the basis and origin over a ladder of maximum displacement cut-offs from 1 to 3~Å, parses the result and checks it. No step of the decomposition is carried out by hand. OUT: \texttt{metadata.json}, listing every irrep with its wave vector $\mathbf k$, order-parameter direction (OPD), amplitudes $A_s$ (supercell-normalized) and $A_p$ (parent-cell-normalized, $A_p=A_s/\sqrt{s}$ for a supercell of $s$ parent cells), the strain block and the isotropy subgroup $H_i$; and the exported mode vectors. $G_1$ then requires $H\subset G$, a basis with $\det=k\,z_H/z_G$, and agreement between the space group implied by the decomposition and the one assigned in block 2 (\SFig{fig:s_index}).

\paragraph{Blocks 4 and 5, Postprocess and Build grid ($f_2$).} The Path Planner turns the decomposition into an audited grid of structures. Block 4 classifies the modes, selects the active set and certifies the mode vectors; block 5 sizes the grid, builds every frame and audits its symmetry. Both are described in detail in the next section (Extended Data Fig.~\ref{fig:planner}, eqs~\eqref{eq:decomp}--\eqref{eq:dmin}). $G_2$ passes the grid on only if all mode-vector checks passed and every frame has its expected space group.

\paragraph{Block 6, Compute ($f_3$ Runner, $f_4$ Reviewer).} IN: the audited frames, the campaign directory, the engine (VASP on an HPC system, or a machine-learned interatomic potential such as MACE~\cite{Batatia2022} locally), the account, the queue and the maximum number of jobs in flight. $f_3$ writes one calculation directory per frame (\texttt{frame\_NNNNN/} with the VASP inputs and a \texttt{point.json} that records the frame's amplitudes), packages the campaign into parts with their submission scripts and a frame map, and runs a preflight that checks every frame against the system profile (species order, $+U$ arrays, magnetic moments, $k$-mesh, POTCARs). It then launches the campaign (VASP through the scheduler, Slurm~\cite{Yoo2003} or PBS; MACE as a batch evaluation), monitors it and resubmits failed parts. While the jobs run, the graph pauses (\texttt{interrupt()}); a job-end hook resumes it, and $G_3$ passes only when no job is queued or running. $f_4$ collects the energies and applies the convergence criteria of Section~\ref{sec:dft}, flagging as SCF-limited (self-consistent-field step limit) the runs that still stop at the step limit after resubmission. $G_4$ passes when every energy is converged or flagged. OUT: \texttt{energies.csv} (frame, amplitudes, energy, convergence flag). For CsSnI$_3$ all 1{,}036 frames converged in 30-minute parts.

\paragraph{Block 7, Analysis ($f_5$ Analyst, $f_6$ Skeptic).} IN: \texttt{energies.csv}, \texttt{grid\_manifest.json} and the \texttt{point.json} files. $f_5$ builds the energy surfaces (planes, rings and axes), cross-checking the interpolant; fits an invariant Landau polynomial, its form fixed by the model of the system or its degree chosen by cross-validation (Section~\ref{sec:landau}); and locates minima, saddles and barriers, including minimum-energy paths. $G_5$ checks the fit against limits set per analysis in the policy file, always with at least ten points per coefficient. For the CsSnI$_3$ plane fit, the 10-fold cross-validation RMS and the fit RMS must not exceed 0.5~meV/f.u., no residual may exceed 2~meV/f.u., and every located minimum must lie at least two grid steps inside the fitted region. For the manganite fits, the limits apply per 30-atom cell to the residuals of the fitted energies (not to the per-formula-unit RMS with $P$ relaxed of Section~\ref{sec:res_mn}): a fit RMS of 5 and 8~meV, a cross-validation RMS of 6 and 10~meV and a largest residual of 15 and 25~meV (12~meV for the InMnO$_3$ plane fit) for InMnO$_3$ and ErMnO$_3$, with the minima at least half a $K_3$ step inside the window. Where a surface is analysed directly without a fit (Ca$_3$Ti$_2$O$_7$), $G_5$ requires a complete grid, a minimum at least two steps inside it, and energies that obey the mirror symmetries of the parent to within 1~meV/f.u. A fit that fails for lack of coverage returns the graph to $f_2$ to refine the grid. $f_6$ then re-derives every number that enters the report from the stored data, independently of $f_5$, and $G_6$ passes only if every number agrees. OUT: figures, the Landau table and the report. For CsSnI$_3$, three interpolants give the same ring corrugation (Section~\ref{sec:landau}).

\paragraph{Gates.} Each gate is a deterministic Python function that returns pass or $\mathrm{fail}(c)$ with an error code $c$ and a detail string:
$G_0$, configuration valid (schema, value ranges, structure files present, budget within limits);
$G_1$, decomposition valid ($H\subset G$, basis $\det=k\,z_H/z_G$, space group implied by the decomposition equal to the assigned one, mode norms $\lVert\mathbf e_i\rVert=A_p\sqrt n$);
$G_2$, grid valid (all four mode-vector checks pass, every frame's space group on the descent tree, no atomic collisions);
$G_3$, jobs finished (none queued or running);
$G_4$, energies usable (converged, or flagged and excluded from gradient fits);
$G_5$, fit acceptable (cross-validation error, residuals, coverage);
$G_6$, claims verified (every reported number re-derived). \STab{tab:gates} lists the checks and error codes.

\paragraph{Retry budget.} The rule table and the sub-agents used by the recovery node are described in the next two paragraphs. After any fix the failing node is re-run and the same gate re-checks the result, within a retry budget of three attempts per node; when the budget is exhausted, the case goes to a human.

\paragraph{Rule table.} The rule table $\mathcal{R}$ (\texttt{recovery.yaml}) is the first and cheapest source of a fix (\STab{tab:rules}). Each rule is a record
\begin{equation*}
r=\bigl(i,\ c,\ \text{condition},\ \text{fix},\ \text{parameters},\ n_{\max},\ \text{source}\bigr),
\end{equation*}
keyed by the step $i$ and error code $c$. The optional condition is a predicate that code evaluates on the error detail (for example $f_T>10^{-6}$, or not converged after NELM steps); the fix is an action that code already implements (resubmit, recalibrate, run the basis search, re-match atoms, flag a run as SCF-limited) with fixed parameters; $n_{\max}$ limits how often the rule may be applied to the same case; and the source records whether the rule was seeded from a past campaign or added from a human decision. When a gate fails, the recovery node takes the most specific rule whose key and condition match; if none matches, or the rule has reached $n_{\max}$, the case falls through to the sub-agent. Rules contain no language-model output and are applied deterministically, so a failure that has been seen once is fixed the same way every time. The table was seeded with the fixes applied by hand during the campaigns reported here: for example, walltime $\to$ resubmit the part; rigid translation in a polar mode $\to$ subtract the mean displacement; SCF step limit $\to$ resubmit with robust mixing and, if that fails, keep the energy as SCF-limited; and no basis from the decomposition $\to$ run the basis search. A human decision is appended as a new rule with its source, date and the case that produced it, and the hash of the table in force is logged with every attempt. The rule table is distinct from the policy (\texttt{policy.yaml}), which holds the scientific choices (the threshold $A_{\mathrm{thr}}$ and rules R1--R5, the budget $B$ and policies D1--D5, and the strain policies P1--P5): policies decide how a landscape is built, rules decide how a failure is repaired.

\paragraph{Sub-agents.} Each gate has one sub-agent, with a menu of the fixes that a rule cannot always choose between (\STab{tab:subagents}): $G_1$, widening the basis search (for example for HfO$_2$ $P4_2/nmc\to Pbcn$ when the stored rule has not fixed it); $G_2$, removing a rigid translation, rebuilding the single-mode check frames on the fixed parent lattice, and the borderline selection of rule R5 (include, exclude or anchor only); $G_4$, resubmitting failed job parts unchanged or with robust mixing, or flagging them and continuing; $G_5$, densifying the grid and the hard-secondary decision of D4; and $G_6$, re-running the analysis. Every option names the error codes it is admissible for, and the recovery node offers the sub-agent and the human only the options admissible for the failing code: a coverage failure whose minimum lies near the grid edge is not offered the densification that addresses a thin grid, and a fit tolerance is not offered the hold that belongs to the D4 judgment call (\SSec{sec:si_eligibility}). Every menu also offers stop, and a failure whose admissible menu is stop alone, among them the menus of $G_0$ and $G_3$, a strain-policy choice or a grid over budget, goes to the human without a model call. A sub-agent is a single language-model call with a fixed contract,
\begin{equation*}
\sigma_i:\ \bigl(c,\ \text{detail},\ S|_i,\ \text{policy},\ \Omega_i(c)\bigr)\ \longmapsto\ (\omega,\ \text{reason}),\qquad \omega\in\Omega_i(c)\cup\{\varnothing\},
\end{equation*}
where $S|_i$ is the slice of the state that the failing gate checked and $\Omega_i(c)$ is the list of admissible fixes that code generates for the failure, each with its parameters already filled in (for example the tolerances on the ladder at which the two symmetry engines agree, or the next-ranked basis from the basis search). The model returns one option identifier $\omega$ and a reason; $\omega=\varnothing$ means that no listed fix applies and sends the case to a human. The answer is requested as structured output constrained by a schema whose only admissible identifiers are those of $\Omega_i(c)$ (JSON-schema or enumeration constraints of the provider's interface), and it is validated again before it is applied: an identifier outside $\Omega_i(c)$, malformed output or a repeated choice that has already failed is rejected, logged as invalid and passed to the human, never re-asked; only transport errors are retried. The same interface serves models of several providers, which makes the sub-agent's model a setting of the run rather than of the code. The model therefore never writes a number into the state: it only selects among actions that code has prepared and that the same gate will check again. Every call is logged with its input, the option list, the choice and the reason. Every option on a menu sets a parameter that the node it returns to reads, which a test checks against the node code. The reference model is Claude Sonnet~5.5 (\texttt{claude-sonnet-5-5}), called at the provider's default temperature; the models compared, the two prompt variants and the protocol are given in Section~\ref{sec:eval}, and the prompts are released with the code.

\paragraph{Role of the language model.} The language model is used in two bounded places: in $f_0$, to turn the request into a draft configuration that $G_0$ validates, and in the sub-agents, to choose one fix from a list the code provides. Both are evaluated (Section~\ref{sec:eval}). It does not compute decompositions, structures, energies or fits, does not decide whether a gate passes, does not route the graph, unlike tool-calling agents in which the model chooses the next action~\cite{Yao2023}, and does not write numbers into the state. These limits follow from errors met during development, in which a model wrote fluent but wrong crystallographic statements, such as wrong space-group symbols or missing zone-boundary irreps, that the deterministic checks of $G_1$ caught (\SSec{sec:si_llm}). Its contribution can therefore be measured by an ablation in which unmatched failures go directly to a human (Section~\ref{sec:autonomy}, Fig.~\ref{fig:agentic}b).

\subsection{Path Planner: from decomposition to an audited grid}\label{sec:planner}
The Path Planner $f_2$ converts the decomposition into a grid of structures in which every frame has a known, checked space group (Extended Data Fig.~\ref{fig:planner}). It receives \texttt{metadata.json} (irreps, amplitudes $A_i$, strain, isotropy subgroups $H_i$) after $G_1$, together with the policy (threshold $A_{\mathrm{thr}}$, rules R1--R5, budget $B$, policies D1--D5 and the strain policy; summarized with examples in \STab{tab:si_policies}), and works in four stages. Checks $c_1$--$c_3$ close the first three stages and $G_2$ closes the last; a failed check goes to the recovery node like any gate failure: the $G_2$ sub-agent settles borderline selections (R5) and the amplitude-audit fixes; strain-policy and grid-budget failures go to the human; and coverage is checked at $G_5$ (\STab{tab:subagents}). Equations~\eqref{eq:decomp}--\eqref{eq:dmin} below are numbered as in Extended Data Fig.~\ref{fig:planner}.

\paragraph{Select modes.} The parent-to-child distortion is written as internal displacements, a sum of irrep modes, together with a homogeneous strain of the lattice,
\begin{equation}\label{eq:decomp}
\Delta\rv=\sum_i A_i\,\eh_i ,\qquad \mathbf L=(\mathbf I+\boldsymbol\varepsilon)\,\mathbf L_0 ,
\end{equation}
where $\eh_i$ is the unit mode vector of irrep $i$ and $A_i$ its amplitude, both in Cartesian coordinates of the parent setting, and $\boldsymbol\varepsilon$ is the strain that takes the parent lattice $\mathbf L_0$ to the child lattice $\mathbf L$. Each irrep is classified as displacive or strain, primary or secondary, and polar or non-polar. For polar $\Gamma$ irreps, whose vector representation contains the acoustic translation, the decomposition fixes the free origin along the polar axis by convention, so its mode vector can carry an arbitrary rigid translation (\SSec{sec:origin}). The amplitude is therefore made translation-free before any threshold is applied,
\begin{equation}\label{eq:tf}
A_i^{\mathrm{tf}}=A_i\,\frac{\|\et_i\|}{\|\e_i\|},
\end{equation}
with $\et_i$ defined in eq.~\eqref{eq:trans}. Five rules then select the active set $\Aset$, whose amplitudes become the axes of the landscape:
R1, strain irreps never become axes and are handled by the strain policy (an irrep that carries both displacive modes and a strain component, such as $\Gamma_5^+$ in $P2_1/c$ or $\Gamma_3^+$ in $Pbcn\to P2_1/c$, remains a displacive irrep and only its strain part goes to the strain policy);
R2, a primary displacive irrep is active if its translation-free amplitude reaches the threshold,
\begin{equation}\label{eq:thr}
i\ \text{primary},\ A_i^{\mathrm{tf}}\ge A_{\mathrm{thr}}\;\Rightarrow\; i\in\Aset ,
\end{equation}
whereas a secondary irrep, one induced by the primaries through their couplings, does not become an axis even above the threshold: it is held at zero on the grid and sampled with a few points where its coupling matters (D4), unless R4 keeps it as the polarization axis. Operationally, the primaries are found by a walk that takes the displacive irreps in order of decreasing translation-free amplitude and adds each only if its isotropy subgroup lowers the symmetry of the running intersection, until the intersection equals $H$; all other displacive irreps are secondaries (\SSec{sec:si_primaries}, \STab{tab:si_primaries});
R3, the active set must generate the child group,
\begin{equation}\label{eq:gen}
\bigcap_{i\in\Aset} H_i = H ,
\end{equation}
and if it does not, $G_2$ fails (\texttt{R3\_FAIL}) and the case goes to the recovery node;
R4, a polar child keeps at least one polar irrep, even below threshold, so that the polarization has an axis;
R5, when the smallest active amplitude exceeds the largest inactive one by less than 0.1~Å, the selection is not clear-cut and the irreps near the threshold go to the borderline sub-agent, which includes them, excludes them or keeps them as anchor points only; the answer is applied inside the rules (include keeps the borderline active irreps as axes; exclude removes them and holds them at zero, failing if the subgroup is no longer generated; anchor only adds one frame per removed irrep at the child amplitudes, whose energy is reported), so that $f_2$ builds the grid on it, $G_2$ checks it and $f_6$ re-derives it.
Check $c_1$ asks whether $\Aset$ generates $H$. \SFig{fig:s_select} shows the selection and the rules applied to every system. All campaigns used $A_{\mathrm{thr}}=0.05$~Å. Irreps are ranked and thresholded on the translation-free supercell-normalized amplitudes, $A_s^{\mathrm{tf}}=A_s\sqrt{1-f_T}$ (eq.~\eqref{eq:fT}). The distinction matters for the ranking: in Ca$_3$Ti$_2$O$_7$ the raw $A_s$ of $\Gamma_5^-$ (1.120~Å) exceeds that of $X_2^+$ (0.847~Å), but 75\% of its squared norm is a rigid translation, and its translation-free value, 0.559~Å, places it below $X_2^+$, so that the walk returns $X_3^-$ and $X_2^+$. In CsSnI$_3$ the secondaries $R_5^+$ and $X_5^+$ pass the threshold (on $A_s$) but are held at zero on the plane by R2, and $X_5^+$ is scanned separately. In the manganites the polar $\Gamma_2^-$ is a secondary that R4 keeps as the polarization axis. In Ca$_3$Ti$_2$O$_7$, R4 would add $\Gamma_5^-$ as a third axis; the policy file instead holds it at zero, so that the plane follows the same two-primary protocol as CsSnI$_3$, and the override is recorded with its reason in the audit log. The trilinear $X_3^-X_2^+\Gamma_5^-$ coupling, and with it the stabilization of the child, therefore lies outside this plane (Section~\ref{sec:res_perov}). For InMnO$_3$, the polar $\Gamma_2^-$ ($A_s^{\mathrm{tf}}=0.099$~Å; 0.101~Å raw) also passes the threshold, but its margin over the largest inactive irrep, $K_1$ (0.023~Å), is 0.076~Å (0.078~Å on raw $A_s$), so R5 applies as well; in the gated rerun the borderline case went to the human, who kept $\Gamma_2^-$ as an axis.

\paragraph{Certify mode vectors.} Four ordered checks act on the vectors that will be used to build structures. First, atoms of the child are matched to atoms of the parent by a species-constrained assignment $\pi$, and the match is accepted only if no atom moves by more than 2~Å,
\begin{equation}\label{eq:match}
\max_j\,\bigl|\rv^{H}_{\pi(j)}-\rv^{G}_j\bigr| < 2~\text{\AA}.
\end{equation}
Second, the rigid translation, the mean displacement over the $N$ atoms of the cell, is removed from every mode vector,
\begin{equation}\label{eq:trans}
\et_{i,j}=\e_{i,j}-\eb_i,\qquad \eb_i=\frac{1}{N}\sum_{j=1}^{N}\e_{i,j},
\end{equation}
and its fraction of the squared norm is recorded,
\begin{equation}\label{eq:fT}
f_T = \frac{N\,|\eb_i|^2}{\|\e_i\|^2}.
\end{equation}
Zone-boundary irreps have $f_T=0$; polar $\Gamma$ irreps do not ($f_T=0.76$ for $\Gamma_4^-$ in BiFeO$_3$, 0.75 for $\Gamma_5^-$ in Ca$_3$Ti$_2$O$_7$, 0.33 for $\Gamma_5^-$ in HfO$_2$ $P4_2/nmc\to Pca2_1$). Third, each mode vector is calibrated so that unit amplitude reproduces the child,
\begin{equation}\label{eq:scale}
s_i = \frac{A_i^{\mathrm{child}}}{A_i^{\mathrm{movie}}},
\end{equation}
where $A_i^{\mathrm{movie}}$ is the amplitude carried by the exported mode vector; for CsSnI$_3$ these vectors carry 0.80 of the child displacement, so $s_i=1.25$ (1.758 against 1.406~Å), and for Ca$_3$Ti$_2$O$_7$ they are $s_i=1.25$ for $X_3^-$ and 2.50 for $X_2^+$, so that the factor differs between irreps and each vector is calibrated separately. The child's projection on each calibrated vector must reproduce its $A_p$ to within 2\%. Equivalently, each vector can be normalized by fixing its norm at $\sqrt{n}$~Å per Å of $A_p$, where $n$ is the number of primitive parent cells in the frame. $n$ need not equal the supercell factor of the decomposition: the 48-atom Ca$_3$Ti$_2$O$_7$ frame contains $n=4$ primitive $I4/mmm$ cells, whereas the decomposition's supercell contains 2, so that $A_s=\sqrt2\,A_p$ there. $G_1$ therefore checks $\lVert\mathbf e_i\rVert=A_p\sqrt{n}$ with $n$ counted from the structure. Fourth, a single-mode frame is built for each active irrep and its space group must equal the isotropy subgroup,
\begin{equation}\label{eq:single}
\SG\bigl(\rv_0+Q\,\eh_i\bigr)=H_i\qquad\forall\, i\in\Aset .
\end{equation}
The single-mode frame is built exactly as the grid builds it, at $Q=A_{p,i}$ on the fixed parent lattice, and the check passes only if spglib returns the space-group number of $H_i$ at a tolerance of $10^{-2}$~Å (default angle tolerance); a match at another tolerance of the ladder counts as a failure (code \texttt{SINGLE\_MODE\_SG}; \SSec{sec:si_primaries}). The order matters: mismatched atoms, a rigid translation and a scale error all leave the space group unchanged, so a symmetry check alone would pass them. Check $c_2$ requires all four checks to pass (\SFig{fig:s_certify}); the symmetry check is made by three independent routes: the isotropy subgroup, the space group of the built frame and the descent tree (\SFig{fig:s_three}).

\paragraph{Size the grid.} The grid dimension counts free order-parameter components, not irreps (D1),
\begin{equation}\label{eq:dim}
n=\sum_{i\in\Aset} d_i^{\,\mathrm{free}} ,
\end{equation}
where $d_i^{\,\mathrm{free}}$ is the number of free parameters in the order-parameter direction of irrep $i$: one for a direction such as $(a,0,0)$ or $(a,a,a)$ with a fixed ratio, two for $K_3$ in the manganites when the phase is sampled ($Q$ and $\Phi$; one on the $\Phi=0$ plane), and three for each of $\Gamma_4^-$ and $R_5^-$ when BiFeO$_3$ is sampled in all directions. With $m_i$ points along axis $i$ over $\pm Q_{\max,i}$ in steps $\Delta Q_i$, the full grid is used when it fits the budget $B$ (D2),
\begin{equation}\label{eq:budget}
m_i = \frac{2Q_{\max,i}}{\Delta Q_i} + 1,\qquad N = \prod_{i=1}^{n} m_i\le B ;
\end{equation}
each axis may have its own range and step (InMnO$_3$ plane: $K_3$ $\pm0.66$~Å in steps of 0.094~Å and $\Gamma_2^-$ $\pm0.235$~Å in steps of 0.034~Å, $N=15\times15$).
Otherwise the grid is reduced: primaries keep axes and coupled pairs share planes (D3, not yet implemented: an over-budget grid fails $G_2$ with \texttt{OVER\_BUDGET} and goes to the human); hard secondaries receive only a few points, a short scan along their own vector at declared points of the primaries, fitted by a quartic that gives the curvature, the linear term induced by the primaries, the relaxed amplitude and the energy released; if that energy exceeds half the smallest energy difference the run interprets, $G_5$ fails (\texttt{SECONDARY\_COUPLED}) and the sub-agent or a human decides whether the secondary stays at zero with the caveat stated (D4); and one DFT point is computed per symmetry orbit of the grid $\mathcal{Q}$ under the parent group, the result being unfolded to every node of the orbit with the parent operations (D5). Each orbit is labelled once by its canonical node, the lexicographically largest of its images, the number of orbits is checked against Burnside's count (the orbit-counting lemma)~\cite{Burnside1911,Neumann1979}, $|\mathcal{Q}/G|=|G|^{-1}\sum_{g\in G}|\mathcal{Q}^{g}|$, where $\mathcal{Q}^{g}$ is the set of grid nodes fixed by $g$, and any cut on the grid, such as a minimum interatomic contact, is applied per orbit, since it is an orbit invariant,
\begin{equation}\label{eq:orbit}
N_{\mathrm{DFT}} = |\mathcal{Q}/G| .
\end{equation}
The descent tree then gives the expected space group of every node,
\begin{equation}\label{eq:tree}
\SG(\Q)=\bigcap_{i:\,Q_i\neq 0} H_i,\qquad \SG(\mathbf{0})=G .
\end{equation}
Check $c_3$ requires $N\le B$ and that every subgroup on the tree between $G$ and $H$ appears on the grid (\SFig{fig:s_dim}). CsSnI$_3$ gives $n=2$ and a full plane (D2) with $m=31$ and $N=961$; for BiFeO$_3$ ($n=6$), D5 reduces 786{,}410 grid nodes at two volumes to 9{,}597 symmetry orbits, computed in 10{,}054 runs with the refinement points and controls. In the rerun configurations, $B=1{,}100$ frames for CsSnI$_3$ and $1{,}400$ for Ca$_3$Ti$_2$O$_7$.

\paragraph{Build and audit.} Each frame is the parent plus a superposition of certified unit modes, on the lattice set by the strain policy,
\begin{equation}\label{eq:build}
\rv(\Q)=\rv_0+\sum_{i\in\Aset} Q_i\,\eh_i,\qquad \mathbf{L}=\mathbf{L}_0 ,
\end{equation}
where $\eh_i$ is translation-free and calibrated, so $Q_i$ is in Å of translation-free $A_p$ and the child lies at $Q_i=A_{p,i}^{\mathrm{tf}}$ on each axis (for InMnO$_3$ $\Gamma_2^-$, $f_T=0.0396$ and $A_p^{\mathrm{tf}}=0.0573$~Å against $A_p=0.0585$~Å). Five strain policies govern the lattice: P1, by default one fixed lattice $\mathbf{L}_0$ for every frame, with strain irreps set to zero; P2, the cell used as the fixed lattice of P1 must be stated explicitly: the parent cell (CsSnI$_3$, Ca$_3$Ti$_2$O$_7$, BiFeO$_3$) or the child cell (manganites); P3, when phase ordering or barriers are asked for, or the strain is large, the lattice is interpolated with a single strain coordinate $s$ (the HfO$_2$ paths, Section~\ref{sec:paths}); P4, cell-relaxed anchors (ISIF~3) are added at the phase nodes to size the strain contribution; and P5, the report states what the fixed lattice biases (well depths, but usually not the shape of the brim or ring). P1--P3 are implemented in the workflow; in the campaigns reported here the relaxed anchors of P4 were computed by script and the bias statement of P5 was written by hand, and neither is yet produced by the workflow itself. \SFig{fig:s_strain} shows the choice for each system. $f_2$ audits every frame with the spglib engine: its space group must equal the expected one, and no two atoms may come closer than $d_{\mathrm{cut}}$,
\begin{align}
\SG_{\mathrm{spglib}}\bigl(\rv(\Q)\bigr)&=\SG(\Q)\qquad\forall\,\Q,\label{eq:audit}\\
d_{\min}(\Q)&\ge d_{\mathrm{cut}} . \label{eq:dmin}
\end{align}
OUT: the audited frames (one POSCAR per node), \texttt{grid\_manifest.json} with the amplitudes $Q_i$ of every node, and the expected and found space group of every frame. $G_2$ passes the grid to the Runner ($f_3$, block~6). For CsSnI$_3$ all 961 plane frames match the tree: 1~$\times$~No.~221, 30~$\times$~No.~127, 30~$\times$~No.~74 and 900~$\times$~No.~62. The collision distance is $d_{\mathrm{cut}}=1.0$~Å by default and for the HfO$_2$ paths, and 1.6~Å for the BiFeO$_3$ grid.

\subsection{Validation of the structure generation by interpolation}\label{sec:interp}
To validate the frames independently (Fig.~\ref{fig:perovskite}c,f), every frame of the CsSnI$_3$ and Ca$_3$Ti$_2$O$_7$ planes was rebuilt from displacement fields between whole structures, without the mode vectors. Two fields were taken, both starting from the parent: one to a structure with only the first irrep ($\Delta\mathbf R_1$; $P4/mbm$ for CsSnI$_3$, $Cmcm$ for Ca$_3$Ti$_2$O$_7$) and one to the child restricted to the two primaries ($\Delta\mathbf R_2$; $Pnma$, and for Ca$_3$Ti$_2$O$_7$ an $A2_1am$ structure at $1/\sqrt2$ of the child amplitudes). Because these fields contain only the two primary irreps, a frame at amplitudes $(x,y)$ is reached by $\mathbf R=\mathbf R_0+\alpha_1\Delta\mathbf R_1+\alpha_2\Delta\mathbf R_2$, with $(\alpha_1,\alpha_2)=(x,y)\,M^{-1}$ and $M=\bigl(\begin{smallmatrix}A_1^{(1)}&0\\A_1^{(2)}&A_2^{(2)}\end{smallmatrix}\bigr)$ built from the amplitudes of the two end structures (CsSnI$_3$: $M=\bigl(\begin{smallmatrix}1.006&0\\0.8431&0.2286\end{smallmatrix}\bigr)$; Ca$_3$Ti$_2$O$_7$: $\bigl(\begin{smallmatrix}0.581&0\\0.581&0.424\end{smallmatrix}\bigr)$). The two constructions were compared on dense grids (CsSnI$_3$: $\pm1.2$~Å in steps of 0.024~Å; Ca$_3$Ti$_2$O$_7$: 10{,}201 nodes over $\pm1.27$~Å in steps of 0.0254~Å) with vectors and $M^{-1}$ stored to nine decimal places. The two constructions agree to the rounding of the stored vectors at every node, with the same space groups (Section~\ref{sec:res_perov}; scaling of the residual with the number of stored digits in \SSec{sec:si_interp}).

\subsection{DFT calculations}\label{sec:dft}
$f_3$ computed all energies with VASP~\cite{Kresse1996a} and PAW potentials~\cite{Blochl1994,Kresse1999}, using PBE~\cite{Perdew1996} (CsSnI$_3$, Ca$_3$Ti$_2$O$_7$, HfO$_2$), PBEsol~\cite{Perdew2008}$+U$~\cite{Dudarev1998} (BiFeO$_3$) and LDA~\cite{Ceperley1980,Perdew1981}$+U$~\cite{Liechtenstein1995,Dudarev1998} (manganites), with the settings of \STab{tab:dft}; the live BaTiO$_3$ run used PBE for the paths and PBEsol for the plane (Section~\ref{sec:livemode}). Landscapes are static calculations on the lattice set by the strain policy (P1 unless stated). Energies are taken only from runs whose last energy change is below EDIFF within NELM steps. Runs that still stop at the step limit after resubmission with robust mixing (\STab{tab:rules}) are kept as SCF-limited energies and excluded from gradient fits. Relaxed anchors (strain policy P4) relax both cell and ions.

\subsection{Landau models}\label{sec:landau}
Energies per formula unit (per 30-atom cell for the manganites) relative to the parent at the same lattice are fitted with polynomials built from the invariants of the active order parameters under the parent group. The degree is fixed by the model form for CsSnI$_3$ and the manganites and chosen by cross-validation and a blind test for BiFeO$_3$, and minima, saddles and barriers are located on the fitted surface and checked against DFT points.

\paragraph{CsSnI$_3$.} The ($M_3^+$, $R_4^+$) plane, with amplitudes $M$ and $R$, is fitted to
\begin{equation}\label{eq:cssni3F}
F=\tfrac{a_1}{2}M^2+\tfrac{b_1}{4}M^4+\tfrac{c_1}{6}M^6+\tfrac{a_2}{2}R^2+\tfrac{b_2}{4}R^4+\tfrac{c_2}{6}R^6+\tfrac{\lambda}{2}M^2R^2+\tfrac{\mu}{2}\bigl(M^4R^2+M^2R^4\bigr)
\end{equation}
by linear least squares on the 545 plane nodes with $\sqrt{M^2+R^2}\le1.05$~Å (RMS 0.10~meV/f.u.), giving $a_1=-0.149$, $a_2=-0.155$~eV\,Å$^{-2}$, $b_1=0.267$, $b_2=0.261$ and $\lambda=0.255$~eV\,Å$^{-4}$ (all coefficients in \STab{tab:cssni3_landau}). The ring is also analysed directly on the DFT surface of the $31\times31$ grid (step 0.08~\AA). Because the grid locates each minimum only to within one step, the DFT ring radius spans 0.72--0.80~\AA\ and the DFT corrugation 1.5--1.6~meV/f.u.\ (1.50, 1.54 and 1.50~meV/f.u.\ with bilinear, bicubic and spherical-spline interpolants at $r=0.74$~\AA); the fit, which is smooth between the nodes, gives a single radius of 0.76~\AA\ and 1.4~meV/f.u. Both place the minimum on the $R_4^+$ axis and give the same $\cos2\theta$ form, and the 0.1--0.2~meV/f.u.\ difference is of the size expected from locating the minima to within one grid step. The Ca$_3$Ti$_2$O$_7$ plane, computed without $\Gamma_5^-$, is analysed directly on the DFT grid and not fitted.

\paragraph{Hexagonal manganites.} The ($Q$, $\Phi$, $P$) energies are fitted to the form of Artyukhin \textit{et al.}~\cite{Artyukhin2014} and Meier \textit{et al.}~\cite{Meier2020},
\begin{equation}
\label{eq:meier}
F=\tfrac{a}{2}Q^2+\tfrac{b}{4}Q^4+\tfrac{Q^6}{6}\bigl(c+c'\cos6\Phi\bigr)-gQ^3P\cos3\Phi+\tfrac{g'}{2}Q^2P^2+\tfrac{a_P}{2}P^2 ,
\end{equation}
by linear least squares, in supercell-normalized amplitudes ($A_s=\sqrt3A_p$) and eV per 30-atom cell, for comparison with Table~I of Meier \textit{et al.} Minimizing over $P$ gives $P^*=gQ^3\cos3\Phi/(a_P+g'Q^2)$ and the hat $F(Q,\Phi)=F(Q,\Phi,P^*)$ of Fig.~\ref{fig:manganite}. Each compound was sampled on two grids on its fixed child lattice (\STab{tab:mn_grids}): a $15\times15$ ($K_3$, $\Gamma_2^-$) plane at $\Phi=0$ (InMnO$_3$: $|K_3|\le0.66$~Å, $|P|\le0.235$~Å, the 225 structures of \SSec{sec:si_mn}; ErMnO$_3$: $|K_3|\le0.755$~Å, $|P|\le0.158$~Å) and a ($Q$, $\Phi$, $P$) grid with five radii (InMnO$_3$ 0.35--0.70~Å, ErMnO$_3$ 0.40--0.80~Å), $\Phi=0$--$30^\circ$ in $5^\circ$ steps (a wedge that covers the whole brim, because the energy obeys $E(\Phi)=E(-\Phi)=E(\Phi+60^\circ)$ with $P\to-P$; \SSec{sec:si_k3}) and five polar amplitudes (InMnO$_3$ $|P|\le0.10$~Å, ErMnO$_3$ $|P|\le0.20$~Å), 175 structures. For ErMnO$_3$, 11 structures on the $\Gamma_2^-$ axis at $Q=0$ ($|P|\le0.5$~Å) fix $a_P=0.583$~eV\,Å$^{-2}$ per cell, the harmonic coefficient of a polynomial in $P^2$, $P^4$ and $P^6$ fitted to all 11 energies relative to the parent, and $a_P$ is held at that value in the joint fit; other windows ($|P|\le0.2$--0.5~Å) and orders give $a_P=0.51$--0.58 but change the hat and brim energies by less than 0.1~meV/f.u.\ (\STab{tab:mn_aP}, \SFig{fig:s_mn_pstar}). The coefficients of Fig.~\ref{fig:manganite} and \STab{tab:mn} come from a joint fit of all seven coefficients, including $c'$ (six for ErMnO$_3$, whose $a_P$ is fixed), to the plane and the ($Q$, $\Phi$, $P$) grid (\STab{tab:mn_fig4}). The plane-fit coefficients of Extended Data Table~\ref{tab:meier} come from the plane alone, on which $\cos6\Phi=1$ and $c'$ cannot be separated from $c$; for these fits the window is $|Q|\le0.58$~Å ($A_p$) and $|P|\le0.075$~Å (translation-free). For validation, the DFT energy at each node of both grids is minimized over $P$ with a quartic in $P$, and the hat depth, brim barrier, $Q^*$ and $P^*$ are compared with the model with $P$ relaxed (\SFig{fig:s_mn_validation}).

\paragraph{BiFeO$_3$.} The six-component landscape is fitted with invariant polynomials in $\mathbf P$ and $\mathbf R$. The invariants are orbit sums of monomials under the 96 operations that act on ($\mathbf P$, $\mathbf R$); there are 2, 9, 27, 68, 152 and 315 invariants up to degrees 2, 4, 6, 8, 10 and 12, respectively (2, 7, 18, 41, 84 and 163 at each degree, as Molien's formula requires, and none of odd degree). The six components are obtained from the single-direction decomposition by splitting $\Gamma_4^-$ into its Cartesian parts, with the rigid translation removed, and $R_5^-$ (Bi at the origin) into its rotations about the three axes; the six vectors sum back to the decomposed fields to $3\times10^{-16}$~Å, and the 96 operations are derived on the folded ten-atom cell as signed permutations of $(\mathbf P,\mathbf R)$. The DFT gradients, projected on the six vectors, are checked against the energies by a Simpson line integral along the grid lines, $E(c)-E(a)=\tfrac h3(g_a+4g_b+g_c)$, which gives a slope of 1.0003 over 15{,}774 triples below 1~eV/f.u. The fit uses energies and all six energy gradients (gradient length scale 0.1~Å), with each orbit counted once, within a window $E<1$~eV/f.u.\ (seven data per point). Degree 12 was chosen using a validation model fitted to the 3{,}906 grid orbits alone: degree 10 leaves a blind error of 3.4~meV/f.u.\ on a set of 361 refinement points not used in the fit (274 off the grid lattice, 87 coinciding with grid nodes), against 1.15~meV/f.u.\ for degree 12 (20\% orbit hold-out: 0.48~meV/f.u.; 253 later validation points: 0.57~meV/f.u.). The final model is fitted to all 4{,}482 points at the $R3c$ volume (V1, $a=3.9407$~Å; V0 is the relaxed cubic volume) in the window, with grid, refinement and validation points together (20\% hold-out 0.51~meV/f.u.; 10-fold cross-validation 0.50~meV/f.u.); the 253 validation points were chosen adaptively from a fit to the earlier data (interpolation gaps, plane checks and path images) and form a second round rather than part of the planned grid. The final model shifts every energy by at most 0.5~meV/f.u.\ relative to the validation model. At fourth order the polynomial contains three mixed invariants, $P^2R^2$, $(\mathbf P\cdot\mathbf R)^2$ and $\sum_iP_i^2R_i^2$, the three independent components of the polar--tilt coupling tensor of effective Hamiltonians~\cite{Kornev2006}. The fit statistics are given in \STab{tab:bfo_fits} and the second- and fourth-order coefficients in \STab{tab:bfo_coef}. Stationary points are found from 60 random starts and classified by the index of the six-dimensional Hessian; switching paths are minimum-energy paths from the string method~\cite{E2002String} with 41 images. As an independent check, a second surface is built from the same data: a degree-8 invariant trend fitted to the energies plus a thin-plate radial-basis interpolant~\cite{Duchon1977} of the residual on the symmetry-unfolded data. This surface passes through every DFT point, validation points included (\STab{tab:bfo_rbf}), and the 17 planes of \SFig{fig:s_bfo_17planes} are drawn from it.

\subsection{Polymorph paths in HfO$_2$}\label{sec:paths}
\paragraph{IRIS paths.} $f_1$ decomposed each transition T1--T5 in its own setting, that is, with respect to the reference phase of that transition (\STab{tab:hfo2_set}), and a straight path was built in amplitude and strain space,
\begin{equation}\label{eq:path}
A_i(\lambda)=\lambda\,A_i^{\mathrm{child}},\qquad s(\lambda)=\lambda,\qquad \lambda=k/9,\ k=0,\dots,9 ,
\end{equation}
where $s$ is a single strain coordinate snapped to the child metric (strain policy P3), so that $\lambda=1$ reproduces the child. The cell matrix is interpolated linearly, $\mathbf L(\lambda)=\mathbf L_0+\lambda(\mathbf L_1-\mathbf L_0)$, with the child cell oriented onto the lattice of the decomposition; for the two monoclinic children this orientation includes a rigid rotation of $4.9^\circ$ (T3) and $5.0^\circ$ (T5), which leaves the end points unchanged and changes the intermediate cells by at most 0.19\% in volume (at $\lambda=\tfrac12$) relative to a rotation-free path. Static calculations with the HfO$_2$ settings of \STab{tab:dft} and ISYM~0 (the $5\times6\times5$ mesh of T4 and T5 follows from their $Pbcn$-parent cells) give the energy profile; the frozen-path barrier is its maximum on the grid of ten points and is an upper bound to the true barrier (a spline through the points raises it by at most 0.15~meV/f.u.). Two consistency checks were applied, the agreement of the same crystal built in the settings of different transitions and the closure of end-point energies across paths (T2 then T4 against T1, T2 then T5 against T3); because the closure gaps equal the sums of the mesh differences, both measure $k$-point convergence at the 0.06~meV/f.u.\ level (values in \SSec{sec:si_hfo2} and \STab{tab:hfo2}). T6 ($P2_1/c\to Pca2_1$), whose end phases are not group--subgroup related, is the concatenation of T5 reversed and T4, so it passes through $Pbcn$.

\paragraph{Expert paths.} The expert paths compared in Extended Data Table~\ref{tab:expert} and Fig.~\ref{fig:hafnia}b were constructed in a single common representation, the tetragonal $P4_2/nmc$ (No.~137) parent. The symmetry-adapted distortions of the $P2_1/c$, $Pca2_1$ and $Pbcn$ phases were first expressed in this common setting and compared through the overlaps of their normalized displacement vectors, which define a seven-dimensional reduced order-parameter basis: common $\Gamma_1^+$ and $M_3$ coordinates, three phase-specific $M_1$ coordinates ($M_1^{(14)}$, $M_1^{(29)}$ and $M_1^{(60)}$, for $P2_1/c$, $Pca2_1$ and $Pbcn$), and a $\Gamma_5^+$ coordinate for $P2_1/c$ and a $\Gamma_5^-$ coordinate for $Pca2_1$. For each transition the mode coordinates of the two end points were interpolated linearly, and the intermediate atomic positions were rebuilt from the common parent-mode basis. The lattice was treated separately: from the two end cells $\mathbf L_A$ and $\mathbf L_B$ of the path, the six lattice parameters $(a,b,c,\alpha,\beta,\gamma)$ of an intermediate structure were interpolated with inverse-square weights,
\begin{equation}\label{eq:expertlat}
\mathbf L=\frac{d_A^{-2}\,\mathbf L_A+d_B^{-2}\,\mathbf L_B}{d_A^{-2}+d_B^{-2}},
\end{equation}
where $d_A$ and $d_B$ are the structural distances of the intermediate configuration to the two end points. Because both end-point distortions are embedded in the same parent setting, a generic intermediate keeps only the symmetry operations common to the two embeddings. For T1--T3 and T5 the path therefore keeps the symmetry of its lower end phase, whereas for T4 and T6 only the operations of $P2_1$ are common to the two embeddings, and the intermediates have $P2_1$ symmetry (Extended Data Table~\ref{tab:expert}). The two constructions therefore differ in three ways: the reference in which the modes are defined (each transition's own reference phase for IRIS, one common parent for the expert paths), the route between phases that are not group--subgroup related (through $Pbcn$ for IRIS, directly for the expert paths), and the lattice path (one strain coordinate linear in $\lambda$ for IRIS, distance-weighted interpolation of the end cells for the expert paths). Each expert path has seven intermediate images, at path fractions 0.033, 0.2, $1/3$, $1/2$, $2/3$, 0.8 and 0.967 between the two end structures (nine points per path, against ten evenly spaced points in $\lambda$ for IRIS); at these images the weight of the second end cell in eq.~\eqref{eq:expertlat} is 0.001, 0.06, 0.2, 0.5, 0.8, 0.94 and 0.999, so the cell follows the atoms with a lag. All expert-path energies are static calculations with the same electronic-structure settings as the IRIS paths (\STab{tab:dft}).

\subsection{Evaluation of the agentic layer}\label{sec:eval}
All evaluations run in replay mode on the stored DFT energies, so that no calculation is repeated and every outcome can be compared with a reference. Faults are injected only through the run configuration (never by altering data or code) and recorded by $f_0$; the fault list (F001--F053) and the benchmark cases are fixed in an evaluation set that was frozen, with a hash, before any language model saw a case.

\paragraph{Benchmark cases.} Each case is the exact prompt that a sub-agent receives at a real gate failure, captured from a replay together with the option menu of that failure. The 53 cases cover all six systems (HfO$_2$ 14, CsSnI$_3$ 12, ErMnO$_3$ 8, InMnO$_3$ 7, Ca$_3$Ti$_2$O$_7$ 7, BiFeO$_3$ 5) and five classes: failures that a stored rule would fix, with the rule table switched off (29); failures for which a rule was tried and the failure returned (6); failure codes without a rule (5); decoys whose correct answer is to stop (11); and judgment calls (2: the borderline InMnO$_3$ selection of rule R5 and the CsSnI$_3$ $X_5^+$ secondary of policy D4). The BiFeO$_3$ cases follow the failure episodes of the real campaign (walltime kills resubmitted unchanged, SCF failures rerun with robust mixing, SCF-limited near-metallic frames). Sixteen cases (30\%) are held out: they were added after the menus and prompts were final and are balanced over the systems, rather than drawn at random. Every answer is scored as correct (the reference fix), safe (stop or another escalation to the human, which damages nothing but costs an intervention) or harmful (a fix that changes the result wrongly); invalid outputs are counted separately. A policy that always stops scores 22 of 53 correct and none harmful. The reference labels were set by the authors before the set was frozen: for a failure that a stored rule fixes, the fix of that rule; for a decoy, stop; for the two judgment calls, the decision taken in the published campaign, with stop counted as safe and the opposite decision as harmful.

\paragraph{Models and protocol.} Nine models from three providers, three tiers each (Anthropic, OpenAI and Google; identifiers taken from each provider's API model listing on 17 October 2026 and pinned in the evaluation set, listed in \STab{tab:si_models}, with the version returned by the API logged with every call; reference model Claude Sonnet~5.5~\cite{ModelSonnet55}, and Claude Opus~5.5~\cite{ModelOpus55}, GPT-6.1 Sol~\cite{ModelGPT61Sol} and Gemini~4 Argon~\cite{ModelGemini4Argon} among the others), answer every case in two prompt variants, with (guided) and without (bare) the one-line guidance that states the fix knowledge of the rule table, ten times each (9{,}540 calls, all on 17 October 2026). Each model was first checked with a dummy prompt, before any project data was sent. None of the providers publishes a dated snapshot for these identifiers, so the dates of use are part of their identification. Answers are constrained to the menu identifiers by the providers' structured-output interfaces; no temperature is sent and reasoning or effort settings are left at each provider's default (for the Anthropic models, effort \texttt{high} for Sonnet~5.5 and \texttt{medium} for Opus~5.5, the defaults of each model at the time; extended thinking is on for both), because one provider's current models reject the temperature parameter, so that variation is sampled by the repeats. The option order is shuffled for every case and repeat, identically for all models and variants, so that the comparison is paired and the known sensitivity of language models to option order~\cite{Zheng2024MCQ} does not bias it. Rates are reported with 95\% confidence intervals from a bootstrap~\cite{Efron1979} over cases in which the repeats of a case are resampled together; models are compared with exact McNemar tests~\cite{McNemar1947}, against the reference model and pairwise, with Holm correction~\cite{Holm1979} within each variant and split, and with bootstrap intervals of the differences in correct and harmful rates. The McNemar tests pair individual calls and so treat the ten repeats of a case as independent, which overstates significance; statements about differences between models rest on the bootstrap intervals over cases. We also report the consistency of each model across repeats, the change in its correct rate when the correct option is listed first or last (position bias), and its tokens, cost and latency. Three trivial policies serve as reference points: always stop, the first listed option and a uniformly random option.

\paragraph{Workflow-level experiments.} End-to-end replays of all thirteen system configurations measure completion and the human decisions needed without faults. Two checks behave differently in replay: $G_3$ stops a frame that has no stored structure to match, and $f_6$ reads the energies of its second route from the stored campaign files and compares some quantities directly with them. A catch by either check is therefore reported as replay-only when it would not occur in production. The closure and cross-setting checks of $G_5$ for the HfO$_2$ paths use the energies of the other paths, which is valid in production when the paths are computed in the same campaign, as here. Each fault is classed as an error, which corrupts a result if it passes undetected, or a limit fault, which only makes a threshold stricter on correct data, and the outcomes are reported by class. The fault set is run with all gates (gated), with no gates (ungated) and with one gate removed at a time; for each fault we record the gate that catches it, how it is resolved (rule, sub-agent or human) and, from the submission records of $f_3$, how many DFT runs would have been submitted before it was detected. Human interventions per run are compared with the model off (rules and human only) and with the reference model as sub-agent. For the full recovery stack (Fig.~\ref{fig:agentic}e), each choice of the recovery policy (the stored rule where one applies, otherwise the reference model; ten repeats) was executed in replay on the 53 recovery cases of the benchmark (built from the faulted runs), with the gates on and with them off. Kill-and-resume tests interrupt runs after $f_2$, $f_3$ and $f_5$ and resume them in a new process.

\paragraph{Deterministic baselines.} Three policies without a language model were written after the aggregate model results were known, and frozen, with a hash, before they were scored, using only the prompts and the frozen cases: the rules alone (the stored rule if one applies, otherwise stop); a decision tree that picks an option from the gate, error code, detail fields and the guidance line of the prompt (unknown cases: stop); and retrieval, which returns the reference option of the most similar other case (same gate and code, then the closest detail text by TF--IDF cosine~\cite{Salton1988}; leave-one-out, and a variant that remembers only the seed cases). They were scored on the 53 cases with the same scorer, and registered as sub-agents and run through the 53 faulted runs. Hand-offs count every question to the human and every run that a sub-agent stops. A fault scripts a limited number of job failures; a run that completed only because the scripted failures had run out after a resubmission is counted as not completed for every policy.

\paragraph{Executing every option.} For each of the 53 cases the checkpoint at which the gate failed was loaded, and one branch was forked per menu option (208 branches), the option being injected as the answer to a single human question at that point, so that no rule or sub-agent could act first. Each branch ran to the end with all gates and rules on and the same retry budget; later human questions were answered with the published decision or with stop. Final numbers were compared with the fault-free replay at the precision of every reported quantity (716 quantities), fixed before the branches were run, and each branch was classed as completed correctly, completed after a change of SCF mixing only (whose identical energies are assumed by replay), refused by the next gate and then repaired, stopped by a later gate, stopped by choice, or completed wrongly. The 124 branches whose option is neither stop nor the reference were run again with the gates switched off after the fork. Branches whose resubmitted runs take their stored energies are flagged, and the labels of the benchmark were left unchanged.

\paragraph{Intake.} Fifty requests were frozen with a hash before any model saw them: each of the 13 run configurations asked for in three styles (expert, naming the transition, analysis, cell and profile; terse; non-specialist), 5 underspecified requests whose correct answer is a question (or, for U01, the only stored analysis), 3 out-of-scope requests (no group--subgroup route, an unsupported engine, molecular dynamics) and 3 requests for materials with no stored calculations (PbTiO$_3$, YInO$_3$, Ca$_3$Mn$_2$O$_7$), whose correct answer is a valid configuration flagged as needing new calculations without the settings of another material. For each request the reference is the stored configuration, with defensible alternatives listed in advance; a field the request does not state is correct if it is left to the default. Four models (Claude Sonnet~5.5, the reference~\cite{ModelSonnet55}; Claude Opus~5.5~\cite{ModelOpus55}; GPT-6.1 Sol~\cite{ModelGPT61Sol}; Gemini~4 Argon~\cite{ModelGemini4Argon}) answered each request five times (primary run, 18--19 October 2026; profile-description rerun, 19 October 2026) through the structured-output interface, and two deterministic baselines answered once each: a keyword lookup, built from the tables of system names, formulas, space groups and transitions only, and a rule-based parser (\texttt{intake/parser.py}) with a vocabulary for every field of the schema (element-name readings of the formulas, Hermann--Mauguin symbols and crystal-system words resolved on the transition table, phrases for the analyses, strain policies, functionals and magnetic orders of the profiles, budgets and outputs, out-of-scope terms, and chemical formulas not in the table), written from the schema, the tables and the system prompt without consulting the requests and frozen with a hash (\texttt{intake/PARSER.sha256}) before scoring; a second version with one declared fix, made after the first scoring, is reported alongside (\SSec{sec:si_parser}). Every configuration with all critical fields right was run end to end in replay. A rerun in which only the descriptions of the system profiles in the prompt were changed, to name the functional, the $+U$ scheme and the magnetic order, was decided after the first results and is reported as a sensitivity check; the first run is the primary result.

\subsection{Audit and reproducibility}\label{sec:audit}
Every structure, input file and decomposition is stored with a hash; every attempt at every node is logged with its input hashes, gate result, error code, fix, who chose it (rule, sub-agent or human) and retry count; and every energy is linked to the frame and amplitudes that produced it through \texttt{point.json} and \texttt{grid\_manifest.json}. All reported numbers are re-derived from these files by $f_6$ before they are accepted. An example of the resulting audit trail, for two CsSnI$_3$ replays, is given in \SSec{sec:si_audit} and \STab{tab:si_audit}.

\subsection{Live mode}\label{sec:livemode}
Live mode (\texttt{run.mode: live}) runs the same graph, gates and recovery as replay, with three nodes connected to external services instead of stored files; all other nodes are unchanged.

\paragraph{Intake and structures.} The request is typed into a chat; $f_0$ passes it to the intake model, which returns the configuration fields described in Section~\ref{sec:eval}, and $G_0$ checks them as in replay. The parent and child structures are then fetched from the Materials Project~\cite{Jain2013} through its API: for each requested space group, the entry with the lowest energy above the convex hull whose structure has that space group is taken. Because Materials Project structures are stored to a finite precision, each candidate is symmetrized at a tolerance of 0.1~\AA\ and the result is verified to have the requested space group at $10^{-3}$~\AA\ before it is accepted; entries whose symmetry does not survive this test are skipped. The accepted structures, their identifiers and energies above the hull are written to a provenance file that the audit hashes.

\paragraph{Decomposition.} $f_1$ calls FINDSYM and ISODISTORT on the ISOTROPY server with the same sequence of steps as the stored decompositions (\SSec{sec:si_stages}); the raw server responses are saved with the run, so the decomposition can be re-derived offline, and $G_1$ is unchanged.

\paragraph{Calculations.} $f_3$ writes, for every frame, POSCAR, INCAR, KPOINTS and POTCAR. The settings used for the BaTiO$_3$ paths are PBE (the plane was run with PBEsol from an older configuration template, \SSec{sec:si_live}), a plane-wave cut-off of 600~eV, an electronic convergence of $10^{-7}$~eV, Gaussian smearing of 0.05~eV, no spin polarization, at most 200 electronic steps and no ionic relaxation (static calculations); the $\Gamma$-centred $k$-mesh is set from the parent cell by $k_i=\lceil |\mathbf{b}_i|/0.20\,\text{\AA}^{-1}\rceil$ and kept identical for every frame of a path, and the PAW potentials are Ba\_sv, Ti and O. A live executor, a separate process, groups the frames into parts of five static calculations, writes one PBS script per part with the cluster's module and MPI lines (one node of 32 cores, 1~h wall-time limit), submits them with \texttt{qsub}, polls the queue every 30~s and, when every part has ended, collects the energies (\texttt{OSZICAR}, \texttt{OUTCAR}) into the campaign's energy table at full precision and resumes the graph with a job-end event; $G_3$ and $G_4$ then run as in replay. Two recovery rules exist for live faults, both seeded from the campaigns of this work: an electronic step limit (code \texttt{NELM}) resubmits the whole path with robust mixing (AMIX 0.2, BMIX $10^{-4}$, NELM 300) so that every frame of a path keeps one INCAR, as $G_4$ requires, and a wall-time kill (\texttt{WALLTIME}) resubmits the unfinished frames of the affected parts. Neither fired in the BaTiO$_3$ campaign. The calculations ran on the Carbon cluster of the Center for Nanoscale Materials, Argonne National Laboratory.

\paragraph{Two-component plane.} For the $(P_z,P_x)$ plane, $f_2$ takes the first component $\mathbf{e}_1$ as the child's $\Gamma_4^-$ displacement field per \AA\ of amplitude, as for a path, and builds the second, $\mathbf{e}_2$, as the image of $\mathbf{e}_1$ under the parent's four-fold rotation about $y$, applied about an atomic site so that the rotation maps the frame onto itself whatever origin ISODISTORT chose; $G_2$ checks that every atom maps onto an equal atom and that $|\mathbf{e}_2|=|\mathbf{e}_1|$. The grid is $(2m-1)\times(2m-1)$ points over $[-q_\mathrm{max},q_\mathrm{max}]^2$ with $q_\mathrm{max}=1.2$ times the child's amplitude ($m=11$ for the reported run), at the fixed parent cell. The plane has $4mm$ symmetry ($Q_1\to-Q_1$, $Q_2\to-Q_2$, $Q_1\leftrightarrow Q_2$), so only one representative of each orbit is computed and carries its orbit multiplicity; $G_2$ checks the number of orbits both by Burnside's lemma and by brute-force enumeration, and that the multiplicities sum to the grid size. $f_5$ fits the $4mm$-invariant expansion $E=a(Q_1^2+Q_2^2)+b(Q_1^4+Q_2^4)+cQ_1^2Q_2^2+d(Q_1^6+Q_2^6)+e(Q_1^4Q_2^2+Q_1^2Q_2^4)$ by least squares weighted by multiplicity, with $E(0,0)=0$, and locates the minimum numerically in the symmetry-unique wedge; $G_5$ checks coverage, the fit residuals and that the minimum lies inside the sampled range. $f_6$ re-derives the amplitude, the second component, the orbit counts, every frame's space group and shortest interatomic distance, every energy from the collected files, the fit and the minimum.

\paragraph{Chat interface.} A browser page served by the run shows the chat on the left and, on the right, the workflow board with each node's state, gate result and wall time, the run plan, the PBS job table and, when available, the energy profile or the plane as a heat map with the fitted contours. The language model is used only in $f_0$, to write the configuration from the request, and as the gate sub-agents of the recovery node; it does not route the workflow, submit calculations or accept results. The only human input in the reported run was the confirmation to submit, which is not a decision of the workflow.

\paragraph{Recording.} Supplementary Video~1 was made by capturing the screen that showed the chat page and the cluster terminal every 2~s (\texttt{screencapture}, macOS) and assembling the frames with ffmpeg at 10 frames per second, that is, 20 times real time; no frame was removed or reordered and nothing else was edited. At the end of the campaign the SHA-256 of every file in the run directories (inputs, outputs, audit logs, checkpoints, chat transcripts) was recorded; the list is deposited with the data.

\section*{Data availability}
The frozen evaluation set (53 recovery cases, 53 faults, 50 intake requests, with their SHA-256 freeze files), the raw model answers and the results behind every evaluation table and figure are in the IRIS repository (\url{https://github.com/sukritimanna/IRIS}, directories \texttt{eval/}, \texttt{intake/} and \texttt{results/}; \SSec{sec:si_repo}). All input structures, generated frames, DFT energies, fitted coefficients and audit logs of the six stored campaigns, together with the complete run directories of the live BaTiO$_3$ campaign (Materials Project provenance, ISODISTORT responses, VASP inputs and outputs, PBS scripts, audit logs, checkpoints, chat transcripts and the SHA-256 list of every file) and the screen recording (Supplementary Video~1), will be deposited in a public Zenodo archive with a DOI on publication.

\section*{Code availability}
IRIS, including the gate definitions, recovery rules, sub-agent prompts, the deterministic deciders, the option replay, the intake with its keyword lookup and rule-based parser, the leave-one-system-out analysis, the test suite and the scripts of the live BaTiO$_3$ campaign, is at \url{https://github.com/sukritimanna/IRIS} under the MIT licence (\SSec{sec:si_repo}); the version used for this work will be archived on Zenodo under a DOI on publication.

\bibliographystyle{unsrtnat}
\bibliography{refs}

\section*{Acknowledgements}
This work was supported by the U.S. Department of Energy (DOE), Office of Science, Basic Energy Sciences, under the AI-Pathfinder project - Materials Discovery Cloud. Work performed at the Center for Nanoscale Materials, a U.S. DOE Office of Science User Facility, was supported by the U.S. DOE, Office of Science, Office of Basic Energy Sciences, under Contract No.~DE-AC02-06CH11357. A portion of this research was conducted at the Center for Nanophase Materials Sciences, a DOE Office of Science User Facility operated by Oak Ridge National Laboratory under Contract No.~DE-AC05-00OR22725. This research used resources of the National Energy Research Scientific Computing Center (NERSC), a DOE Office of Science User Facility supported by the Office of Science of the U.S. DOE under Contract No.~DE-AC02-05CH11231, using NERSC awards. We also acknowledge the computing facilities of the Laboratory Computing Resource Center (LCRC) at Argonne National Laboratory. Language models (Claude, Anthropic) assisted with writing and debugging code and with editing the text; the authors checked all content and take full responsibility for it.

\section*{Author contributions}
S.M., P.G., P.D. and S.K.R.S.S. conceived the project. S.M. and P.G. developed the initial version of the workflow. S.M. designed and implemented IRIS: the gated state graph and its seven agents ($f_0$--$f_6$), the deterministic gates ($G_0$--$G_6$), the mode-selection and dimensionality policies, mode-vector certification, the recovery stack (stored rules, menu-limited sub-agents and human hand-off) and the chat interface. S.M. performed the DFT campaigns for CsSnI$_3$, Ca$_3$Ti$_2$O$_7$, InMnO$_3$, ErMnO$_3$, BiFeO$_3$ and HfO$_2$, the Landau fits and the comparison with published models, and designed and ran the evaluation (end-to-end replays, fault injection and ablation, the recovery benchmark across nine language models, and the intake evaluation). S.A. and P.D. acted as human experts, answering the decisions IRIS handed over and reviewing the results. R.R. contributed to the discussion of the ferroelectric and multiferroic results. S.M. and P.D. analysed the results. S.M. prepared the figures and the Supplementary Information and wrote the manuscript with input from P.D. and S.K.R.S.S. All authors reviewed and approved the final manuscript.

\section*{Competing interests}
The authors declare no competing interests.

\section*{Extended Data}\label{sec:ED}
\addcontentsline{toc}{section}{Extended Data}
\setcounter{figure}{0}\setcounter{table}{0}
\renewcommand{\figurename}{Extended Data Fig.}\renewcommand{\tablename}{Extended Data Table}
\renewcommand{\theHfigure}{ED\arabic{figure}}\renewcommand{\theHtable}{ED\arabic{table}}

\begin{table}[!htbp]
\centering\footnotesize
\caption{Blocks, owner agents, products and gates.}
\label{tab:blocks}
\begin{tabularx}{\linewidth}{@{}>{\raggedright}p{3.0cm}>{\raggedright}p{3.0cm}>{\raggedright}X>{\raggedright\arraybackslash}X@{}}
\toprule
Block & Agent & Produces & Gate at hand-off\\
\midrule
0 Intake & $f_0$ Intake (LLM) & config, intent & $G_0$ config valid\\
1--3 Input, preprocess, decompose & $f_1$ Crystallographer & $G\supset H$, basis, decomposition & $G_1$ decomposition valid\\
4--5 Postprocess, build grid & $f_2$ Path Planner & active modes, certified vectors, audited frames & $G_2$ grid valid\\
6 Compute & $f_3$ Runner, $f_4$ Reviewer & DFT or machine-learned interatomic potential (MLIP) energies & $G_3$ jobs done; $G_4$ converged or flagged\\
7 Analysis & $f_5$ Analyst, $f_6$ Skeptic & Landau model, barriers, report & $G_5$ fit acceptable; $G_6$ claims verified\\
\bottomrule
\end{tabularx}
\end{table}

\begin{landscape}
\begin{figure}[p]
  \centering
  \includegraphics[width=\linewidth,height=0.74\textheight,keepaspectratio]{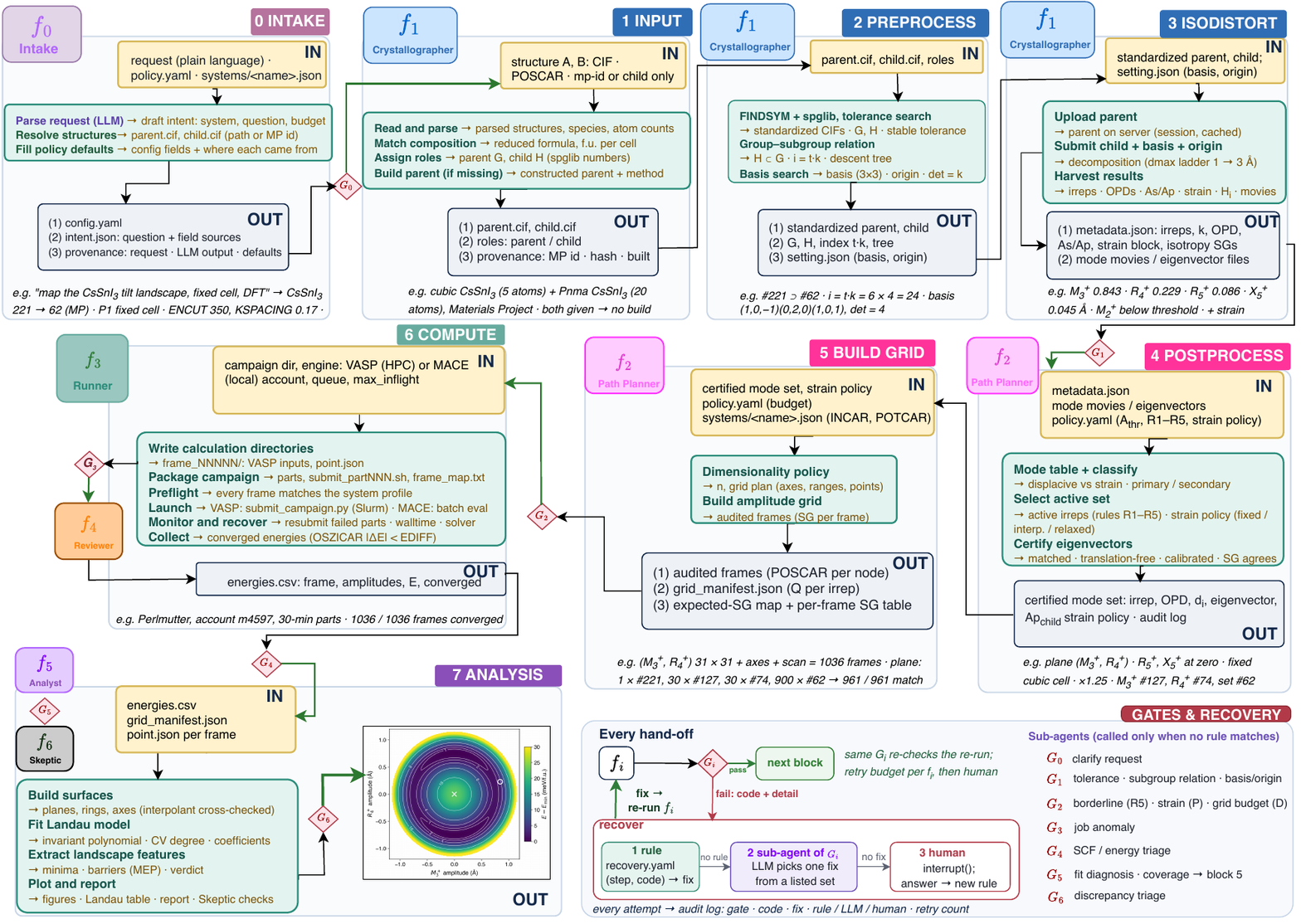}
  \caption{\textbf{Detailed IRIS workflow, shown for CsSnI$_3$.}
The eight blocks of the state graph of Fig.~\ref{fig:pipeline}, each with its owner agent (chip), inputs (IN), steps (bold, step $\to$ product), outputs (OUT) and the CsSnI$_3$ example (italic); gates $G_0$--$G_6$ (red diamonds) sit on the hand-offs.
The blocks are 0 Intake ($f_0$); 1 Input and 2 Preprocess ($f_1$: structures, space groups, group--subgroup relation and supercell basis); 3 Decompose ($f_1$; labelled with the name of its decomposition engine, Methods block~3); 4 Postprocess and 5 Build grid ($f_2$: mode selection, certification, dimensionality and frame audit); 6 Compute ($f_3$ Runner, $f_4$ Reviewer); and 7 Analysis ($f_5$ Analyst, $f_6$ Skeptic). Bottom right: gates and recovery, with the sub-agents of each gate (called only when no stored rule matches), the retry budget per node and the audit log of every attempt; inset in block~7, the CsSnI$_3$ landscape of Fig.~\ref{fig:perovskite}b.
For CsSnI$_3$, the ($M_3^+$, $R_4^+$) plane on the fixed cubic lattice, with axes and scans, gives 1{,}036 frames; all 961 plane frames have the expected space group and all 1{,}036 converged.
Every block is described step by step in Methods (Section~\ref{sec:agent}) and, with the complete CsSnI$_3$ example, in \SSec{sec:si_leg_ed1}.}
  \label{fig:workflow}
\end{figure}
\end{landscape}
\begin{figure}[!tp]
  \centering
  \includegraphics[width=\textwidth,height=0.70\textheight,keepaspectratio]{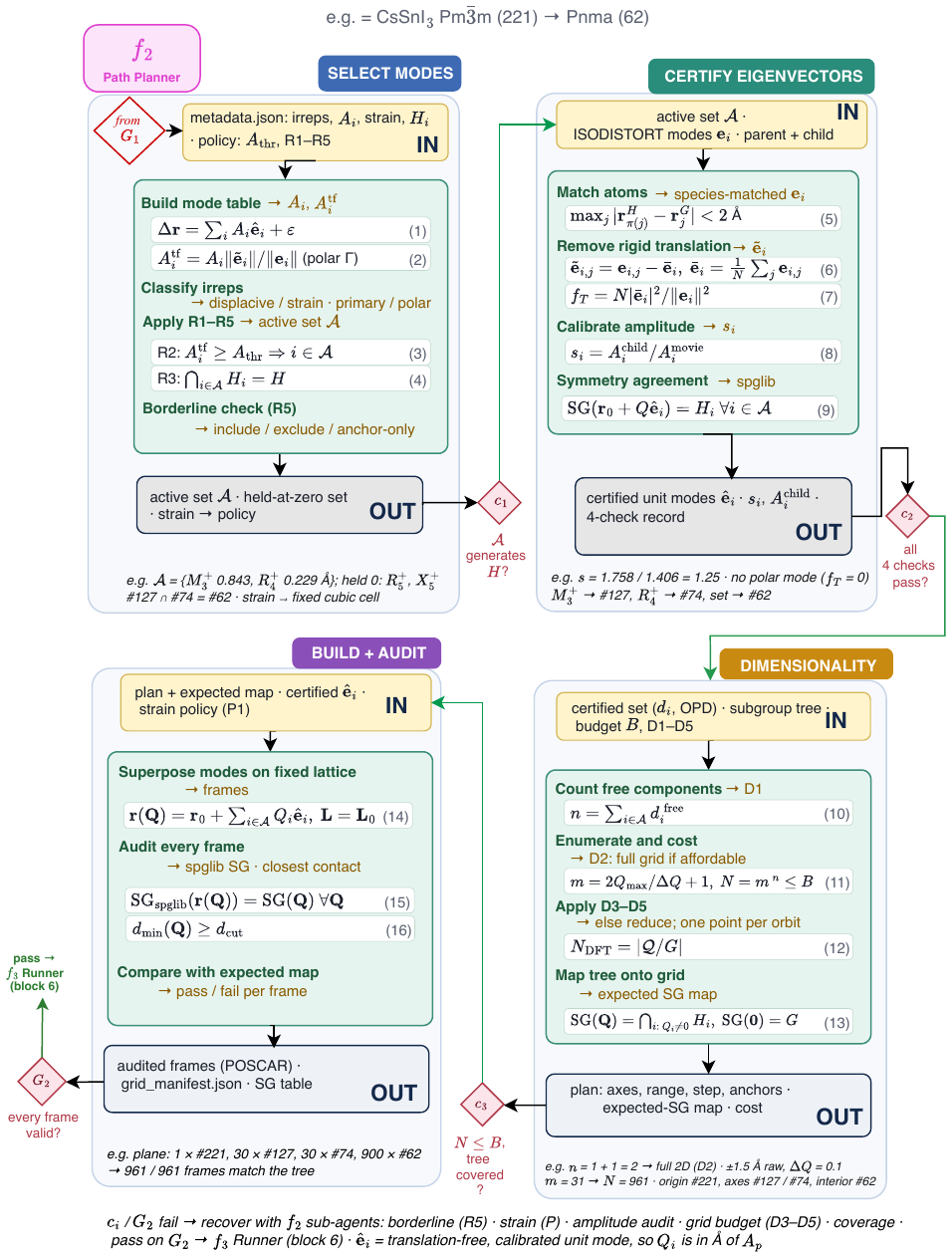}
  \caption{\textbf{Inside the Path Planner ($f_2$): from decomposition to an audited grid, shown for CsSnI$_3$.}
The Planner receives the decomposition after $G_1$ and runs four stages, read clockwise from the top left, each with inputs (IN), steps with their equations (numbered as in Methods, Section~\ref{sec:planner}) and outputs (OUT): Select modes (eqs~\eqref{eq:decomp}--\eqref{eq:gen}, rules R1--R5, check $c_1$); Certify eigenvectors (Methods: Certify mode vectors; eqs~\eqref{eq:match}--\eqref{eq:single}, check $c_2$); Dimensionality (Methods: Size the grid; eqs~\eqref{eq:dim}--\eqref{eq:tree}, policies D1--D5, of which D3 is not yet implemented; check $c_3$); and Build and audit (eqs~\eqref{eq:build}--\eqref{eq:dmin}, gate $G_2$).
A failed check goes to the recovery node; of the sub-agent boxes in the figure (borderline, strain, amplitude audit, grid budget, coverage), strain and grid-budget failures go to the human as implemented and coverage is checked at $G_5$ (\STab{tab:subagents}). Coordinates $Q_i$ are in \AA\ of translation-free $A_p$.
For CsSnI$_3$, the active set is $\{M_3^+, R_4^+\}$ and the $31\times31$ plane gives $N=961$ frames, all matching the descent tree.
The stage-by-stage description and the complete CsSnI$_3$ values are given in \SSec{sec:si_leg_ed2}.}
  \label{fig:planner}
\end{figure}

\clearpage

\begin{table}[!htbp]
\centering\footnotesize
\caption{Systems, active modes, grids and key results.}
\label{tab:systems}
\begin{tabularx}{\linewidth}{@{}>{\raggedright}p{1.8cm}>{\raggedright}p{2.5cm}>{\raggedright}X>{\raggedright}p{2.2cm}>{\raggedright\arraybackslash}X@{}}
\toprule
System & Transition & Active modes (\AA) & Grid & Key result\\
\midrule
CsSnI$_3$ & $Pm\bar{3}m\to Pnma$ & $M_3^+$ 0.843, $R_4^+$ 0.229 & $31\times31$, 1{,}036 frames & ring at $r=0.76$~\AA\ (fit; 0.72--0.80~\AA\ on the DFT grid), depth $\approx-22$~meV/f.u., corrugation 1.4~meV/f.u.\ (fit; 1.5--1.6 on the grid)\\
Ca$_3$Ti$_2$O$_7$ & $I4/mmm\to A2_1am$ & $X_3^-$ 0.822, $X_2^+$ 0.599 ($\Gamma_5^-$ 0.395 tf, held at zero) & $37\times37$, 1{,}372 frames & child 51~meV/f.u.\ above the plane minimum ($\Gamma_5^-=0$)\\
InMnO$_3$ & $P6_3/mmc\to P6_3cm$ & $K_3$ 0.528, $\Gamma_2^-$ 0.057 (tf) & $15\times15$ plane + $5\times7\times5$ ($Q$, $\Phi$, $P$); 400 frames & hat 43~meV/f.u., brim barrier 3.3~meV/f.u.; $\tilde b=0.70$--0.72 vs 0.73 (Meier); $g$ within 4\% (plane fits)\\
ErMnO$_3$ & $P6_3/mmc\to P6_3cm$ & $K_3$ 0.604, $\Gamma_2^-$ 0.126 (tf) & as InMnO$_3$ + 11 on the $\Gamma_2^-$ axis; 411 frames & hat 170~meV/f.u., brim barrier 34.3~meV/f.u.; single-well polar axis (improper); $\tilde b=0.91$ vs 0.91 (Meier), $a$ within 3\%\\
BiFeO$_3$ & $Pm\bar{3}m\to R3c$ & $\Gamma_4^-$ 0.526 (tf), $R_5^-$ 0.694 & 6D, 9{,}597 orbits (10{,}054 runs) & $R3c$ $-950.0$~meV/f.u.\ at (0.519, 0.698)~\AA; switching via $Ima2$, 101.3~meV/f.u.\\
HfO$_2$ & four polymorphs, T1--T6 & $M_1$, $M_3$, $\Gamma_5^\pm$; $\Gamma_2^-$, $\Gamma_3^+$; $\Gamma_1^+$ & five 10-point paths & barriers 6.43, 24.68, 43.95, 80.50 and 52.37~meV/f.u.\ (T1--T5), 151.5~meV/f.u.\ (T6)\\
\bottomrule
\end{tabularx}
\end{table}

\begin{table}[!htbp]
\centering\footnotesize
\caption{Landau coefficients of InMnO$_3$ and ErMnO$_3$ from IRIS and from Meier \textit{et al.}~\cite{Meier2020}, eq.~\eqref{eq:meier}, in eV\,Å$^{-n}$ ($n$, the total power of the amplitudes in the term) per 30-atom cell with amplitudes $A_s$. Plane fit: the nodes of the $15\times15$ ($K_3$, $\Gamma_2^-$) plane at $\Phi=0$ within $|Q|\le0.58$~Å ($A_p$) and $|P|\le0.075$~Å (Methods, Section~\ref{sec:landau}), on which $c'$ is not defined, in three magnetic and $+U$ settings. Joint fit: the plane and the ($Q$, $\Phi$, $P$) grid (A-type antiferromagnetic (AFM) order, Dudarev scheme for both compounds; for ErMnO$_3$ $a_P$ fixed from the $Q=0$ axis), the coefficients behind Fig.~\ref{fig:manganite} and \STab{tab:mn}. Meier \textit{et al.}: Table~I of ref.~\cite{Meier2020}, where the InMnO$_3$ values are taken from ref.~\cite{Smabraten2018}. RMS: plane fit, fitted energies (meV per cell); joint fit, all nodes with $P$ relaxed (meV/f.u.). Derived quantities in \STab{tab:mn_red} and \STab{tab:mn_fig4}.}
\label{tab:meier}
\setlength{\tabcolsep}{3.6pt}
\begin{tabular}{@{}lccccc@{\hspace{10pt}}cc@{}}
\toprule
 & \multicolumn{5}{c}{InMnO$_3$} & \multicolumn{2}{c}{ErMnO$_3$}\\
\cmidrule(lr){2-6}\cmidrule(l){7-8}
 & \multicolumn{3}{c}{IRIS, plane fit} & IRIS & Meier & IRIS & Meier\\
\cmidrule(lr){2-4}
 & \shortstack{A-AFM\\Dudarev} & \shortstack{A-AFM\\Liechtenstein} & \shortstack{$\uparrow\uparrow\downarrow$\\Liechtenstein} & \shortstack{joint\\Dudarev} & \textit{et al.} & joint & \textit{et al.}\\
\midrule
$a$     & $-1.211$ & $-1.221$ & $-1.357$ & $-1.206$ & $-0.82$ & $-3.708$ & $-3.83$\\
$b$     & $1.216$  & $1.253$  & $1.341$  & $1.206$  & $1.13$  & $3.973$  & $6.20$\\
$c$     & $0.764$  & $0.710$  & $0.728$  & $0.700$  & $0.81$  & $0.427$  & $1.06$\\
$c'$    & --       & --       & --       & $0.068$  & $0.04$  & $0.040$  & $0.06$\\
$a_P$   & $2.711$  & $2.687$  & $2.444$  & $2.663$  & $3.48$  & $0.583$  & $0.54$\\
$g$     & $0.986$  & $0.988$  & $1.011$  & $0.962$  & $1.02$  & $2.266$  & $3.76$\\
$g'$    & $5.375$  & $5.339$  & $5.545$  & $5.310$  & $4.85$  & $10.330$ & $17.43$\\
RMS     & 3.9 & 3.9 & 3.9 & 3.5 & -- & 6.5 & --\\
\bottomrule
\end{tabular}
\end{table}

\begin{table}[!htbp]
\centering\footnotesize
\caption{\textbf{Gated end-to-end reruns on the stored DFT energies.} Frames: built frames matched to stored structures. Re-derived: numbers of the final report re-derived by $f_6$. Human: decisions taken by a human during the run; all three are the two published decisions, R5 (InMnO$_3$, met in two reruns) and D4 (CsSnI$_3$). The HfO$_2$ row combines the five path configurations T1--T5, giving the 13 run configurations of Section~\ref{sec:autonomy}.}
\label{tab:reruns}
\begin{tabularx}{\linewidth}{@{}>{\raggedright}p{2.9cm}>{\raggedright}p{2.1cm}>{\raggedright}X>{\raggedright}p{1.4cm}>{\raggedright\arraybackslash}p{1.15cm}@{}}
\toprule
Rerun & Frames & Gates & Re-derived & Human\\
\midrule
CsSnI$_3$ plane & 962/962$^{a}$ & all pass & 43/43 & 0\\
CsSnI$_3$ $X_5^+$ scans (D4) & 39/39 & $G_5$: secondary coupled $\to$ human & 53/53 & 1\\
Ca$_3$Ti$_2$O$_7$ plane & 1{,}372/1{,}372 & all pass ($\Gamma_5^-$ held at zero by policy) & 38/38 & 0\\
InMnO$_3$ plane & 225/225 & $G_2$: R5 borderline $\to$ human & 39/39 & 1\\
ErMnO$_3$ plane and $P$ axis & 236/236 & all pass & 40/40 & 0\\
InMnO$_3$ plane and hat, joint fit & 400/400 & $G_2$: R5 borderline $\to$ human & 37/37 & 1\\
ErMnO$_3$ plane, $P$ axis and hat, joint fit & 411/411 & all pass & 38/38 & 0\\
BiFeO$_3$ six-component grid, fit & 10{,}307/10{,}307 & all pass & 84/84 & 0\\
HfO$_2$ paths T1--T5 & 50/50 & all pass & 344/344 & 0\\
\bottomrule
\end{tabularx}
\begin{flushleft}\footnotesize $^{a}$961 plane frames and the $P4/mbm$ anchor.\end{flushleft}
\end{table}

\begin{table}[!htbp]
\centering\footnotesize
\caption{Errors caught by each check in the real campaigns and, for the last row, in the option replay.}
\label{tab:errors}
\begin{tabularx}{\linewidth}{@{}>{\raggedright}p{2.8cm}>{\raggedright}p{2.6cm}>{\raggedright}X>{\raggedright\arraybackslash}X@{}}
\toprule
Check (owner) & System & Error caught & Consequence without it\\
\midrule
Basis search ($f_1$) & HfO$_2$ 137$\to$60 & no basis offered at $+11.3\%$ strain & no decomposition\\
Atom matching ($f_2$) & Ca$_3$Ti$_2$O$_7$ & atoms permuted, vector norms 40 and 28~\AA & wrong mode vectors, same space group\\
Translation removal ($f_2$) & Ca$_3$Ti$_2$O$_7$, BiFeO$_3$ & 75--76\% of $\|\e\|^2$ a rigid shift & polar axis overstated about twofold\\
Scale calibration ($f_2$) & CsSnI$_3$, Ca$_3$Ti$_2$O$_7$ & exported mode vectors at $0.80\times$; in Ca$_3$Ti$_2$O$_7$ a factor 2 between $X_3^-$ and $X_2^+$ & $Q=1$ falls short of the child by 20\% or more\\
Symmetry agreement ($f_2$) & CsSnI$_3$ & raw frame No.~55 where No.~127 expected & wrong subgroup on an axis\\
Cell count ($f_1$) & Ca$_3$Ti$_2$O$_7$ & mode norms set for the 2-cell supercell of the decomposition, not the 4 primitive cells of the frame & axes off by $\sqrt2$; child misplaced on the plane\\
SCF check ($f_4$) & BiFeO$_3$ & 20 runs not converged & noisy corner energies in the fit\\
Missing frames ($f_4$, $f_5$) & option replay (CsSnI$_3$, BiFeO$_3$, ErMnO$_3$) & frames of a failed job part dropped and the fit continued (7 of 22 wrong results in the first replay) & landscape fitted on fewer frames than planned, unnoticed\\
\bottomrule
\end{tabularx}
\end{table}

\begin{table}[!htbp]
\centering\footnotesize

\caption{\textbf{Construction of the HfO$_2$ transition paths by IRIS and by an expert.} IRIS decomposes each transition with respect to its own reference phase, whereas the expert construction expresses every structure in a common $P4_2/nmc$ (No.~137) order-parameter basis. Both constructions include the change of lattice along the path in addition to the symmetry-mode distortions, written as direct sums ($\oplus$) of irreps with their order-parameter directions. In the IRIS column, $\lambda\in[0,1]$ is the fraction along the path: every mode amplitude and the strain $\varepsilon$, decomposed into the irreps listed in parentheses, are scaled together by $\lambda$ from the reference ($\lambda=0$) to the child ($\lambda=1$) (eq.~\eqref{eq:path}; strain policy P3), so that the lattice follows the atoms. In the expert construction the lattice is interpolated separately between the two end cells (eq.~\eqref{eq:expertlat}), and superscripts (29), (60) and (14) mark the components specific to $Pca2_1$, $Pbcn$ and $P2_1/c$ in the shared basis; modes without a superscript are common to the phases. The intermediate symmetry is the space group of the structures along the path, with the number of point operations in parentheses; a path with fewer operations has left the symmetry manifold of its end phases. $\Gamma_1^+$, $\Gamma_2^-$ and $\Gamma_3^+$ are one-dimensional with direction $(a)$; $M_1$ and $M_3$ are two-dimensional with direction $(a,0)$ or $(0,a)$; $\Gamma_5^+$ and $\Gamma_5^-$ are two-dimensional with directions $(a,a)$ and $(a,-a)$.}
\label{tab:expert}
\renewcommand{\arraystretch}{1.0}\footnotesize
\begin{tabularx}{\textwidth}{@{}L{0.55cm}L{2.2cm}L{4.5cm}>{\raggedright\arraybackslash}X L{2.2cm}L{1.45cm}@{}}
\toprule
& & \multicolumn{2}{l}{Modes along the path} & \multicolumn{2}{l}{Intermediate symmetry}\\
\cmidrule(lr){3-4}\cmidrule(l){5-6}
Path & Transition & IRIS (own reference) & Expert (common $P4_2/nmc$ basis) & IRIS & Expert\\
\midrule
T1 & $P4_2/nmc\to Pca2_1$ & $P4_2/nmc$: $\lambda[M_1(0,a)\oplus M_3(a,0)\oplus\Gamma_5^-(a,-a)\oplus\Gamma_1^+(a)]\oplus\lambda\varepsilon(\Gamma_1^+,\Gamma_4^+)$ & $M_1^{(29)}\oplus M_3\oplus\Gamma_5^{-(29)}\oplus\Gamma_1^+$ & $Pca2_1$ (4) & $Pca2_1$ (4)\\
T2 & $P4_2/nmc\to Pbcn$ & $P4_2/nmc$: $\lambda[M_1(a,0)\oplus\Gamma_1^+(a)]\oplus\lambda\varepsilon(\Gamma_1^+,\Gamma_4^+)$ & $M_1^{(60)}\oplus\Gamma_1^+$ & $Pbcn$ (8) & $Pbcn$ (8)\\
T3 & $P4_2/nmc\to P2_1/c$ & $P4_2/nmc$: $\lambda[M_1(0,a)\oplus M_3(0,a)\oplus\Gamma_5^+(a,a)\oplus\Gamma_1^+(a)]\oplus\lambda\varepsilon(\Gamma_1^+,\Gamma_4^+,\Gamma_5^+)$ & $M_1^{(14)}\oplus M_3\oplus\Gamma_5^{+(14)}\oplus\Gamma_1^+$ & $P2_1/c$ (4) & $P2_1/c$ (4)\\
T4 & $Pbcn\to Pca2_1$ & $Pbcn$: $\lambda[\Gamma_2^-(a)\oplus\Gamma_1^+(a)]\oplus\lambda\varepsilon(\Gamma_1^+)$ & $M_1^{(60)}\oplus M_1^{(29)}\oplus M_3\oplus\Gamma_5^{-(29)}\oplus\Gamma_1^+$ & $Pca2_1$ (4) & $P2_1$ (2)\\
T5 & $Pbcn\to P2_1/c$ & $Pbcn$: $\lambda[\Gamma_3^+(a)\oplus\Gamma_1^+(a)]\oplus\lambda\varepsilon(\Gamma_1^+,\Gamma_3^+)$ & $M_1^{(60)}\oplus M_1^{(14)}\oplus M_3\oplus\Gamma_5^{+(14)}\oplus\Gamma_1^+$ & $P2_1/c$ (4) & $P2_1/c$ (4)\\
T6 & $P2_1/c\to Pca2_1$ & via $Pbcn$: T5 reversed, then T4 & $M_1^{(14)}\oplus M_1^{(29)}\oplus M_3\oplus\Gamma_5^{+(14)}\oplus\Gamma_5^{-(29)}\oplus\Gamma_1^+$ & $P2_1/c$ (4) $\to$ $Pbcn$ (8) $\to$ $Pca2_1$ (4) & $P2_1$ (2)\\
\bottomrule
\end{tabularx}
\end{table}
\clearpage

\clearpage
\IRISsinumbering
\newcommand{\IRISsitoc}{\etocsettocstyle{\section*{Contents of the Supplementary Information}}{}\localtableofcontents\clearpage}
\begin{center}
{\LARGE\textbf{Supplementary Information}}\\[6pt]
{\large \IRIStitle}\\[6pt]
\parbox{\textwidth}{\centering \IRISauthors}
\end{center}
\addcontentsline{toc}{part}{Supplementary Information}
\makeatletter
\let\IRISmaincite\cite
\protected\def\cite#1{\def\IRIS@keys{}\@for\IRIS@k:=#1\do{\ifx\IRIS@keys\@empty\edef\IRIS@keys{SI-\IRIS@k}\else\edef\IRIS@keys{\IRIS@keys,SI-\IRIS@k}\fi}\expandafter\IRISmaincite\expandafter{\IRIS@keys}}
\makeatother
\noindent This document follows the order of the main text. Supplementary Section~\ref{sec:si_legends} gives the complete descriptions of the figures whose legends are shortened in the main text. Supplementary Sections~\ref{sec:si_gates}--\ref{sec:si_dec} cover the workflow: the gates, error codes, rule table and sub-agents, and the language-model errors met during development (Section~\ref{sec:si_gates}); detailed views of the crystallographic stages of $f_1$ and $f_2$ (Section~\ref{sec:si_stages}); and the symmetry-mode decompositions of all systems (Section~\ref{sec:si_dec}). Section~\ref{sec:si_dft} lists the electronic-structure settings of every campaign. Supplementary Sections~\ref{sec:si_perov}--\ref{sec:si_hfo2} then follow the results in the order of the main text, each opening with the full account of the corresponding main-text section: the perovskite tilt transitions CsSnI$_3$ and Ca$_3$Ti$_2$O$_7$ (Section~\ref{sec:si_perov}); the two-dimensional $K_3$ order parameter and the Landau models of the hexagonal manganites (Sections~\ref{sec:si_k3} and~\ref{sec:si_mn}); the six-dimensional landscape of BiFeO$_3$ (Section~\ref{sec:si_bfo}); and the HfO$_2$ paths (Section~\ref{sec:si_hfo2}). Section~\ref{sec:origin} gives the polar-origin convention used throughout, Section~\ref{sec:si_eval} the evaluation of the agentic layer, with the full account of the gated replays and the extended discussion of limitations, Section~\ref{sec:si_live} the prospective BaTiO$_3$ campaign, and Section~\ref{sec:si_repo} the layout of the code and data repository. Equation, figure and table numbers without an S refer to the main text.

\IRISsitoc

\section{Complete figure descriptions}\label{sec:si_legends}
The legends of main-text Figs~\Mref{fig:pipeline} and~\Mref{fig:perovskite} and Extended Data Figs~\Mref{fig:workflow} and~\Mref{fig:planner} are shortened to the journal length; their complete descriptions, with every panel element, convention and value, are given here.

\subsection{Main-text Fig.~\Mref{fig:pipeline}: overview of IRIS}\label{sec:si_leg_fig1}
\noindent \textbf{Overview of IRIS as a gated state graph.}
IRIS is a state graph that runs from the human (left) to END through seven agent nodes, $f_0$--$f_6$, each followed by a gate $G_i$ on its outgoing edge.
$f_0$ Intake, the only node that calls a language model directly, turns a plain-language request and two crystal structures into a validated configuration.
$f_1$ Crystallographer assigns the space groups of the parent and child structures and decomposes the distortion into irrep modes.
$f_2$ Path Planner selects the active modes, certifies their vectors and builds the grid of frames.
$f_3$ Runner submits the DFT calculations; while the jobs run, the graph (LangGraph~\cite{LangGraph}) pauses with \texttt{interrupt()}, the process exits, and the graph resumes from its checkpoint when Slurm reports the jobs finished (orange).
$f_4$ Reviewer checks every energy, $f_5$ Analyst fits the Landau model and locates minima and barriers, and $f_6$ Skeptic re-derives every reported number.
The gates are deterministic code (table, top right) that return pass ($\checkmark$) or $\mathrm{fail}(c)$ with an error code $c$; they pass a hand-off only if the configuration ($G_0$), decomposition ($G_1$) and grid ($G_2$) are valid, the jobs are finished ($G_3$), the energies are usable ($G_4$), the fit is acceptable ($G_5$) and the claims are verified ($G_6$).
The table gives, for each gate, the question it answers and what it checks; for $G_2$, the four mode-vector checks of Methods (atom match, rigid translation removed, amplitude calibration, single-mode symmetry) and the space group of every frame on the descent tree.
Black arrows lead into a gate and green arrows ($\checkmark$) leave a passed gate.
A failure travels along the fail bus (red) to the recovery node (bottom), which tries a stored rule $\mathcal{R}(i,c)$, then the gate's sub-agent $\sigma_i$, a language-model call that must choose one fix from the listed set $\Omega_i(c)$, then a human $h$; the fix is applied, $f_i$ is re-run and $G_i$ re-checks the result (purple).
The human decides through \texttt{interrupt()}, the graph resumes with \texttt{Command(decision)}, and the answer is stored as a new rule, $\mathcal{R}\leftarrow\mathcal{R}\cup\{(i,c)\mapsto h\}$.
A coverage failure at $G_5$ returns directly to $f_2$ to refine the grid (loop above the nodes).
The human works with IRIS through a chat interface (top left), shown for CsSnI$_3$: (1) the plain-language request passes from the human to $f_0$, which returns the resolved configuration for confirmation; (2) when recovery reaches the human, the failure and its listed fixes appear in the chat, here the $M_3^+$-only frame found in No.~55 instead of No.~127 at check $c_2$, shown as it would reach the human if the stored rule \texttt{f2\_single\_mode\_sg\_fixed\_lattice} did not resolve it, and the answer resumes the graph through \texttt{Command(decision)} (dashed purple arrow between the human node and the chat); (3) the report is returned after $G_6$.
The output of this run (right, arrow from END) is the CsSnI$_3$ energy landscape on the $(M_3^+,R_4^+)$ plane of main-text Fig.~\Mref{fig:perovskite}b.
The chat panel shows the intended interface; the intake step was evaluated on text requests (main-text Section~\Mref{sec:autonomy}), and the front end itself was not.

\subsection{Main-text Fig.~\Mref{fig:perovskite}: perovskite irrep planes}\label{sec:si_leg_fig2}
\noindent \textbf{IRIS energy landscapes on irrep planes and validation of the structure generation.}
Each IRIS frame is the parent plus certified irrep mode vectors; for validation, every frame was also rebuilt by linear interpolation along the displacement fields from the parent to two phases (CsSnI$_3$: $P4/mbm$ and $Pnma$; Ca$_3$Ti$_2$O$_7$: $Cmcm$ and $A2_1am$), which contain only the two primary irreps.
\textbf{a},~CsSnI$_3$ structures viewed along the doubled $b$ axis: the cubic parent ($Pm\bar{3}m$, $a^0a^0a^0$), $M_3^+$ alone (in-phase rotation $a^0a^0c^+$, $P4/mbm$), $R_4^+$ alone (anti-phase tilt $a^-a^-c^0$, $Imma$) and both ($a^-b^+a^-$, $Pnma$), drawn at 1.2~\AA\ per irrep to make the tilts visible. Cs, teal; Sn, lilac; I, purple; the equatorial I atoms of the second octahedral layer are shown in a lighter shade to separate the two layers, which rotate in the same sense under $M_3^+$ and in opposite senses under $R_4^+$.
\textbf{b},~Energy landscape on the ($M_3^+$, $R_4^+$) plane at the fixed parent lattice, $E-E_{\min}$ per formula unit; contours at 1, 2, 4, 8 and 16~meV/f.u.
\textbf{c},~Norm of the atomic-position difference $\lVert\Delta R\rVert$ between the IRIS frame and the interpolated structure on a $101\times101$ grid spanning the plane in \textbf{b}, in n\AA\ ($10^{-9}$~\AA); the maximum full-cell norm, 8.7~n\AA\ (3.1~n\AA\ per atom), is the rounding of the stored nine-decimal vectors.
\textbf{d},~Ca$_3$Ti$_2$O$_7$ structures on the fixed 48-atom $I4/mmm$ cell: the parent ($I4/mmm$, constructed from the child), $X_3^-$ alone (tilt $a^-a^-c^0$, $Cmcm$), $X_2^+$ alone (rotation $a^0a^0c^+$, $Cmce$) and both ($A2_1am$). Ca, grey; Ti, blue; O, red.
\textbf{e},~Energy landscape on the ($X_3^-$, $X_2^+$) plane at the fixed parent lattice, with the polar $\Gamma_5^-$ mode and strain held at zero (0.071~Å steps); $E-E_{\min}$ per formula unit, contours every 50~meV/f.u.
\textbf{f},~As \textbf{c}, on a $101\times101$ grid spanning the plane in \textbf{e} (maximum per-atom difference 1.0~n\AA\ within the disk shown, 1.1~n\AA\ over the full $\pm1.27$~Å grid).
In \textbf{b}, \textbf{c}, \textbf{e} and \textbf{f}, amplitudes are $A_p$ in \AA; the white $\times$ marks the parent and the white $\circ$ the child (CsSnI$_3$: $Pnma$ at (0.843, 0.229)~\AA; Ca$_3$Ti$_2$O$_7$: $A2_1am$ at (0.822, 0.599)~\AA). Energies are DFT (PBE; CsSnI$_3$ on a $31\times31$ grid over $\pm1.2$~\AA, Ca$_3$Ti$_2$O$_7$ on a $37\times37$ grid over $\pm1.27$~\AA; Supplementary Table~\ref{tab:dft}), shown within a disk inscribed in the grid; colour scales are capped at 30 (\textbf{b}) and 600~meV/f.u.\ (\textbf{e}).
The blue bubbles above \textbf{b} and \textbf{e} give, verbatim, the non-specialist forms of the tested human requests for these two landscapes, E03 (``Why does cubic CsSnI3 distort into its orthorhombic low-temperature phase? Compute the energy surface of the tilting.'') and E09 (``Ca3Ti2O7 is said to be a hybrid improper ferroelectric. Compute the energy map of the two octahedral rotations that make it polar.''; Supplementary Table~\ref{tab:si_requests}); for each, $f_0$ returned the configuration behind the panel in 20 of 20 model calls (four models, five repeats), and that configuration reproduces the panel in replay (main-text Section~\Mref{sec:autonomy}). The campaigns themselves were not launched from these requests.
The agreement in \textbf{c} and \textbf{f} is an algebraic closure test of the frame construction, not a test of the energies.

\subsection{Extended Data Fig.~\Mref{fig:workflow}: detailed workflow}\label{sec:si_leg_ed1}
\noindent \textbf{Detailed IRIS workflow, shown for CsSnI$_3$.}
The eight blocks (0--7) into which the seven nodes of main-text Fig.~\Mref{fig:pipeline} are divided, each with its owner agent (chip), inputs (IN), steps (bold step $\to$ product), outputs (OUT) and the CsSnI$_3$ example (italic); gates $G_0$--$G_6$ (red diamonds) sit on the hand-offs.
\textbf{0 Intake} ($f_0$, the only language-model step) parses the request, resolves the two structures (file or Materials Project~\cite{Jain2013} identifier) and fills the remaining fields from the policy, recording where each came from; $G_0$ validates the configuration.
\textbf{1 Input} ($f_1$) reads the structures, matches their compositions, assigns the parent $G$ and child $H$, and builds the parent if only the child is given.
\textbf{2 Preprocess} ($f_1$) assigns and standardizes the space groups of both structures over a tolerance search, establishes $H\subset G$, the index $i=t\cdot k$ and the descent tree, and searches for a supercell basis with $\det=k$ when the decomposition offers none.
\textbf{3 Decompose} ($f_1$; labelled in the figure with the name of the decomposition engine of main-text Methods, Block~3) decomposes the distortion with this basis and origin over a ladder of displacement cut-offs (1.0, 1.5, 2.0~Å, then the basis search) and harvests the irreps, order-parameter directions, amplitudes $A_s$ and $A_p$, strain, isotropy subgroups $H_i$ and mode vectors; $G_1$ checks the decomposition.
\textbf{4 Postprocess} ($f_2$) classifies the modes, selects the active set with rules R1--R5 and the threshold $A_{\mathrm{thr}}$, sets the strain policy (fixed, interpolated or relaxed) and certifies the mode vectors (matched, translation-free, calibrated, space group agrees).
\textbf{5 Build grid} ($f_2$) sets the grid dimension and plan with the dimensionality policy, builds the frames and audits the space group of every frame against the expected map; $G_2$ passes the grid on.
\textbf{6 Compute} ($f_3$ Runner, $f_4$ Reviewer) writes and packages the calculations, checks each against the system profile, launches them (VASP~\cite{Kresse1996a} through Slurm~\cite{Yoo2003}, or MACE~\cite{Batatia2022} locally), resubmits failed parts and collects converged energies; $G_3$ waits for every job and $G_4$ admits only usable energies.
\textbf{7 Analysis} ($f_5$ Analyst, $f_6$ Skeptic) builds the surfaces with cross-checked interpolants, fits an invariant Landau polynomial with the degree chosen by cross-validation, locates minima and barriers, and reports; $G_5$ checks the fit and sends coverage failures back to block~5, and $G_6$ verifies every reported number.
Gates and recovery (bottom right): a failed gate returns an error code to a rule table, then to that gate's sub-agent, which picks one fix from a listed set, then to a human, whose answer becomes a new rule; the same gate re-checks the re-run, and every attempt is logged.
CsSnI$_3$: $Pm\bar{3}m$ (No.~221) $\to$ $Pnma$ (No.~62), $i=24$ with basis $(1,0,\bar1)(0,2,0)(1,0,1)$, $\det=4$; $M_3^+$ 0.843, $R_4^+$ 0.229, $R_5^+$ 0.086, $X_5^+$ 0.045~Å, $M_2^+$ below threshold, plus strain; the ($M_3^+$, $R_4^+$) plane on the fixed cubic lattice ($31\times31$) plus axes and scans gives 1036 frames; all 961 plane frames have the expected space group (1~$\times$~No.~221, 30~$\times$~No.~127, 30~$\times$~No.~74, 900~$\times$~No.~62), and all 1036 converged.

\subsection{Extended Data Fig.~\Mref{fig:planner}: Path Planner}\label{sec:si_leg_ed2}
\noindent \textbf{Inside the Path Planner ($f_2$): from decomposition to an audited grid, shown for CsSnI$_3$.}
The Planner receives the decomposition after $G_1$ and runs four stages, read clockwise from the top left, each with inputs (IN), steps with their equations (numbers as in the Methods text) and outputs (OUT); checks $c_1$--$c_3$ close the first three stages and $G_2$ the last.
\textbf{Select modes}: the distortion is written as a sum of irrep modes (eq.~1), polar $\Gamma$ amplitudes are made translation-free (eq.~2), and rules R1--R5 select the active set $\mathcal{A}$ (R2 amplitude threshold on primary irreps, eq.~3; R3 the active subgroups must intersect to $H$, eq.~4; R5 borderline irreps are included, excluded or kept as anchors only); $c_1$ asks whether $\mathcal{A}$ generates $H$.
\textbf{Certify eigenvectors}: atoms are matched by species (eq.~5), the rigid translation is removed and its fraction recorded (eqs.~6, 7), each mode is calibrated to the child amplitude (eq.~8), and each single-mode frame must have its isotropy subgroup $H_i$ (eq.~9); $c_2$ requires all four checks to pass.
\textbf{Dimensionality}: the grid dimension counts free order-parameter components (D1, eq.~10), the full grid is used if it fits the budget $B$ (D2, eq.~11), otherwise it is reduced with one DFT point per symmetry orbit (D3--D5, eq.~12), and the descent tree gives the expected space group of every node (eq.~13); $c_3$ requires $N\le B$ and a covered tree.
\textbf{Build and audit}: frames are built by superposing the certified unit modes $\hat{\mathbf{e}}_i$ on the lattice set by the strain policy (P1, fixed parent lattice; eq.~14), and every frame must have its expected space group (eq.~15) and no atoms closer than $d_{\mathrm{cut}}$ (eq.~16); $G_2$ passes the audited grid to the Runner ($f_3$, block~6).
A failed $c_i$ or $G_2$ goes to the recovery node (sub-agent boxes in the figure: borderline, strain, amplitude audit, grid budget, coverage); as implemented, the $G_2$ sub-agent chooses among the options of Supplementary Table~\ref{tab:subagents} (remove a translation, rebuild the check frames on the fixed lattice, R5), an over-budget grid goes to the human and coverage is checked at $G_5$.
Because $\hat{\mathbf{e}}_i$ is translation-free and calibrated, $Q_i$ is in Å of translation-free $A_p$, $A_p^{\mathrm{tf}}=A_p\sqrt{1-f_T}$, which equals $A_p$ for every irrep without a rigid translation.
CsSnI$_3$: $\mathcal{A}=\{M_3^+\ 0.843,\ R_4^+\ 0.229$~Å$\}$ with $R_5^+$ and $X_5^+$ held at zero and $\#127\cap\#74=\#62$; $s_i=1.758/1.406=1.25$ and no polar mode ($f_T=0$); $n=2$, $m=31$, $N=961$; all 961 frames match the tree (1~$\times$~No.~221, 30~$\times$~No.~127, 30~$\times$~No.~74, 900~$\times$~No.~62).

\section{Gates, error codes and sub-agents}\label{sec:si_gates}

Each gate $G_i$ returns pass or $\mathrm{fail}(c)$ with an error code $c$. Supplementary Table~\ref{tab:gates} lists what each gate checks and what its sub-agent may choose when no stored rule matches; Supplementary Table~\ref{tab:rules} lists the rule table, seeded from fixes that were applied during the campaigns reported in the main text, and Supplementary Table~\ref{tab:subagents} gives the menu of each sub-agent as implemented. The evaluation of the sub-agents and of the gates is reported in Supplementary Section~\ref{sec:si_eval}.

\begin{table}[!tp]
\centering\footnotesize\renewcommand{\arraystretch}{0.96}
\caption{Gates, their checks and error codes, and the options of the sub-agent called by the recovery node when no rule matches (Supplementary Table~\ref{tab:subagents}). Any node that crashes returns \texttt{NODE\_ERROR}; a test that is not implemented returns \texttt{NOT\_IMPLEMENTED} and can never pass silently. The codes listed are the principal ones; the complete list, predicate by predicate, is in Section~\ref{sec:si_gates_formal}.}
\label{tab:gates}
\begin{tabularx}{\linewidth}{@{}L{0.7cm}L{1.6cm}XL{3.6cm}@{}}
\toprule
Gate & After & Check (pass condition); error codes & Sub-agent options besides \texttt{stop}\\
\midrule
$G_0$ & $f_0$ Intake & configuration matches the schema; values within ranges; structure files present and readable, unchanged (pinned SHA-256) and of the expected space group (spglib~\cite{Togo2024}, $10^{-3}$~Å); budget within limits; injected faults only in replay mode; writes only inside the run directory; \texttt{SCHEMA}, \texttt{CIF\_MISSING}, \texttt{CIF\_UNREADABLE}, \texttt{CIF\_SG}, \texttt{CIF\_CHANGED}, \texttt{BUDGET}, \texttt{FAULTS}, \texttt{PATHS}; intake (main-text Methods, Block~0): \texttt{AMBIGUOUS}, \texttt{OUT\_OF\_SCOPE}, \texttt{NEEDS\_NEW\_DFT} & none: \texttt{stop} only, no model call\\
$G_1$ & $f_1$ Crystallographer & $H\subset G$ with index $t\cdot k$; supercell basis with $\det=k\,z_H/z_G$ ($z$, lattice points per conventional cell); space group stable on the tolerance ladder and equal in both symmetry engines; an independent mapping agrees with the decomposition's maximum displacement to 0.01~Å (the atom-matching basis search for a non-polar child; for a polar child, whose origin along the polar axis is free, the mapping with the mean displacement removed against the decomposition's shift-free value); mode norms $\lVert\mathbf e_i\rVert=A_p\sqrt n$; \texttt{NOT\_SUBGROUP}, \texttt{SG\_LADDER}, \texttt{SG\_DISAGREE}, \texttt{BASIS\_DET}, \texttt{NO\_BASIS}, \texttt{BASIS\_DISAGREE}, \texttt{DECOMP\_NORM} & widen the basis search\\
$G_2$ & $f_2$ Path Planner & selection rules R3--R5; four mode-vector checks pass (eqs.~\Mref{eq:match}--\Mref{eq:single}); grid within budget and covering the subgroup tree; every frame's space group equals the expected one (eq.~\Mref{eq:audit}); no collisions (eq.~\Mref{eq:dmin}); \texttt{R3\_FAIL}, \texttt{R4\_NO\_POLAR}, \texttt{R5\_BORDERLINE}, \texttt{ATOM\_MATCH}, \texttt{TRANSLATION}, \texttt{CALIBRATION}, \texttt{SINGLE\_MODE\_SG}, \texttt{OVER\_BUDGET}, \texttt{TREE\_COVERAGE}, \texttt{FRAME\_SG}, \texttt{D\_MIN}; for strain-interpolated paths (P3), rigid rotation of the child cell within its limit: \texttt{PATH\_LATTICE} & remove the translation; rebuild check frames on the fixed lattice; R5: include, exclude or anchor only\\
$G_3$ & $f_3$ Runner & no job queued or running; in replay mode, every frame matched to a stored structure (fractional coordinates within $10^{-4}$ for the two-irrep grids, Cartesian positions within $10^{-3}$~Å for paths and the six-component grid; optionally after removing a rigid shift of at most 0.05~Å, which leaves the energy unchanged); \texttt{JOBS\_PENDING}, \texttt{JOBS\_CANCELLED}, \texttt{UNMATCHED}, \texttt{RIGID\_SHIFT} & none: \texttt{stop} only, no model call\\
$G_4$ & $f_4$ Reviewer & every energy converged ($|\Delta E|<$ EDIFF within NELM) or flagged as SCF-limited; every frame accounted for; one set of inputs across frames (INCAR compared as tag--value sets, KPOINTS without the comment line); energies within a physical range; magnetic moments within the limit of the system profile; no frame dropped from the planned grid without a recorded human authorisation (\texttt{MISSING\_FRAMES}, also tested at $G_5$ after the fit); formula units per cell consistent with the composition; no more SCF-limited runs than declared; \texttt{ACCOUNTING}, \texttt{WALLTIME}, \texttt{NELM}, \texttt{NO\_ORIGIN}, \texttt{ENERGY\_RANGE}, \texttt{DFT\_INCONSISTENT}, \texttt{MAGNETIC}, \texttt{PER\_FU}, \texttt{SCF\_LIMITED} & resubmit; robust mixing; flag and continue\\
$G_5$ & $f_5$ Analyst & cross-validation error, residuals and coverage of the fit within the limits of main-text Methods, Block~7 (Section~\Mref{sec:agent}); for a surface analysed without a fit, complete grid and parent mirror symmetry; \texttt{NO\_FIT}, \texttt{CV\_ERROR}, \texttt{RESIDUALS}, \texttt{THIN\_GRID}, \texttt{COVERAGE}, \texttt{NO\_SURFACE}, \texttt{MIRROR}, \texttt{SCRIPT\_FAILED}, \texttt{SCRIPT\_MISMATCH}; HfO$_2$ paths, closure of the end-point energies across paths, agreement of the same crystal across settings and, in live mode, agreement of a plane's origin with the paths of its campaign: \texttt{CLOSURE}, \texttt{CROSS\_SETTING}, \texttt{NETWORK}, \texttt{SETTINGS\_MISMATCH} (the last two added after the evaluation freeze, Section~\ref{sec:si_live_settings}); hard-secondary
scans (D4): \texttt{SECONDARY\_COUPLED} & densify (returns to $f_2$); D4: hold the secondary at zero with a caveat\\
$G_6$ & $f_6$ Skeptic & every reported number re-derived from stored data by an independent route; \texttt{NOT\_REDERIVED}, \texttt{MISMATCH} & re-run the analysis\\
\bottomrule
\end{tabularx}
\end{table}

\begin{table}[!htbp]
\centering\scriptsize\renewcommand{\arraystretch}{0.92}
\caption{The rule table $\mathcal{R}$ (\texttt{recovery.yaml}) as seeded from fixes applied during the campaigns of the main text. Each rule is keyed by the step and error code, may carry a condition evaluated by code on the error detail, and is applied at most the stated number of times per run (or per frame for compute rules) before the case falls through to the sub-agent; in addition, each node has a retry budget of three attempts, after which the case goes to the human. A fix configures the node that runs next: compute failures return to $f_3$, which resubmits only the failed parts. Rules marked $^{\dagger}$ are implemented in \texttt{recovery.yaml} and were active in the end-to-end reruns and the evaluation of main-text Section~\Mref{sec:autonomy}, together with \texttt{f0\_schema\_reask}, which needs the intake model and was inactive; the others record fixes applied by script during the campaigns and are not yet rules of the implementation; \texttt{MISSING\_FRAMES} is a gate test added after the option replay, listed here for its stop-and-authorize rule.}
\label{tab:rules}
\begin{tabularx}{\linewidth}{@{}L{0.7cm}L{3.35cm}L{2.6cm}XL{0.55cm}L{1.9cm}@{}}
\toprule
Step & Error code & Condition & Fix & Max & Seeded from\\
\midrule
$f_1$ & \texttt{DECOMP\_RATE\_LIMIT} & HTTP 403 or 429 & wait with exponential back-off, retry from the session cache & 5 & all systems\\
$G_1$ & \texttt{NO\_BASIS}$^{\dagger}$ & decomposition lists no basis for $H\subset G$ & widen the atom-matching cut-off $d_{\max}$ (1.0, 1.5, 2.0~Å); in the campaign, then the basis search ($\det=k$, scored by atom matching) & 3 & HfO$_2$ 137$\to$60\\
$G_1$ & \texttt{SG\_LADDER} & assignment changes along the ladder & adopt the tightest tolerance at which the two symmetry engines agree & 1 & all systems\\
$G_2$ & \texttt{ATOM\_MATCH} & $\max_j|\Delta\mathbf r_j|>2$~Å & re-match atoms by species-constrained assignment & 1 & Ca$_3$Ti$_2$O$_7$\\
$G_2$ & \texttt{TRANSLATION}$^{\dagger}$ & $f_T>10^{-6}$ & subtract the mean displacement; rescale $A^{\mathrm{tf}}$ & 1 & BiFeO$_3$, Ca$_3$Ti$_2$O$_7$, HfO$_2$, InMnO$_3$\\
$G_2$ & \texttt{CALIBRATION} & $|s_i-1|>0.01$ & calibrate with $s_i$ (eq.~\Mref{eq:scale}) & 1 & CsSnI$_3$, Ca$_3$Ti$_2$O$_7$, InMnO$_3$\\
$G_2$ & \texttt{R4\_NO\_POLAR} & $H$ polar, no polar irrep active & add the largest polar irrep (R4), unless the policy holds it at zero & 1 & -- (policy rule)\\
$G_2$ & \texttt{SINGLE\_MODE\_SG}$^{\dagger}$ & $H_i$ found only at a tolerance other than $10^{-2}$~Å & rebuild the single-mode frame on the fixed parent lattice and re-check & 1 & CsSnI$_3$ (No.~55)\\
$f_3$ & \texttt{LDAU\_ORDER} & $+U$ arrays do not follow the POSCAR species order & rebuild LDAUL/U/J from the species order & 1 & ErMnO$_3$\\
$f_3$ & \texttt{POTCAR\_MISSING} & POTCAR absent for a species & stop the part; escalate & 0 & all systems\\
$G_3$ & \texttt{ALLOCATION\_EXHAUSTED} & account balance zero & escalate (choice of account) & 0 & BiFeO$_3$\\
$G_4$ & \texttt{WALLTIME}$^{\dagger}$ & job killed at the time limit & return to $f_3$; resubmit the failed parts unchanged & 1 & BiFeO$_3$ (50 runs, 45 recovered)\\
$G_4$ & \texttt{NELM}$^{\dagger}$ & not converged within NELM & return to $f_3$; resubmit the failed parts with AMIX 0.2, BMIX $10^{-4}$, NELM 300 (campaign: also LMAXMIX 4, AMIX\_MAG 0.8, BMIX\_MAG $10^{-4}$) & 1 & BiFeO$_3$ (15 runs, none recovered$^{\ddagger}$)\\
$G_4$ & \texttt{NELM\_AFTER\_ROBUST} & still not converged after resubmission & stop, keep $E_0$ as SCF-limited, exclude from gradient fits & -- & BiFeO$_3$ (20 runs: 15 after robust mixing, 5 after a walltime resubmission)\\
$G_4$, $G_5$ & \texttt{MISSING\_FRAMES} & a frame of the planned grid has no energy in the fit and no human has authorised the smaller grid (\texttt{policy.\allowbreak{}fit.\allowbreak{}allow\_\allowbreak{}missing\_\allowbreak{}frames}) & stop; resubmit the parts, or the human authorises the smaller grid, which every report then states & 0 & -- (added after the option replay, Section~\ref{sec:si_replay_found})\\
$G_5$ & \texttt{COVERAGE} & stationary point at the edge of the sampled region & return to $f_2$ with $Q_{\max}$ extended & 1 & --\\
$G_5$ & \texttt{THIN\_GRID}$^{\dagger}$ & fewer than ten points per coefficient & return to $f_2$; halve the grid step (subject to the budget) & 1 & --\\
\bottomrule
\end{tabularx}
\begin{flushleft}\footnotesize $^{\ddagger}$None of the 15 converged on this resubmission; all were then flagged by \texttt{NELM\_AFTER\_ROBUST}. The rule is kept as the first remedy because it costs one resubmission and leaves the frame usable either way.\end{flushleft}
\end{table}

Each implemented rule is stored as one entry of \texttt{recovery.yaml}; the rule for SCF failures reads (abridged)
\begin{verbatim}
f4_nelm_robust_mixing:
  step: G4
  code: NELM
  goto: f3                  # resubmit only the failed parts
  fix: {resubmit_failed: true,
        incar: {AMIX: 0.2, BMIX: 1.0e-4, NELM: 300}}
  source: seeded            # seeded | human
  origin: bifeo3_6d campaign, 15 runs
\end{verbatim}
A decision taken by a human is appended in the same form with \texttt{source: human}, the date and the identifier of the case that produced it, so that its provenance can be traced in the audit log. The hash of the rule table in force is recorded with every attempt.

\begin{table}[!htbp]
\centering\scriptsize
\caption{Recovery menus $\Omega_i(c)$ as implemented (\texttt{recovery.yaml}). For each gate, the stored rules are tried first; if none applies or the failure returns, the sub-agent of that gate receives the error code and detail, the slice of the state the gate checked and this menu, and returns one option identifier and a reason. Each option sets a parameter (the configuration patch) that the node it returns to reads, which a test checks against the node code; parameters are fixed in code, never written by the language model. A menu with only \texttt{stop} ($G_0$, $G_3$) goes to the human without a model call. Options that changed nothing a node reads were removed before the evaluation. Since the eligibility conditions of Section~\ref{sec:si_eligibility} (added after the evaluation freeze), every non-stop option also carries the error codes it is admissible for (\texttt{for:}): \texttt{widen\_basis\_search} \texttt{NO\_BASIS}; \texttt{include\_borderline}, \texttt{exclude\_borderline}, \texttt{anchor\_only} \texttt{R5\_BORDERLINE}; \texttt{rebuild\_fixed\_lattice} \texttt{SINGLE\_MODE\_SG}; \texttt{remove\_translation} \texttt{TRANSLATION}; \texttt{resubmit} \texttt{WALLTIME}; \texttt{robust\_mixing} \texttt{NELM}; \texttt{flag\_and\_continue} \texttt{WALLTIME}, \texttt{NELM}; \texttt{densify} \texttt{THIN\_GRID}; \texttt{accept\_secondary\_hold} \texttt{SECONDARY\_COUPLED}; \texttt{rerun\_analysis} \texttt{MISMATCH}, \texttt{NOT\_REDERIVED}; a failure whose admissible menu is \texttt{stop} alone is decided in code.}
\label{tab:subagents}
\begin{tabularx}{\linewidth}{@{}L{0.55cm}L{2.9cm}L{2.75cm}X L{1.0cm}@{}}
\toprule
Gate & Rules tried first (code) & Option & What it does (configuration patch) & Returns to\\
\midrule
$G_0$ & \texttt{f0\_\allowbreak{}schema\_\allowbreak{}reask} (\texttt{SCHEMA}) & \texttt{stop} & stop and ask the human for a corrected configuration & human\\
\midrule
$G_1$ & \texttt{f1\_\allowbreak{}no\_\allowbreak{}basis\_\allowbreak{}dmax\_\allowbreak{}ladder} (\texttt{NO\_\allowbreak{}BASIS}) & \texttt{widen\_\allowbreak{}basis\_\allowbreak{}search} & widen the atom-matching basis search to $d_{\max}=2.0$~\AA\ (\texttt{basis\_\allowbreak{}dmax\_\allowbreak{}A: 2.0}) & $f_1$\\
 & & \texttt{stop} & stop and ask the human & human\\
\midrule
$G_2$ & \texttt{f2\_\allowbreak{}translation} (\texttt{TRANSLATION}); \texttt{f2\_\allowbreak{}single\_\allowbreak{}mode\_\allowbreak{}sg\_\allowbreak{}fixed\_\allowbreak{}lattice} (\texttt{SINGLE\_\allowbreak{}MODE\_\allowbreak{}SG}) & \texttt{include\_\allowbreak{}borderline} & R5: include the borderline irreps as axes (\texttt{r5\_\allowbreak{}decision: include}) & $f_2$\\
 & & \texttt{exclude\_\allowbreak{}borderline} & R5: exclude the borderline irreps (\texttt{exclude}) & $f_2$\\
 & & \texttt{anchor\_\allowbreak{}only} & R5: keep the borderline irreps as anchor points only (\texttt{anchor\_\allowbreak{}only}) & $f_2$\\
 & & \texttt{rebuild\_\allowbreak{}fixed\_\allowbreak{}lattice} & rebuild the single-mode check frames on the fixed parent lattice (\texttt{single\_\allowbreak{}mode\_\allowbreak{}check\_\allowbreak{}on: fixed\_\allowbreak{}parent\_\allowbreak{}lattice}) & $f_2$\\
 & & \texttt{remove\_\allowbreak{}translation} & subtract the rigid translation (mean displacement) from the polar mode vectors (\texttt{subtract\_\allowbreak{}mean\_\allowbreak{}displacement}) & $f_2$\\
 & & \texttt{stop} & stop and ask the human & human\\
\midrule
$G_3$ & -- & \texttt{stop} & stop and ask the human & human\\
\midrule
$G_4$ & \texttt{f4\_\allowbreak{}walltime\_\allowbreak{}resubmit} (\texttt{WALLTIME}); \texttt{f4\_\allowbreak{}nelm\_\allowbreak{}robust\_\allowbreak{}mixing} (\texttt{NELM}) & \texttt{resubmit} & resubmit the failed parts unchanged (\texttt{resubmit\_\allowbreak{}failed}) & $f_3$\\
 & & \texttt{robust\_\allowbreak{}mixing} & resubmit the failed parts with robust mixing (\texttt{AMIX} 0.2, \texttt{BMIX} $10^{-4}$, \texttt{NELM} 300) & $f_3$\\
 & & \texttt{flag\_\allowbreak{}and\_\allowbreak{}continue} & flag the unconverged frames and continue without them (\texttt{flag\_\allowbreak{}unconverged}) & $f_4$\\
 & & \texttt{stop} & stop and ask the human & human\\
\midrule
$G_5$ & \texttt{f5\_\allowbreak{}thin\_\allowbreak{}grid\_\allowbreak{}densify} (\texttt{THIN\_\allowbreak{}GRID}) & \texttt{densify} & return to $f_2$ and densify the grid (\texttt{densify: 2}) & $f_2$\\
 & & \texttt{accept\_\allowbreak{}secondary\_\allowbreak{}hold} & D4: keep the secondary at zero and report the energy it would release as a caveat (\texttt{d4\_\allowbreak{}accept}) & $f_5$\\
 & & \texttt{stop} & stop and ask the human & human\\
\midrule
$G_6$ & -- & \texttt{rerun\_\allowbreak{}analysis} & re-run $f_5$ and check again & $f_5$\\
 & & \texttt{stop} & stop and ask the human & human\\
\bottomrule
\end{tabularx}
\end{table}

\begingroup\sloppy
\providecommand{\gc}[1]{[\texttt{#1}]}
\providecommand{\pr}[2]{p_{#1,#2}}

\subsection{Formal definition of the gates}\label{sec:si_gates_formal}

\paragraph{Common form.}
A gate is the function \texttt{run\_gate(G\_i, S)} of \texttt{iris\_agent/gates.py}, which evaluates an ordered list of named tests on the state $S$ (a dictionary, checkpointed after every node); each test is a pure Python function returning \texttt{None} on pass or a pair (code, detail) on fail. With $\pr{i}{k}$ the $k$-th test of gate $i$ and $c_{i,k}$ its code,
\begin{equation}\label{eq:si_gate}
G_i(S)=\bigwedge_{k=1}^{n_i}\pr{i}{k}\bigl(S|_i\bigr),\qquad
G_i=\mathrm{fail}(c_{i,k^\ast}),\quad k^\ast=\min\{k:\ \pr{i}{k}(S|_i)=\mathrm{false}\},
\end{equation}
where $S|_i$ is the slice that node $f_i$ wrote: $S|_0=$ \texttt{config}, \texttt{inputs}; $S|_1=$ \texttt{decomposition}; $S|_2=$ \texttt{selection}, \texttt{grid}; $S|_3=$ \texttt{jobs}; $S|_4=$ \texttt{energies} (and \texttt{grid}, for the frame count); $S|_5=$ \texttt{model} (and \texttt{energies}, \texttt{grid}); $S|_6=$ \texttt{report}. All $n_i$ tests are evaluated and logged; the gate's code is that of the first failing test in list order, the order given below.\footnote{The main text says that a gate ``returns'' its first failure; the code evaluates every test, so the audit log of a failed gate also records the later ones. The outcome is the same.} A test that raises fails with \texttt{TEST\_ERROR}; one returning the sentinel \texttt{SKIP} fails with \texttt{NOT\_IMPLEMENTED} in every mode but \texttt{skeleton}. The gate neither modifies nor rolls back the state: the node's output stays in $S$, the pair $(i,c)$ goes to the recovery node (main-text Section~\Mref{sec:agent}), the fix is a patch to the overrides the re-run node reads, and the same gate is evaluated again. \texttt{tests\_for(gate, state)} picks the list from the policy: each standard list below (the two-irrep grid of CsSnI$_3$, Ca$_3$Ti$_2$O$_7$ and the manganites) is followed by its variants for a path (\texttt{policy.path}), a two-component plane (\texttt{policy.plane2}), the six-component grid (\texttt{policy.grid.six}) and the analyses named by \texttt{policy.analysis}. $\SG_\epsilon(X)$ is the space-group \emph{number} spglib~\cite{Togo2024} returns for the Cartesian structure $X$ at \texttt{symprec} $=\epsilon$ (Å) with its default angle tolerance (pymatgen's \texttt{SpacegroupAnalyzer}; \texttt{spglib.get\_symmetry\_dataset} directly for the six-component grid). \texttt{policy.*}, \texttt{run.*}, \texttt{expected.*} are configuration keys (\texttt{docs/CONFIG\_SCHEMA.md}), campaign values in parentheses; other numbers are constants of \texttt{gates.py} or of the node named.

\paragraph{$G_0$, configuration valid (after $f_0$).}
With $\mathcal I=$ \texttt{config.intake} (present only when $f_0$ parsed a request with the language model; $\pr{0}{1}$--$\pr{0}{4}$ hold when absent) and $X_r$, $r\in\{G,H\}$, the two structure files,
\begin{equation}\label{eq:si_G0}
G_0=\textstyle\bigwedge_{k=1}^{11}\pr{0}{k}:
\end{equation}
$\pr{0}{1}$--$\pr{0}{4}$: $\mathcal I.\mathrm{status}\notin\{$\texttt{out\_of\_scope}, \texttt{ambiguous}, \texttt{needs\_new\_dft}$\}$, $\mathcal I.\mathrm{checks}=\varnothing$ \gc{OUT\_OF\_SCOPE}, \gc{AMBIGUOUS}, \gc{SCHEMA}, \gc{NEEDS\_NEW\_DFT};
$\pr{0}{5}$: the schema validator of \texttt{config.py} returns no error (every key present, of its type and range; \texttt{run.allow\_live\_submission} iff \texttt{run.mode} $=$ \texttt{live}) \gc{SCHEMA};
$\pr{0}{6}$: both files exist \gc{CIF\_MISSING};
$\pr{0}{7}$: both parse \gc{CIF\_UNREADABLE} and $\SG_{10^{-3}}(X_r)=$ \texttt{expected.$r$\_sg} \gc{CIF\_SG};
$\pr{0}{8}$: $\mathrm{SHA256}(X_r)=$ \texttt{expected.$r$\_sha256} for every pinned digest \gc{CIF\_CHANGED};
$\pr{0}{9}$: $B=$ \texttt{run.budget\_runs} $\in\mathbb Z$, $0<B\le20{,}000$ (\texttt{MAX\_BUDGET}) \gc{BUDGET};
$\pr{0}{10}$: \texttt{config.faults} $=\varnothing$ or \texttt{run.mode} $=$ \texttt{replay} \gc{FAULTS};
$\pr{0}{11}$: \texttt{paths.agent\_root} is non-empty and differs from \texttt{paths.repo\_root} \gc{PATHS}.\footnote{Supplementary Table~\ref{tab:gates} words this as ``writes only inside the run directory''; the gate tests only that the two roots differ. Confinement of every write to \texttt{agent\_root} is enforced at write time by \texttt{paths.Paths.out}, not by $G_0$.}

\paragraph{$G_1$, decomposition valid (after $f_1$).}
Let $G$, $H$ be the expected group numbers; $t=|P_G|/|P_H|$ the ratio of point-group orders (integer, else undefined); $k$, $i$, $\mathbf B$ the klassengleiche index (the supercell factor printed by ISODISTORT), the total index and the $3\times3$ basis; $z_G$, $z_H$ the lattice points per conventional cell ($P$ 1; $A,B,C,I$ 2; $F$ 4; $R$ on hexagonal axes 3); $\ell_r(\epsilon)=\SG_\epsilon(X_r)$ the ladder $f_1$ computes at $\epsilon\in\{10^{-4},10^{-3},10^{-2},10^{-1},0.3\}$~Å; and $\mathbf e_i$ the Cartesian field of irrep $i$ at the child amplitude on the $N$-atom frame, which holds $n=N/N_{\mathrm{prim}}(G)$ primitive parent cells.
\begin{equation}\label{eq:si_G1}
G_1=\pr{1}{1}\wedge\pr{1}{2}\wedge\pr{1}{3}\wedge\pr{1}{4}\wedge\pr{1}{5}:
\end{equation}
$\pr{1}{1}$: $H\le G$ by pymatgen's \texttt{SpaceGroup.is\_subgroup}, a walk of the maximal-subgroup tables of the International Tables, Vol.~A, keyed by number, and $t\cdot k=i$ \gc{NOT\_SUBGROUP}; $\ell_r(10^{-3})=\ell_r(10^{-2})=$ \texttt{expected.$r$\_sg} for both structures \gc{SG\_LADDER};\footnote{The ladder has five rungs, $10^{-4}$ to $0.3$~Å (main-text Methods, Block~2); the gate requires stability on the two middle rungs, the others feeding the recovery menu.} the stored FINDSYM~\cite{Stokes2005} number, when present, equals the expected one \gc{SG\_DISAGREE}.
$\pr{1}{2}$: $\det\mathbf B=k\,z_H/z_G>0$ \gc{BASIS\_DET}; unless switched off, the atom-matching basis search (\texttt{iris/lib/basis\_search.py}) finds a right-handed basis with largest displacement below \texttt{policy.basis\_dmax\_A} (2.0~Å; the \texttt{NO\_BASIS} rule tries 1.0, 1.5, 2.0) \gc{NO\_BASIS}, with $\det\in\{s,\ s\,z_H/z_G\}$ \gc{BASIS\_DET}, and its displacement $\Delta_{\mathrm{search}}<2.0$~Å agrees with the decomposition to 0.01~Å, against the printed or translation-free value for a non-polar child, or, for a polar child, the centre-of-mass-fixed mapping of $f_2$ against the translation-free value (Section~\ref{sec:origin}) \gc{BASIS\_DISAGREE}.
$\pr{1}{3}$: $\{H_{\mathrm{ISODISTORT}},\ \SG_{10^{-3}}(X_H),\ \SG_{10^{-2}}(X_0+\sum_i\mathbf e_i)\}=\{H\}$, the last being the child rebuilt from the stored mode matrix on the parent lattice \gc{SG\_DISAGREE}.
$\pr{1}{4}$ (eq.~\Mref{eq:decomp}): $n$ recorded and $\bigl|\lVert\mathbf e_i\rVert-|A_{p,i}|\sqrt n\bigr|\le2\times10^{-4}\sqrt n$~Å for every irrep \gc{DECOMP\_NORM}.
$\pr{1}{5}$ (\texttt{policy.components}, BiFeO$_3$): the six component fields partition the stored $\Gamma_4^-$ and $R_5^-$ fields to $10^{-9}$~Å, each carries $A_p^{\mathrm{tf}}/\sqrt3$ to \texttt{components.tol\_A} ($10^{-4}$~Å), and the 48 point operations about Bi times the odd parent translation give 96 distinct signed-permutation matrices with closure residual $\le$ \texttt{tol\_A} \gc{COMPONENTS}; these equal the campaign toolkit's matrices, the fields folded onto the ten-atom grid cell equal its basis with $\lVert\mathbf B_k\rVert^2=2$, and the child lies on a grid node to $10^{-3}$ \gc{BASIS\_DISAGREE}.

\paragraph{$G_2$, grid valid (after $f_2$).}
Let $\Aset$ be the active set, $T(\mathcal S)$ the lowest-index node of the ISODISTORT subgroup tree whose displacive order parameters contain $\mathcal S$ ($T(\varnothing)=G$), $\mathcal F$ the built frames, and $\hat{\mathbf e}_i$ the calibrated, translation-free unit vectors (eqs~\Mref{eq:trans}--\Mref{eq:scale}).
\begin{equation}\label{eq:si_G2}
G_2=\pr{2}{1}\wedge\pr{2}{2}\wedge\pr{2}{3}\wedge\pr{2}{4}\wedge\pr{2}{5}\wedge\pr{2}{6}\wedge\pr{2}{7}\wedge\pr{2}{8}:
\end{equation}
$\pr{2}{1}$ ($c_1$, eq.~\Mref{eq:gen}): \texttt{selection.code} $=$ \texttt{OK}, else the gate returns that code (\texttt{R3\_FAIL}: the primaries, walked in decreasing $A_s^{\mathrm{tf}}$, never reach $H$, or $\Aset$ does not generate it; \texttt{R4\_NO\_POLAR}; \texttt{R5\_BORDERLINE}: $\min_{i\in\Aset}A_i-\max_{i\notin\Aset}A_i<$ \texttt{policy.r5\_margin} (0.1~Å) with no recorded decision; \texttt{R5\_EXCLUDE\_BREAKS\_H}); and $T(\Aset)=H$ by number and index \gc{R3\_FAIL}.\footnote{$\bigcap_{i\in\Aset}H_i$ in eq.~\Mref{eq:gen} is not computed group-theoretically: $f_2$ reads it from the subgroup tree ISODISTORT returns, as the highest-symmetry node whose order-parameter set contains $\Aset$. The built-frame route of $\pr{2}{8}$ tests the same statement on structures, and $f_6$ repeats the lookup.}
$\pr{2}{2}$: a recorded R5 decision $\delta$ (\texttt{overrides.f2.r5\_decision}) with borderline irreps satisfies \texttt{selection.r5\_decision} $=\delta$, the grid axes equal \texttt{selection.active}, and for \texttt{anchor\_only} the anchor frames are exactly the removed irreps \gc{R5\_NOT\_APPLIED}.
$\pr{2}{3}$ (eq.~\Mref{eq:match}): the child maps onto the parent supercell (\texttt{StructureMatcher}, then a species-constrained Hungarian assignment $\pi$, centre of mass fixed) with $\max_j|\rv^H_{\pi(j)}-\rv^G_j|<2.0$~Å \gc{ATOM\_MATCH}.
$\pr{2}{4}$ (eqs~\Mref{eq:trans}, \Mref{eq:fT}): for each $i\in\Aset$ the mean was subtracted or $f_{T,i}\le10^{-6}$, and the stored vector has $\max_\alpha|\frac1N\sum_je_{i,j\alpha}|\le10^{-6}$ \gc{TRANSLATION}.
$\pr{2}{5}$: with the child field $\Delta\rv^H$ aligned to the domain of largest overlap with $\sum_iA^{\mathrm{tf}}_{p,i}\hat{\mathbf e}_i$ under the parent operations, $\bigl|\langle\Delta\rv^H,\hat{\mathbf e}_i\rangle/\lVert\hat{\mathbf e}_i\rVert^2-A^{\mathrm{tf}}_{p,i}\bigr|\le0.02\,A^{\mathrm{tf}}_{p,i}$ for every $i\in\Aset$ \gc{CALIBRATION}.\footnote{$s_i$ of eq.~\Mref{eq:scale} is computed and recorded (\texttt{s\_matrix\_over\_movie}) but is not a predicate: the gate tests the 2\,\% projection condition stated in the main text. The condition $|s_i-1|>0.01$ in Supplementary Table~\ref{tab:rules} describes the campaign-time fix, not the test the gate evaluates.}
$\pr{2}{6}$ (eq.~\Mref{eq:single}): $\SG_{10^{-2}}\bigl(\rv_0+A^{\mathrm{tf}}_{p,i}\hat{\mathbf e}_i\bigr)=H_i$ for every $i\in\Aset$, $H_i$ the isotropy-subgroup number ISODISTORT lists, the frame built on the fixed parent lattice (\texttt{policy.single\_mode\_check\_on}); the ladder $\{10^{-4},\dots,10^{-1}\}$ is recorded and a match at another rung is still a failure \gc{SINGLE\_MODE\_SG}.
$\pr{2}{7}$ ($c_3$, eqs~\Mref{eq:budget}, \Mref{eq:tree}): the grid was built, $\prod_im_i+n_{\mathrm{anchors}}\le B$ (polar, D4 and R5-anchor frames included), $|\mathcal F|\le B$ \gc{OVER\_BUDGET}, and $\{T(\mathcal S):\mathcal S\subseteq\Aset\}\subseteq\{\SG(\Q):\Q\in\mathcal F\}$ \gc{TREE\_COVERAGE}.
$\pr{2}{8}$ (eqs~\Mref{eq:audit}, \Mref{eq:dmin}): $\SG_{10^{-2}}(\rv(\Q))=T(\{i:Q_i\neq0\})$ for every frame (a polar $(Q,\Phi)$ frame expects the group declared for its $\Phi$ class) \gc{FRAME\_SG}, and $d_{\min}(\Q)\ge$ \texttt{policy.d\_cut} (1.0~Å; BiFeO$_3$ 1.6~Å) \gc{D\_MIN}.\footnote{The frame audit of the two-irrep grid, the paths and the plane is made at \texttt{SYMPREC} $=10^{-2}$~Å (\texttt{nodes/f2\_path\_planner.py}), the tolerance of the strict single-mode check; only the six-component grid uses $10^{-3}$~Å (main-text Methods, Block~2).}

\emph{Variants.} Path (\texttt{G2\_path}; eq.~\Mref{eq:path}): the $\lambda=1$ cell has the child metric, $\max|\mathbf L_1\mathbf L_1^{\mathsf T}-\mathbf C\mathbf C^{\mathsf T}|\le10^{-6}$~Å$^2$, and its rigid rotation against $\mathbf L_0$ is below \texttt{max\_rigid\_rotation\_deg} of \texttt{policy.path} when that key is set \gc{PATH\_LATTICE};\footnote{No HfO$_2$ configuration sets this limit, so the $4.9^\circ$ and $5.0^\circ$ rotations of T3 and T5 are recorded, not gated.} the $\lambda=1$ frame matches the child file with \texttt{StructureMatcher} RMS $\le$ \texttt{path.child\_rms\_tol} ($10^{-3}$) \gc{PATH\_END}; $\SG_{10^{-2}}=G$ at $\lambda=0$, $=H$ at $\lambda>0$, $d_{\min}\ge d_{\mathrm{cut}}$ \gc{FRAME\_SG}, \gc{D\_MIN}; in replay the path equals the stored \texttt{path.json} \gc{PLAN\_DISAGREE}. Plane (\texttt{G2\_plane2}): $\lVert\hat{\mathbf e}_2\rVert=\lVert\hat{\mathbf e}_1\rVert$ and $\langle\hat{\mathbf e}_1,\hat{\mathbf e}_2\rangle=0$ to $10^{-6}$ \gc{COMPONENTS}; the orbit representatives number $\frac18(n^2+4n+3)$, Burnside's count for $4mm$ on the $n\times n$ grid, with multiplicities summing to $n^2$, and $|\mathcal F|\le B$ \gc{ORBITS}, \gc{BUDGET}; origin frame in $G$, axis frames in $H$, $d_{\min}\ge d_{\mathrm{cut}}$ \gc{FRAME\_SG}, \gc{D\_MIN}. Six-component grid (\texttt{G2\_grid6}; eq.~\Mref{eq:orbit}): $\Aset=\{$\texttt{components.polar}, \texttt{components.rotation}$\}$ \gc{R3\_FAIL}; per dataset the orbit count equals $|G|^{-1}\sum_g|\mathcal Q^g|$ over the 96 matrices, multiplicities sum to the node count and divide 96, and the counts before and after the per-orbit contact cut ($d_{\min}\ge1.6$~Å) equal \texttt{grid.six.expected} \gc{ORBITS}; orbits, contacts, planned points and controls equal the stored campaign \gc{ORBITS\_DISAGREE}, \gc{CONTACT\_CUT}, \gc{PLAN\_DISAGREE}, \gc{CONTROLS}; $\SG_{10^{-3}}$ of every frame equals the stored \texttt{point\_info} and the special nodes their expected groups \gc{SG\_DISAGREE}, \gc{SINGLE\_MODE\_SG}; $|\mathcal F|\le B$ \gc{OVER\_BUDGET}.

\paragraph{$G_3$, jobs finished (after $f_3$).}
With $J$ the job book,
\begin{equation}\label{eq:si_G3}
\begin{aligned}
G_3=\ &\bigl[J.\mathrm{queued}=J.\mathrm{running}=0\bigr]\wedge\bigl[J.\mathrm{cancelled}=0\bigr]\\
&\wedge\bigl[J.\mathrm{unmatched}=\varnothing,\ n_{\mathrm{matched}}=|\mathcal F|\bigr]\wedge\bigl[\textstyle\max_f\lVert\mathbf t_f\rVert\le t_{\max}\bigr],
\end{aligned}
\end{equation}
with codes \texttt{JOBS\_PENDING}, \texttt{JOBS\_CANCELLED}, \texttt{UNMATCHED}, \texttt{RIGID\_SHIFT}. In replay a frame is matched to a stored structure when, after the one atom permutation found at $\Q=\mathbf0$, species agree, $\max|\Delta\mathbf L|\le$ \texttt{run.replay.tol\_lattice\_A} ($10^{-4}$~Å) and the largest coordinate deviation is within $10^{-4}$ (fractional; two-irrep grid) or \texttt{run.replay.tol\_A} ($10^{-3}$~Å; paths and six-component grid). The last term applies only with \texttt{run.replay.match\_modulo\_translation}, $\mathbf t_f$ being the removed mean displacement and $t_{\max}=$ \texttt{max\_rigid\_shift\_A} (0.05~Å). The menus of $G_0$ and $G_3$ are \texttt{stop} alone.

\paragraph{$G_4$, energies usable (after $f_4$).}
Let $E=\{E_f\}$ be the linked energies per cell, $\Phi$ the flagged frames with optional job code $\kappa_f\in\{$\texttt{WALLTIME}, \texttt{NELM}, \texttt{NODE\_FAIL}, \texttt{OOM}, \texttt{UNKNOWN}$\}$ and an \emph{accepted} bit set only by the recorded decision \texttt{flag\_and\_continue}, $E_0$ the energy at $\Q=\mathbf0$, $z=$ \texttt{policy.fit.per\_fu}, and $\mathcal D=\{f\in\Phi:\mathrm{accepted}_f,\ \neg\,\mathrm{energy\_kept}_f,\ f\notin E\}$ the frames dropped from the plan.
\begin{equation}\label{eq:si_G4}
\begin{aligned}
G_4=\ &\bigl[|E|+|\Phi|=|\mathcal F|\bigr]\wedge\bigl[\forall f\in\Phi:\kappa_f=\varnothing\vee\mathrm{accepted}_f\bigr]\wedge\bigl[E_0\ \text{defined}\bigr]\\
&\wedge\bigl[1\le\tfrac{10^3}{z}(\max E-\min E)\le10^4\ \text{meV}\bigr]\wedge\bigl[\text{one INCAR/KPOINTS set}\bigr]\\
&\wedge\bigl[\mathcal D=\varnothing\ \vee\ \texttt{policy.fit.allow\_missing\_frames}\bigr],
\end{aligned}
\end{equation}
with codes \texttt{ACCOUNTING}, $\kappa_f$ itself, \texttt{NO\_ORIGIN}, \texttt{ENERGY\_RANGE}, \texttt{DFT\_INCONSISTENT} (INCARs compared as tag--value sets, KPOINTS without its comment line; one set per D4 scan allowed under \texttt{policy.d4.settings: own\_per\_scan}) and \texttt{MISSING\_FRAMES}, which $G_5$ tests again with every planned frame lacking an energy counted. The convergence rule of main-text Section~\Mref{sec:dft}, last $|\Delta E|<$ EDIFF within NELM steps, is applied by the collector that writes the stored energies; in replay, where OSZICAR files are absent, a two-irrep-grid frame counts as converged iff the stored CSV holds its energy.\footnote{For paths and planes $G_4$ re-evaluates the rule from each point's stored \texttt{result.json}, $n_{\mathrm{SCF}}<$ NELM and $|\Delta E_{\mathrm{last}}|<$ EDIFF \gc{NOT\_CONVERGED}; for the two-irrep grid it does not.}

\emph{Variants.} Path or plane (\texttt{G4\_path}): accounting with the per-point convergence rule \gc{ACCOUNTING}, \gc{NOT\_CONVERGED}, \gc{NO\_ORIGIN}, \gc{ENERGY\_RANGE}; $z$ equals the formula units counted from the composition \gc{PER\_FU}; one CSV row per frame \gc{NO\_ENERGY}; $E/z$ and $\Delta E(\lambda)$ re-derived from the cell energies agree with \texttt{result.json} and the CSV to $6\times10^{-4}$~meV/f.u.\ \gc{ENERGY\_MISMATCH}; stored group $=$ frame group \gc{FRAME\_SG}; $\max_f|m_f|\le$ \texttt{policy.qc.mag\_max\_muB} ($0.01\,\mu_B$/cell) \gc{MAGNETIC}; one INCAR/KPOINTS set equal to the campaign template by hash, one POTCAR set \gc{DFT\_INCONSISTENT}; \texttt{MISSING\_FRAMES}. Six-component grid (\texttt{G4\_grid6}; limits in \texttt{policy.qc} of \texttt{bifeo3.yaml}): at most \texttt{max\_scf\_limited} (20) SCF-limited runs \gc{SCF\_LIMITED}; energies, labels and groups re-derived against the CSV \gc{ENERGY\_MISMATCH}, \gc{LABEL\_MISMATCH}, \gc{SG\_DISAGREE}; the 95 control repeats at images $g\mathbf q$ reproduce $E(\mathbf q)$ to \texttt{control\_dE\_meV} (0.01) and $g\nabla E(\mathbf q)$ to \texttt{control\_dgrad\_meV\_per\_A} (0.5) \gc{CONTROLS}; the slope of $E(c)-E(a)$ on the Simpson integral $\frac h3(g_a+4g_b+g_c)$ of the projected gradients along grid lines is within \texttt{line\_slope\_tol} (0.02) of 1 \gc{GRADIENTS}; $|m_{\mathrm{Fe}}|\in[3.5,4.5]\,\mu_B$ with opposite signs, $|M_{\mathrm{tot}}|\le0.05\,\mu_B$ \gc{MAGNETISM}; one input set \gc{DFT\_INCONSISTENT}; \texttt{MISSING\_FRAMES}.

\paragraph{$G_5$, fit acceptable (after $f_5$).}
For the Landau plane (\texttt{landau\_plane}; eq.~\Mref{eq:cssni3F}) let $\sigma$, $\rho$ be the training RMS and largest residual (meV/f.u.) on the $n_{\mathrm{fit}}$ nodes with $r\le$ \texttt{policy.fit.rmax}; $\sigma_{\mathrm{CV}}$ the RMS of the $k$-fold cross-validation ($k=$ \texttt{fit.cv\_folds} $=10$, fold of node $j$ is $j\bmod k$, each fold re-solving the same least-squares problem on the others); $r^{\max}_{\mathrm{ring}}$ the largest radius of the ring of minima; $\Delta q=$ \texttt{grid.dq}; and $\tau_\sigma,\tau_{\mathrm{CV}},\tau_\rho,\nu$ the limits \texttt{fit.max\_rms\_meV}, \texttt{max\_cv\_rms\_meV}, \texttt{max\_abs\_res\_meV}, \texttt{min\_points\_per\_coef} (0.5, 0.5, 2.0, 10 for CsSnI$_3$).
\begin{equation}\label{eq:si_G5}
\begin{aligned}
G_5=\ &\bigl[\mathcal D'=\varnothing\vee\texttt{allow\_missing\_frames}\bigr]\wedge\bigl[\text{fitted}\bigr]\wedge\bigl[\sigma_{\mathrm{CV}}\le\tau_{\mathrm{CV}}\bigr]\wedge\bigl[\sigma\le\tau_\sigma,\ \rho\le\tau_\rho\bigr]\\
&\wedge\bigl[n_{\mathrm{fit}}\ge8\nu\bigr]\wedge\bigl[r^{\max}_{\mathrm{ring}}\le r_{\max}-2\Delta q\bigr]\wedge\bigl[\text{D4 secondaries hard, interior, within their share}\bigr],
\end{aligned}
\end{equation}
with codes \texttt{MISSING\_FRAMES} ($\mathcal D'$: planned frames without an energy in the fit), \texttt{NO\_FIT}, \texttt{CV\_ERROR}, \texttt{RESIDUALS}, \texttt{THIN\_GRID} (eight coefficients), \texttt{COVERAGE},\footnote{For the plane fit the margin of two steps and the count of eight coefficients are constants of \texttt{gates.py}; \texttt{policy.fit.edge\_margin\_steps} is read by the \texttt{meier} and \texttt{surface} variants only.} and under \texttt{policy.d4} (CsSnI$_3$ $X_5^+$): every quartic scan has positive curvature \gc{SECONDARY\_SOFT}, no scan minimum sits at an end of its range \gc{SECONDARY\_SCAN\_EDGE}, and the largest released energy is $\le$ \texttt{d4.max\_fraction} (0.5) $\times$ the quantity named by \texttt{d4.compare\_to} (the fit's ring corrugation), unless \texttt{accept\_secondary\_hold} is recorded \gc{SECONDARY\_COUPLED}.

\emph{Variants.} Surface without a fit (\texttt{surface}, Ca$_3$Ti$_2$O$_7$): every node and anchor has an energy \gc{THIN\_GRID}, \gc{COVERAGE}; the grid minimum lies $\ge$ \texttt{policy.surface.edge\_margin\_steps} (2) steps from the edge \gc{COVERAGE}; $\max|E(x,y)-E(-x,y)|$ and $\max|E(x,y)-E(x,-y)|$, the two parent mirrors of the plane, are $\le$ \texttt{surface.mirror\_tol\_meV} (1.0~meV/f.u.) \gc{MIRROR}; the campaign's stand-alone script reproduces them to $10^{-3}$ and the phase energies to 0.01~meV \gc{SCRIPT\_FAILED}, \gc{SCRIPT\_MISMATCH}. Manganites (\texttt{meier\_QP}, \texttt{meier\_joint}; eq.~\Mref{eq:meier}; per 30-atom cell, \texttt{inmno3.yaml}: 5, 6 and 12~meV for the plane fit, 15 for the joint fit; ErMnO$_3$ 8, 10, 25~meV; $\nu=10$, \texttt{edge\_margin\_steps} 0.5): $\sigma_{\mathrm{CV}}\le\tau_{\mathrm{CV}}$; $\sigma\le\tau_\sigma$, $\rho\le\tau_\rho$; $n_{\mathrm{fit}}\ge\nu\,n_{\mathrm{coef}}$ ($n_{\mathrm{coef}}$ fitted coefficients, a fixed $a_P$ excluded); $\max(Q^\ast,Q_{\mathrm{coupled}})\le q_{\max}-$ \texttt{edge\_margin\_steps} $\cdot\Delta q$ \gc{CV\_ERROR}, \gc{RESIDUALS}, \gc{THIN\_GRID}, \gc{COVERAGE}. Six-component fit (\texttt{landau6d}; 1.0, 1.0, 5.0~meV/f.u., $\nu=10$): the invariants built number $\sum_{d=2}^{d_{\max}}m_d$, with $m_d$ the coefficients of Molien's series, none of odd degree \gc{BASIS\_INCOMPLETE}; 10-fold CV and 20\,\% hold-out RMS $\le\tau_{\mathrm{CV}}$ \gc{CV\_ERROR}; residuals \gc{RESIDUALS}; points per coefficient $\ge\nu$ \gc{THIN\_GRID}; the lowest stationary point on the symmetric lines is \texttt{fit.six.child\_phase} ($R3c$) with Hessian index 0, and the lowest of 60 random-start minimizations agrees with it to 0.5~meV/f.u.\ \gc{WRONG\_GROUND\_STATE}. Path (\texttt{path}; \texttt{policy.path.network}): one energy per point \gc{THIN\_GRID}; every end structure that \texttt{StructureMatcher} (\texttt{ltol} 0.02, \texttt{stol} 0.05, $1^\circ$) identifies with an end of another stored path has $|E^{\mathrm{own}}-E^{\mathrm{other}}|/z\le$ \texttt{cross\_setting\_meV} (0.1 for HfO$_2$; 1.0 in the BaTiO$_3$ live configuration) \gc{CROSS\_SETTING}; every declared chain connects this path's ends by structure and $\bigl|\Delta E^{\mathrm{own}}_{\mathrm{end}}-\sum_{T\in\mathrm{chain}}\Delta E^{T}_{\mathrm{end}}\bigr|\le$ \texttt{closure\_meV} (0.1~meV/f.u.) \gc{CLOSURE}. Plane (\texttt{plane2}): complete orbits \gc{THIN\_GRID}, residuals \gc{FIT\_RMS}, \gc{FIT\_RESIDUAL}, an interior minimum \gc{MIN\_ON\_BORDER}; the INCAR tags that fix the functional and the numerical settings and the POTCAR data sets of the plane's origin equal those of $\lambda=0$ of every finished path of the campaign, read from the written inputs \gc{SETTINGS\_MISMATCH} (metadata only; added 31 October 2026, Section~\ref{sec:si_live_settings}); and the plane's origin agrees with $E(\lambda=0)$ of every finished path to \texttt{plane2.network\_meV} (1.0~meV/f.u.) \gc{NETWORK} (added after the evaluation freeze). Neither cross-run test was active during the prospective BaTiO$_3$ campaign.

\paragraph{$G_6$, claims verified (after $f_6$).}
$f_6$ writes rows $(\text{name},v,v',\tau)$: a reported quantity $v$, its value $v'$ by a route independent of the node that produced it, and a tolerance $\tau$ ($\tau=\varnothing$: exact comparison of integers, labels, lists).
\begin{equation}\label{eq:si_G6}
G_6=\bigl[\text{rows}\neq\varnothing\bigr]\wedge\bigwedge_{(v,v',\tau)}\Bigl[(\tau=\varnothing\wedge v=v')\vee(\tau\neq\varnothing\wedge|v-v'|\le\tau)\Bigr],
\end{equation}
with codes \texttt{NOT\_REDERIVED}, \texttt{MISMATCH}. For the two-irrep grid (\texttt{f6\_skeptic.py}) the rows are: $A_p$ and $A_s$ of every irrep from $\lVert\mathbf e_i\rVert/\sqrt n$ and $A_p\sqrt s$ ($2$, $3\times10^{-4}$~Å); the active set, primaries, held-at-zero set and $c_1$ group from a second run of the selection rules (exact); for a polar $(Q,\Phi)$ grid, $\hat{\mathbf e}_2$ against $(\mathcal T\hat{\mathbf e}_1-\mathcal T^2\hat{\mathbf e}_1)/\sqrt3$, $\mathcal T$ the lost translation ($10^{-9}$); the frames per space group from frames rebuilt as $\rv_0+\sum_iQ_i\hat{\mathbf e}_i$ and re-assigned by spglib (exact); every linked energy re-read from the CSV by stored identifier, and the formula units from the system file (exact); each Landau coefficient and the RMS from a second solution by QR factorization on a fit set rebuilt from the frames ($10^{-6}$ relative); ring radius, depth and corrugation by a bounded scalar minimization per angle instead of the script's grid search (0.011~Å, 0.05~meV); under D4, each scan's curvature, relaxed amplitude and released energy from a second quartic fit. For a path (\texttt{f6\_path.py}) the space group and $d_{\min}$ are recounted on the \emph{stored} POSCARs, the energies re-read from the collector's CSV rather than \texttt{result.json}, and the cross-setting differences and closure gaps re-derived with the identity $\mathrm{gap}=\Delta_{\mathrm{end}}-\Delta_{\mathrm{start}}-\sum\Delta_{\mathrm{junction}}$ ($10^{-9}$~meV); \texttt{f6\_plane2.py} and \texttt{f6\_grid6.py} do the same for their analyses.

\paragraph{How symmetry is evaluated programmatically.}
Three operations carry every symmetry statement above. First, a space group is always a \emph{number} returned by spglib for a Cartesian structure at a stated \texttt{symprec}; settings, origins and Hermann--Mauguin symbols are never compared, so $Pnma$ and $Pbnm$ are one group and a setting cannot fail a gate; the tolerances in force are listed with each predicate above. Second, $H\subset G$ is pymatgen's \texttt{SpaceGroup.is\_subgroup}, a walk of the maximal-subgroup tables keyed by number, with the index relation $t\cdot s=i$; the gate neither constructs the operations of $H$ nor verifies that a particular basis realizes the relation, which is left to $\det\mathbf B=s\,z_H/z_G$ and the atom-matching basis search. Third, the irrep projection is never verified by applying representation matrices inside a gate; the gates test its \emph{consequences} on structures: (a) each irrep's field has the norm its amplitude implies ($\pr{1}{4}$); (b) the child rebuilt from the sum of all fields, and the child file, have the group $H$ ($\pr{1}{3}$); (c) a frame with irrep $i$ alone at the child amplitude has exactly the isotropy subgroup $H_i$ that ISODISTORT lists, at $10^{-2}$~Å and at no other rung (eq.~\Mref{eq:single}, $\pr{2}{6}$); (d) every superposition $\rv_0+\sum_iQ_i\hat{\mathbf e}_i$ has the group the descent tree assigns to its non-zero set (eqs~\Mref{eq:tree}, \Mref{eq:audit}; $\pr{2}{8}$), which for a $K_3$ pair sampled in $(Q,\Phi)$ means $P6_3cm$, $P\bar3c1$ or $P3c1$ by the class of $\Phi$; (e) the child's field, mapped atom by atom and aligned to the matching domain, projects onto each unit vector to its own amplitude ($\pr{2}{5}$). Operations act on vectors inside a gated computation only in the six-component case, where $f_1$ derives the 96 operations of $Pm\bar3m$ at the $R$ point as the matrices mapping the six component fields onto $\pm$ one another and $\pr{1}{5}$ demands 96 distinct signed permutations with residual $\le10^{-4}$~Å.\footnote{The check of the $K_3$ representation matrices for all 72 parent operations (main-text Section~\Mref{sec:res_mn}, Section~\ref{sec:si_k3}) was made by the campaign script that built the second component; it is not a test of any gate. The gated statements on the $K_3$ pair are (c), (d) and the $G_6$ row that rebuilds $\hat{\mathbf e}_2$ from the lost translation.} $\bigcap_{i\in\Aset}H_i=H$ is likewise a lookup in the tree ISODISTORT returns, tested on structures by route (d) and repeated by $f_6$. A wrong irrep label, isotropy subgroup or mode vector therefore fails a gate only through a structure whose spglib group differs from the expected one, a norm that differs from the amplitude, or a projection that misses the child, the three failure modes that the development errors of Section~\ref{sec:si_llm} exercised.

\paragraph{What no gate checks.}
The formulation is complete with respect to \texttt{gates.py}, and its limits should be stated. No gate checks the exchange--correlation functional, Hubbard $U$, magnetic order, cut-off, $k$-mesh or PAW set: $G_4$ requires them to be the same across a campaign (and equal to the template, for a path), not appropriate; the system profile is an input. No gate checks the physical plausibility of a landscape: the energy range must lie between 1~meV and 10~eV per formula unit, the parent mirrors must hold and a fit must reproduce and cross-validate its data, but a landscape wrong for a physical reason (a wrong magnetic ground state, an unconverged basis, the fixed-lattice bias that strain policy P5 states by hand) passes every gate if it is self-consistent. No gate verifies the decomposition engine: ISODISTORT's irrep labels, directions, isotropy subgroups and subgroup tree are trusted once consequences (a)--(e) hold, as are pymatgen's subgroup tables for $H\subset G$. No gate checks the correctness of a stored rule or a human decision: a rule of \texttt{recovery.yaml} is applied whenever its key and condition match, and an R5, D4 or \texttt{allow\_missing\_frames} decision is applied as recorded; the gates verify that it was applied, not that it was right. Nor are $A_{\mathrm{thr}}$, the R5 margin, $Q_{\max}$ beyond the coverage margin or the fit limits themselves checked; they are policy values that the configuration states and the audit log hashes. The comparisons with stored campaigns (\texttt{PLAN\_DISAGREE}, \texttt{ORBITS\_DISAGREE}, \texttt{CONTACT\_CUT}, \texttt{SCRIPT\_MISMATCH}) are replay-only regression tests. Finally, the gates are deterministic code, so an error in a test recurs identically in every run; the re-derivation of $G_6$ and the three-way symmetry agreement of $G_2$ are where a mistake in one route is caught by another, and nothing checks a quantity that both routes compute from the same wrong input.

\endgroup

\subsection{Language-model errors caught during development}\label{sec:si_llm}
The design rule that a language model never decides whether a step succeeded comes from errors met while building the workflow, in which a model wrote fluent and plausible crystallographic statements that deterministic checks contradicted. In the HfO$_2$ campaigns, model-written text named No.~137 ``$P4_2/mnm$'' (which is No.~136) and No.~60 ``$Pbcm$'' (No.~57); claimed a four-fold volume increase for a transition of index 4, whereas the index is $t\cdot k$ and only $k$ multiplies the cell; stated that $P4_2/nmc\to Pbcn$ involves no $M$-point modes, whereas it is $t2\cdot k2$ with $M_1$ as its primary irrep (Supplementary Table~\ref{tab:dec_hfo2}); and stated that $Pbcn\to P2_1/c$ doubles the cell with $M$ modes, whereas it is $t2$ with $\Gamma$-point irreps only. A model also classified a HfO$_2$ transition as connecting symmetry-equivalent ferroelectric variants, whereas the end points belong to a distinct non-polar subgroup. Each error was caught by the space-group and decomposition checks of $G_1$; in the final design such text is produced, if at all, only after $G_6$ has passed and is not used by any gate.

\subsection{An example of the audit trail}\label{sec:si_audit}
Every run leaves an audit trail: each structure, input file and decomposition is stored with a hash; each attempt at each node is logged with its input hashes, gate result, error code, fix, who chose the fix (stored rule, sub-agent or human) and retry count; each energy is linked to the frame and amplitudes that produced it through \texttt{point.json} and \texttt{grid\_manifest.json}; and $f_6$ re-derives every reported number from these files (main-text Methods, Section~\Mref{sec:audit}). Supplementary Table~\ref{tab:si_audit} shows the summary that this trail gives for two CsSnI$_3$ replays (main-text Section~\Mref{sec:autonomy}): the fault-free run, and the run with fault F001, in which one job part ends at the walltime limit. The rows are taken from the run records of the replays; the full per-attempt logs, with the hashes of every input, are released with the code.

\begin{table}[!htbp]
\centering\small
\caption{\textbf{The audit trail of two CsSnI$_3$ replays.} Fault-free run (thread \texttt{E1\_cssni3}) and the run with injected fault F001 (job part 3 ends with WALLTIME; Supplementary Table~\ref{tab:si_faultlist}). Gate trail: the gates passed in order. Who fixed: the source of each fix (rule, sub-agent or human). $G_6$: reported quantities re-derived by an independent route and matching the stored values.}
\label{tab:si_audit}
\begin{tabularx}{\linewidth}{@{}>{\raggedright}p{3.2cm}>{\raggedright}X>{\raggedright\arraybackslash}X@{}}
\toprule
Record & Fault-free run & Fault F001 (walltime)\\
\midrule
Gate trail & $G_0$ OK $>$ $G_1$ OK $>$ $G_2$ OK $>$ $G_3$ OK $>$ $G_4$ OK $>$ $G_5$ OK $>$ $G_6$ OK & $G_4$ fail(\texttt{WALLTIME}), then all gates pass\\
Gate failures & 0 & 1 ($G_4$, \texttt{WALLTIME})\\
Fix and who chose it & none & stored rule \texttt{f4\_walltime\_resubmit} $\to$ $f_3$\\
Sub-agent fixes / human interventions & 0 / 0 & 0 / 0\\
Job submissions & not recorded in the summary & 2 submissions; 962 runs before detection, 1{,}002 in total\\
Final state & completed at $f_6$ & completed\\
$G_6$ re-derivation & 43 of 43 quantities match & numbers identical to the stored fit\\
\bottomrule
\end{tabularx}
\end{table}

\section{Detailed views of the crystallographic stages}\label{sec:si_stages}

\begin{figure}[!htbp]
\centering
\includegraphics[width=0.92\textwidth]{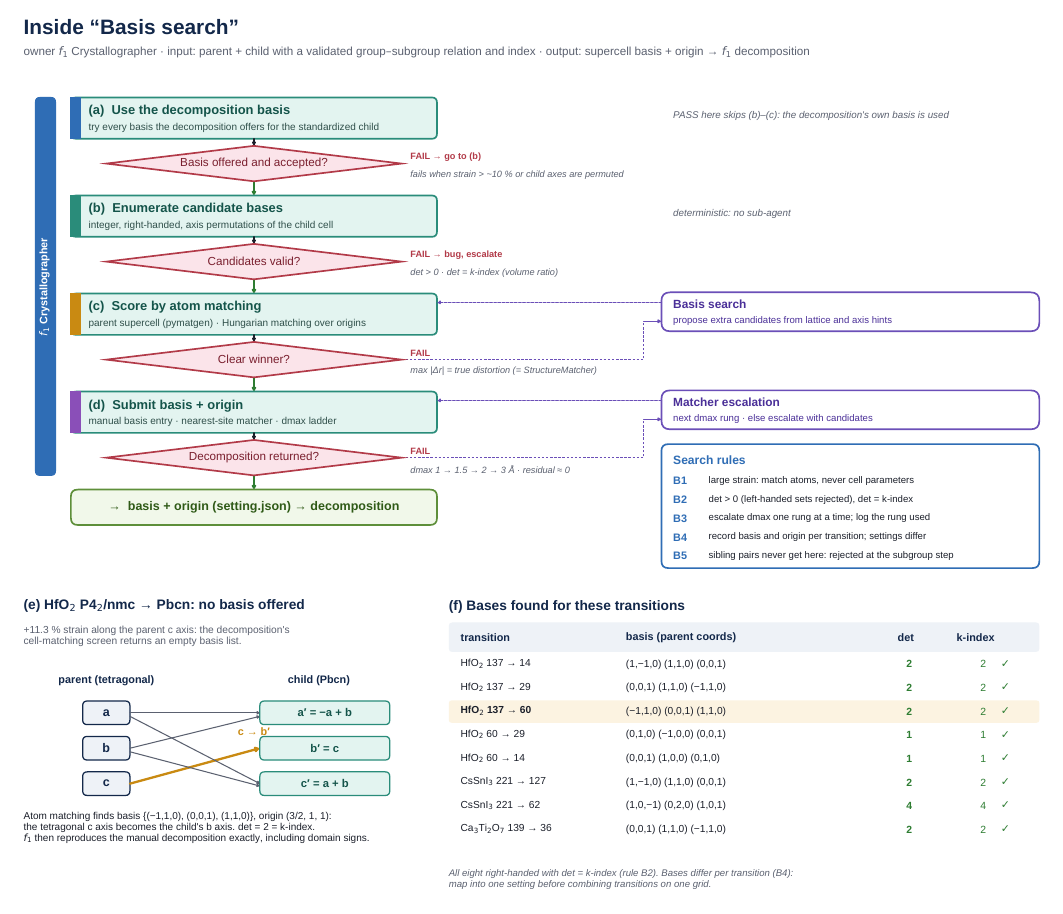}
\caption{\textbf{Basis search ($f_1$).} (a) The bases offered by the decomposition step are tried first. (b,c) If none fits, candidate integer bases with $\det=k$ are enumerated and scored by atom matching. (d) The chosen basis and origin are submitted for the decomposition. Coloured boxes are the steps, in order; pink diamonds are the checks after each step, with the FAIL route (red text) and its condition beneath; purple boxes are the sub-agents a failure is passed to (basis search, matcher escalation), and the blue box lists the search rules B1--B5. (e) HfO$_2$ $P4_2/nmc\to Pbcn$, for which the decomposition offered no basis because the transition carries 11.3\% strain along the tetragonal $c$ axis: atom matching finds the basis $(\bar1,1,0)(0,0,1)(1,1,0)$ with $\det=2=k$. (f) Bases found for the transitions of this work, in parent coordinates, with their determinant and $k$-index (highlighted row, the case in \textbf{e}). Space groups are given by number.}
\label{fig:s_basis}
\end{figure}

\begin{figure}[!htbp]
\centering
\includegraphics[width=0.92\textwidth]{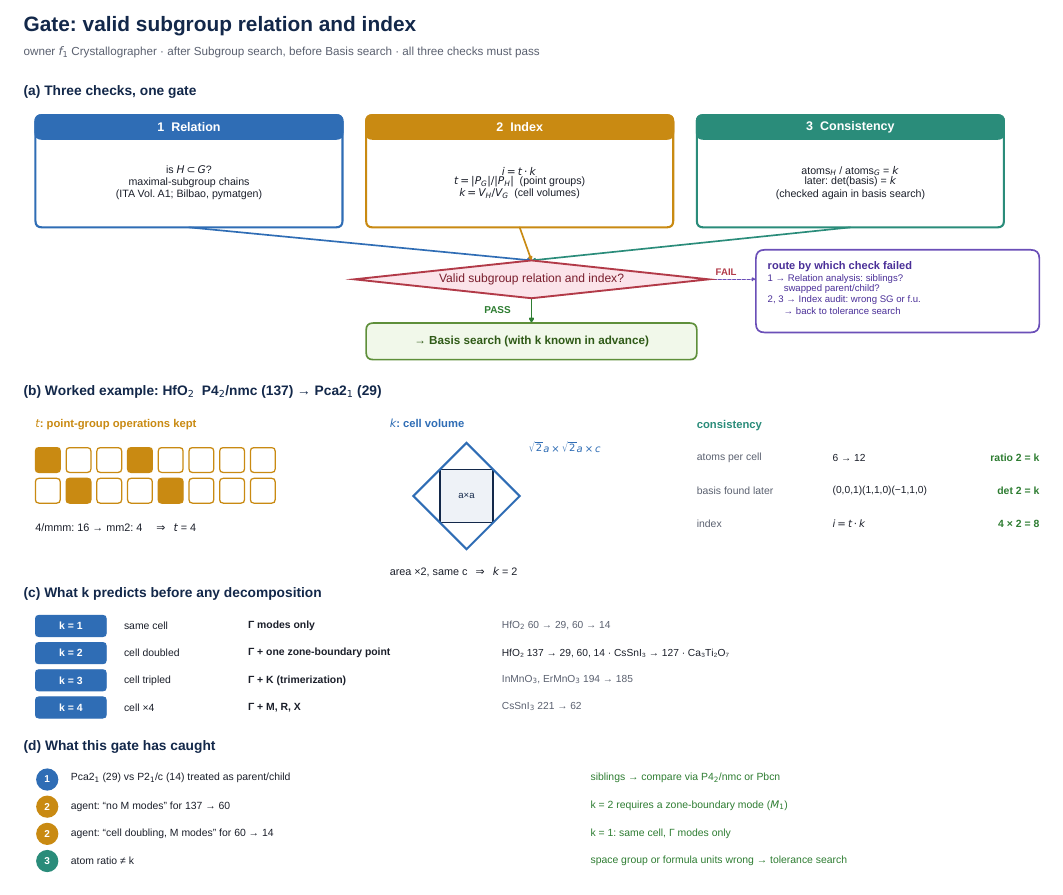}
\caption{\textbf{Gate on the subgroup relation and index ($G_1$).} (a) Three checks: relation $H\subset G$, index $i=t\cdot k$, and agreement between the space group implied by the decomposition and the one assigned by the two symmetry engines of $f_1$. (b) Worked example, HfO$_2$ $P4_2/nmc$ (No.~137) $\to$ $Pca2_1$ (No.~29): the point group falls from $4/mmm$ (16 operations) to $mm2$ (4), so $t=4$; the cell volume doubles, so $k=2$; hence $i=t\cdot k=8$, consistent with the atom ratio and the basis determinant. (c) What $k$ predicts before any decomposition: $k=1$, $\Gamma$ modes only; $k=2$, $\Gamma$ and one zone-boundary point; $k=3$, $\Gamma$ and $K$ (trimerization); $k=4$, $\Gamma$, $M$, $R$ and $X$, with examples from this work. (d) Errors this gate caught during development (numbered), with the check that caught each. Failed checks are routed to the relation analysis, index audit or tolerance search (purple box); see Supplementary Table~\ref{tab:gates} for the error codes.}
\label{fig:s_index}
\end{figure}

\begin{figure}[!htbp]
\centering
\includegraphics[width=0.92\textwidth]{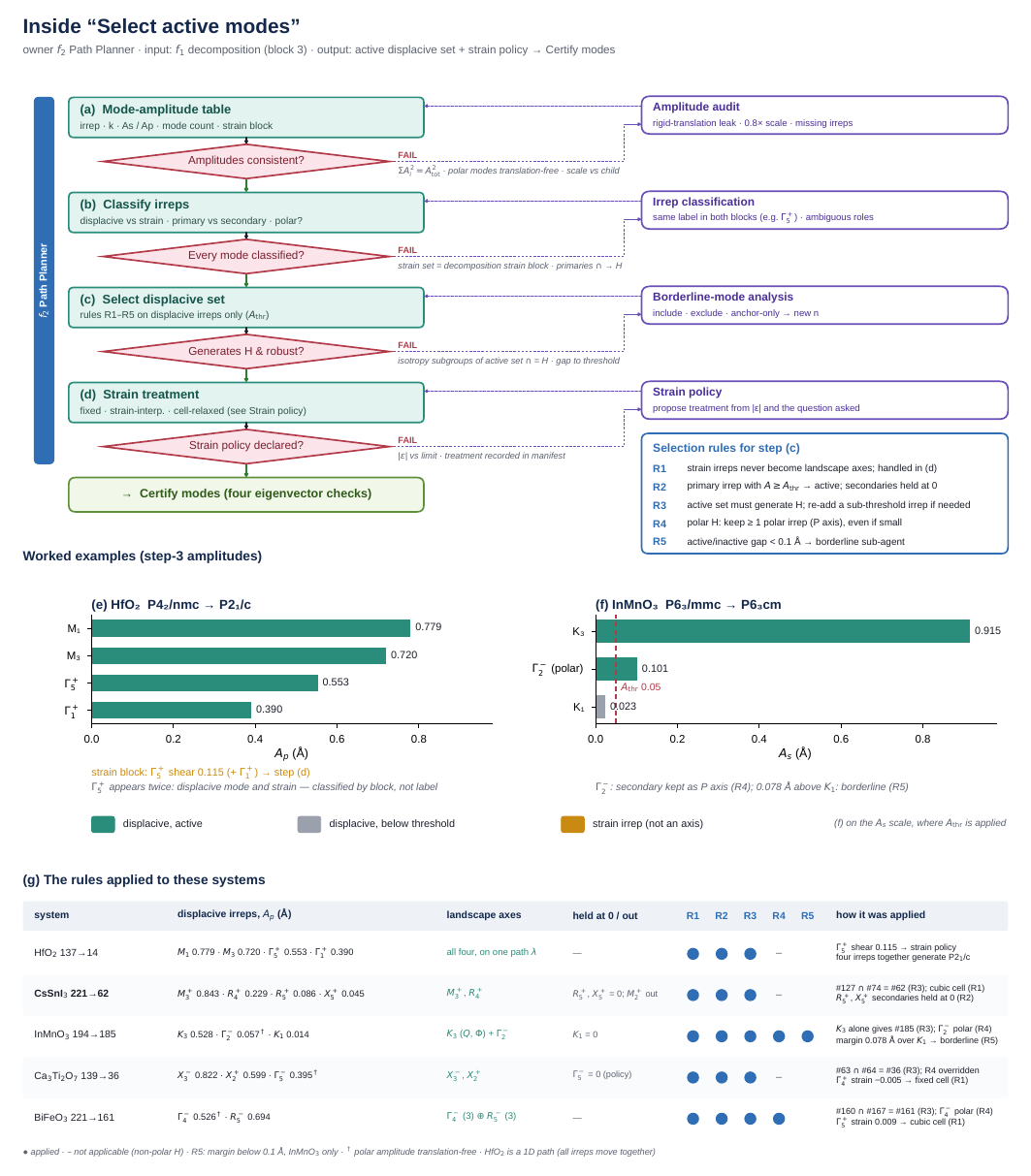}
\caption{\textbf{Selection of the active modes ($f_2$).} (a) Mode-amplitude table. (b) Classification of irreps. (c) Selection of the displacive set by rules R1--R5. (d) Strain treatment. Coloured boxes are the steps, in order; pink diamonds are the checks after each step, with the FAIL route (red text) and its condition beneath; purple boxes are the sub-agents a failure is passed to (amplitude audit, irrep classification, borderline-mode analysis, strain policy), and the blue box lists rules R1--R5. (e) HfO$_2$ $P4_2/nmc\to P2_1/c$, amplitudes $A_p$ in \AA: $M_1$ 0.779, $M_3$ 0.720, $\Gamma_5^+$ 0.553 and $\Gamma_1^+$ 0.390 are displacive and active (green); the $\Gamma_5^+$ shear strain (0.115) goes to step (d) (orange). (f) InMnO$_3$ $P6_3/mmc\to P6_3cm$, where the polar $\Gamma_2^-$, a secondary kept as the polarization axis by R4, is above threshold but only 0.076~Å above the largest inactive irrep ($K_1$; 0.078~Å on raw $A_s$, the value printed in the panel), so the selection is borderline (R5). The dashed red line is $A_{\mathrm{thr}}=0.05$~\AA. (g) The rules applied to each system; in Ca$_3$Ti$_2$O$_7$ the polar $\Gamma_5^-$ is held at zero by a recorded policy override of R4.}
\label{fig:s_select}
\end{figure}

\begin{figure}[!htbp]
\centering
\includegraphics[width=0.92\textwidth]{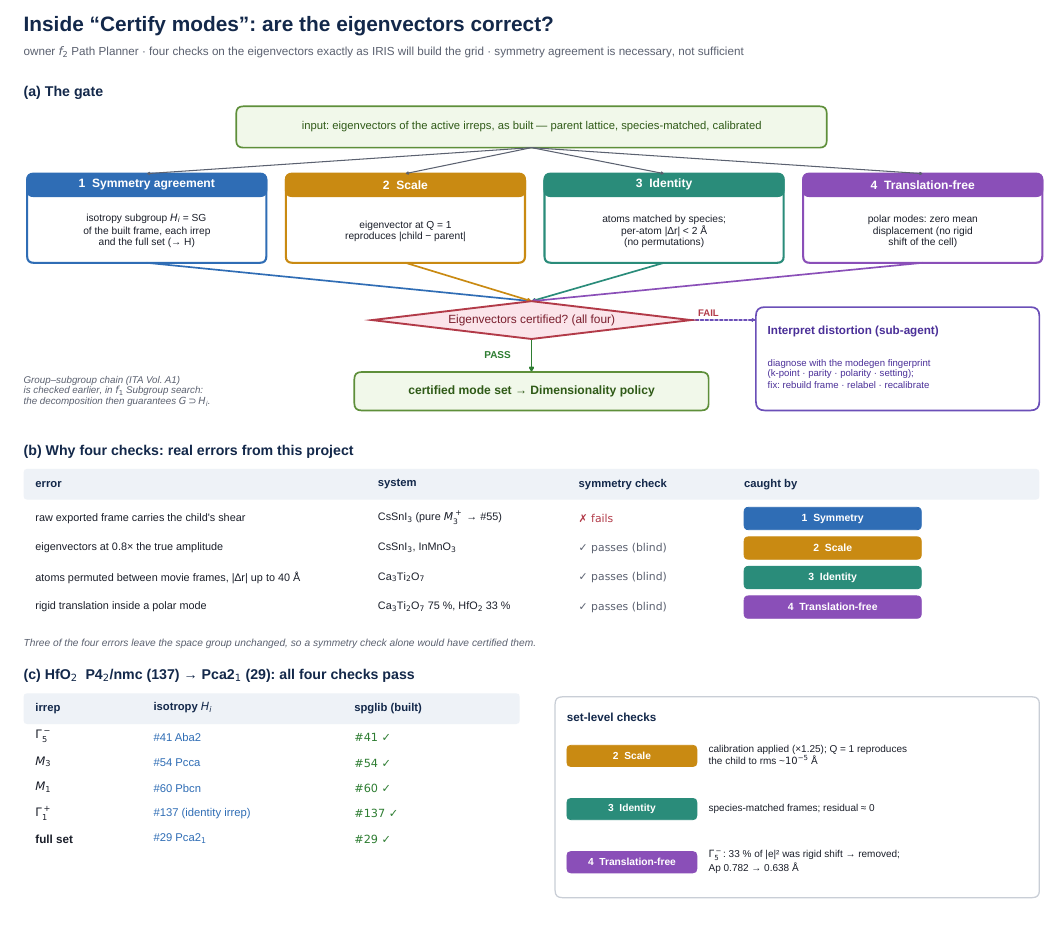}
\caption{\textbf{Certification of the mode vectors ($f_2$).} (a) The four checks of eqs.~\Mref{eq:match}--\Mref{eq:single}. (b) The real errors from this work that make each check necessary. (c) HfO$_2$ $P4_2/nmc\to Pca2_1$: all four checks pass; each irrep's single-mode frame has the space group of its isotropy subgroup (spglib), the calibration factor is 1.25 and the polar $\Gamma_5^-$ carries a rigid shift of 33\% of its squared norm, which is removed. Check colours: symmetry agreement, blue; scale, orange; identity, green; translation-free, purple. A failed check goes to the distortion-interpretation sub-agent (purple box). Three of the four errors in \textbf{b} leave the space group unchanged, so a symmetry check alone would have certified them.}
\label{fig:s_certify}
\end{figure}

\begin{figure}[!htbp]
\centering
\includegraphics[width=0.92\textwidth]{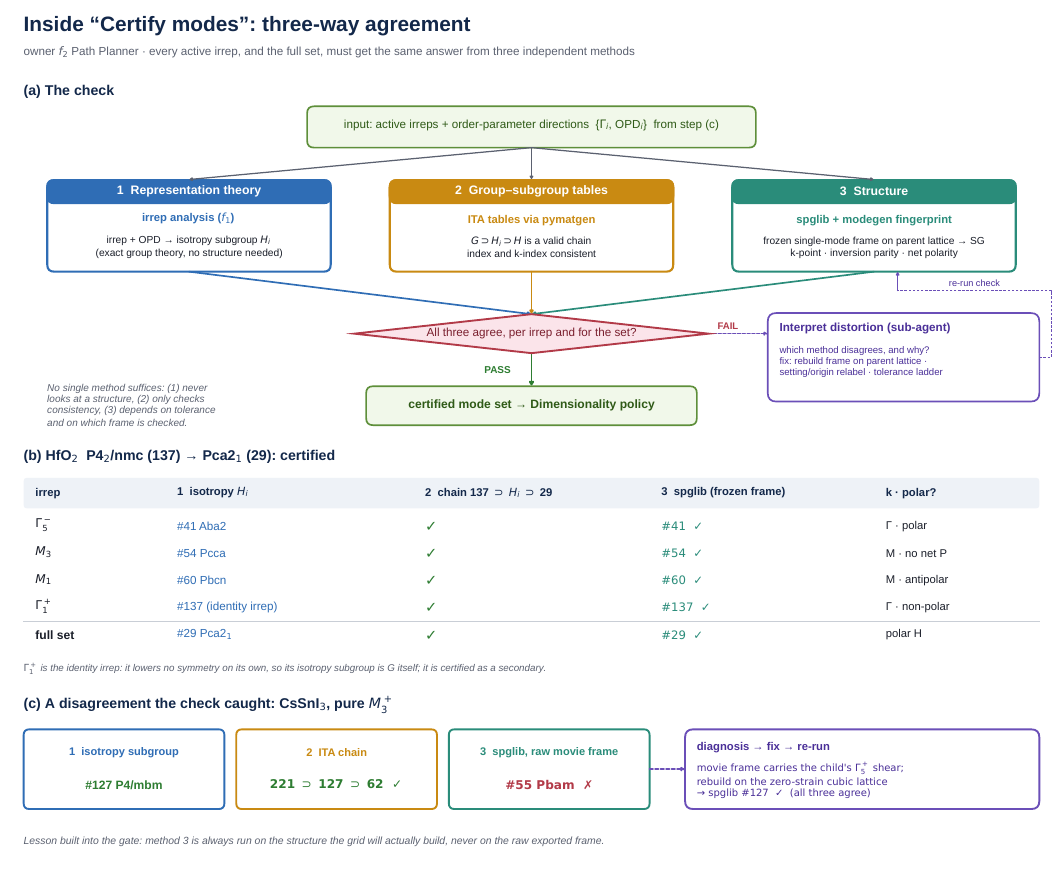}
\caption{\textbf{Three-way symmetry agreement ($f_2$).} (a) Every active irrep, and the full set, must give the same space group from three independent routes (isotropy subgroup, the space group of the built frame, and the descent tree). (b) HfO$_2$ $P4_2/nmc\to Pca2_1$, certified. (c) A disagreement caught for CsSnI$_3$ with $M_3^+$ alone, where the raw frame came out as No.~55 ($Pbam$) instead of No.~127 ($P4/mbm$): the raw frame carried the child's $\Gamma_5^+$ shear, and rebuilding it on the cubic parent lattice restored agreement (Supplementary Section~\ref{sec:si_primaries}). In \textbf{a} and \textbf{b}, method 1 is representation theory (isotropy subgroup $H_i$), method 2 the group--subgroup chain from the International Tables for Crystallography, Vol.~A~\cite{ITA2016}, and method 3 the space group of the built frame (spglib with a modes fingerprint); $\Gamma_1^+$ is the identity irrep, whose isotropy subgroup is the parent. A disagreement goes to the distortion-interpretation sub-agent (purple box).}
\label{fig:s_three}
\end{figure}

\begin{figure}[!htbp]
\centering
\includegraphics[width=0.92\textwidth]{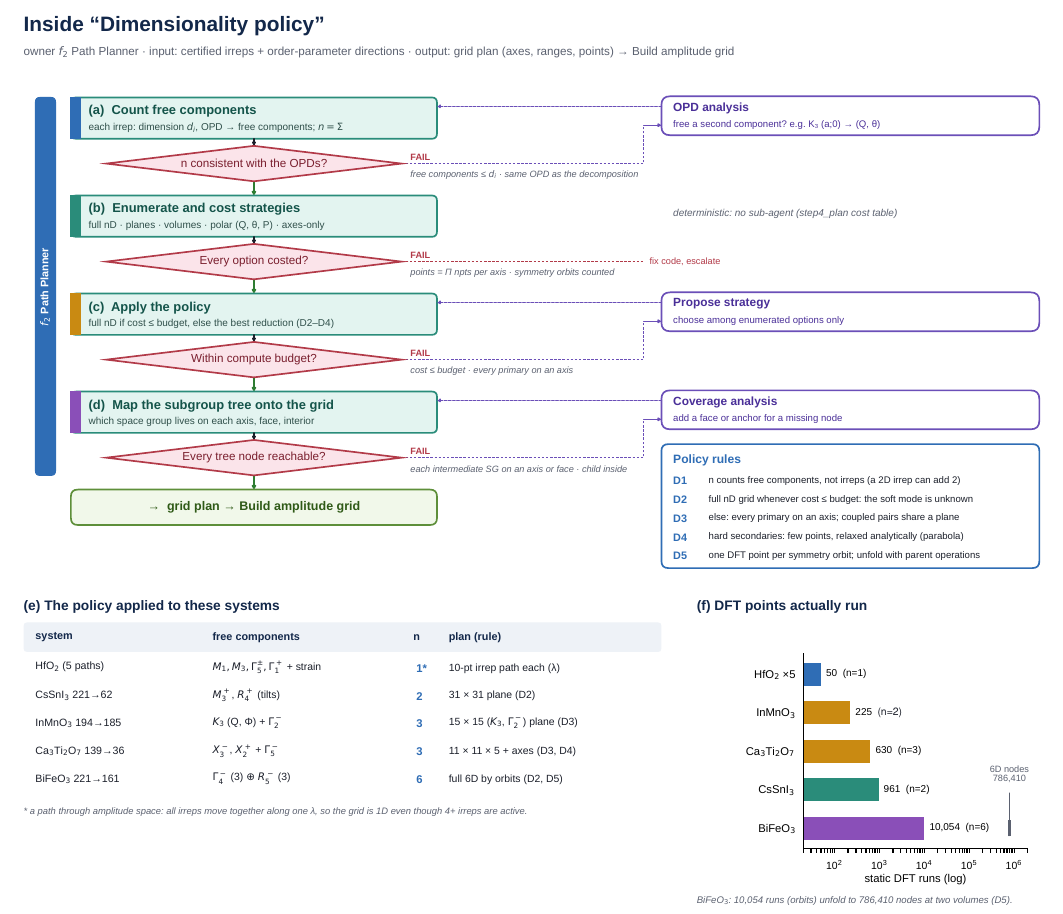}
\caption{\textbf{Dimensionality policy ($f_2$).} (a) Free order-parameter components are counted irrep by irrep (D1). (b) Strategies are enumerated and costed. (c) The policy D1--D5 is applied. (d) The descent tree is mapped onto the grid to give the expected space group of every node. Coloured boxes are the steps, pink diamonds the checks with their FAIL routes (red), purple boxes the sub-agents (order-parameter-direction analysis, strategy proposal, coverage analysis), and the blue box the policy rules D1--D5. (e) The policy applied to each system: free components, grid dimension $n$ and plan (the Ca$_3$Ti$_2$O$_7$ row of \textbf{e} and bar of \textbf{f} show the earlier 630-frame ($X_3^-$, $X_2^+$, $\Gamma_5^-$) grid, superseded by the $37\times37$ plane of 1{,}372 frames with $\Gamma_5^-$ at zero, Supplementary Section~\ref{sec:si_ca3ti2o7_dft}; the D3 labels mark the intended reduction, which is not yet implemented). (f) DFT points actually run, on a logarithmic axis, with the grid dimension $n$; for BiFeO$_3$, 9{,}597 symmetry orbits (10{,}054 runs with refinement points and controls) unfold to 786{,}410 nodes (D5).}
\label{fig:s_dim}
\end{figure}

\begin{figure}[!htbp]
\centering
\includegraphics[width=0.92\textwidth]{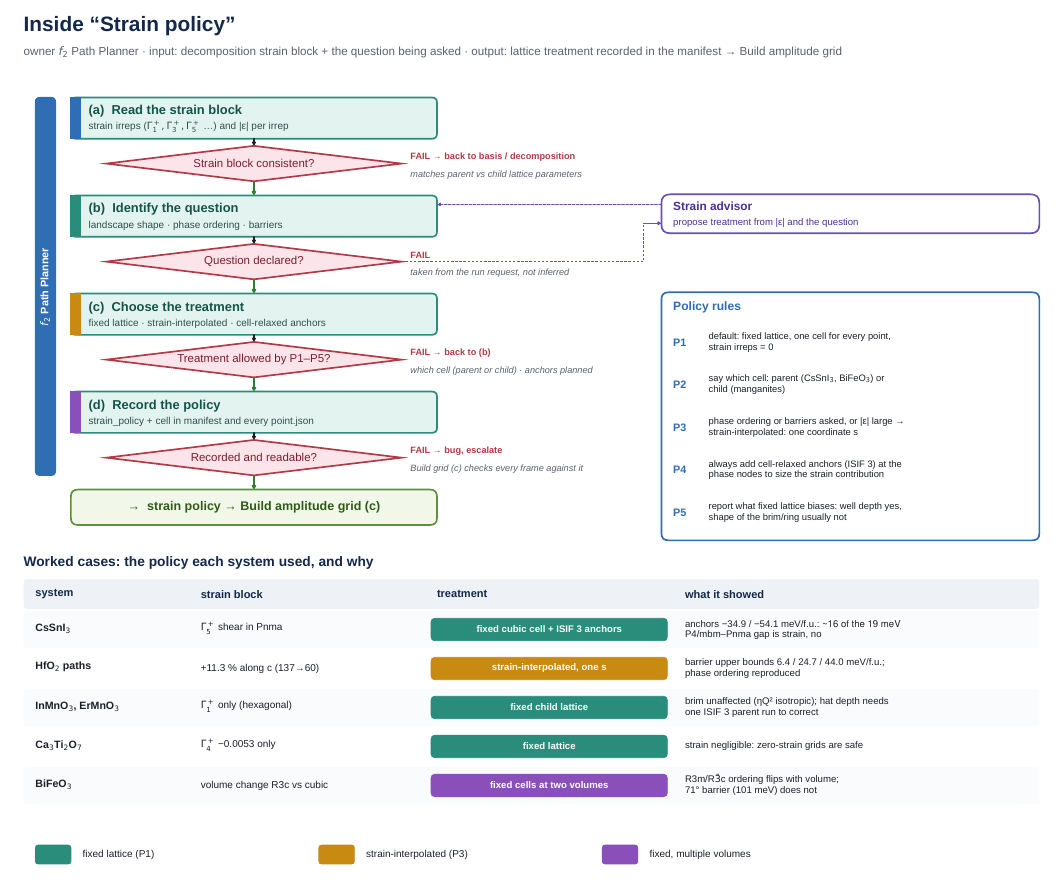}
\caption{\textbf{Strain policy ($f_2$).} (a) The strain block of the decomposition is read. (b) The scientific question is identified. (c) A lattice treatment is chosen by policies P1--P5: fixed lattice, stated cell, strain-interpolated, relaxed anchors, and a statement of what the fixed lattice biases (P4 and P5 are not yet produced by the workflow itself; main-text Methods). (d) The policy is recorded in the grid manifest, and the build stage checks every frame against it. Coloured boxes are the steps, pink diamonds the checks with their FAIL routes (red), the purple box the strain advisor sub-agent and the blue box the policy rules P1--P5. Bottom: the policy each system used and what it showed, coloured by treatment (green, fixed lattice, P1; orange, strain-interpolated, P3; purple, fixed cells at two volumes); values as in main-text Sections~\Mref{sec:res_perov}--\Mref{sec:res_hfo2}.}
\label{fig:s_strain}
\end{figure}

\begin{figure}[!htbp]
\centering
\includegraphics[width=0.92\textwidth]{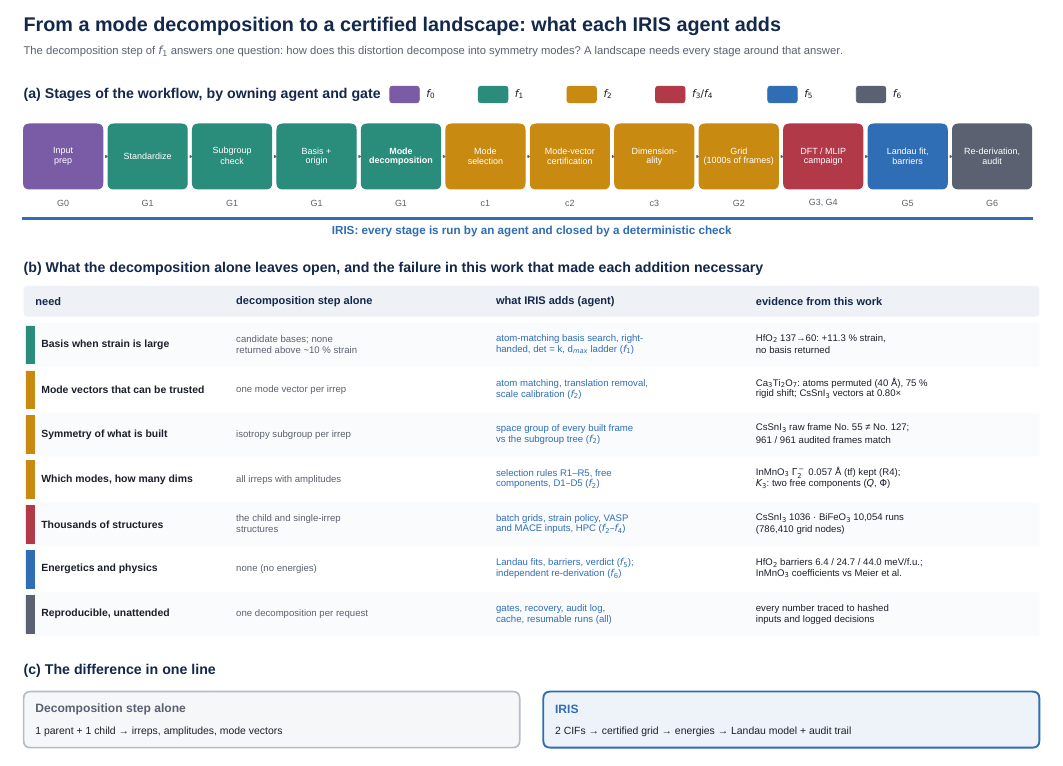}
\caption{\textbf{From a mode decomposition to a certified landscape: what each IRIS agent adds.} (a) The stages of the workflow coloured by the agent that runs them ($f_0$ purple, $f_1$ teal, $f_2$ orange, $f_3$/$f_4$ red, $f_5$ blue, $f_6$ grey), with the gate ($G_i$) or Planner check ($c_i$) that closes each one beneath; the mode decomposition is one stage of $f_1$. (b) For each requirement of a landscape: what the decomposition step returns on its own, what IRIS adds and which agent adds it, and the failure in the campaigns of this work that made the addition necessary (Extended Data Table~\Mref{tab:errors}). (c) Summary: one decomposition per parent--child pair against an audited grid, its energies and a Landau model from two CIFs.}
\label{fig:s_iso}
\end{figure}

Supplementary Figs.~\ref{fig:s_basis}--\ref{fig:s_iso} expand the stages of $f_1$ and $f_2$ summarized in Extended Data Figs~\Mref{fig:workflow} and~\Mref{fig:planner}, each with the systems where the stage mattered: the basis search and the subgroup gate of $f_1$ (Supplementary Figs.~\ref{fig:s_basis} and~\ref{fig:s_index}); mode selection, the four mode-vector checks and the three-way symmetry agreement of $f_2$ (Supplementary Figs.~\ref{fig:s_select}--\ref{fig:s_three}); the dimensionality and strain policies (Supplementary Figs.~\ref{fig:s_dim} and~\ref{fig:s_strain}); and a map of what each agent adds around the decomposition step, with the evidence from this work (Supplementary Fig.~\ref{fig:s_iso}).

Supplementary Table~\ref{tab:si_policies} lists the policies of the Path Planner, the mode-selection rules R1--R5, the dimensionality policies D1--D5 and the strain policies P1--P5, with their implementation status and the case in this work where each one mattered.

\begin{table}[!tp]
\centering\scriptsize
\caption{\textbf{Policies of the Path Planner ($f_2$), with the cases in this work where each mattered.} Mode-selection rules R1--R5 decide which irreps become axes; dimensionality policies D1--D5 decide the dimension and sampling of the grid; strain policies P1--P5 decide the lattice. Definitions and equations in main-text Methods, Section~\Mref{sec:planner}; flowcharts in Supplementary Figs~\ref{fig:s_select}, \ref{fig:s_dim} and~\ref{fig:s_strain}. $A_{\mathrm{thr}}=0.05$~\AA\ in every campaign; amplitudes are translation-free.}
\label{tab:si_policies}
\renewcommand{\arraystretch}{1.05}
\begin{tabularx}{\linewidth}{@{}>{\raggedright}p{0.5cm}>{\raggedright}p{6.2cm}>{\raggedright}p{2.0cm}>{\raggedright\arraybackslash}X@{}}
\toprule
& Rule & Status & Example in this work\\
\midrule
\multicolumn{4}{@{}l}{\textit{Mode selection: which irreps become axes}}\\
R1 & Strain irreps never become axes; an irrep with both displacive and strain parts stays a displacive irrep and only its strain part goes to the strain policy. & implemented & HfO$_2$ T3: the displacive $\Gamma_5^+$ is a mode on the path, its shear-strain part is handled by P3.\\
R2 & A primary displacive irrep is active if $A^{\mathrm{tf}}\ge A_{\mathrm{thr}}$; a secondary, induced by the primaries, is held at zero even above the threshold. & implemented & CsSnI$_3$: $M_3^+$ (0.843~\AA) and $R_4^+$ (0.229~\AA) active; the secondaries $R_5^+$ and $X_5^+$ held at zero. Ca$_3$Ti$_2$O$_7$: ranking on translation-free amplitudes places $\Gamma_5^-$ ($A_s^{\mathrm{tf}}=0.559$~\AA; $X_2^+$ 0.847~\AA) below $X_2^+$.\\
R3 & The active set must generate the child, $\bigcap_i H_i=H$; otherwise $G_2$ fails (\texttt{R3\_FAIL}). & implemented & CsSnI$_3$: $P4/mbm$ (No.~127) $\cap$ $Imma$ (No.~74) $=$ $Pnma$ (No.~62).\\
R4 & A polar child keeps at least one polar irrep, even below threshold, as the polarization axis. & implemented & Manganites: the secondary $\Gamma_2^-$ kept as the $P$ axis. Ca$_3$Ti$_2$O$_7$: the policy file overrides R4 and holds $\Gamma_5^-$ at zero, recorded in the audit log.\\
R5 & If the smallest active amplitude exceeds the largest inactive one by less than 0.1~\AA, the borderline irreps go to the sub-agent or human: include, exclude or anchor only. & implemented (answer applied) & InMnO$_3$: $\Gamma_2^-$ ($A_s^{\mathrm{tf}}=0.099$~\AA) exceeds $K_1$ by only 0.076~\AA\ (0.078~\AA\ on raw $A_s$); the human kept it as an axis.\\
\midrule
\multicolumn{4}{@{}l}{\textit{Dimensionality: how many axes and how densely sampled}}\\
D1 & The grid dimension $n$ counts free order-parameter components, not irreps. & implemented & CsSnI$_3$ $n=2$; manganites $Q$, $\Phi$ and $P$; BiFeO$_3$ $n=6$ (three components each of $\mathbf P$ and $\mathbf R$).\\
D2 & The full grid is used if it fits the budget, $N=\prod_i m_i\le B$. & implemented & CsSnI$_3$ $31\times31=961$ frames ($B=1{,}100$); InMnO$_3$ plane $15\times15$.\\
D3 & An over-budget grid is reduced: primaries keep axes, coupled pairs share planes. & not implemented (over budget: $G_2$ fails, human decides) & ---\\
D4 & Hard secondaries get short scans along their own vector at declared points of the primaries; if the energy released by relaxing one exceeds half the smallest energy difference the run interprets, $G_5$ fails (\texttt{SECONDARY\_COUPLED}) and the sub-agent or human decides whether it stays at zero, with the caveat stated. & implemented & CsSnI$_3$ $X_5^+$: relaxing it at 0.8 of the child primaries releases 1.9~meV/f.u., more than the 1.4~meV/f.u.\ ring corrugation; kept at zero with the caveat (Supplementary Fig.~\ref{fig:s_x5}).\\
D5 & One DFT point per symmetry orbit of the grid, unfolded to every node with the parent operations. & implemented & BiFeO$_3$: 786{,}410 grid nodes reduced to 9{,}597 orbits, computed in 10{,}054 runs with refinement points and controls.\\
\midrule
\multicolumn{4}{@{}l}{\textit{Strain: which lattice the frames use}}\\
P1 & One fixed lattice for every frame, strain irreps at zero (default). & implemented & Every landscape except the HfO$_2$ paths.\\
P2 & The fixed cell is stated explicitly: the parent cell or the child cell. & implemented & Parent cell: CsSnI$_3$, Ca$_3$Ti$_2$O$_7$, BiFeO$_3$; child cell: manganites.\\
P3 & The lattice is interpolated with a single strain coordinate along the path. & implemented & HfO$_2$ paths T1--T6 (main-text eq.~\Meqref{eq:path}).\\
P4 & Cell-relaxed anchors (ISIF~3) at the phase nodes size the strain contribution. & by script, not yet in the workflow & CsSnI$_3$ anchors: strain gives about 16 of the 19.2~meV/f.u.\ between $Pnma$ and $P4/mbm$; BiFeO$_3$ relaxed anchors.\\
P5 & The report states what the fixed lattice biases (well depths, but usually not the shape of the brim or ring). & by hand, not yet in the workflow & The fixed-lattice caveats in the main-text Results and Limitations.\\
\bottomrule
\end{tabularx}
\end{table}

\subsection{Choice of the primary irreps and the strict single-mode check}\label{sec:si_primaries}

\paragraph{Primaries and secondaries.} Rule R2 (main-text Methods) makes only primary irreps axes of the landscape. The primaries are found from the amplitude table and the isotropy subgroups $H_i$: the displacive irreps are taken in order of decreasing translation-free amplitude $A_s^{\mathrm{tf}}$, and each is added only if its $H_i$ lowers the symmetry of the running intersection, until the intersection equals $H$. Every other displacive irrep is a secondary; it is held at zero on the grid and sampled separately where its coupling matters (D4), unless rule R4 keeps it as the polarization axis. If the list is exhausted before the intersection reaches $H$, check $c_1$ fails (rule R3, code \texttt{R3\_FAIL} at $G_2$) and the case goes to the recovery node. An irrep whose $H_i$ already contains the running intersection, such as $\Gamma_1^+$, is never added. Supplementary Table~\ref{tab:si_primaries} applies the procedure to every system; it reproduces the selection used in this work in each case. Ranking on the raw amplitude would fail for Ca$_3$Ti$_2$O$_7$ (amplitudes in main-text Methods, Section~\Mref{sec:planner}): the walk would end at $X_3^-+\Gamma_5^-$ instead of $X_3^-+X_2^+$.

\begin{table}[!htbp]
\centering\small
\caption{\textbf{Primary irreps selected by the walk over decreasing amplitude.} $A_p$ in \AA\ (tf, translation-free); ranking and threshold $A_{\mathrm{thr}}=0.05$~\AA\ use $A_s^{\mathrm{tf}}$. Space-group numbers of the isotropy subgroups in parentheses.}
\label{tab:si_primaries}
\begin{tabularx}{\linewidth}{@{}L{2.1cm}L{4.9cm}L{2.2cm}X@{}}
\toprule
System & Walk & Primaries & Secondaries\\
\midrule
CsSnI$_3$ & $M_3^+$ 0.843 (127); $+R_4^+$ 0.229 (74) $\to$ No.~62 $=H$ & $M_3^+$, $R_4^+$ & $R_5^+$ 0.086, $X_5^+$ 0.045: above $A_{\mathrm{thr}}$ on $A_s$, held at zero (R2); $X_5^+$ scanned separately; $M_2^+$ 0.003 below threshold\\
Ca$_3$Ti$_2$O$_7$ & $X_3^-$ 0.822 (63); $+X_2^+$ 0.599 (64) $\to$ No.~36 $=H$ & $X_3^-$, $X_2^+$ & $\Gamma_5^-$ 0.395 tf (42), held at zero by a recorded policy override of R4\\
InMnO$_3$ & $K_3$ 0.528 (185) $=H$ & $K_3$ & $\Gamma_2^-$ 0.057 tf, kept by R4; borderline margin over $K_1$ (R5)\\
ErMnO$_3$ & $K_3$ 0.604 (185) $=H$ & $K_3$ & $\Gamma_2^-$ 0.126 tf, kept by R4\\
BiFeO$_3$ & $R_5^-$ 0.694 (167); $+\Gamma_4^-$ 0.526 tf (160) $\to$ No.~161 $=H$ & $R_5^-$, $\Gamma_4^-$ & none\\
HfO$_2$ paths & all irreps move together along the path coordinate $\lambda$ (strain policy P3) & -- & $\Gamma_1^+$ never lowers the symmetry\\
\bottomrule
\end{tabularx}
\end{table}

\paragraph{The CsSnI$_3$ No.~55 case.} The single-mode check (fourth certification check, main-text eq.~\Meqref{eq:single}) must be made on the structure that the grid will actually contain. In the first CsSnI$_3$ pass the check was made on the raw single-mode frame exported by the decomposition step. That frame carries the shear strain of the child, so a pure $M_3^+$ distortion was found in $Pbam$ (No.~55) at the tighter tolerances of the ladder and in $P4/mbm$ (No.~127) only at the looser ones, and the check at that time accepted the expected group at any tolerance (Supplementary Fig.~\ref{fig:s_three}c). The disagreement was found in the log before any grid was built and corrected by script: the check structure is now built exactly as the grid builds it, from the parent fractional coordinates plus the displacement on the fixed lattice, and a strict criterion replaces the any-tolerance pass: spglib must return the expected group at a tolerance of $10^{-2}$~\AA\ (default angle tolerance), and a match at another tolerance of the ladder ($10^{-4}$, $10^{-3}$, $10^{-1}$~\AA) does not count. With this construction $M_3^+$ alone gives No.~127 at every tolerance, and the axis and plane grids were built after the correction. In the gated workflow a match only at a loose tolerance is a gate failure (code \texttt{SINGLE\_MODE\_SG}), and the rule table rebuilds the check structure on the fixed parent lattice before escalating.

\section{Symmetry-mode decompositions}\label{sec:si_dec}

Amplitudes are $A_p$ in Å (parent-cell normalized, $A_p=A_s/\sqrt{s}$ for a supercell of $s$ parent cells). Polar $\Gamma$ amplitudes are given both as returned by the decomposition ($f_1$) and translation-free (tf; Supplementary Section~\ref{sec:origin}). Supplementary Table~\ref{tab:dec_perov} covers the perovskite tilt transitions (main-text Fig.~\Mref{fig:perovskite}a,d), Supplementary Table~\ref{tab:dec_mn} the manganites (Fig.~\Mref{fig:manganite}a), Supplementary Table~\ref{tab:dec_bfo} BiFeO$_3$ (Fig.~\Mref{fig:bfo}a; isotropy subgroups along the directions sampled in Fig.~\Mref{fig:bfo}b--e in Supplementary Table~\ref{tab:bfo_iso}) and Supplementary Table~\ref{tab:dec_hfo2} the HfO$_2$ transitions (Fig.~\Mref{fig:hafnia}a).

\begin{table}[!htbp]
\centering
\caption{\textbf{Symmetry-mode decomposition of the perovskite children.} Order-parameter components along the direction of each irrep returned by the decomposition of $f_1$ (child setting), the total amplitude $Q$ of each irrep and the norm $Q_{\rm sec}$ of all secondaries, as $A_p$ in \AA; for an $n$-component direction the components are $\pm Q/\sqrt{n}$. Strains are dimensionless. (a)~CsSnI$_3$, $Pm\bar{3}m\to Pnma$, 20-atom $\sqrt2\times2\times\sqrt2$ cell ($s=4$); directions $R_4^+(a,-a,0)$, $M_3^+(0;a;0)$, $R_5^+(a,a,0)$, $X_5^+(a,a;0,0;0,0)$, $M_2^+(0;a;0)$, strains $\Gamma_1^+(a)$, $\Gamma_3^+(a,-\sqrt3a)$, $\Gamma_5^+(0,0,a)$. (b)~Ca$_3$Ti$_2$O$_7$, $I4/mmm\to A2_1am$, 48-atom cell ($s=2$); directions $X_3^-(0;a)$, $X_2^+(0;a)$, $\Gamma_5^-(a,-a)$, strain $\Gamma_4^+(a)$. $^\dagger$Translation-free value after removal of the rigid translation contained in the raw $\Gamma_5^-$ vector (75\% of its squared norm).}
\label{tab:dec_perov}
\footnotesize
\setlength{\tabcolsep}{3.5pt}
(a) CsSnI$_3$\\[2pt]
\resizebox{\textwidth}{!}{\begin{tabular}{l ccc ccc ccc cc c ccc ccc}
\toprule
 & \multicolumn{3}{c}{$R_4^+$} & \multicolumn{3}{c}{$M_3^+$} & \multicolumn{3}{c}{$R_5^+$} & \multicolumn{2}{c}{$X_5^+$} & $M_2^+$ & \multicolumn{3}{c}{strain} & & & \\
\cmidrule(lr){2-4}\cmidrule(lr){5-7}\cmidrule(lr){8-10}\cmidrule(lr){11-12}\cmidrule(lr){13-13}\cmidrule(lr){14-16}
 & (1) & (2) & (3) & (1) & (2) & (3) & (1) & (2) & (3) & (1) & (2) & (2) & $\Gamma_1^+$ & $\Gamma_3^+$ & $\Gamma_5^+$ & $Q_{R_4^+}$ & $Q_{M_3^+}$ & $Q_{\rm sec}$ \\
\midrule
$Pnma$ child   & $+0.162$ & $-0.162$ & 0 & 0 & 0.843 & 0 & $+0.061$ & $+0.061$ & 0 & $+0.032$ & $+0.032$ & 0.003 & $-0.022$ & $+0.020$ & $+0.004$ & 0.229 & 0.843 & 0.097 \\
$P4/mbm$       & 0 & 0 & 0 & 0 & 1.006 & 0 & 0 & 0 & 0 & 0 & 0 & 0 & -- & -- & -- & 0 & 1.006 & 0 \\
\bottomrule
\end{tabular}}\\[8pt]
(b) Ca$_3$Ti$_2$O$_7$\\[2pt]
\resizebox{\textwidth}{!}{\begin{tabular}{l cc cc cc c c ccc}
\toprule
 & \multicolumn{2}{c}{$X_3^-$} & \multicolumn{2}{c}{$X_2^+$} & \multicolumn{2}{c}{$\Gamma_5^-$} & $\Gamma_1^+$ & $\Gamma_4^+$ (strain) & & & \\
\cmidrule(lr){2-3}\cmidrule(lr){4-5}\cmidrule(lr){6-7}
 & (1) & (2) & (1) & (2) & $P_x$ & $P_y$ & & & $Q_{X_3^-}$ & $Q_{X_2^+}$ & $P_{\Gamma_5^-}$ \\
\midrule
$A2_1am$ child & 0 & 0.822 & 0 & 0.599 & $+0.560$ ($+0.279^\dagger$) & $-0.560$ ($-0.279^\dagger$) & 0.000 & $-0.0053$ & 0.822 & 0.599 & 0.792 ($0.395^\dagger$) \\
\bottomrule
\end{tabular}}
\end{table}

\begin{table}[!htbp]
\centering\small
\caption{Decompositions of the hexagonal manganites ($P6_3/mmc\to P6_3cm$, $i=6$, $s=3$).}
\label{tab:dec_mn}
\begin{tabularx}{\linewidth}{@{}L{2.0cm}L{1.3cm}L{1.5cm}L{2.6cm}X@{}}
\toprule
System & Irrep & OPD & $A_s$ / $A_p$ (Å) & Notes\\
\midrule
InMnO$_3$ & $K_3$ & $(a,0)$ & 0.9146 / 0.5280 & two-dimensional; In $A_{2u}$ 0.267, O1 $E$ $-0.298$, O2 $A_2''$ $-0.344$ (Å, $A_p$)\\
 & $\Gamma_2^-$ & $(a)$ & -- / 0.0585, 0.0573 tf & polar secondary, kept by R4; 4\% rigid translation; above threshold ($A_s^{\mathrm{tf}}=0.099$~Å; 0.101~Å raw), borderline margin 0.076~Å over $K_1$ (R5)\\
 & $K_1$ & $(a,0)$ & -- / 0.0135 & set to zero\\
\midrule
ErMnO$_3$ & $K_3$ & $(a,0)$ & -- / 0.604 & two-dimensional\\
 & $\Gamma_2^-$ & $(a)$ & -- / 0.127, 0.1263 tf & polar\\
\bottomrule
\end{tabularx}
\end{table}

\begin{table}[!htbp]
\centering\small
\caption{\textbf{Symmetry-mode decomposition of BiFeO$_3$} ($Pm\bar3m\to R3c$). Amplitudes $A_p$ in Å; tf, translation-free (Supplementary Section~\ref{sec:origin}).}
\label{tab:dec_bfo}
\begin{tabularx}{\linewidth}{@{}L{1.3cm}L{2.0cm}L{2.6cm}X@{}}
\toprule
Irrep & OPD & $A_p$ (Å) & Notes\\
\midrule
$\Gamma_4^-$ & $(a,a,a)$ & 1.073 / 0.526 tf & polar; 76\% of $\|\mathbf e\|^2$ rigid translation ($f_T=0.76$); alone $R3m$ (160)\\
$R_5^-$ & $(a,a,a)$ & 0.694 & anti-phase tilt; alone $R\bar3c$ (167)\\
\bottomrule
\end{tabularx}
\end{table}

\begin{table}[!htbp]
\centering\small
\caption{Decompositions of the HfO$_2$ transitions. $f_1$ decomposes each transition in its own setting.}
\label{tab:dec_hfo2}
\begin{tabularx}{\linewidth}{@{}L{2.9cm}XL{4.2cm}@{}}
\toprule
Transition & Irreps, $A_p$ (Å) & Strain\\
\midrule
T1 $P4_2/nmc\to Pca2_1$ & $M_1(0,a)$ 0.6254; $M_3(a,0)$ 0.5327 (crossed arms); $\Gamma_5^-(a,-a)$ 0.7818, 0.6379 tf (33\% rigid translation); $\Gamma_1^+$ 0.1731 & $\Gamma_1^+$, $\Gamma_4^+$\\
T2 $P4_2/nmc\to Pbcn$ & $M_1(a,0)$ 1.0141; $\Gamma_1^+$ 0.6160 & $\Gamma_1^+$ 0.0027/0.1130; $\Gamma_4^+$ 0.0482\\
T3 $P4_2/nmc\to P2_1/c$ & $M_1(0,a)$ 0.7793; $M_3(0,a)$ 0.7198 (aligned arms); $\Gamma_5^+(a,a)$ 0.5527; $\Gamma_1^+$ 0.3902 & includes $\Gamma_5^+$ shear 0.1147\\
T4 $Pbcn\to Pca2_1$ & $\Gamma_2^-$ 1.3866 (raw); $\Gamma_1^+$ 0.9405 & $\Gamma_1^+$ only ($+0.028$/$-0.090$/$-0.034$)\\
T5 $Pbcn\to P2_1/c$ & $\Gamma_3^+$ 1.3588; $\Gamma_1^+$ 0.6216 & $\Gamma_1^+$ ($+0.058$/$-0.084$/$-0.026$); $\Gamma_3^+$ shear $-0.1143$\\
\bottomrule
\end{tabularx}
\end{table}

\section{Electronic-structure settings}\label{sec:si_dft}
Supplementary Table~\ref{tab:dft} gives the electronic-structure settings of the six systems (main-text Methods, Section~\Mref{sec:dft}); each is stored as the system profile that $f_0$ selects, and the same settings are used for every frame of a campaign.

\begin{table}[!htbp]
\centering\small
\caption{Electronic-structure settings of the six systems (VASP~\cite{Kresse1996a}, PAW datasets~\cite{Blochl1994,Kresse1999}). Cut-off: plane-wave energy cut-off; magnetism: the collinear order imposed in every calculation.}
\label{tab:dft}
\begin{tabularx}{\linewidth}{@{}>{\raggedright}p{1.9cm}>{\raggedright}p{3.4cm}>{\raggedright}p{2.0cm}>{\raggedright}p{1.4cm}>{\raggedright\arraybackslash}X@{}}
\toprule
System & Functional & Magnetism & Cut-off (eV) & $k$-points; notes\\
\midrule
CsSnI$_3$ & PBE~\cite{Perdew1996}, no spin--orbit coupling & spin-polarized (zero moment) & 350 & $5\times3\times5$ (20-atom cell; KSPACING 0.17~Å$^{-1}$); fixed cubic cell\\
Ca$_3$Ti$_2$O$_7$ & PBE~\cite{Perdew1996} & non-magnetic & 520 & $\Gamma$ $2\times6\times6$; 48-atom cell, fixed $I4/mmm$ lattice, EDIFF $10^{-8}$, LREAL Auto\\
InMnO$_3$ & LDA~\cite{Ceperley1980,Perdew1981}$+U$, $U=8$, $J=0.88$~eV (Mn $d$); Dudarev scheme~\cite{Dudarev1998} for the joint fit, Liechtenstein scheme~\cite{Liechtenstein1995} also for the benchmark plane fits & A-type AFM & 820 & $\Gamma$ $6\times6\times2$; relaxed end cells; PAW datasets of the LDA set (In\_d; Mn\_pv; O)\\
ErMnO$_3$ & LDA$+U$ (Dudarev~\cite{Dudarev1998}), $U=8$, $J=0.88$~eV (Mn $d$) & A-type AFM & 820 & $\Gamma$ $6\times6\times2$; relaxed end cells; PAW datasets of the PBE set (Er\_3, Er $4f$ in core; Mn\_pv; O) for all 419 runs (an LDA functional on PBE-generated PAW datasets; the InMnO$_3$ runs used the LDA set) (the 411 grid structures and 8 cell relaxations)\\
BiFeO$_3$ & PBEsol~\cite{Perdew2008}$+U$~\cite{Dudarev1998}, $U_{\mathrm{eff}}=4$~eV (Fe $d$) & G-type AFM & 600 & $\Gamma$ $6\times6\times6$ (10-atom cell; grids at $a=3.9407$ and 3.8813~\AA); $7\times7\times3$ (30-atom cell; relaxed anchors)\\
HfO$_2$ & PBE~\cite{Perdew1996} & non-magnetic & 600 & $\Gamma$ $5\times5\times5$ (T1--T3), $5\times6\times5$ (T4, T5); EDIFF $10^{-8}$\\
\bottomrule
\end{tabularx}
\end{table}

\section{Perovskite tilt transitions: CsSnI$_3$ and Ca$_3$Ti$_2$O$_7$}\label{sec:si_perov}

This section supports main-text Section~\Mref{sec:res_perov}, Fig.~\Mref{fig:perovskite} and Supplementary Table~\ref{tab:perov}. It opens with the full account of the perovskite results (Supplementary Section~\ref{sec:si_perov_ext}), then shows why the irrep basis and the difference basis give the same frames in both systems (Supplementary Section~\ref{sec:si_interp}), gives the CsSnI$_3$ Landau fit and energy ladder (Supplementary Section~\ref{sec:si_cssni3}), and finally the constructed parent, site-resolved decomposition, closure and DFT settings of Ca$_3$Ti$_2$O$_7$ (Supplementary Sections~\ref{sec:si_ca3ti2o7} and~\ref{sec:si_ca3ti2o7_dft}).

\subsection{Full account of the perovskite results}\label{sec:si_perov_ext}
This subsection gives the complete account of the perovskite transitions summarized in main-text Section~\Mref{sec:res_perov}, with every value and check.

\begin{figure}[!htbp]
\centering
\includegraphics[width=\textwidth]{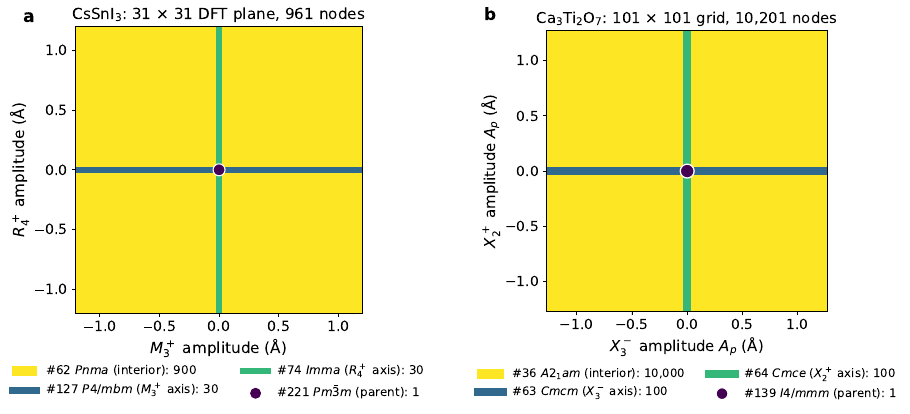}
\caption{\textbf{Space group of every frame on the two planes}, identical in the irrep and difference bases at every node: the parent at the origin, the single-irrep groups on the axes and the child everywhere else. \textbf{a},~CsSnI$_3$, $31\times31$ DFT plane (961 nodes, $\pm1.2$~Å); \textbf{b},~Ca$_3$Ti$_2$O$_7$, $101\times101$ node grid (10{,}201 nodes, $\pm1.27$~Å). Amplitudes are $A_p$ in \AA; space groups from spglib, labelled by number and symbol; colours and counts per group are given in the legends (CsSnI$_3$: Nos~62, 127, 74, 221 with 900, 30, 30, 1 nodes; Ca$_3$Ti$_2$O$_7$: Nos~36, 63, 64, 139 with 10{,}000, 100, 100, 1).}
\label{fig:s_census}
\end{figure}

The first question is whether the frames are what IRIS says they are. CsSnI$_3$ and Ca$_3$Ti$_2$O$_7$ answer it on transitions whose child is known and whose two primary irreps span a plane. For CsSnI$_3$ ($Pm\bar{3}m\to Pnma$, index 24, $\sqrt2\times2\times\sqrt2$ cell of 20 atoms), IRIS finds the in-phase rotation $M_3^+$ (0.843~\AA, direction $(0;a;0)$, one arm) and the anti-phase tilt $R_4^+$ (0.229~\AA, direction $(a,-a,0)$), i.e.\ the Glazer pattern~\cite{Glazer1972} $a^-b^+a^-$ (structures in main-text Fig.~\Mref{fig:perovskite}a; full decomposition in Supplementary Table~\ref{tab:dec_perov}). Both order parameters are three-dimensional ($R_4^+$: one arm, three components; $M_3^+$: three arms, one component each), so the mapped plane is the two-dimensional section of a six-dimensional order-parameter space spanned by the two symmetry-fixed rays whose isotropy subgroups are $P4/mbm$ (No.~127) and $Imma$ (No.~74); their intersection is $Pnma$ (No.~62) (rule R3 of the mode selection, Methods; Supplementary Fig.~\ref{fig:s_select}g), so every interior point of the plane has the symmetry of the ground state. The secondaries are small: $R_5^+$ (0.086~\AA, Cs rumpling), $X_5^+$ (0.045~\AA, antipolar Cs shift) and $M_2^+$ (0.003~\AA), 0.097~\AA\ in total against 0.874~\AA\ for the primaries, and as secondaries they are held at zero on the plane (rule R2; Supplementary Table~\ref{tab:si_primaries}); there is no polar mode, so no origin correction is needed. The ($M_3^+$, $R_4^+$) plane on the fixed cubic lattice ($31\times31$ points, $\pm1.2$~\AA) gives 1{,}036 frames with the axes and scans, and all 961 plane frames have the expected space group (1, 30, 30 and 900 frames of Nos.~221, 127, 74 and 62; Supplementary Fig.~\ref{fig:s_census}a). Before certification, the raw single-mode frame for $M_3^+$ alone came out in No.~55 instead of No.~127; the three-way symmetry agreement of $f_2$ caught this before any calculation (Supplementary Fig.~\ref{fig:s_three}c). The raw frame carried the child's shear strain; the check is now made on the structure the grid builds, with a strict tolerance criterion (Supplementary Section~\ref{sec:si_primaries}). For Ca$_3$Ti$_2$O$_7$ ($I4/mmm\to A2_1am$, index 8), the database holds only the child (Materials Project~\cite{Jain2013}, mp-4163), so IRIS constructs the parent by symmetrizing it: with increasing tolerance the space group climbs from No.~36 through $Cmcm$ (No.~63) to $I4/mmm$ (No.~139), the first body-centred parent in the set, with every Ti six-coordinated (Supplementary Table~\ref{tab:ca3ti2o7_parent}). The active irreps are the tilt $X_3^-$ (0.822~Å) and the rotation $X_2^+$ (0.599~Å; main-text Fig.~\Mref{fig:perovskite}d). Both are two-dimensional, and both condense on the same arm of the $X$ star, $(0;a)$; this is what allows their trilinear coupling to the polar $\Gamma_5^-$, which condenses along $(a,-a)$, that is, along the parent $[1\bar{1}0]$~\cite{Benedek2011,Nowadnick2016}. A rigid translation made up 75\% of the squared norm of the raw $\Gamma_5^-$ mode vector returned by the decomposition (half of its amplitude), which $f_2$ removes (0.395~Å translation-free; Supplementary Fig.~\ref{fig:s_certify}b). Because the parent is derived from the child, the fully symmetric $\Gamma_1^+$ is exactly zero and the only strain is $\Gamma_4^+$ ($-0.005$; full decomposition in Supplementary Table~\ref{tab:dec_perov} and, site by site, Supplementary Table~\ref{tab:dec_ca3ti2o7}). Here 16 of the 48 atoms move under both irreps, a stricter test than CsSnI$_3$, whose two irreps move disjoint displacement components ($M_3^+$ the equatorial I in the $xz$ plane; $R_4^+$ the equatorial I along $y$ and the apical I along $z$; Supplementary Section~\ref{sec:si_interp}). Besides this rigid translation, certification caught three further errors in this system that symmetry checks alone cannot detect. First, the exported mode vectors list the atoms in a different order for different irreps; taken in list order, the $\Gamma_5^-$ and $X_2^+$ vectors became permutations with norms of 40 and 28~Å while every space-group check still passed, and the atom-matching check of $f_2$ (main-text eq.~\Meqref{eq:match}) rejected them. Second, the exported amplitudes do not match the decomposition: the $X_3^-$ and $X_2^+$ vectors carry 0.80 and 0.40 of the child amplitudes (calibration factors 1.25 and 2.50). IRIS therefore normalizes each mode vector to $\sqrt{n}$~Å per Å of $A_p$, with $n=4$ primitive parent cells in this 48-atom frame. Third, the decomposition's own supercell contains only two, and a first normalization to $\sqrt2$~Å per unit had put the plane axes in $A_s=\sqrt2\,A_p$; the norm check of $G_1$ caught this in the gated rerun (Extended Data Table~\Mref{tab:errors}). With the vectors fixed, all 10{,}201 nodes of a $101\times101$ check grid have the expected space group (1, 100, 100 and 10{,}000 frames of Nos.~139, 63, 64 and 36; Supplementary Fig.~\ref{fig:s_ca3ti2o7_coords}d). Supplementary Table~\ref{tab:perov} summarizes both systems.

\begin{figure}[!htbp]
\centering
\includegraphics[width=0.62\textwidth,trim=0 0 380bp 0,clip]{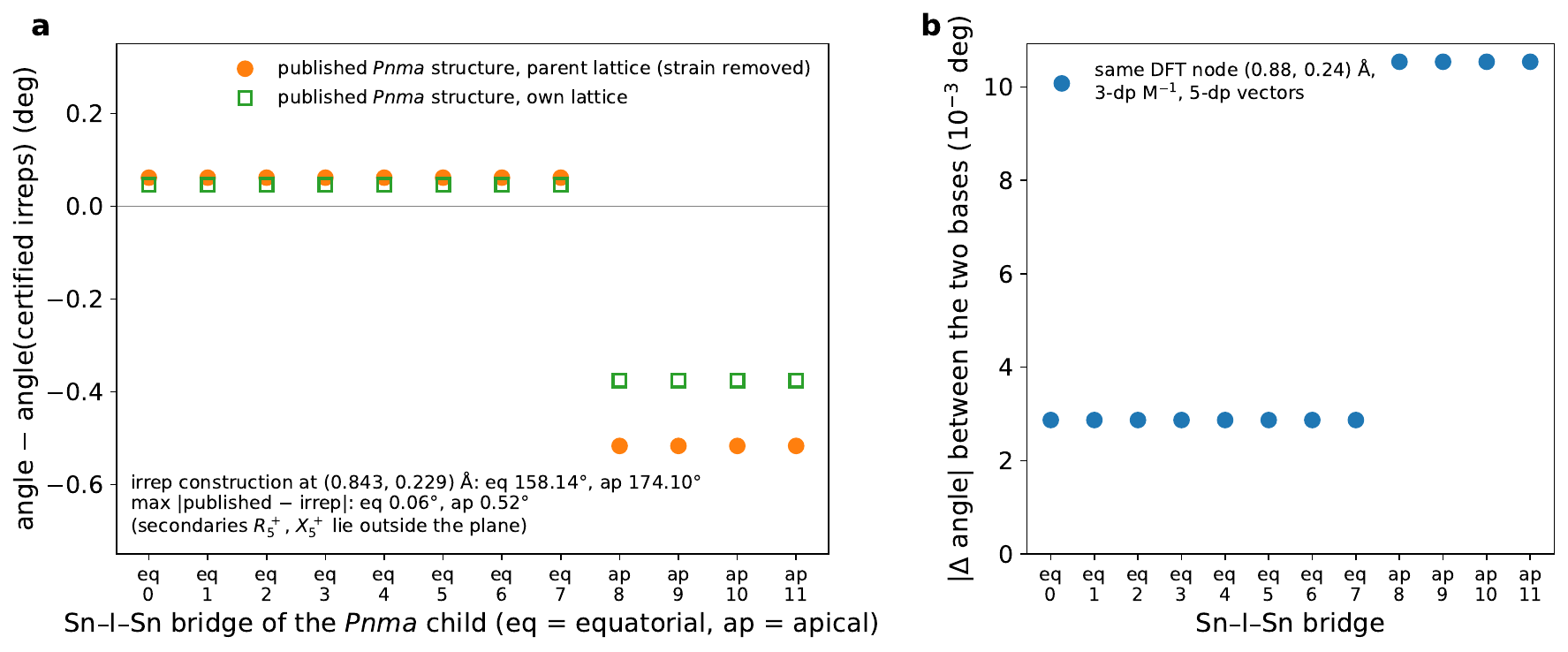}
\caption{\textbf{What a published structure adds that the irrep plane does not contain.} Sn--I--Sn angles of the published $Pnma$ structure of CsSnI$_3$ (on the parent lattice with strain removed, and on its own lattice) relative to the certified two-irrep construction at the child amplitudes, bridge by bridge (Supplementary Table~\ref{tab:cssni3_angles}): the eight equatorial bridges agree to $0.06^\circ$, the four apical bridges differ by $0.52^\circ$ ($0.38^\circ$ on the structure's own lattice); this difference is the part of the published distortion carried by the secondaries $R_5^+$ and $X_5^+$, which lie outside the $(M_3^+,R_4^+)$ plane. Orange circles, published structure on the parent lattice (strain removed); green squares, on its own lattice; eq, equatorial; ap, apical; reference angles of the irrep construction at (0.843, 0.229)~\AA: 158.14$^\circ$ (eq) and 174.10$^\circ$ (ap). A precision check (not plotted) gives the absolute difference of each Sn--I--Sn angle between the irrep and difference bases at the plane node (0.88, 0.24)~\AA, with $\mathsf{M}^{-1}$ stored to three decimals and the vectors to five, as at most $3\times10^{-3}$ (eq) and $1.05\times10^{-2}$ degrees (ap), rounding only (Supplementary Table~\ref{tab:cssni3_audit}).}
\label{fig:s_angles}
\end{figure}

To validate the frames independently, we rebuilt each one by linear interpolation along the displacement fields from the parent to two phases ($P4/mbm$ and $Pnma$ for CsSnI$_3$; $Cmcm$ and $A2_1am$ for Ca$_3$Ti$_2$O$_7$). These fields contain only the two primary irreps, so the interpolation is a linear change of coordinates within the irrep plane, $(\alpha_1,\alpha_2)\mapsto(Q_1,Q_2)=(\alpha_1,\alpha_2)\,\mathsf{M}$ with $\mathsf{M}$ built from the amplitudes of the two end structures (Supplementary Section~\ref{sec:si_interp}), and must reproduce every IRIS frame. The two constructions agree to within 3.1~n\AA\ per atom for CsSnI$_3$ (full-cell norm $\lVert\Delta R\rVert\le8.7$~n\AA, median 3.8~n\AA; 10{,}201 nodes) and 1.1~n\AA\ per atom for Ca$_3$Ti$_2$O$_7$ (10{,}201 nodes; main-text Fig.~\Mref{fig:perovskite}c,f), the rounding level of the stored vectors, with the same space group at every node (Supplementary Fig.~\ref{fig:s_census}), and the two end structures are reproduced to $4\times10^{-15}$~\AA\ when sixteen decimals are stored (Supplementary Table~\ref{tab:cssni3_audit}; the two bases are drawn on the plane in Supplementary Fig.~\ref{fig:s_dgrid}). The agreement is a consequence of symmetry rather than of numerical care: because each difference field lies entirely in the irrep plane, any interpolation along it, or any combination of the two, is by construction a point of that plane with the correct irrep content. In Ca$_3$Ti$_2$O$_7$ the two fields are $36^\circ$ apart ($15^\circ$ in CsSnI$_3$) and share the O1 atoms, yet the Ti--O--Ti angles of the two constructions agree to $5\times10^{-8}{}^\circ$ at every node; the residual falls from 1.1~n\AA\ with nine stored decimals to $4\times10^{-15}$~\AA\ with sixteen, so the agreement is limited only by stored precision (Supplementary Table~\ref{tab:ca3ti2o7_audit}). The 1{,}372 DFT inputs of main-text Fig.~\Mref{fig:perovskite}e are themselves reproduced from the two fields to $8\times10^{-15}$~Å, so one set of energies serves both coordinate systems (Supplementary Fig.~\ref{fig:s_ca3ti2o7_coords}a; Supplementary Fig.~\ref{fig:s_ca3ti2o7_energy}). The same test fails when structures are taken as found: the published $Pnma$ structure is the $(Q_{M_3^+},-Q_{R_4^+})$ domain and carries 0.097~\AA\ of secondaries. Its apical Sn--I--Sn angles therefore differ from the certified construction by 0.52$^\circ$ (Supplementary Section~\ref{sec:si_interp}), and a difference basis built from the published structures shifts the angles at plane nodes by up to 2.1$^\circ$ (Supplementary Table~\ref{tab:cssni3_angles}, Supplementary Fig.~\ref{fig:s_angles}). Interpolating between such structures without first fixing the parent setting, domain and origin mixes irreps and lowers the symmetry of intermediate frames (Supplementary Section~\ref{sec:si_interp}).

The CsSnI$_3$ landscape (main-text Fig.~\Mref{fig:perovskite}b) is a nearly isotropic ring of tilt minima, at $r=0.76$~\AA\ in the Landau fit and within one grid step of it (0.72--0.80~\AA, step 0.08~\AA) on the DFT grid, about 22~meV/f.u.\ below the cubic phase. Its corrugation is only 1.4~meV/f.u.\ in the fit and 1.5--1.6~meV/f.u.\ on the DFT grid, has $\cos2\theta$ form and places the minimum on the $R_4^+$ ($Imma$) axis. A two-irrep Landau fit (RMS 0.10~meV/f.u.; Supplementary Table~\ref{tab:cssni3_landau}, Supplementary Fig.~\ref{fig:s_landau}) explains the near-degeneracy: the harmonic coefficients are almost equal ($a_1=-0.149$, $a_2=-0.155$~eV\,\AA$^{-2}$), and the biquadratic coupling ($\lambda=0.255$~eV\,\AA$^{-4}$) is within 5\% of the quartic coefficients, the condition for an isotropic ring. The child lies outside the ring (0.87~\AA), so the $Pnma$ ground state is selected not by the two tilts but by the $X_5^+$ secondary and by strain: fully relaxed anchors place $P4/mbm$ and $Pnma$ at $-34.9$ and $-54.1$~meV/f.u., against $-36$ and $-56$~meV/f.u.\ reported by Grote and Berger with the same functional (PBE)~\cite{Grote2015} (Supplementary Table~\ref{tab:cssni3_anchors}). On the fixed lattice at 0.8 of the child primaries, relaxing $X_5^+$ lowers the energy by only 1.9~meV/f.u.\ (to 0.155~Å; 0.045~Å in the fully relaxed child, where strain and $R_5^+$ also relax); the scan is asymmetric in $Q_{X_5^+}$ there (Supplementary Fig.~\ref{fig:s_x5}a), the signature of the trilinear $M_3^+R_4^+X_5^+$ invariant, and symmetric at the parent and on the $M_3^+$ axis, where one of the two primaries vanishes. Of the 19.2~meV/f.u.\ by which the relaxed $Pnma$ lies below the relaxed $P4/mbm$, about 3~meV/f.u.\ comes from internal modes (1.4 from the ring corrugation and 1.9 from $X_5^+$) and about 16~meV/f.u.\ from strain (Supplementary Fig.~\ref{fig:s_x5}b), so the tilt pattern of CsSnI$_3$ is selected by the lattice rather than by the octahedral network. In Ca$_3$Ti$_2$O$_7$ (main-text Fig.~\Mref{fig:perovskite}e), the plane is computed with $\Gamma_5^-$ and strain held at zero, so the hybrid-improper trilinear term $X_3^-X_2^+\Gamma_5^-$ is absent. On this plane (1{,}372 frames: $37\times37$ points over $\pm1.27$~Å in steps of 0.071~Å of $A_p$, plus the $Cmcm$, $Cmce$ and $A2_1am$ points), $X_3^-$ is the dominant instability, and the landscape is about twenty times deeper than in CsSnI$_3$. The parent lies 489~meV/f.u.\ above four isolated minima at $(Q_{X_3^-},Q_{X_2^+})\approx(\pm0.92,\pm0.21)$~Å, where CsSnI$_3$ has a continuous ring. The minima sit in valleys elongated along $X_2^+$: once the tilt is large, the rotation changes the energy little, so the two primaries compete rather than cooperate. The $X_2^+$-only wells lie about 175~meV/f.u.\ above the minimum, and the $Cmcm$ and $Cmce$ anchors, at $1/\sqrt2$ of the child amplitudes, lie 189 and 324~meV/f.u.\ above it. The primaries-only child, at (0.822, 0.599)~Å, sits 51~meV/f.u.\ above the minimum, and the plane minimum has about 12\% more tilt and a third of the rotation of the child. As in CsSnI$_3$, the two primaries alone do not select the child; its amplitudes must be fixed by terms outside the plane, namely the trilinear coupling to $\Gamma_5^-$, the secondaries and strain. The landscape keeps $E(Q_{X_2^+})=E(-Q_{X_2^+})$ to 0.013~meV/f.u.\ and $E(Q_{X_3^-})=E(-Q_{X_3^-})$ to 0.42~meV/f.u. The parent symmetry requires both; the larger $X_3^-$ residual is numerical (real-space projection, \texttt{LREAL = Auto}) and is 0.1\% of the landscape depth (Supplementary Fig.~\ref{fig:s_ct_mirror}, Supplementary Section~\ref{sec:si_ca3ti2o7_dft}). Mapping the $\Gamma_5^-$ direction, where the trilinear term acts, is left for a separate grid.

\begin{table}[!htbp]
\centering\small
\caption{\textbf{The two perovskite tilt transitions at a glance.} Amplitudes are $A_p$ in Å; tf, translation-free. Energies per formula unit on the fixed parent lattice unless stated; the census counts frames of the parent, the two single-irrep subgroups and the child (CsSnI$_3$: DFT plane; Ca$_3$Ti$_2$O$_7$: $101\times101$ check grid). Full decompositions in Supplementary Table~\ref{tab:dec_perov}; closure in Supplementary Table~\ref{tab:cssni3_audit} and Supplementary Table~\ref{tab:ca3ti2o7_audit}.}
\label{tab:perov}
\renewcommand{\arraystretch}{1.25}
\begin{tabularx}{\textwidth}{@{}L{3.3cm}>{\raggedright\arraybackslash}X>{\raggedright\arraybackslash}X@{}}
\toprule
& CsSnI$_3$ & Ca$_3$Ti$_2$O$_7$\\
\midrule
Transition, index, cell & $Pm\bar3m\to Pnma$, 24; 20 atoms ($s=4$) & $I4/mmm\to A2_1am$, 8; 48 atoms (4 primitive parent cells; $s=2$ in the decomposition)\\
Parent & from the database & constructed by symmetrizing the child (mp-4163)\\
Primary irreps (direction) & $M_3^+$ $(0;a;0)$ 0.843; $R_4^+$ $(a,-a,0)$ 0.229 & $X_3^-$ $(0;a)$ 0.822; $X_2^+$ $(0;a)$ 0.599\\
Single-irrep subgroups & $P4/mbm$ (127), $Imma$ (74) & $Cmcm$ (63), $Cmce$ (64)\\
Secondaries & $R_5^+$ 0.086, $X_5^+$ 0.045, $M_2^+$ 0.003 (0.097 in total); no polar mode & $\Gamma_5^-$ 0.792 raw, 0.395 tf; strain $\Gamma_4^+$ $-0.005$; $\Gamma_1^+$ zero\\
Caught at certification & $M_3^+$ frame in No.~55, not 127; vectors at $0.80\times$ & atoms permuted (norms 40 and 28~Å); vectors at $0.80\times$ and $0.40\times$ ($s_i=1.25$, 2.50); 75\% of $\Gamma_5^-$ a rigid translation; vector norms $\sqrt2$ too small (decomposition supercell of 2 cells, frame of 4)\\
Plane grid & $31\times31$, $\pm1.2$~Å; 1{,}036 frames with axes and scans & $37\times37$, $\pm1.27$~Å; 1{,}372 frames with anchors\\
Space-group census & 1 / 30 / 30 / 900 & 1 / 100 / 100 / 10{,}000\\
Irrep vs difference basis & fields $15^\circ$ apart; 8.7~nÅ (full cell) & fields $36^\circ$ apart, 16 atoms shared; 1.1~nÅ (per atom)\\
Landscape & ring of minima at $r=0.76$~Å (fit; 0.72--0.80~Å on the DFT grid), $-22$~meV/f.u.; corrugation 1.4~meV/f.u.\ (fit; 1.5--1.6 on the grid) & four minima at $(\pm0.92,\pm0.21)$~Å, 489~meV/f.u.\ below the parent; child 51~meV/f.u.\ above\\
Landau fit & $a_1=-0.149$, $a_2=-0.155$~eV\,Å$^{-2}$; $\lambda=0.255$~eV\,Å$^{-4}$; RMS 0.10~meV/f.u. & not fitted (plane without $\Gamma_5^-$)\\
What selects the child & $X_5^+$ ($-1.9$~meV/f.u.\ at 0.8 of the child primaries, fixed lattice) and strain (about 16 of the 19.2~meV/f.u.\ relaxed gap) & terms outside the plane: trilinear coupling to $\Gamma_5^-$, secondaries, strain\\
\bottomrule
\end{tabularx}
\end{table}

\subsection{Irrep basis versus difference basis}
\label{sec:si_interp}

\paragraph{Construction.}
Let $R_C$ be the parent in the child cell and $e_1$, $e_2$ the certified mode vectors of the two primary irreps, normalized so that $|e_i|=\sqrt{n}$~\AA\ per \AA\ of $A_p$ ($n$ the number of primitive parent cells in the frame; $n=4$ for both CsSnI$_3$, 20/5 atoms, and Ca$_3$Ti$_2$O$_7$, 48/12 atoms). An IRIS frame is $R(Q_1,Q_2)=R_C+Q_1e_1+Q_2e_2$. The two reference phases are $R_A=R_C+Q_1^Ae_1$ (single irrep: $P4/mbm$ with $Q_1^A=1.006$~\AA; $Cmcm$ with $0.581$~\AA) and $R_B=R_C+Q_1^Be_1+Q_2^Be_2$ (CsSnI$_3$: the child, $(0.843,0.229)$~\AA; Ca$_3$Ti$_2$O$_7$: an $A2_1am$ structure on the child direction at $1/\sqrt2$ of the child amplitudes, $(0.581,0.424)$~\AA). The difference fields are $\Delta R_A=R_A-R_C$ and $\Delta R_B=R_B-R_C$, computed by plain subtraction of Cartesian coordinates under the minimum-image convention, with no projection on any mode. A difference-basis frame is $R(\alpha_1,\alpha_2)=R_C+\alpha_1\Delta R_A+\alpha_2\Delta R_B$. Because $\Delta R_A$ and $\Delta R_B$ are linear combinations of $e_1$ and $e_2$,
\begin{equation}
(Q_1,Q_2)=(\alpha_1,\alpha_2)\,\mathsf{M},\qquad
\mathsf{M}=\begin{pmatrix}Q_1^A & 0\\ Q_1^B & Q_2^B\end{pmatrix},
\end{equation}
so the two bases are related by the invertible matrix $\mathsf{M}$ (CsSnI$_3$: $\mathsf{M}^{-1}=\bigl(\begin{smallmatrix}0.994036&0\\-3.666105&4.374453\end{smallmatrix}\bigr)$; Ca$_3$Ti$_2$O$_7$: $\sqrt2\times\bigl(\begin{smallmatrix}1.216545&0\\-1.669449&1.669449\end{smallmatrix}\bigr)$). Every difference-basis frame is an irrep-basis frame and vice versa; the two constructions can differ only by the arithmetic used to store the vectors and $\mathsf{M}^{-1}$. The angle between $\Delta R_A$ and $\Delta R_B$ is $15.2^\circ$ for CsSnI$_3$ and $36.1^\circ$ for Ca$_3$Ti$_2$O$_7$; the difference basis is oblique but complete.

\paragraph{Why the symmetry is exact.}
In CsSnI$_3$ the two primaries move disjoint displacement components: $M_3^+(0;a;0)$ displaces the eight equatorial I atoms within the $xz$ plane (in-phase rotation about $y$, the long $b$ axis of the 20-atom cell), while $R_4^+(a,-a,0)$ displaces the equatorial I atoms along $y$ and the four apical I atoms along $z$ (anti-phase tilt about $a$, apical/equatorial ratio $\sqrt2$); Cs and Sn are fixed. The two vectors are therefore orthogonal, $e_1\cdot e_2=0$, and any combination $Q_1e_1+Q_2e_2$ with $Q_1Q_2\neq0$ has the isotropy subgroup of the sum, $P4/mbm\cap Imma=Pnma$. In Ca$_3$Ti$_2$O$_7$ the two irreps share 16 of the 48 atoms ($X_3^-$ moves all 48, $X_2^+$ moves 16 O), so the vectors are not disjoint; they are still orthogonal because they transform as different irreps of $I4/mmm$, and the same argument gives $Cmcm\cap Cmce=A2_1am$ on every interior node. Supplementary Table~\ref{tab:cssni3_audit} lists the numerical agreement; the space-group census of the two bases is identical at every node of the $101\times101$ check grids (CsSnI$_3$: 1, 100, 100 and 10{,}000 nodes of Nos.~221, 127, 74 and 62; Ca$_3$Ti$_2$O$_7$: 1, 100, 100 and 10{,}000 of Nos.~139, 63, 64 and 36).

\paragraph{Where the two bases differ.}
Supplementary Table~\ref{tab:cssni3_audit} also shows how the residual scales with the stored precision: with nine-decimal vectors and $\mathsf{M}^{-1}$ it is at the n\AA\ level, and with sixteen decimals it falls to $10^{-15}$~\AA. The residual is rounding, not physics. The situation is different when the reference structures are taken as published rather than rebuilt from certified irreps. The Materials Project $Pnma$ structure of CsSnI$_3$ is the $(Q_{M_3^+},-Q_{R_4^+})$ domain and contains 0.097~\AA\ of $R_5^+$, $X_5^+$ and $M_2^+$ (Supplementary Table~\ref{tab:dec_perov}). The certified two-irrep construction reproduces its equatorial Sn--I--Sn angle to $0.06^\circ$ but its apical angle only to $0.52^\circ$ (Supplementary Table~\ref{tab:cssni3_angles}), because the apical displacement carried by the secondaries lies outside the $(M_3^+,R_4^+)$ plane. A difference field built from the published structure and placed on a grid with the certified $P4/mbm$ field yields interior frames in which the secondaries are slaved to one axis with a fixed ratio, and whose space group is lower than $Pnma$ wherever the sign of $\alpha_2$ is reversed, since $-\Delta R_B$ is not a domain of the child. For CsSnI$_3$ this effect is below the fitting noise (the trilinear $X_5^+M_3^+R_4^+$ term is small at 0.045~\AA\ of $X_5^+$); for Ca$_3$Ti$_2$O$_7$, with a polar $\Gamma_5^-$ of 0.395~\AA\ and the hybrid-improper trilinear coupling, it would not be, which is why both of its difference fields are built from the certified irreps of the constructed parent (Supplementary Section~\ref{sec:si_ca3ti2o7}).

\paragraph{Displacement patterns and distances.} CsSnI$_3$: in the 20-atom child cell the unit mode vectors have $|\mathbf e|=\sqrt n=2$~Å per Å of $A_p$ ($n=4$ primitive parent cells). $\Delta\mathbf R_{CT}$ (parent $\to P4/mbm$) moves the eight equatorial I atoms by $(\pm0.503,0,\pm0.503)$~Å, an in-phase rotation about $y$; $\Delta\mathbf R_{CO}$ (parent $\to Pnma$, primaries only) moves the equatorial I by $(\pm0.42155,\pm0.11426,\pm0.42155)$~Å and the apical I by $(0,0,\pm0.16170)$~Å; Cs and Sn do not move. The two irreps therefore move disjoint displacement components. The Frobenius distances over the 20-atom cell are $|\mathbf R_T-\mathbf R_C|=2.012000$, $|\mathbf R_O-\mathbf R_C|=1.747084$ and $|\mathbf R_O-\mathbf R_T|=0.561407$~Å (1.006, 0.873542 and 0.280703~Å after division by $\sqrt n=2$).

\paragraph{Structures taken as found.} The agreement above requires that both end structures be in the parent setting, in one domain and with the polar origin fixed. Interpolating linearly between the Materials Project $P4/mbm$ and $Pnma$ cells of CsSnI$_3$ as found gives $P1$ intermediates, with the cell volume collapsing towards zero at the midpoint and atoms 0.3--0.7~Å apart. Differences between atoms paired by index give $|\Delta\mathbf R|$ of tens of Å and about 80\% $P1$ frames; differences between matched atoms in Cartesian coordinates carry the rigid rotation between the two cell settings and give $P\bar1$ and $P2/m$ frames, a third of them unphysical. Only matched atoms with fractional-coordinate differences in a common parent frame reproduce the symmetric grid, that is, a construction that has already rebuilt the parent frame of IRIS. A linear change of basis within the irrep plane, by contrast, cannot change any symmetry: a grid on the difference-vector lattice of CsSnI$_3$ (961 frames) contains only Nos.~62, 127 and 221. Paths built from certified modes also differ in energy: switching $+M_3^+\to-M_3^+$ along a straight line through the parent costs 21.1~meV/f.u.\ on the DFT grid nodes (the fitted $M_3^+$-axis minimum lies 21.3~meV/f.u.\ below the parent, Supplementary Table~\ref{tab:cssni3_landau}), but 0.0~meV/f.u.\ along the ring of main-text Fig.~\Mref{fig:perovskite}b, on which every frame is a certified subgroup.

\begin{table}[!htbp]
\centering
\caption{\textbf{Agreement between irrep-basis and difference-basis frames.} Maximum norm $\lVert\Delta R\rVert$ (or maximum per-atom displacement where stated) between the same frame built in the two bases, the maximum change in the B--X--B angle, and the space-group agreement. The precision rows use the same 10{,}201-node grids with the vectors and $\mathsf{M}^{-1}$ stored to the given number of decimals.}
\label{tab:cssni3_audit}
\small
\begin{tabularx}{\linewidth}{@{}l >{\raggedright\arraybackslash}X r r l@{}}
\toprule
system & frames & $\max\lVert\Delta R\rVert$ & $\max\Delta\angle$ & space groups \\
\midrule
CsSnI$_3$ & 10{,}201 nodes, 9-dp vectors and $\mathsf{M}^{-1}$ (main-text Fig.~\Mref{fig:perovskite}c) & 8.7 n\AA\ (3.1 n\AA$^{a}$) & $10^{-7\,\circ}$ & identical \\
CsSnI$_3$ & exact child points $P4/mbm$, $Pnma$, 9 dp & 1.4 n\AA & -- & identical \\
CsSnI$_3$ & same, 16 dp & $4\times10^{-15}$ \AA & -- & identical \\
Ca$_3$Ti$_2$O$_7$ & 10{,}201 nodes, 9-dp read-back (main-text Fig.~\Mref{fig:perovskite}f) & 1.1 n\AA$^{a}$ & $5\times10^{-8\,\circ}$ & identical \\
Ca$_3$Ti$_2$O$_7$ & 1{,}372 DFT frames ($37\times37$ plane + anchors), 16 dp & $7.6\times10^{-15}$ \AA & -- & identical \\
Ca$_3$Ti$_2$O$_7$ & child endpoints $Cmcm$, $A2_1am$, 16 dp & $1.2\times10^{-15}$ \AA & -- & identical \\
\bottomrule
\end{tabularx}
\begin{flushleft}\footnotesize $^{a}$maximum per-atom displacement.\end{flushleft}
\end{table}

\begin{table}[!htbp]
\centering
\caption{\textbf{Sn--I--Sn angles in CsSnI$_3$} (degrees; equatorial / apical) for the $P4/mbm$ and $Pnma$ structures built from the certified irreps at the child amplitudes, for the published structures, and for the nearest plane nodes in the irrep basis and in a difference basis built from the published structures. The irrep construction reproduces the published $P4/mbm$ angle to $0.006^\circ$ and the published $Pnma$ equatorial angle to $0.06^\circ$; the $0.5^\circ$ apical gap is the part of the published $Pnma$ distortion carried by the secondaries ($R_5^+$, $X_5^+$), which lie outside the $(M_3^+,R_4^+)$ plane.}
\label{tab:cssni3_angles}
\small
\begin{tabular}{l cc}
\toprule
structure & $P4/mbm$ (eq / ap) & $Pnma$ (eq / ap) \\
\midrule
irrep construction at child amplitudes & 154.498 / 180 & 158.142 / 174.100 \\
published structure, parent frame & 154.492 / 180 & 158.204 / 173.583 \\
published structure, own lattice & 154.492 / 180 & 158.188 / 173.724 \\
plane node (1.04, 0) / (0.88, 0.24), irrep basis & 153.666 / 180 & 157.204 / 173.817 \\
same nodes, difference basis from published structures & 154.640 / 180 & 159.335 / 174.394 \\
\bottomrule
\end{tabular}
\end{table}

\begin{figure}[!htbp]
\centering
\includegraphics[width=\textwidth]{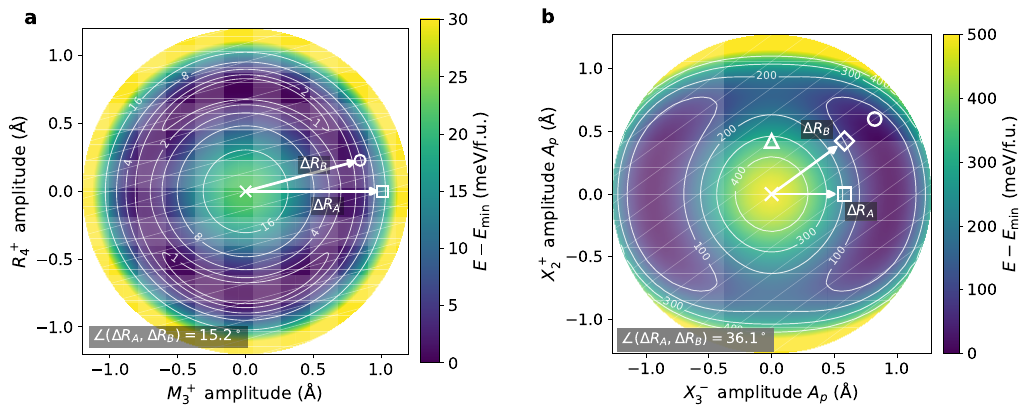}
\caption{\textbf{The difference basis drawn on the irrep plane.} \textbf{a},~CsSnI$_3$; \textbf{b},~Ca$_3$Ti$_2$O$_7$. The energy landscapes of main-text Fig.~\Mref{fig:perovskite} with the two difference fields $\Delta R_A$ (parent $\to$ single-irrep phase) and $\Delta R_B$ (parent $\to$ child) as arrows and the lines $\alpha_1,\alpha_2={\rm const}$ (step 0.5) of the difference basis. The two bases are related by $(Q_1,Q_2)=(\alpha_1,\alpha_2)\,\mathsf{M}$ (Supplementary Section~\ref{sec:si_interp}); the fields are oblique ($15.2^\circ$ and $36.1^\circ$, angle given in each panel) but span the plane. Colour, DFT energy $E-E_{\min}$ per formula unit (capped at 30 and 500~meV/f.u.); white lines, energy contours; grey lines, constant $\alpha_1$ or $\alpha_2$; $\times$, parent. Amplitudes are $A_p$ in \AA. For CsSnI$_3$, $\Delta R_A$ ends at the $P4/mbm$ point ($\square$, (1.006, 0)~Å) and $\Delta R_B$ at the child ($\circ$, (0.843, 0.229)~Å). For Ca$_3$Ti$_2$O$_7$, $\Delta R_A$ and $\Delta R_B$ end at the $Cmcm$ ($\square$) and $A2_1am$ ($\diamond$) anchors at $1/\sqrt2$ of the child amplitudes, (0.581, 0) and (0.581, 0.424)~Å, $\triangle$ marks the $Cmce$ anchor (0, 0.424)~Å, and $\circ$ the child at (0.822, 0.599)~Å.}
\label{fig:s_dgrid}
\end{figure}

\subsection{CsSnI$_3$: Landau fit, secondary and energy ladder}\label{sec:si_cssni3}
\begin{figure}[!htbp]
\centering
\includegraphics[width=\textwidth]{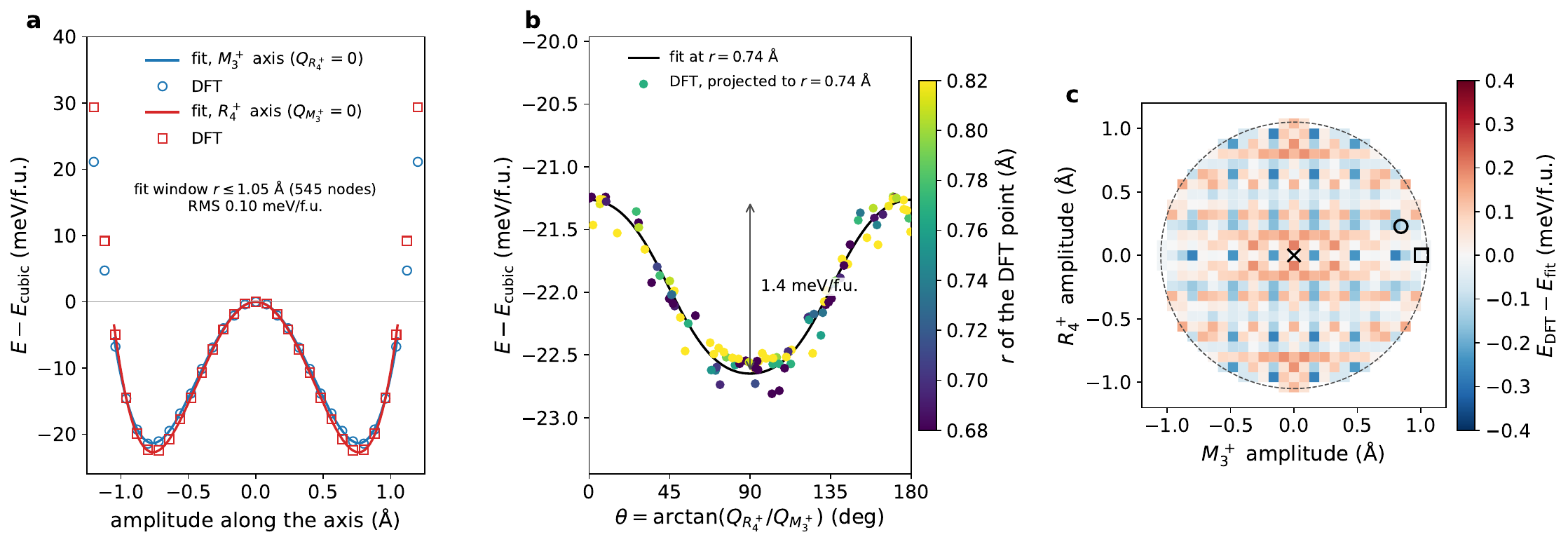}
\caption{\textbf{Two-irrep Landau fit of the CsSnI$_3$ plane.} \textbf{a},~DFT energies along the $M_3^+$ axis (circles) and the $R_4^+$ axis (squares) with the fitted polynomial of Supplementary Table~\ref{tab:cssni3_landau} (lines; fit window $r\le1.05$~\AA, 545 nodes, RMS 0.10~meV/f.u.). \textbf{b},~Energy around the ring at $r=0.74$~\AA\ (the grid radius nearest the fitted 0.76~\AA) from the fit (line) and the 100 random DFT points near the ring, projected onto $r=0.74$~\AA\ along the fit (dots, coloured by their $r$); the corrugation is 1.4~meV/f.u.\ with a $\cos2\theta$ form and the minimum on the $R_4^+$ ($Imma$) axis, $\theta=\arctan(Q_{R_4^+}/Q_{M_3^+})=90^\circ$; colour scale, $r$ of each DFT point (0.68--0.82~\AA). \textbf{c},~Fit residual $E_{\mathrm{DFT}}-E_{\mathrm{fit}}$ over the plane (meV/f.u.); dashed circle, fit window $r=1.05$~\AA; $\times$, parent; $\circ$, child (0.843, 0.229)~\AA; $\square$, $P4/mbm$ point (1.006, 0)~\AA; the $\pm0.3$~meV/f.u.\ alternation between neighbouring nodes is the numerical noise of the single-point calculations.}
\label{fig:s_landau}
\end{figure}

\begin{figure}[!htbp]
\centering
\includegraphics[width=\textwidth]{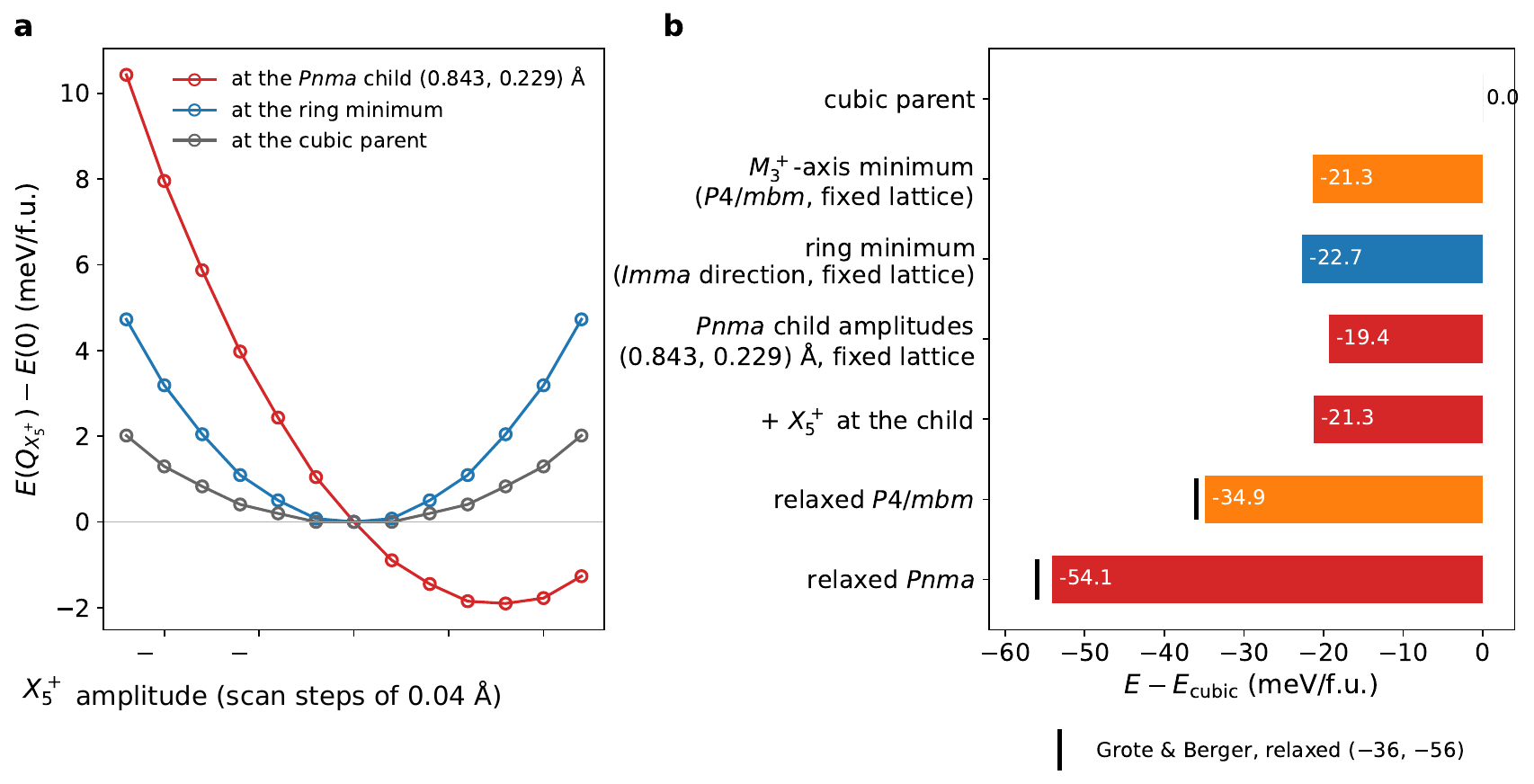}
\caption{\textbf{The $X_5^+$ secondary and the energy ladder of CsSnI$_3$.} \textbf{a},~Energy $E(Q_{X_5^+})-E(0)$ along the $X_5^+$ mode at the cubic parent (grey), on the $M_3^+$ axis at the grid node nearest the ring radius ($M_3^+=0.720$~\AA, $R_4^+=0$, one grid step inside the fitted axis minimum at 0.764~\AA, Supplementary Table~\ref{tab:cssni3_anchors}; blue, labelled ring minimum in the panel) and at 0.8 of the child primaries (red) ($M_3^+=0.672$, $R_4^+=0.184$~\AA; the scan was built from the uncalibrated mode vectors, so its labels 0.84 and 0.23 correspond to these amplitudes). With both primaries present the scan is asymmetric, reflecting the linear term of the trilinear $M_3^+R_4^+X_5^+$ invariant ($-24$~meV\,Å$^{-1}$/f.u.); a quartic fit gives a curvature of 150~meV\,Å$^{-2}$/f.u., a relaxed $X_5^+$ of 0.155~Å and an energy gain of 1.9~meV/f.u. At the parent and on the $M_3^+$ axis one of the two primaries is zero and the scan is even (curvature 59 and 155~meV\,Å$^{-2}$/f.u.). The relaxed 0.155~Å is a fixed-lattice value at 0.8 of the primaries, with $R_5^+$ and strain held at zero; the fully relaxed child has 0.045~Å of $X_5^+$ in its own cell, with 0.086~Å of $R_5^+$. Each scan has 13 points over $\pm0.24$~\AA\ of $X_5^+$ in steps of 0.04~\AA\ (the horizontal axis is in units of this step), computed with ENCUT 350~eV and spin polarization; energies are relative to the scan's own $X_5^+=0$ point, so only differences within a scan are used. \textbf{b},~Energies relative to the cubic parent (Supplementary Table~\ref{tab:cssni3_anchors}): fixed-lattice minima along the $M_3^+$ axis and on the ring, the $Pnma$ child amplitudes on the plane with and without $X_5^+$, and the fully relaxed $P4/mbm$ and $Pnma$ phases (bar colours: orange, $P4/mbm$-type; blue, $Imma$ direction; red, $Pnma$), with the relaxed values of Grote and Berger~\cite{Grote2015} ($-36$ and $-56$~meV/f.u.) as black ticks. All energies are DFT (PBE).}
\label{fig:s_x5}
\end{figure}

Supplementary Table~\ref{tab:cssni3_landau} and Supplementary Fig.~\ref{fig:s_landau} give the two-irrep Landau fit of the CsSnI$_3$ plane, which explains the near-isotropic ring of main-text Fig.~\Mref{fig:perovskite}b; Supplementary Fig.~\ref{fig:s_x5} and Supplementary Table~\ref{tab:cssni3_anchors} show how the $X_5^+$ secondary and strain select the $Pnma$ child.

\begin{table}[!htbp]
\centering
\caption{\textbf{Two-irrep Landau fit of the CsSnI$_3$ plane.} $E=\tfrac{a_1}{2}Q_1^2+\tfrac{b_1}{4}Q_1^4+\tfrac{c_1}{6}Q_1^6+\tfrac{a_2}{2}Q_2^2+\tfrac{b_2}{4}Q_2^4+\tfrac{c_2}{6}Q_2^6+\tfrac{\lambda}{2}Q_1^2Q_2^2+\tfrac{\mu}{2}\,(Q_1^4Q_2^2+Q_1^2Q_2^4)$ with $Q_1=Q_{M_3^+}$, $Q_2=Q_{R_4^+}$ in \AA\ ($A_p$) and $E$ per formula unit relative to the cubic parent; least-squares fit to the 545 plane nodes with $r=\sqrt{Q_1^2+Q_2^2}\le1.05$~\AA\ (Supplementary Fig.~\ref{fig:s_landau}).}
\label{tab:cssni3_landau}
\small
\begin{tabular}{l r l}
\toprule
coefficient & value & unit \\
\midrule
$a_1$ ($M_3^+$) & $-0.1486$ & eV\,\AA$^{-2}$ \\
$b_1$ & $0.2669$ & eV\,\AA$^{-4}$ \\
$c_1$ & $-0.0211$ & eV\,\AA$^{-6}$ \\
$a_2$ ($R_4^+$) & $-0.1554$ & eV\,\AA$^{-2}$ \\
$b_2$ & $0.2606$ & eV\,\AA$^{-4}$ \\
$c_2$ & $0.0129$ & eV\,\AA$^{-6}$ \\
$\lambda$ & $0.2552$ & eV\,\AA$^{-4}$ \\
$\mu$ & $0.0108$ & eV\,\AA$^{-6}$ \\
RMS residual & 0.098 & meV/f.u. \\
\midrule
minimum on the $R_4^+$ axis ($Imma$), fit & $0.761$~\AA, $-22.7$ & \AA, meV/f.u. \\
ring radius on the DFT grid (one grid step) & 0.72--0.80 & \AA \\
minimum on the $M_3^+$ axis ($P4/mbm$), fit & $0.764$~\AA, $-21.3$ & \AA, meV/f.u. \\
ring corrugation $E(\theta=0)-E(\theta=90^\circ)$, fit & 1.4 & meV/f.u. \\
ring corrugation on the DFT grid & 1.5--1.6 & meV/f.u. \\
fit at the $Pnma$ child amplitudes (0.843, 0.229)~\AA & $-19.4$ & meV/f.u. \\
fit at the $P4/mbm$ child amplitude (1.006, 0)~\AA & $-10.5$ & meV/f.u. \\
\bottomrule
\end{tabular}
\end{table}

\begin{table}[!htbp]
\centering
\caption{\textbf{Energy ladder of CsSnI$_3$} (meV per formula unit relative to the cubic parent; Supplementary Fig.~\ref{fig:s_x5}b). Fixed-lattice values are the Landau fit of the plane at the stated point; the $X_5^+$ gain is from the DFT scan at 0.8 of the child primaries (Supplementary Fig.~\ref{fig:s_x5}a) and is likely larger at the child, where the trilinear term is stronger; relaxed values are full ISIF~3 relaxations. Literature values are from Table~1 of ref.~\cite{Grote2015} (DFT-PBE, PAW, 500~eV cut-off, no spin--orbit coupling), the same functional as here.}
\label{tab:cssni3_anchors}
\small
\begin{tabular}{l r}
\toprule
state & $E-E_{\rm cubic}$ \\
\midrule
$M_3^+$-axis minimum, 0.764~\AA\ ($P4/mbm$ symmetry, fixed cubic lattice) & $-21.3$ \\
ring minimum, 0.761~\AA\ on the $R_4^+$ axis ($Imma$, fixed cubic lattice) & $-22.7$ \\
$Pnma$ child amplitudes (0.843, 0.229)~\AA, fixed lattice & $-19.4$ \\
\quad $X_5^+$ relaxed at 0.8 of the child primaries (gain) & $-1.9$ \\
fully relaxed $P4/mbm$ (this work / Grote \& Berger~\cite{Grote2015}) & $-34.9$ / $-36$ \\
fully relaxed $Pnma$ (this work / Grote \& Berger~\cite{Grote2015}) & $-54.1$ / $-56$ \\
\midrule
$E(Pnma)-E(P4/mbm)$, relaxed & $-19.2$ \\
\quad internal modes: ring corrugation 1.4 $+$ $X_5^+$ 1.9 & $\approx-3$ \\
\quad strain & $\approx-16$ \\
\bottomrule
\end{tabular}
\end{table}

\subsection{Ca$_3$Ti$_2$O$_7$: constructed parent, site-resolved decomposition and closure}\label{sec:si_ca3ti2o7}
\begin{figure}[!htbp]
\centering
\includegraphics[width=\textwidth]{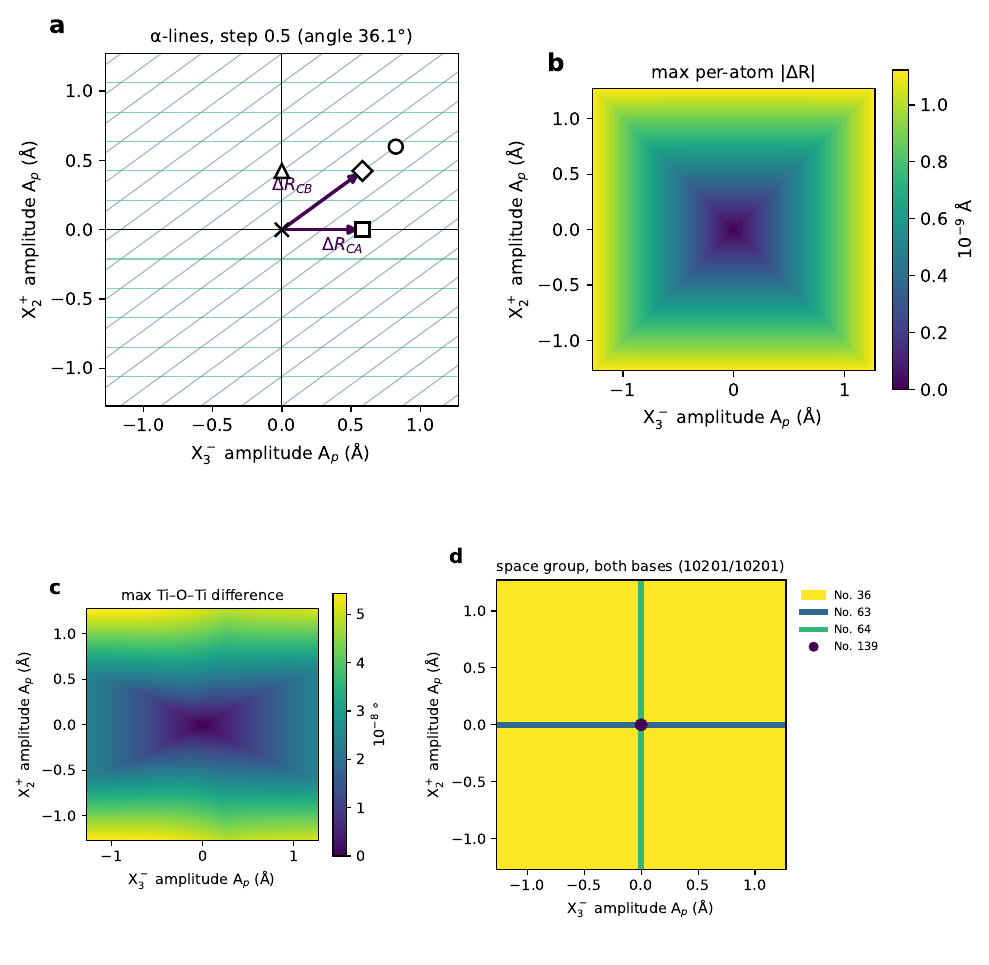}
\caption{\textbf{Ca$_3$Ti$_2$O$_7$: the irrep plane in two coordinate systems, structures.} Amplitudes are $A_p$ ($\pm1.27$~Å); the anchors $Cmcm$ ($\square$), $Cmce$ ($\triangle$) and $A2_1am$ ($\diamond$) lie at $1/\sqrt2$ of the child amplitudes, $\circ$ marks the child at (0.822, 0.599)~Å and $\times$ the parent.
\textbf{a},~Lines of constant $\alpha_1$ and $\alpha_2$ (difference basis, steps of 0.5) drawn on the ($X_3^-$, $X_2^+$) irrep plane (green, constant $\alpha_2$; grey, constant $\alpha_1$); arrows are $\Delta R_{CA}$ (parent$\to Cmcm$ anchor) and $\Delta R_{CB}$ (parent$\to A2_1am$ anchor), 36.1° apart.
\textbf{b},~Largest per-atom position difference between the two constructions on the $101\times101$ grid ($10^{-9}$~Å).
\textbf{c},~Largest Ti--O--Ti angle difference ($10^{-8}$°).
\textbf{d},~Space group per node, identical in both bases (No.~36 interior, No.~63 on the $X_3^-$ axis, No.~64 on the $X_2^+$ axis, No.~139 at the origin). Energies in the two coordinate systems are in Supplementary Fig.~\ref{fig:s_ca3ti2o7_energy}.}
\label{fig:s_ca3ti2o7_coords}
\end{figure}

\begin{figure}[!htbp]
\centering
\includegraphics[width=0.92\textwidth]{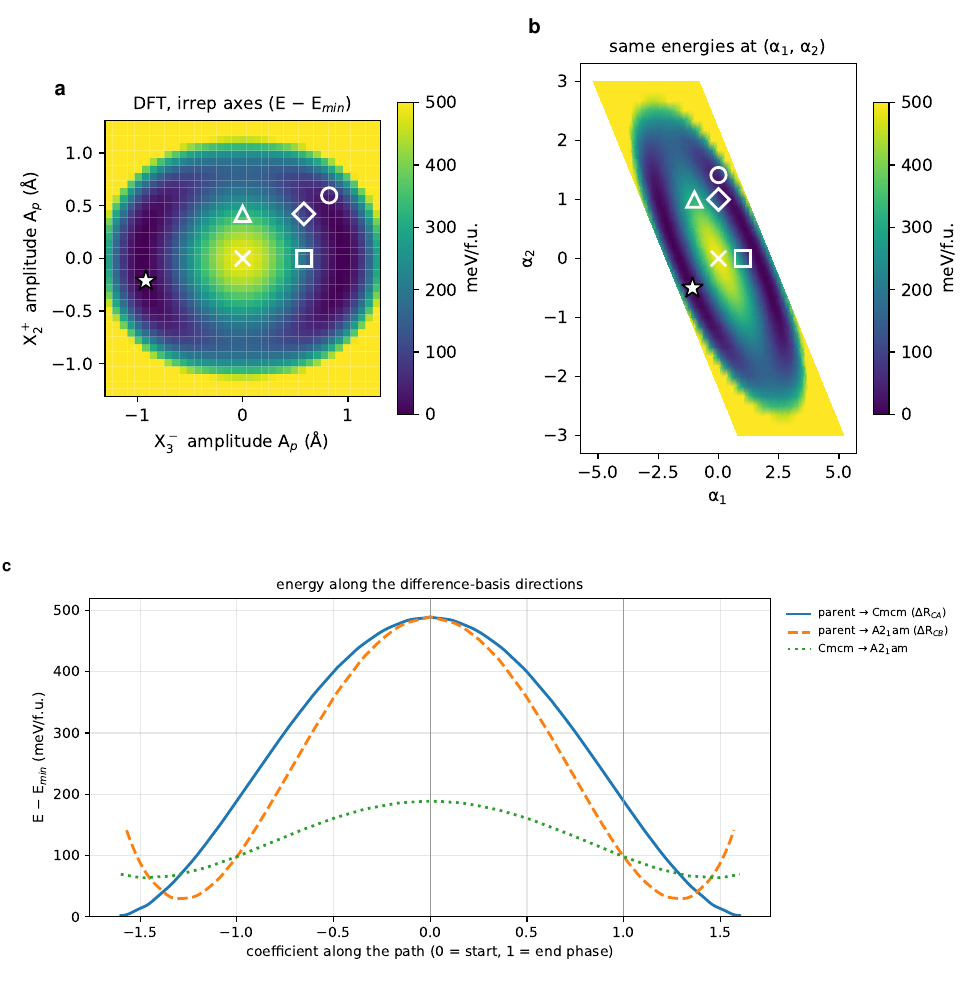}
\caption{\textbf{Ca$_3$Ti$_2$O$_7$: the irrep plane in two coordinate systems, energies.} The same DFT (PBE) energies in the irrep basis and in the difference basis of Supplementary Fig.~\ref{fig:s_ca3ti2o7_coords}a. Amplitudes are $A_p$; markers as in Supplementary Fig.~\ref{fig:s_ca3ti2o7_coords}, with $\star$ one of the four plane minima, $(-0.92,-0.21)$~Å. Colour scales in \textbf{a} and \textbf{b}: $E-E_{\min}$ per formula unit, capped at 500~meV/f.u.
\textbf{a},~DFT energies of the 1{,}369 plane points plotted against the irrep axes and \textbf{b},~the same energies placed at their $(\alpha_1,\alpha_2)=(x,y)M^{-1}$; the square grid becomes a parallelogram.
\textbf{c},~Energy along the difference-basis directions: parent$\to Cmcm$ (solid), parent$\to A2_1am$ (dashed) and $Cmcm\to A2_1am$ (dotted), against the coefficient along the path (0, start; 1, end phase; extended beyond both ends), from a spline of the DFT plane.}
\label{fig:s_ca3ti2o7_energy}
\end{figure}

The database holds only the $A2_1am$ child, so IRIS constructs the parent (Supplementary Table~\ref{tab:ca3ti2o7_parent}); Supplementary Table~\ref{tab:dec_ca3ti2o7} resolves the decomposition by Wyckoff site, Supplementary Table~\ref{tab:ca3ti2o7_audit} gives the closure between the irrep basis and the difference basis, and Supplementary Figs.~\ref{fig:s_ca3ti2o7_coords} and~\ref{fig:s_ca3ti2o7_energy} show the plane in both coordinate systems, as structures and as energies. The irreps $X_3^-$ and $X_2^+$ are two-dimensional (star of $X$: $(\tfrac12,\tfrac12,0)$ and $(\tfrac12,\tfrac12,1)$, one-dimensional small irrep per arm) and condense on the same arm, $(0;a)$, which allows the trilinear $X_3^-X_2^+\Gamma_5^-$ coupling; the mode counts (6 for $X_3^-$, 2 for $X_2^+$, 7 for $\Gamma_5^-$) are copies of each irrep on the Wyckoff sites, not order-parameter dimensions. The subgroup tree is $I4/mmm$ (139) $\to$ $Fmmm$ (69, $\Gamma_4^+$) $\to$ $Cmcm$ (63, $X_3^-$), $Cmce$ (64, $X_2^+$) and $Fmm2$ (42, $\Gamma_5^-$) $\to$ $A2_1am$ (36).

\begin{table}[!htbp]
\centering
\caption{\textbf{Constructed $I4/mmm$ parent of Ca$_3$Ti$_2$O$_7$.} The database holds only the $A2_1am$ child (mp-4163, 48 atoms). The parent is the spglib-refined structure of the child at increasing tolerance; the ladder passes through the single-tilt $Cmcm$ before reaching $I4/mmm$.}
\label{tab:ca3ti2o7_parent}
\small
\begin{tabular}{ll}
\toprule
Tolerance (Å) & Space group \\
\midrule
0.05--0.6 & $A2_1am$ (No.~36; $Cmc2_1$ in the standard setting) \\
0.7--0.8 & $Cmcm$ (No.~63) \\
0.9 & $I4/mmm$ (No.~139) \\
\midrule
\multicolumn{2}{l}{Parent: $a=3.844$, $c=19.443$~Å; 24 atoms (conventional), 12 (primitive)} \\
\multicolumn{2}{l}{Ca 2a (0,0,0), Ca 4e (0,0,0.1875), Ti 4e (½,½,0.0995)} \\
\multicolumn{2}{l}{O 8g (0,½,0.0981), O 4e (½,½,0.1976), O 2b (½,½,0)} \\
\multicolumn{2}{l}{Ti six-coordinated (Ti--O $\le$ 2.4~Å); min.\ distance 1.907~Å; child/parent volume 2.000} \\
\bottomrule
\end{tabular}
\end{table}

\begin{table}[!htbp]
\centering
\caption{\textbf{Symmetry-mode decomposition of Ca$_3$Ti$_2$O$_7$} ($I4/mmm\to A2_1am$, index 8). Order-parameter (OP) directions are those returned by the decomposition of $f_1$; $A_s=\sqrt{2}\,A_p$ in this 48-atom cell. $\Gamma_5^-$ is given as reported and, in parentheses, after removing the rigid translation (75\% of the squared norm of the raw vector, half of its amplitude). $\Gamma_1^+$ vanishes because the parent was constructed by symmetrizing the child.}
\label{tab:dec_ca3ti2o7}
\resizebox{\textwidth}{!}{%
\begin{tabular}{lllcccccccc}
\toprule
Irrep & OP direction & dim & Ca1 (4e) & Ca2 (2a) & Ti (4e) & O1 (8g) & O2 (4e) & O3 (2b) & $A_p$ (Å) & alone \\
\midrule
$X_3^-$ & $(0;a)$ & 2 & $-0.064$ & $-0.023$ & $-0.001$ & $-0.536$ & $+0.484$ & $-0.386$ & 0.822 & $Cmcm$ (63) \\
$X_2^+$ & $(0;a)$ & 2 & -- & -- & -- & $-0.599$ (B$_2$), $-0.002$ (B$_1$) & -- & -- & 0.599 & $Cmce$ (64) \\
$\Gamma_5^-$ & $(a,-a)$ & 2 & 0.000 & $-0.411$ & $-0.254$ & $-0.303$, $-0.307$ & $-0.438$ & $-0.123$ & 0.792 (0.395) & $Fmm2$ (42) \\
$\Gamma_1^+$ & $(a)$ & 1 & 0 & -- & 0 & 0 & 0 & -- & 0 & $I4/mmm$ \\
$\Gamma_4^+$ & $(a)$ & 1 & \multicolumn{6}{c}{strain $-0.0053$} & -- & $Fmmm$ (69) \\
\bottomrule
\end{tabular}}
\end{table}

\begin{table}[!htbp]
\centering
\caption{\textbf{Irrep basis versus interpolation along displacement fields, Ca$_3$Ti$_2$O$_7$.} $C=I4/mmm$, $A=Cmcm=C+0.581\,e_{X_3^-}$, $B=A2_1am=C+0.581\,e_{X_3^-}+0.424\,e_{X_2^+}$ (primaries only, $1/\sqrt2$ of the child amplitudes), fixed 48-atom parent lattice; $\lvert e\rvert=2$~Å per Å of $A_p$ (four primitive cells). The stored fields were built with $\lvert e\rvert=\sqrt2$~Å per unit, i.e.\ in $A_s$; all amplitudes are given here in $A_p=A_s/\sqrt2$. Fields $\Delta R_{CA}$, $\Delta R_{CB}$ written to and read back from text files; $(x,y)=\alpha M$.}
\label{tab:ca3ti2o7_audit}
\small
\begin{tabular}{p{0.46\textwidth}p{0.48\textwidth}}
\toprule
Quantity & Value \\
\midrule
$\lVert R_A-R_C\rVert$, $\lVert R_B-R_C\rVert$, $\lVert R_B-R_A\rVert$ & 1.162484, 1.438391, 0.847114~Å \\
Angle between $\Delta R_{CA}$ and $\Delta R_{CB}$ & 36.081° (CsSnI$_3$: 15.17°) \\
$M$; $M^{-1}$ & $\begin{pmatrix}0.581&0\\0.581&0.424\end{pmatrix}$; $\sqrt2\times\begin{pmatrix}1.216545012&0\\-1.669449082&1.669449082\end{pmatrix}$ \\
Atoms moved by $X_3^-$ / $X_2^+$ / both & 48 / 16 / 16 (CsSnI$_3$: disjoint) \\
$C+\Delta R_{CA}$, $C+\Delta R_{CB}$ vs irrep endpoints & $5\times10^{-16}$, $1.2\times10^{-15}$~Å \\
$101\times101$ grid ($\pm1.27$~Å, step 0.025~Å), POSCAR files, 9-decimal read-back & max $1.1\times10^{-9}$~Å; Ti--O--Ti $\le5.4\times10^{-8}$° \\
Space groups (both bases) & identical at 10{,}201/10{,}201 nodes: 10{,}000 No.~36, 100 No.~63, 100 No.~64, 1 No.~139 \\
Midpoints and 5 random $(\alpha_1,\alpha_2)$ & $\le4.4\times10^{-15}$~Å \\
1{,}372 DFT inputs rebuilt from $\Delta R_{CA}$, $\Delta R_{CB}$ & $7.6\times10^{-15}$~Å (16 decimals), $1.1\times10^{-9}$~Å (9 decimals) \\
\midrule
\multicolumn{2}{p{0.95\textwidth}}{Precision (max per-atom error over the grid, Å), by decimals kept in the vectors / in $M^{-1}$:} \\
9 / 9: $1.1\times10^{-9}$; 16 / 9: $9.4\times10^{-11}$ & 16 / 16: $4.1\times10^{-15}$ \\
\midrule
\multicolumn{2}{p{0.95\textwidth}}{Ti--O--Ti (in-plane mean / apical), at the parent and the three anchors: $C$ 178.34/180.00°; $Cmcm$ 168.74/163.96°; $Cmce$ 167.32/180.00°; $A2_1am$ 163.14/163.96°} \\
\bottomrule
\end{tabular}
\end{table}

\subsection{DFT on the Ca$_3$Ti$_2$O$_7$ irrep plane}\label{sec:si_ca3ti2o7_dft}
\begin{figure}[!htbp]
\centering
\includegraphics[width=\textwidth]{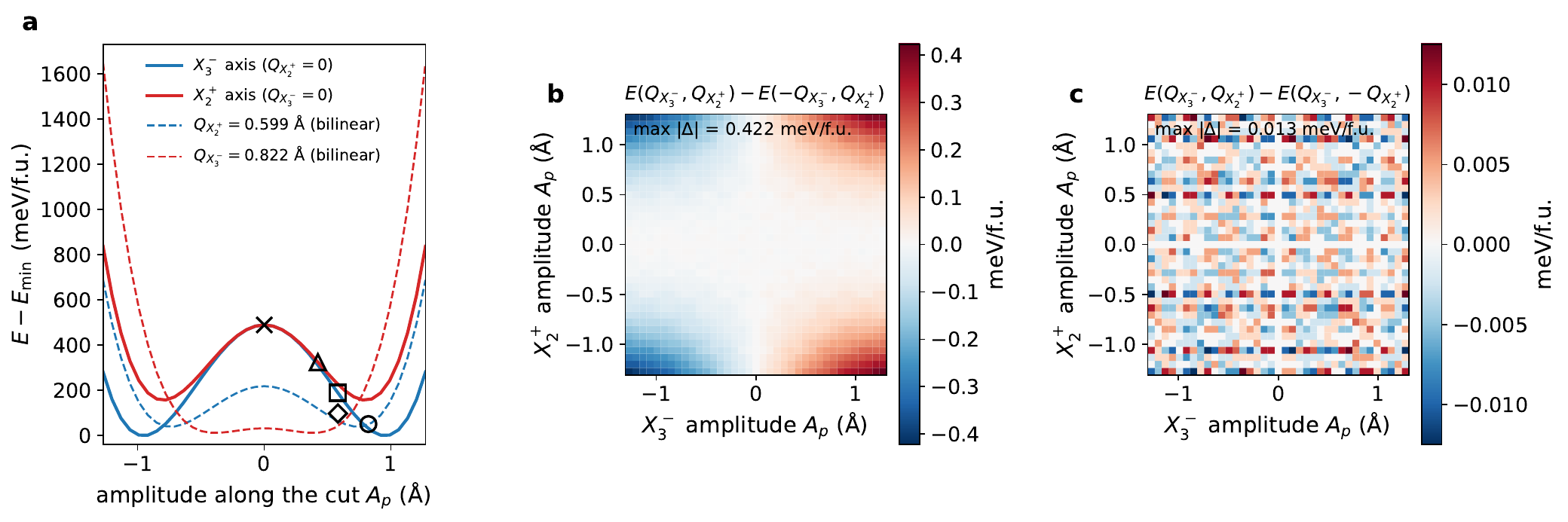}
\caption{\textbf{Ca$_3$Ti$_2$O$_7$ plane with $\Gamma_5^-$ and strain held at zero.} DFT (PBE) on the $37\times37$ plane of main-text Fig.~\Mref{fig:perovskite}e. \textbf{a},~$E-E_{\min}$ per formula unit along cuts on the $X_3^-$ axis (blue solid) and $X_2^+$ axis (red solid) and through the child amplitudes, $Q_{X_2^+}=0.599$~Å (blue dashed) and $Q_{X_3^-}=0.822$~Å (red dashed), the off-node cuts bilinearly interpolated from the grid; markers: parent $\times$, the $Cmcm$ $\square$, $Cmce$ $\triangle$ and $A2_1am$ $\diamond$ anchors at $1/\sqrt2$ of the child amplitudes, and the child $\circ$. All amplitudes are $A_p$. \textbf{b},~$E(Q_{X_3^-},Q_{X_2^+})-E(-Q_{X_3^-},Q_{X_2^+})$, maximum 0.42~meV/f.u., and \textbf{c},~$E(Q_{X_3^-},Q_{X_2^+})-E(Q_{X_3^-},-Q_{X_2^+})$, maximum 0.013~meV/f.u.; both are required to vanish by the parent symmetry, and the residual in \textbf{b} is the real-space projection error of the single-point calculations (\texttt{LREAL = Auto}), negligible on the 50~meV contour scale of main-text Fig.~\Mref{fig:perovskite}e.}
\label{fig:s_ct_mirror}
\end{figure}

Static VASP calculations (PBE; PAW Ca\_pv, Ti, O; 520~eV cut-off; $\Gamma$-centred $2\times6\times6$ $k$-mesh; Gaussian smearing; \texttt{EDIFF}~$=10^{-8}$~eV; \texttt{PREC = Accurate}; \texttt{LREAL = Auto}) were run on the fixed 48-atom $I4/mmm$ cell (in the $A2_1am$ child setting, $a=19.443$ and $b=c=5.436$~Å $=\sqrt2\times3.844$~Å), with settings identical to those of the earlier 630-frame grid described below. The plane holds $37\times37$ points over $\pm1.27$~Å in 0.071~Å steps of $A_p$, plus three anchors at $1/\sqrt2$ of the child amplitudes, $Cmcm$ (0.581, 0), $Cmce$ (0, 0.424) and $A2_1am$ (0.581, 0.424)~Å: 1{,}372 frames, all converged (mean 5.7~min per frame). The campaign normalized its mode vectors to $\sqrt2$~Å per unit, so its stored coordinates are $A_s$ ($\pm1.8$~Å in 0.1~Å steps); they are converted here to $A_p=A_s/\sqrt2$, which is independent of the scale of the exported mode vectors. On this plane the minima lie at $(\pm0.919,\pm0.212)$~Å and the child, at (0.822, 0.599)~Å, lies 51~meV/f.u.\ above them (bilinear interpolation); the closest interatomic contact on the plane is 1.88~Å. The parent symmetry requires $E$ to be even in each amplitude. The $X_2^+$ mirror holds to 0.013~meV/f.u.\ and the $X_3^-$ mirror to 0.42~meV/f.u.; the latter is attributed to real-space projection and will be removed with \texttt{LREAL = .FALSE.} before any sub-meV Landau fit.
Supplementary Fig.~\ref{fig:s_ct_mirror} shows cuts through the plane and both mirror residuals.

An earlier 630-frame ($X_3^-$, $X_2^+$, $\Gamma_5^-$) grid built from the unnormalized exported vectors (calibration factors $s_i=1.25$ for $X_3^-$ and 2.50 for $X_2^+$, so that $Q=1$ fell short of the child by 20\% and 60\%) did not reach the child along $X_2^+$; it is superseded by the plane above and is the example of a scale error that space-group checks cannot detect.

\section{The two-dimensional $K_3$ order parameter}\label{sec:si_k3}

This section details how IRIS builds the second $K_3$ basis vector of the hexagonal manganites (main-text Section~\Mref{sec:res_mn}).

The decomposition by $f_1$ gives $P6_3/mmc\to P6_3cm$ with the $K_3$ order-parameter direction $(a,0)$. The little-group irrep at $k=(\tfrac13,\tfrac13,0)$ is complex and one-dimensional, so a real displacement needs it and its conjugate at $-k$: the physical irrep is two-dimensional, with a real basis $(K_{3a},K_{3b})=(Q\cos\Phi,Q\sin\Phi)$ given by the real and imaginary parts of the complex mode. Operations that keep $k$ act as a common phase; operations that send $k\to-k$ act as complex conjugation, $\mathrm{diag}(\pm1,\mp1)$ (Supplementary Table~\ref{tab:k3}).

\begin{table}[!htbp]
\centering\small
\caption{Action of the parent operations on the $K_3$ pair $(a,b)$, from the irrep matrices of $P6_3/mmc$ harvested by $f_1$.}
\label{tab:k3}
\begin{tabularx}{\linewidth}{@{}L{4.2cm}X@{}}
\toprule
Matrix on $(a,b)$ & Parent operations\\
\midrule
$\mathrm{diag}(1,1)$ & $1$, $3$, $3^{-1}$, $-2_{[210]}$, $-2_{[120]}$, $-2_{[1\bar10]}$\\
$\mathrm{diag}(1,-1)$ & $2_{[001]}$, $6$, $6^{-1}$ ($6_3$ family); $-2_{[100]}$, $-2_{[110]}$, $-2_{[010]}$\\
$\mathrm{diag}(-1,1)$ & $2_{[210]}$, $2_{[120]}$, $2_{[1\bar10]}$; $\bar1$; $\bar3$, $\bar3^{-1}$\\
$\mathrm{diag}(-1,-1)$ & $2_{[100]}$, $2_{[110]}$, $2_{[010]}$; $-2_{[001]}$; $\bar6$, $\bar6^{-1}$\\
$\begin{pmatrix}-\tfrac12&\tfrac{\sqrt3}{2}\\-\tfrac{\sqrt3}{2}&-\tfrac12\end{pmatrix}$ & parent lattice translations $(x+1,y,z)$, $(x,y+1,z)$\\
\bottomrule
\end{tabularx}
\end{table}

Three consequences follow. (i) The $6_3$ screw axis and the $c$ glides map $b\to-b$, which forces $K_{3b}=0$ in the $P6_3cm$ ground state; the stabilizers give $P6_3cm$ (No.~185) for $(a,0)$, $P\bar3c1$ (No.~165, centrosymmetric, so $P=0$ by symmetry) for $(0,a)$, and $P3c1$ (No.~158) for a general $(a,b)$. (ii) A parent lattice translation rotates the pair by $120^\circ$, which is how IRIS builds the second basis vector from the first,
\begin{equation}
\mathbf e_2=\frac{T\mathbf e_1-\cos120^\circ\,\mathbf e_1}{\sin120^\circ};
\end{equation}
the sign of the rotation only maps $\Phi\to-\Phi$, and the Landau energy $F$ (main-text eq.~\Meqref{eq:meier}) is even in $\Phi$. (iii) $a^3-3ab^2=Q^3\cos3\Phi$ is odd under $\bar1$ and $2_{[100]}$, like $P_z$, so $Q^3P\cos3\Phi$ is an invariant; $\mathrm{Re}(a+ib)^6=Q^6\cos6\Phi$ is invariant as well.

For InMnO$_3$ the pair built this way was checked against all 72 operations of the parent in the child cell (24 point operations times three lost translations). The rebuilt $\mathbf e_2$ has an overlap of $+1.000000$ with the stored vector; $\mathbf e_1\cdot\mathbf e_2=6\times10^{-8}$; $|\mathbf e_1|=|\mathbf e_2|=0.91457$ ($A_s$ of $K_3$); and all 72 operations close on $(\mathbf e_1,\mathbf e_2)$. The space group of the parent plus $\cos\Phi\,\mathbf e_1+\sin\Phi\,\mathbf e_2$ is No.~185 at $\Phi=0^\circ$ and $60^\circ$, No.~158 at $15^\circ$ and $45^\circ$, and No.~165 at $30^\circ$ and $90^\circ$.

\section{Additional results for the hexagonal manganites}\label{sec:si_mn}

This section collects the grids, the coefficient sets and the validation behind main-text Section~\Mref{sec:res_mn}, Fig.~\Mref{fig:manganite}, Extended Data Table~\Mref{tab:meier} and Supplementary Table~\ref{tab:mn}. Two coefficient sets are used. The InMnO$_3$ benchmark against Meier \textit{et al.}~\cite{Meier2020} (plane-fit columns of Extended Data Table~\Mref{tab:meier}) is fitted to the $\Phi=0$ ($K_3$, $\Gamma_2^-$) plane alone, in three magnetic and $+U$ settings, to test the sensitivity of the comparison to the magnetic order and the $+U$ scheme; $c'$ is not defined on that plane. The landscapes of Fig.~\Mref{fig:manganite} use a joint fit to the plane and the ($Q$, $\Phi$, $P$) grid, which also determines $c'$; these coefficients, for both InMnO$_3$ and ErMnO$_3$, are the joint-fit columns of Extended Data Table~\Mref{tab:meier} and are repeated in Supplementary Table~\ref{tab:mn_fig4} with the landscape they predict. The same table gives the landscape predicted by the coefficients of Meier \textit{et al.}\ for both compounds, on which the comparison of the main text rests. Supplementary Table~\ref{tab:mn_grids} lists the grids, Supplementary Table~\ref{tab:mn_red} the quantities derived from the InMnO$_3$ plane fits, and Supplementary Figs.~\ref{fig:s_mn_plane}--\ref{fig:s_mn_pstar} compare the models with DFT.

\subsection{Full account of the manganite results}\label{sec:si_mn_ext}
This subsection gives the complete account of the manganite results summarized in main-text Section~\Mref{sec:res_mn}, with every value and comparison.

The hexagonal manganites $R$MnO$_3$ test two steps the perovskites do not need: building an order parameter that the decomposition gives only in part, and benchmarking the resulting Landau model against an independent one. In $P6_3/mmc\to P6_3cm$ (index 6, 30-atom cell) the primary irrep $K_3$ is two-dimensional: its components $(K_{3a},K_{3b})=(Q\cos\Phi,\,Q\sin\Phi)$ are the real and imaginary parts of the complex mode at $k=(\tfrac13,\tfrac13,0)$. The decomposition by $f_1$ returns only the direction $(a,0)$ realized in the child (Supplementary Table~\ref{tab:dec_mn}). IRIS generates the second component by applying a parent lattice translation that is lost in the child, which rotates the $K_3$ pair by $120^\circ$, and verifies that the pair carries the $K_3$ representation matrices for all 72 parent operations (Supplementary Section~\ref{sec:si_k3}, Supplementary Table~\ref{tab:k3}). Frames then have space group $P6_3cm$ at $\Phi=n\cdot60^\circ$, $P\bar{3}c1$ at $\Phi=30^\circ+n\cdot60^\circ$ and $P3c1$ otherwise; main-text Fig.~\Mref{fig:manganite}a shows the corresponding structures. The polar $\Gamma_2^-$ enters only through the trilinear invariant $Q^3P\cos3\Phi$~\cite{Fennie2005,Artyukhin2014}. In InMnO$_3$ the child has $K_3$ 0.528~Å and $\Gamma_2^-$ 0.057~Å (translation-free, after removing a 4\% rigid translation); the polar mode, a secondary kept as the polarization axis by rule R4, also passes the amplitude threshold ($A_s^{\mathrm{tf}}=0.099$~Å, 0.101~Å raw, against $A_{\mathrm{thr}}=0.05$~Å), but only 0.076~Å of $A_s^{\mathrm{tf}}$ (0.078~Å raw) separates it from the largest inactive irrep ($K_1$, $A_s=0.023$~Å), so the selection is flagged as not clear-cut and handled as borderline (rule R5, Methods; Supplementary Fig.~\ref{fig:s_select}f). ErMnO$_3$ has $K_3$ 0.604~Å and $\Gamma_2^-$ 0.126~Å (Supplementary Table~\ref{tab:mn}).

\begin{figure}[!htbp]
\centering
\includegraphics[width=\linewidth]{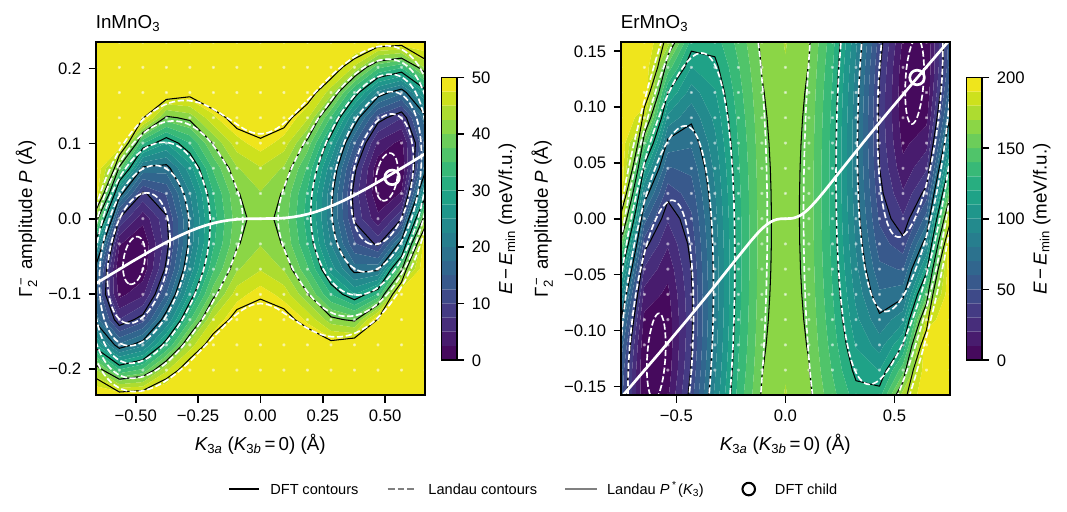}
\caption{DFT energy on the ($K_3$, $\Gamma_2^-$) plane at $\Phi=0$ ($K_{3b}=0$; negative $K_{3a}$ is the $\Phi=180^\circ$ domain) for InMnO$_3$ and ErMnO$_3$, relative to the minimum (colour and black contours), with the Landau model of Supplementary Table~\ref{tab:mn_fig4} (white dashed contours) and its optimal polarization $P^*(K_3)$ (white line). The circle marks the DFT child. DFT on the $15\times15$ plane of each compound (settings in Supplementary Table~\ref{tab:dft}); amplitudes $A_p$ in Å; colour scales capped at 50 (InMnO$_3$) and 200~meV/f.u.\ (ErMnO$_3$). The two wells are related by $(K_3,P)\to(-K_3,-P)$, which the DFT energies obey to 0.1~meV; the tilt of the wells is the trilinear coupling $-gQ^3P\cos3\Phi$.}
\label{fig:s_mn_plane}
\end{figure}

\begin{figure}[!htbp]
\centering
\includegraphics[width=\linewidth]{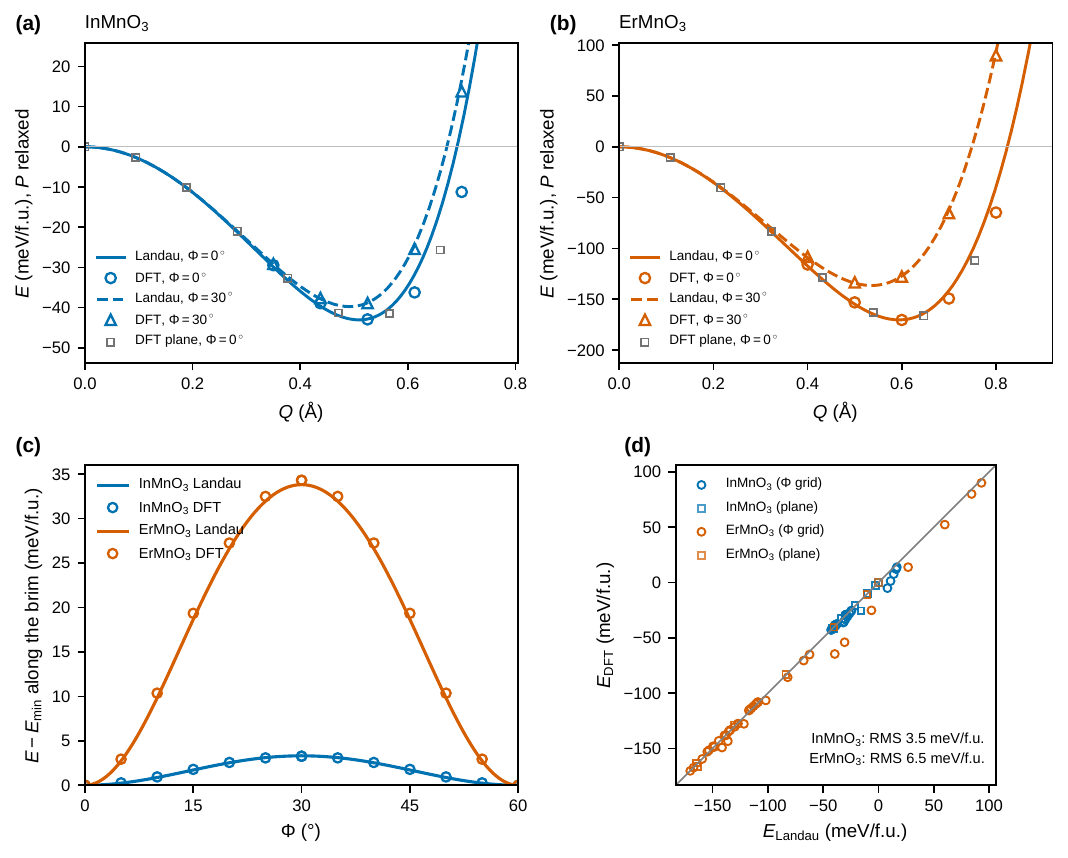}
\caption{Validation of the Landau models against DFT, $P$ relaxed at every point; blue, InMnO$_3$; orange, ErMnO$_3$; energies per formula unit, $Q$ in Å of $A_p$. (a,b)~Radial cuts through the minimum ($\Phi=0$, circles and solid line) and the $P\bar{3}c1$ saddle ($\Phi=30^\circ$, triangles and dashed line); squares, ($K_3$, $\Gamma_2^-$) plane. (c)~Energy along the brim, relative to its minimum, against $\Phi$ (circles, DFT; lines, Landau): the domain-wall barrier is 3.3 (InMnO$_3$) and 34.3~meV/f.u.\ (ErMnO$_3$) in DFT, 3.3 and 33.8 in the model. (d)~DFT against Landau energies for all relaxed nodes (circles, ($Q$, $\Phi$, $P$) grid; squares, plane); RMS 3.5 (InMnO$_3$) and 6.5~meV/f.u.\ (ErMnO$_3$). The largest deviations are at $Q$ beyond the brim, where terms above sixth order, outside eq.~\Meqref{eq:meier}, contribute.}
\label{fig:s_mn_validation}
\end{figure}

For each compound IRIS computes two grids on the fixed child lattice (Supplementary Table~\ref{tab:mn}, Supplementary Table~\ref{tab:mn_grids}): a $15\times15$ ($K_3$, $\Gamma_2^-$) plane at $\Phi=0$ (225 structures), which fixes the radial and polar coefficients, and a ($Q$, $\Phi$, $P$) grid of five radii around the brim, seven angles from $0$ to $30^\circ$ in $5^\circ$ steps and five polar amplitudes (175 structures), which fixes the angular term $c'$. The symmetry of the $K_3$ pair reduces $\Phi$ to this interval: $F$ is even in $\Phi$, and a shift by $60^\circ$ maps $P\to-P$ (Supplementary Section~\ref{sec:si_k3}). For ErMnO$_3$, 11 further structures on the $\Gamma_2^-$ axis at $Q=0$ fix $a_P$. IRIS fits both grids jointly to the Landau form of Artyukhin \textit{et al.}~\cite{Artyukhin2014} and Meier \textit{et al.}~\cite{Meier2020}, main-text eq.~\Meqref{eq:meier}, and minimizes over $P$ to obtain $F(Q,\Phi)$ (main-text Fig.~\Mref{fig:manganite}b,c; coefficients in Supplementary Table~\ref{tab:mn_fig4}). With $P$ relaxed at every node, the model reproduces the DFT energies to 3.5~meV/f.u.\ (InMnO$_3$) and 6.5~meV/f.u.\ (ErMnO$_3$) RMS; the largest deviations lie beyond the brim, where terms above sixth order contribute (Supplementary Fig.~\ref{fig:s_mn_plane}, Supplementary Fig.~\ref{fig:s_mn_validation}).

\begin{figure}[!htbp]
\centering
\includegraphics[width=0.48\linewidth]{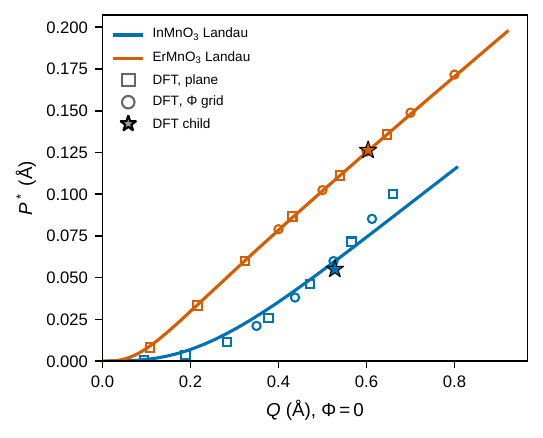}\hfill
\includegraphics[width=0.48\linewidth]{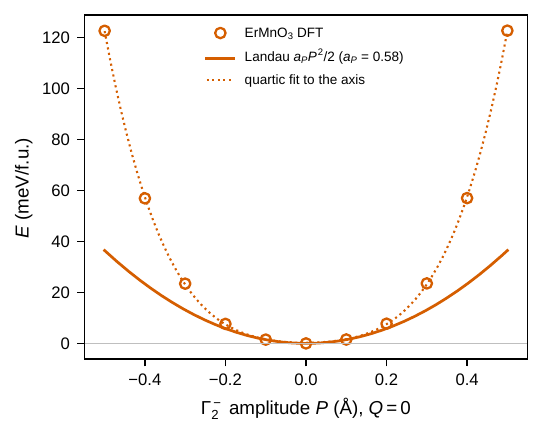}
\caption{Left: optimal polar amplitude $P^*$ against $Q$ at $\Phi=0$ (both in Å of $A_p$; blue, InMnO$_3$; orange, ErMnO$_3$), Landau model (lines) and DFT (squares, plane; circles, $\Phi$ grid; stars, DFT child). $P^*$ grows as $Q^3$ at small $Q$ and tends to $(g/g')Q$ at large $Q$. Right: energy along $\Gamma_2^-$ at $Q=0$ for ErMnO$_3$ (circles, DFT; meV/f.u.): a single well, so the polarization is not unstable on its own and appears only through its coupling to $K_3$ (improper ferroelectricity). Solid, $a_PP^2/2$ with $a_P$ of Supplementary Table~\ref{tab:mn_fig4}; dotted, an earlier quartic fit to the axis ($a_P=0.49$, $b_P=3.9$~eV\,Å$^{-4}$ per cell), shown for comparison only; the same quartic over $|P|\le0.5$~Å, refitted in the replay, gives $a_P=0.507$ (Supplementary Table~\ref{tab:mn_aP}) and fits the axis poorly, so it is not used. The dependence of $a_P$ on the fit window and order is given in Supplementary Table~\ref{tab:mn_aP}.}
\label{fig:s_mn_pstar}
\end{figure}

Both compounds show a Mexican hat: a maximum at the parent, a brim of radius $Q^*$, and six $P6_3cm$ minima at $\Phi=n\cdot60^\circ$ separated by $P\bar{3}c1$ saddles. The hat is 43~meV/f.u.\ deep in InMnO$_3$ and 170~meV/f.u.\ in ErMnO$_3$ (Landau 43.1 and 170.3, DFT 42.9 and 170.4~meV/f.u.), and the brim radius, $Q^*=0.510$ and 0.594~Å in the model and 0.523 and 0.605~Å in DFT, lands on the relaxed child (0.528 and 0.604~Å). The brim is ten times more strongly corrugated in ErMnO$_3$: the barrier between neighbouring domains along the brim is 3.3 and 34.3~meV/f.u.\ in DFT, against 3.3 and 33.8~meV/f.u.\ in the model (Supplementary Fig.~\ref{fig:s_mn_validation}c). The corrugation comes from the polar coupling, not from the sixth-order angular term. With $P$ relaxed, the trilinear term lowers the energy by $(gQ^3\cos3\Phi)^2/[2(a_P+g'Q^2)]$, which is largest at $\Phi=0$ and vanishes at $\Phi=30^\circ$; without $c'$ it would give barriers of 4.94 and 35.84~meV/f.u., and the fitted $c'>0$ (0.068 and 0.040~eV\,Å$^{-6}$ per cell) lowers them by 1.62 and 2.06~meV/f.u., to 3.32 and 33.78~meV/f.u.\ (minimum over $Q$ and $P$ at $\Phi=0$ and $30^\circ$; a 2{,}000- and 4{,}000-point scan in $Q$ and a continuous minimization agree to $10^{-5}$~meV/f.u.); on its own (with $g=0$) it would favour the $P\bar3c1$ saddles, by 1.3 and 1.4~meV/f.u. ErMnO$_3$ has a three times stronger harmonic instability ($a=-3.71$ against $-1.21$~eV\,Å$^{-2}$), hence the deeper hat, and it is more corrugated because its trilinear coupling is 2.4 times larger ($g=2.27$ against 0.96~eV\,Å$^{-4}$) and its polar axis 4.6 times softer ($a_P=0.58$ against 2.66~eV\,Å$^{-2}$; Extended Data Table~\Mref{tab:meier}). The polar axis alone is nevertheless a single well (Supplementary Fig.~\ref{fig:s_mn_pstar}, right): polarization appears only through its coupling to $K_3$, the signature of improper ferroelectricity~\cite{Fennie2005,Artyukhin2014}. The optimal polarization follows $P^*=gQ^3\cos3\Phi/(a_P+g'Q^2)$ (main-text Fig.~\Mref{fig:manganite}d), growing as $Q^3$ at small $Q$ and tending to $(g/g')Q$ beyond the brim (Supplementary Fig.~\ref{fig:s_mn_pstar}, left). It is largest in the $P6_3cm$ wells (Landau 0.056 and 0.124~Å, DFT 0.059 and 0.127~Å, child 0.057 and 0.126~Å), vanishes at the centrosymmetric saddle and changes sign, so neighbouring domains carry opposite polarization, and the DFT values follow the model at every $\Phi$.

Meier \textit{et al.}~\cite{Meier2020,Smabraten2018} give Landau coefficients for both compounds, which provide an independent benchmark (Extended Data Table~\Mref{tab:meier}). The InMnO$_3$ reference set was obtained with the same LDA$+U$ parameters ($U=8$~eV, $J=0.88$~eV) in a different code. To test whether the comparison depends on the magnetic order or the $+U$ scheme, IRIS also fitted the $\Phi=0$ plane alone in three settings; $c'$ is not defined on that plane. The anharmonic and coupling coefficients $b$, $c$, $g$, $g'$ and $a_P$ agree to within a factor of 0.70--1.19, $g$ to within 4\%, and the reduced brim parameter $\tilde b=bQ_0^2/|a|$ is 0.70--0.72 against 0.73 (Supplementary Table~\ref{tab:mn_red}). The one real difference is the harmonic coefficient $a$, 1.48--1.65 times larger in magnitude. It moves the brim out by 14--17\% ($Q^*=0.510$--0.523~Å against 0.447~Å) and doubles its depth ($-43$ to $-51$ against $-22$~meV/f.u.), and the IRIS brim, unlike that of the reference coefficients, lies close to the relaxed child (0.528~Å). Changing the $+U$ scheme, the magnetic order (A-type or the frustrated $\uparrow\uparrow\downarrow$ arrangement) or the convention for the polar axis does not remove the difference. The remaining candidates are the PAW datasets and reference cells of the two codes and, most likely, the direction of the $K_3$ coordinate. $K_3$ occurs on three Wyckoff sites (Supplementary Table~\ref{tab:dec_mn}), so its displacement pattern is not fixed by symmetry: Meier \textit{et al.}\ displaced the atoms along force-constant eigenvectors of the parent~\cite{Meier2020}, whereas IRIS uses the $K_3$ vector of the relaxed child, and the two directions give different energy profiles even at equal amplitude. The joint fit differs from the A-type plane fits by at most 8\% in any shared coefficient ($c$ against the Dudarev fit; 4\% otherwise), so adding the angular grid does not change this comparison. The ErMnO$_3$ reference coefficients were also computed with LDA$+U$ ($U=8$~eV, $J=0.88$~eV) and PAW datasets, in ABINIT~\cite{Gonze2016,Meier2020}. For ErMnO$_3$ the pattern is reversed: the harmonic coefficient agrees to 3\% ($-3.71$ against $-3.83$~eV\,Å$^{-2}$) and $a_P$ to 8\%, whereas $b$, $c$, $g$ and $g'$ are 0.40--0.64 times the reference values. Unlike the InMnO$_3$ runs, which use LDA PAW datasets, the ErMnO$_3$ runs combine the LDA functional with PAW datasets generated for PBE (Supplementary Table~\ref{tab:dft}), a further difference from the reference calculation besides the $K_3$ direction. The reduced brim parameter is again the same, $\tilde b=0.91$ in both, but the softer anharmonic terms move the IRIS brim out to $Q^*=0.594$~Å (0.486~Å from the reference coefficients), deepen the hat to 170~meV/f.u.\ (118~meV/f.u.) and raise the brim barrier to 33.8~meV/f.u.\ (25.0~meV/f.u.; Supplementary Table~\ref{tab:mn_fig4}); as for InMnO$_3$, the IRIS brim lies at the relaxed child (0.604~Å). In both compounds, therefore, the two calculations agree on the shape of the hat, measured by $\tilde b$, and differ in its scale, and the benchmark identifies which coefficients carry the difference: $a$ in InMnO$_3$, the anharmonic and coupling terms in ErMnO$_3$.

\begin{table}[!htbp]
\centering\small
\caption{\textbf{The two hexagonal manganites at a glance.} Amplitudes are $A_p$ in Å; tf, translation-free. Coefficients of main-text eq.~\Meqref{eq:meier} in eV\,Å$^{-n}$ per 30-atom cell with amplitudes $A_s=\sqrt3A_p$, from the joint fit to both grids that underlies main-text Fig.~\Mref{fig:manganite}; energies per formula unit. DFT values have $P$ relaxed at every node; the brim barrier is the energy of the $P\bar3c1$ saddle above the $P6_3cm$ minimum along the brim. Decompositions in Supplementary Table~\ref{tab:dec_mn}; grids in Supplementary Table~\ref{tab:mn_grids}; coefficients and validation in Supplementary Table~\ref{tab:mn_fig4} and Supplementary Fig.~\ref{fig:s_mn_validation}.}
\label{tab:mn}
\renewcommand{\arraystretch}{1.25}
\begin{tabularx}{\textwidth}{@{}L{3.6cm}>{\raggedright\arraybackslash}X>{\raggedright\arraybackslash}X@{}}
\toprule
& InMnO$_3$ & ErMnO$_3$\\
\midrule
Transition, cell & \multicolumn{2}{>{\raggedright\arraybackslash}p{\dimexpr\textwidth-3.6cm-2\tabcolsep\relax}@{}}{$P6_3/mmc\to P6_3cm$, index 6; 30 atoms ($s=3$); LDA$+U$ (Dudarev, $U=8$, $J=0.88$~eV), A-type AFM}\\
Child lattice & $a=5.9075$, $c=11.4058$~Å & $a=6.0396$, $c=11.3269$~Å\\
Order parameters in the child & $K_3$ $(a,0)$ 0.528; $\Gamma_2^-$ 0.057 tf (4\% rigid translation removed; active, borderline margin 0.076~Å on $A_s^{\mathrm{tf}}$) & $K_3$ $(a,0)$ 0.604; $\Gamma_2^-$ 0.126 tf (1\% rigid translation)\\
Second $K_3$ component & \multicolumn{2}{>{\raggedright\arraybackslash}p{\dimexpr\textwidth-3.6cm-2\tabcolsep\relax}@{}}{parent translation, $120^\circ$ rotation of the pair; $K_3$ matrices verified for all 72 parent operations}\\
Frame symmetry & \multicolumn{2}{>{\raggedright\arraybackslash}p{\dimexpr\textwidth-3.6cm-2\tabcolsep\relax}@{}}{$P6_3cm$ at $\Phi=n\cdot60^\circ$; $P\bar3c1$ at $30^\circ+n\cdot60^\circ$; $P3c1$ otherwise}\\
Grids & $15\times15$ plane + $5\times7\times5$ ($Q$, $\Phi$, $P$); 400 structures & $15\times15$ plane + $5\times7\times5$ ($Q$, $\Phi$, $P$) + 11 on the $\Gamma_2^-$ axis; 411 structures\\
$a$, $b$, $c$, $c'$ & $-1.206$, 1.206, 0.700, 0.068 & $-3.708$, 3.973, 0.427, 0.040\\
$a_P$, $g$, $g'$ & 2.663, 0.962, 5.310 & 0.583 ($a_P$ fixed from the $Q=0$ axis), 2.266, 10.330\\
Fit RMS (meV/f.u.) & 3.5 & 6.5\\
Hat depth, Landau / DFT (meV/f.u.) & 43.1 / 42.9 & 170.3 / 170.4\\
Brim barrier, Landau / DFT (meV/f.u.) & 3.3 / 3.3 & 33.8 / 34.3\\
$Q^*$, Landau / DFT / child & 0.510 / 0.523 / 0.528 & 0.594 / 0.605 / 0.604\\
$P^*$, Landau / DFT / child & 0.056 / 0.059 / 0.057 & 0.124 / 0.127 / 0.126\\
Polar axis at $Q=0$ & single well, $a_P>0$ & single well ($a_P=0.58$~eV\,Å$^{-2}$ per cell from the axis; 0.51--0.58 for other fit windows); improper\\
Benchmark against Meier \textit{et al.}~\cite{Meier2020} (Extended Data Table~\Mref{tab:meier}) & plane fits, $c'$ omitted: $b$, $c$, $g$, $g'$, $a_P$ within 0.70--1.19; $a$ 1.48--1.65$\times$; $\tilde b$ 0.70--0.72 vs 0.73; $Q^*$ 0.510--0.523 vs 0.447; hat 43--51 vs 22 & joint fit: $a$ within 3\%, $a_P$ 1.08$\times$; $b$, $c$, $g$, $g'$ 0.40--0.64$\times$; $\tilde b$ 0.91 vs 0.91; $Q^*$ 0.594 vs 0.486; hat 170 vs 118\\
\bottomrule
\end{tabularx}
\end{table}

\begin{table}[!htbp]
\centering\small
\caption{DFT grids for the manganite Landau models (fixed child lattice, static LDA$+U$). Amplitudes $A_p$ in Å.}
\label{tab:mn_grids}
\begin{tabularx}{\linewidth}{@{}>{\raggedright}p{2.6cm}>{\raggedright}X>{\raggedright\arraybackslash}X@{}}
\toprule
 & InMnO$_3$ & ErMnO$_3$ \\
\midrule
($K_3$, $\Gamma_2^-$) plane, $\Phi=0$ & $15\times15$: $K_3\in[-0.66,0.66]$, $P\in[-0.235,0.235]$ & $15\times15$: $K_3\in[-0.755,0.755]$, $P\in[-0.158,0.158]$ \\
($Q$, $\Phi$, $P$) grid & $Q$ 0.35--0.70 (5), $\Phi$ 0--30$^\circ$ (7), $P$ $\pm0.10$ (5) & $Q$ 0.40--0.80 (5), $\Phi$ 0--30$^\circ$ (7), $P$ $\pm0.20$ (5) \\
$\Gamma_2^-$ axis, $Q=0$ & -- & 11 points, $P\in[-0.5,0.5]$ \\
structures & 225 + 175 & 225 + 175 + 11 \\
\bottomrule
\end{tabularx}
\end{table}

\begin{table}[!htbp]
\centering\small
\caption{Landau coefficients used for main-text Fig.~\Mref{fig:manganite} (joint fit to the ($K_3$, $\Gamma_2^-$) plane and the ($Q$, $\Phi$, $P$) grid; for ErMnO$_3$, $a_P$ is fixed from a sixth-order fit to the 11 points of the $Q=0$ axis), the landscape they predict against DFT, and the same quantities from the coefficients of Meier \textit{et al.}~\cite{Meier2020} (Table~I there; InMnO$_3$ from ref.~\cite{Smabraten2018}). Coefficients in eV\,Å$^{-n}$ per 30-atom cell, amplitudes $A_s$. DFT values: $P$ relaxed at every grid node (quartic in $P$), minimum over the sampled radii at each $\Phi$. RMS: relaxed nodes of both grids, model with $P$ relaxed. $Q^*$, $P^*$ in $A_p$; $\tilde b=bQ_0^2/|a|$ with $Q_0$ the radius of the $P=0$ well at $\Phi=0$ ($c'$ included).}
\label{tab:mn_fig4}
\begin{tabular}{@{}lcccc@{}}
\toprule
 & \multicolumn{2}{c}{InMnO$_3$} & \multicolumn{2}{c}{ErMnO$_3$}\\
\cmidrule(lr){2-3}\cmidrule(l){4-5}
 & IRIS & Meier \textit{et al.} & IRIS & Meier \textit{et al.}\\
\midrule
$a$      & $-1.206$ & $-0.82$ & $-3.708$ & $-3.83$\\
$b$      & $1.206$  & $1.13$  & $3.973$  & $6.20$\\
$c$      & $0.700$  & $0.81$  & $0.427$  & $1.06$\\
$c'$     & $0.068$  & $0.04$  & $0.040$  & $0.06$\\
$a_P$    & $2.663$  & $3.48$  & $0.583$  & $0.54$\\
$g$      & $0.962$  & $1.02$  & $2.266$  & $3.76$\\
$g'$     & $5.310$  & $4.85$  & $10.330$ & $17.43$\\
\midrule
$\tilde b$ & 0.69 & 0.72 & 0.91 & 0.91\\
hat depth (meV/f.u.), Landau / DFT    & 43.1 / 42.9 & 21.9 & 170.3 / 170.4 & 117.9\\
brim barrier (meV/f.u.), Landau / DFT & 3.3 / 3.3   & 2.1  & 33.8 / 34.3 & 25.0\\
brim barrier with $c'=0$ (meV/f.u.)   & 4.9 & -- & 35.8 & --\\
$Q^*$ (Å), Landau / DFT / child       & 0.510 / 0.523 / 0.528 & 0.444 & 0.594 / 0.605 / 0.604 & 0.486\\
$P^*$ (Å), Landau / DFT / child       & 0.056 / 0.059 / 0.057 & 0.042 & 0.124 / 0.127 / 0.126 & 0.100\\
RMS, relaxed nodes (meV/f.u.)         & 3.5 & -- & 6.5 & --\\
\bottomrule
\end{tabular}
\end{table}

\begin{table}[!htbp]
\centering\small
\caption{Quantities derived from the plane-fit coefficients of Extended Data Table~\Mref{tab:meier} with main-text eq.~\Meqref{eq:meier} at $\Phi=0$, $P$ minimized ($c'$ omitted in all columns). $Q_0$: radius of the $P=0$ well; $Q^*$, $P^*$: coupled minimum; amplitudes $A_p$ in Å; energies per formula unit relative to the parent. $\tilde b=bQ_0^2/|a|$ (with $Q_0$ in $A_s$).}
\label{tab:mn_red}
\begin{tabular}{@{}lcccc@{}}
\toprule
 & \multicolumn{3}{c}{IRIS (this work)} & Meier \textit{et al.} \\
\cmidrule(lr){2-4}
 & \shortstack{A-AFM\\Dudarev} & \shortstack{A-AFM\\Liechtenstein} & \shortstack{$\uparrow\uparrow\downarrow$\\Liechtenstein} & \\
\midrule
$Q_0$ (Å)                       & 0.481 & 0.482 & 0.492 & 0.419 \\
$Q^*$ (Å)                       & 0.510 & 0.513 & 0.523 & 0.447 \\
$P^*$ (Å)                       & 0.057 & 0.058 & 0.062 & 0.043 \\
$E(Q_0,0)$ (meV/f.u.)           & $-38.5$ & $-38.9$ & $-44.9$ & $-19.6$ \\
$E(Q^*,P^*)$ (meV/f.u.)         & $-43.4$ & $-43.9$ & $-50.7$ & $-22.1$ \\
polar gain (meV/f.u.)           & $-4.8$  & $-5.0$  & $-5.8$  & $-2.5$ \\
$\tilde b$                      & 0.70 & 0.72 & 0.72 & 0.73 \\
\midrule
\shortstack[l]{DFT child (decomposition by $f_1$)\\$K_3$ / $\Gamma_2^-$ (Å)} & \multicolumn{3}{c}{0.528 / 0.057} & -- \\
\bottomrule
\end{tabular}
\end{table}

In the units of Meier \textit{et al.}\ ($A_s$, energies per 30-atom cell), the Dudarev column reads $Q_0=0.833$~Å, $E(Q_0,0)=-231$~meV, coupled minimum $(Q^*,P^*)=(0.883,\,+0.099)$~Å and a hat of 260~meV (Liechtenstein: 0.836, $-233$, (0.889, $+0.100$), 263; $\uparrow\uparrow\downarrow$: 0.852, $-269$, (0.906, $+0.107$), 304; Meier: 0.726, $-118$, (0.774, $+0.074$), 133); the hat measured directly from the DFT plane at $\Phi=0$ for the Dudarev calculation is 246~meV per cell (41~meV/f.u.); at the lowest point of the $(Q,\Phi,P)$ hat grid it is 42.9~meV/f.u.\ (Supplementary Table~\ref{tab:mn}), the value the joint fit reproduces.

All three InMnO$_3$ benchmark calculations used the same 225 structures (the $15\times15$ ($K_3$, $\Gamma_2^-$) plane of Supplementary Table~\ref{tab:mn_grids}) on the relaxed $P6_3cm$ child ($a=5.9075$, $c=11.4058$~Å), with LDA$+U$ ($U=8$, $J=0.88$~eV), ENCUT 820~eV and a $\Gamma$-centred $6\times6\times2$ mesh. The frustrated $\uparrow\uparrow\downarrow$ cell carries a net moment of $8\,\mu_B$ per cell. The plane was built before the polar-origin correction of Supplementary Section~\ref{sec:origin}: its $\Gamma_2^-$ vector contained a rigid translation of $f_T=0.0396$ of its squared norm, and its calibration against the translation-free child therefore gave each unit of the label only $1-f_T=0.960$ of the child's polar content. Converting the labels to translation-free amplitudes multiplies them by $\sqrt{1-f_T}\,(1-f_T)=0.9412$, computed directly as the projection of each structure on the translation-free vector; the structures and energies are unchanged, only $a_P$, $g$ and $g'$ are affected, and the translation-free child amplitude is 0.0573~Å. The ($Q$, $\Phi$, $P$) grid was built after the correction with a translation-free vector, so both InMnO$_3$ grids, and all ErMnO$_3$ grids, share one polar scale. ErMnO$_3$ was computed on its relaxed $P6_3cm$ child ($a=6.0396$, $c=11.3269$~Å, Mn $\pm4.01\,\mu_B$, A-type AFM) and a parent built from it ($a=3.4869$~Å $=a_{\mathrm{child}}/\sqrt3$); its $\Gamma_2^-$ mode carries only 1\% rigid translation.

\begin{table}[!htbp]
\centering\small
\caption{\textbf{Sensitivity of the ErMnO$_3$ polar stiffness to the axis fit.} $a_P$ (eV\,Å$^{-2}$ per 30-atom cell, $A_s$) from the $Q=0$ axis for each window and order, and the hat and brim of the joint fit with $a_P$ fixed at that value. The published choice (bold) fits the axis with $P^2$, $P^4$ and $P^6$ over $|P|\le0.5$~Å; the quartic over the same window fits the axis poorly and is not used.}
\label{tab:mn_aP}
\begin{tabular}{llcccc}
\toprule
$|P|$ window (Å) & powers & points & $a_P$ & axis RMS (meV) & hat / brim (meV/f.u.)\\
\midrule
0.2 & $P^2,P^4$ & 5 & 0.554 & 0.17 & 170.25 / 33.76\\
0.2 & $P^2,P^4,P^6$ & 5 & 0.555 & 0.17 & 170.25 / 33.76\\
0.3 & $P^2,P^4$ & 7 & 0.551 & 0.22 & 170.25 / 33.76\\
0.3 & $P^2,P^4,P^6$ & 7 & 0.555 & 0.21 & 170.25 / 33.76\\
0.4 & $P^2,P^4$ & 9 & 0.554 & 0.24 & 170.25 / 33.76\\
0.4 & $P^2,P^4,P^6$ & 9 & 0.550 & 0.23 & 170.24 / 33.76\\
0.5 & $P^2,P^4$ & 11 & 0.507 & 1.83 & 170.21 / 33.72\\
\textbf{0.5} & $\mathbf{P^2,P^4,P^6}$ & \textbf{11} & \textbf{0.583} & \textbf{0.50} & \textbf{170.27 / 33.78}\\
\bottomrule
\end{tabular}
\end{table}

\section{BiFeO$_3$: the coupled $\Gamma_4^-\oplus R_5^-$ landscape}
\label{sec:si_bfo}
\emergencystretch=2em

This section gives the basis, grid, models, validation and physics behind main-text Section~\Mref{sec:res_bfo}, Fig.~\Mref{fig:bfo} and Supplementary Table~\ref{tab:bfo}: the order parameters, their symmetry and the campaign (Supplementary Section~\ref{sec:si_bfo_grid}); the invariant polynomial and the independent interpolant (Supplementary Tables~\ref{tab:bfo_fits}--\ref{tab:bfo_rbf}); all 17 inequivalent high-symmetry planes (Supplementary Section~\ref{sec:si_bfo_planes}); competition and axis locking (Supplementary Section~\ref{sec:si_bfo_coupling}); the sampling audit and validation set (Supplementary Section~\ref{sec:si_bfo_cov}); switching paths, volume dependence and polarization (Supplementary Tables~\ref{tab:bfo_paths}--\ref{tab:bfo_berry}); relaxed anchors (Supplementary Table~\ref{tab:bfo_lit}); magnetism and domain states (Supplementary Section~\ref{sec:si_bfo_magnet}); and the structure files (Supplementary Section~\ref{sec:si_bfo_struct}).

\subsection{Full account of the BiFeO$_3$ results}\label{sec:si_bfo_ext}
\begin{table}[!htbp]
\centering\small
\caption{\textbf{BiFeO$_3$ at a glance.} Amplitudes are $A_p$ in Å; tf, translation-free. Energies per formula unit at the fixed cubic lattice of the $R3c$ volume (V1, $a_c=3.9407$~Å) unless stated; V0, relaxed cubic volume ($a_c=3.8813$~Å). Decomposition in Supplementary Table~\ref{tab:dec_bfo}; models, planes, paths, polarization and structures in Supplementary Section~\ref{sec:si_bfo}.}
\label{tab:bfo}
\renewcommand{\arraystretch}{1.2}
\begin{tabularx}{\textwidth}{@{}L{3.6cm}>{\raggedright\arraybackslash}X@{}}
\toprule
Transition, cell & $Pm\bar3m\to R3c$; 10-atom fcc$(2a_c)$ cell; PBEsol$+U$ ($U_{\mathrm{eff}}=4$~eV), G-type AFM, collinear, no spin--orbit coupling\\
Order parameters & $\Gamma_4^-$ $(a,a,a)$ 1.073 raw, 0.526 tf ($f_T=0.76$); $R_5^-$ $(a,a,a)$ 0.694; both three-dimensional\\
Symmetry on ($\mathbf P$, $\mathbf R$) & 96 operations; no invariant odd in $\mathbf P\cdot\mathbf R$, so $\mathbf P\parallel\mathbf R$ and $\mathbf P\parallel-\mathbf R$ are degenerate\\
Sampling & one DFT point per orbit: V1 6{,}724 orbits (558{,}685 nodes), V0 2{,}873 (227{,}725); 361 refinement points (274 off the grid, 87 on grid nodes); 253 validation points; 20 SCF-limited runs outside the fitting window\\
Models & degree-12 invariant polynomial (315 invariants): blind RMS 1.15 (max 5.0) on the refine set and 0.57 (max 2.7) on the 253 validation points; final fit to all V1 data, 20\% hold-out 0.51~meV/f.u.; independent degree-8 + RBF interpolant: 1.29 on the validation points, 0.09 along the switching paths\\
Stationary points & $R3c$ $-950.0$ at (0.519, 0.698), the only minimum; $Ima2$ $+101.3$ (index 1); $R3m$ $+211.2$; $R\bar3c$ $+274.2$ above $R3c$\\
Polar--tilt coupling & no trilinear term (forbidden by symmetry); biquadratic $\lambda_1=+12.2$ (competition), $\lambda_2=-5.8$ (axis locking), $\lambda_3=-1.2$~eV\,Å$^{-4}$ (Supplementary Table~\ref{tab:bfo_coef}); penalty $+465$~meV/f.u.\ at $R3c$, $|\mathbf P|$ and $|\mathbf R|$ reduced by 13.4\% and 11.8\%; angular stiffness 1{,}830~meV/f.u.\ per rad$^2$\\
17 inequivalent planes & coupling energy positive on all ($+510$ to $+3{,}090$); joint minima on 6 (\#1, \#2, \#8, \#11, \#14, \#16); single-mode minima on the other 11\\
Switching (V1 / V0) & joint $71^\circ$, $109^\circ$, $180^\circ$ via $Ima2$: 101.3 / 101.6; $\mathbf P$ reversal via $R\bar3c$: 274 / 194; $\mathbf R$ reversal via $R3m$: 211 / 270\\
Domain states (inferred) & joint route: Ederer--Spaldin state 4 ($\mathbf P$, $\mathbf D$, $\mathbf M$ reversed), 101; state 8 ($\mathbf M$ unchanged): 274 direct, 211 via $\mathbf R$ reversal\\
Polarization (Berry phase, $\mathbf R=0$) & 103~$\mu$C\,cm$^{-2}$ at $p=0.526$ (Neaton \textit{et al.}: 95 for $R3c$); $Z^*_{\mathrm{mode}}$ 8.66 at $p\to0$, 6.01 at 0.6~Å; homogeneous $E_c\approx27$~MV\,cm$^{-1}$\\
Relaxed anchors & $R3m$ $+169$, $R\bar3c$ $+225$, $Pm\bar3m$ $+893$ above $R3c$ (Di\'eguez \textit{et al.}: $+169$, $+230$, $+906$)\\
\bottomrule
\end{tabularx}
\end{table}

This subsection gives the complete account of the BiFeO$_3$ results summarized in main-text Section~\Mref{sec:res_bfo}, including the campaign statistics, the coercive-field bounds and the literature comparisons.

BiFeO$_3$ tests the dimensionality policy and the recovery machinery at scale. The transition $Pm\bar{3}m\to R3c$ couples two three-dimensional order parameters, the polar $\Gamma_4^-$ ($\mathbf P$) and the antiferrodistortive octahedral tilt $R_5^-$ ($\mathbf R$). The raw $\Gamma_4^-$ amplitude returned by the decomposition, 1.073~Å, carried a rigid translation that made up 76\% of its squared norm ($f_T=0.76$, 51\% of the amplitude; Supplementary Section~\ref{sec:origin}); the certified, translation-free value is 0.526~Å, beside 0.694~Å for $R_5^-$ (main-text Fig.~\Mref{fig:bfo}a shows single-irrep structures; Supplementary Table~\ref{tab:dec_bfo} and Supplementary Table~\ref{tab:bfo_iso} list the decomposition and isotropy subgroups). Under the 96 operations that act on ($\mathbf P$, $\mathbf R$), the odd translation reverses $\mathbf R$ alone, so no invariant is odd in $\mathbf P\cdot\mathbf R$ and domains with $\mathbf P$ parallel and antiparallel to $\mathbf R$ are exactly degenerate (Supplementary Section~\ref{sec:si_bfo_grid}). A full grid in the six-dimensional ($\mathbf P$, $\mathbf R$) space is unaffordable, so the dimensionality policy computes one DFT point per symmetry orbit. In total, 10{,}054 PBEsol$+U$ calculations on fixed cubic lattices (strain policy P1, Supplementary Fig.~\ref{fig:s_strain}; G-type antiferromagnetic order) cover 786{,}410 grid nodes (Supplementary Fig.~\ref{fig:s_dim}f): 6{,}724 orbits (558{,}685 nodes) at the volume of $R3c$ ($a=3.9407$~Å) and 2{,}873 orbits (227{,}725 nodes) at the relaxed cubic volume ($a=3.8813$~Å), with 361 points on three refinement planes (274 off the grid, 87 coinciding with grid nodes) and 96 control repeats (67 at V1, 29 at V0), of which the 95 converged ones (one V1 control is SCF-limited) reproduce their partners to 0.0015~meV/f.u.\ (Supplementary Section~\ref{sec:si_bfo_grid}). The campaign also exercised failure handling at scale: 65 of the 10{,}054 runs failed on the first pass, and 20 near-metallic runs above the 1~eV fitting window remained SCF-limited after resubmission and were kept for energies only (Supplementary Section~\ref{sec:si_bfo_grid}); these fixes now seed the rule table (Supplementary Table~\ref{tab:rules}). In the campaign these fixes were applied by script; in the gated replay the same failures, injected as faults, are fixed by the stored rules with numbers identical to the published ones (main-text Section~\Mref{sec:autonomy}).

A degree-12 invariant polynomial (315 invariants), fitted to the grid alone, reproduces the 361 refinement energies (274 off the grid lattice, 87 on grid nodes), never used in that fit, with an RMS error of 1.15~meV/f.u.\ (0.48~meV/f.u.\ on a 20\% hold-out of grid orbits), and the 253 validation points computed afterwards with an RMS error of 0.57~meV/f.u.; the final model is fitted to all V1 points within the 1~eV window, 4{,}482 of the 7{,}319 computed (Supplementary Table~\ref{tab:bfo_fits}). Its only minimum is $R3c$, at $|\mathbf P|=0.519$ and $|\mathbf R|=0.698$~Å, 950.0~meV/f.u.\ below the cubic phase; $Ima2$ ($-848.8$~meV/f.u.) is a saddle of index 1, and the single-irrep phases $R3m$ ($-738.9$) and $R\bar{3}c$ ($-675.9$) are unstable towards the missing order parameter (all stationary points in Supplementary Table~\ref{tab:bfo_stat}). Unlike the manganites and Ca$_3$Ti$_2$O$_7$, BiFeO$_3$ has no trilinear coupling, because every invariant is even in both $\mathbf P$ and $\mathbf R$ (Supplementary Section~\ref{sec:si_bfo_planes}). The lowest-order couplings are the three biquadratic invariants of Supplementary Table~\ref{tab:bfo_coef}. The isotropic term $\lambda_1P^2R^2$ ($+12.2$~eV\,Å$^{-4}$ at the cubic point) makes the two order parameters compete: at the minimum each suppresses the other by 12--13\%, at a cost of 465~meV/f.u. The term $\lambda_2(\mathbf P\cdot\mathbf R)^2$ ($-5.8$~eV\,Å$^{-4}$) locks their axes: misaligning them by $10^\circ$ costs 28~meV/f.u., and beyond about $35^\circ$ one order parameter collapses. At $R3c$ the softest transverse mode of the coupled Hessian rotates $\mathbf P$ and $\mathbf R$ together (3.00~eV\,Å$^{-2}$, against 5.25 for $\mathbf P$ and 3.75 for $\mathbf R$ alone), and it is the start of the switching path below (Supplementary Table~\ref{tab:bfo_hess}, Supplementary Section~\ref{sec:si_bfo_coupling}).

Main-text Fig.~\Mref{fig:bfo} shows both effects. On planes where $\mathbf P$ and $\mathbf R$ lie along different axes the polarization vanishes and the minima are pure tilt phases, $Imma$ and $R\bar{3}c$ (main-text Fig.~\Mref{fig:bfo}b,c). With the axes aligned, both order parameters survive, in four minima at $Ima2$ and $R3c$ (main-text Fig.~\Mref{fig:bfo}d,e), which are degenerate because no term linear in $\mathbf P\cdot\mathbf R$ is allowed. The four planes of main-text Fig.~\Mref{fig:bfo} are four of the 17 symmetry-inequivalent planes with $\mathbf P$ and $\mathbf R$ along $\langle001\rangle$, $\langle110\rangle$ or $\langle111\rangle$, to which the 169 axis pairs reduce under the 96 operations (Supplementary Section~\ref{sec:si_bfo_planes}, Supplementary Fig.~\ref{fig:s_bfo_17planes}, Supplementary Table~\ref{tab:bfo_planes}). The coupling energy is positive on all 17, from $+510$~meV/f.u.\ for parallel $\langle110\rangle$ axes to $+3{,}090$~meV/f.u.\ for perpendicular ones; joint minima survive only on six planes (the three aligned ones, the two at $35^\circ$ and the perpendicular $\langle001\rangle$ plane \#2), and on the other 11, one order parameter collapses to the lowest single-mode phase. The low-symmetry $Cc$ minima of the two $35^\circ$ planes ($+183$ and $+208$~meV/f.u.\ above $R3c$) are minima within their planes only, not stationary points of the full space. These results do not depend on the polynomial form: an independent interpolant (a degree-8 invariant trend plus symmetric radial basis functions that reproduces every DFT point exactly) gives the same minimum to 0.5~meV/f.u.\ and the same barrier hierarchy (Supplementary Table~\ref{tab:bfo_rbf}). The 253 further DFT calculations placed between grid nodes, on check planes and along the switching paths agree with this interpolant to 1.3~meV/f.u.\ RMS overall and to 0.09~meV/f.u.\ RMS (maximum 0.29~meV/f.u.) along the paths (Supplementary Table~\ref{tab:bfo_extra}). Every point below the cubic energy on the 17 planes lies within 1.13 grid steps of a DFT point, so no low-energy region is missing from the grid (Supplementary Section~\ref{sec:si_bfo_cov}).

The model also gives homogeneous switching barriers. Rotating $\mathbf P$ by $71^\circ$, $109^\circ$ or $180^\circ$ with $\mathbf R$ following passes only through $Ima2$-type saddles, at 101.3~meV/f.u., whereas reversing $\mathbf P$ with the tilts fixed, the path studied by Xu \textit{et al.}~\cite{Xu2017}, costs 274~meV/f.u.\ through $R\bar{3}c$ (Supplementary Table~\ref{tab:bfo_paths}). The joint barrier does not depend on volume (101.6~meV/f.u.\ at the relaxed cubic volume), whereas the two single reversals do: compression weakens the polar mode much more than the tilt, the order of $R3m$ and $R\bar3c$ reverses, and the $\mathbf P$-reversal barrier falls to 194~meV/f.u.\ while the $\mathbf R$-reversal barrier rises to 270~meV/f.u.\ (Supplementary Table~\ref{tab:bfo_vol}). The routes differ in what they do to the magnetization. The Dzyaloshinskii--Moriya vector that cants the G-type spins is set by the sense of the octahedral rotation, so the weak magnetization $\mathbf M\propto\mathbf D\times\mathbf L$ with $\mathbf D\propto\mathbf R$ reverses with $\mathbf R$ and not with $\mathbf P$~\cite{Ederer2005}. The cheapest route, the joint rotation, therefore reverses $\mathbf P$, $\mathbf R$ and $\mathbf M$ together, whereas reversing $\mathbf P$ alone leaves $\mathbf M$ unchanged but costs 274~meV/f.u.: in the homogeneous limit the locking of the polar and tilt axes is what carries the magnetization with the polarization. In the labels of Ederer and Spaldin, the joint route reaches state 4 ($\mathbf P$, $\mathbf D$ and $\mathbf M$ reversed) across a 101~meV/f.u.\ barrier, whereas state 8 ($\mathbf P$ reversed, $\mathbf M$ unchanged), which they considered the likely outcome because reversing the rotations is costly, needs 274~meV/f.u.\ directly or 211~meV/f.u.\ by reversing $\mathbf R$ first (Supplementary Table~\ref{tab:bfo_states}, Supplementary Section~\ref{sec:si_bfo_magnet}); which state a field selects depends on the field term, not included here. Along the untilted polar axis, Berry-phase calculations give 103~$\mu$C\,cm$^{-2}$ at the $R3c$ polar amplitude, against 95~$\mu$C\,cm$^{-2}$ reported by Neaton \textit{et al.}~\cite{Neaton2005}. They also give a mode effective charge of 8.66~$e$ per formula unit at the cubic point that falls by 30\% towards the $R3m$ minimum, matching the drop of the Born charges of Neaton \textit{et al.}\ from the cubic to the $R3c$ phase (Supplementary Section~\ref{sec:si_berry}, Supplementary Table~\ref{tab:bfo_berry}). The homogeneous coercive field along the untilted line is about 27~MV\,cm$^{-1}$ (about 11~MV\,cm$^{-1}$ in the model with the tilt kept at $R3c$), a 0~K, fixed-cell upper bound, against 8~MV\,cm$^{-1}$ from effective-Hamiltonian dynamics~\cite{Xu2017} and about 12~kV\,cm$^{-1}$ measured in single crystals, where switching proceeds by nucleation~\cite{Lebeugle2007}. Fully relaxed anchors give $R3c<R3m<R\bar{3}c\ll Pm\bar{3}m$ ($+169$, $+225$ and $+893$~meV/f.u.\ above $R3c$), within 5~meV/f.u.\ of the PBEsol$+U$ values of Di\'eguez \textit{et al.}~\cite{Dieguez2011} and Fedorova \textit{et al.}~\cite{Fedorova2022} for $R3m$ and $R\bar{3}c$ and within 16~meV/f.u.\ for the cubic phase (Supplementary Table~\ref{tab:bfo_lit}). The $R3c$ model, written out in the hexagonal setting, puts Fe at $z=0.2243$, against 0.2212 in experiment~\cite{Kubel1990}, with the residual reflecting the fixed cubic metric (Supplementary Section~\ref{sec:si_bfo_struct}, Supplementary Table~\ref{tab:bfo_structs}). The magnetic order is held G-type in all runs: the magnetization is inferred from symmetry, not computed, and the spin canting and the cycloid require spin--orbit calculations that lie outside the present landscape.

\subsection{Order parameters, isotropy subgroups and grid}
\label{sec:si_bfo_grid}

\paragraph{Basis.}
The six coordinates are $\mathbf P=(P_1,P_2,P_3)$ for $\Gamma_4^-$ and $\mathbf R=(R_1,R_2,R_3)$ for $R_5^-$ (Bi at the origin, the setting of the decomposition by $f_1$; $R_5^-$ here is $R_4^+$ in the Fe-origin setting used by Di\'eguez \textit{et al.}\ and Fedorova \textit{et al.}). All coordinates are in \AA\ of $A_p$. The $\Gamma_4^-$ pattern is translation-free (Supplementary Section~\ref{sec:origin}). Per \AA\ of $P_i$, the displacements along $x_i$ are Bi $+0.794$, Fe $+0.120$, O$_\parallel$ $-0.079$ and O$_\perp$ $-0.417$~\AA. $P_i>0$ moves Bi along $+x_i$. $R_i>0$ rotates the octahedron at Fe$(\tfrac12,\tfrac12,\tfrac12)$ right-handedly about $+x_i$. The decomposition of relaxed $R3c$ gives $|\mathbf P|=0.526$~\AA\ (1.073~\AA\ before the rigid translation is removed) and $|\mathbf R|=0.694$~\AA\ (Supplementary Table~\ref{tab:dec_bfo}). The polar direction of the grid overlaps the unstable force-constant eigenvector of the cubic phase to 0.998 at V1 and 0.999 at V0, and is close to the soft polar mode of Di\'eguez \textit{et al.}~\cite{Dieguez2011} (Bi 0.80, Fe 0.06, O $-0.42$/$-0.04$).

\paragraph{Symmetry.}
The 48 point operations of $Pm\bar3m$, combined with the odd translation, give 96 operations on $(\mathbf P,\mathbf R)$. The translation reverses $\mathbf R$ and leaves $\mathbf P$ unchanged. In closed form every operation is
\begin{equation}
G=\begin{pmatrix} A & 0\\ 0 & t\,\det(A)\,A\end{pmatrix},\qquad A\in\{\text{signed }3\times3\text{ permutations}\},\quad t=\pm1 .
\end{equation}
$\mathbf P$ is a polar vector and $\mathbf R$ an axial vector. Because $t$ changes the sign of $\mathbf R$ alone, no invariant is odd in $\mathbf P\cdot\mathbf R$. Domains with $\mathbf P\parallel\mathbf R$ and $\mathbf P$ antiparallel to $\mathbf R$ are therefore exactly degenerate.

\begin{table}[!htbp]
\centering
\caption{\textbf{Isotropy subgroups of $Pm\bar3m$ for $\Gamma_4^-\oplus R_5^-$ in BiFeO$_3$.} Space groups were identified with spglib~\cite{Togo2024} (tolerance $10^{-3}$~\AA) from structures generated in the 10-atom fcc$(2a_c)$ cell. The plane numbers refer to Supplementary Fig.~\ref{fig:s_bfo_17planes} and Supplementary Table~\ref{tab:bfo_planes}. Several entries with the same symbol (e.g.\ $Ima2$) are different subgroups in different settings.}
\label{tab:bfo_iso}
\begin{tabular}{lll}
\toprule
$\Gamma_4^-$ ($\mathbf P$) & $R_5^-$ ($\mathbf R$) & space group \\
\midrule
$(0,0,0)$ & $(0,0,0)$ & $Pm\bar3m$ (221) \\
$(a,0,0)$ & 0 & $P4mm$ (99) \\
$(a,a,0)$ & 0 & $Amm2$ (38) \\
$(a,a,a)$ & 0 & $R3m$ (160) \\
0 & $(b,0,0)$ & $I4/mcm$ (140) \\
0 & $(b,b,0)$ & $Imma$ (74) \\
0 & $(b,b,b)$ & $R\bar3c$ (167) \\
$(0,0,a)$ & $(0,0,b)$ & $I4cm$ (108) \quad[\#1] \\
$(a,a,0)$ & $(b,b,0)$ & $Ima2$ (46) \quad[\#8] \\
$(a,a,a)$ & $(b,b,b)$ & $R3c$ (161) \quad[\#16] \\
$(0,0,a)$ & $(b,0,0)$ & $Fmm2$ \quad[\#2] \\
$(0,0,a)$ & $(b,b,0)$ & $Ima2$ \quad[\#4] \\
$(a,a,0)$ & $(0,0,b)$ & $Ima2$ \quad[\#7] \\
$(a,a,0)$ & $(b,-b,0)$ & $Imm2$ \quad[\#10] \\
$(a,a,0)$ & $(b,-b,b)$ & $C2$ \quad[\#12] \\
\multicolumn{2}{l}{\#3, \#6, \#15} & $Cm$ \\
\multicolumn{2}{l}{\#5, \#11, \#13, \#14, \#17} & $Cc$ \\
$(a,a,0)$ & $(b,0,b)$ & $P1$ \quad[\#9] \\
\bottomrule
\end{tabular}
\end{table}

\paragraph{Grid and campaign.}
The energy surface was sampled with one DFT point per symmetry orbit on a lattice with steps $\Delta P=0.15189$~\AA\ and $\Delta R=0.20037$~\AA. The lattice reaches $|\mathbf P|\le0.911$~\AA\ and $|\mathbf R|\le1.202$~\AA. Points with any interatomic distance below 1.6~\AA\ were not computed. The datasets were:
\begin{itemize}
  \item main\_V1: $a_c=3.9407$~\AA, the volume of relaxed $R3c$; 6{,}724 orbits, which unfold to 558{,}685 grid nodes.
  \item main\_V0: $a_c=3.8813$~\AA, the relaxed cubic volume; 2{,}873 orbits, 227{,}725 nodes.
  \item refine\_V1: 361 points on the three planes with $\mathbf P\parallel\mathbf R$ along $[001]$, $[110]$ and $[111]$, 274 of them off the grid lattice and 87 coinciding with main\_V1 nodes.
  \item extra\_V1: 253 validation points (Supplementary Section~\ref{sec:si_bfo_cov}).
\end{itemize}
In terms of runs, the 10{,}054 grid calculations comprise the 361 refinement points, 6{,}771 converged and 20 SCF-limited runs at V1, and 2{,}902 runs at V0; they cover every orbit inside the contact cut.
All runs used PBEsol$+U$ ($U_{\rm eff}=4$~eV on Fe, Dudarev~\cite{Dudarev1998}), G-type antiferromagnetic order, collinear spins without spin--orbit coupling, a $6\times6\times6$ $k$-mesh and the 10-atom cell. Each run was a static calculation that recorded energy, forces and stress. Energies are referenced to $Pm\bar3m$ at the same volume.

Quality control:
\begin{itemize}
  \item Control repeats agree to 0.0015~meV/f.u.
  \item The 87 points that coincide between refine\_V1 and main\_V1 agree to 0.005~meV/f.u.
  \item $|m_{\rm Fe}|$ lies between 3.91 and 4.14~$\mu_{\rm B}$ throughout.
\end{itemize}

Of the 10{,}054 grid runs, 65 failed on the first pass: 50 hit the walltime (45 converged on resubmission) and 15 hit the SCF step limit (NELM 150) and did not converge with robust mixing (LMAXMIX 4, AMIX 0.2, BMIX $10^{-4}$, AMIX\_MAG 0.8, BMIX\_MAG $10^{-4}$, NELM 300). These 15 and the 5 walltime runs that did not converge on resubmission, 20 SCF-limited runs in all, were stopped after about 80 steps to obtain $E_0$:
\begin{itemize}
  \item They lie at the outer corner of the grid ($|\mathbf P|=0.87$--0.91, $|\mathbf R|=1.17$--1.20~\AA), 1.7--5.6~eV/f.u.\ above the cubic phase, with spin gaps below 0.15~eV.
  \item Their energies are uncertain by up to $\pm15$~meV/f.u. Two independent attempts agree to better than 1~meV for 7 of them and to 10--12.5~meV for 4.
  \item They are kept for energies and excluded from gradient fits. None lies inside the 1~eV fitting window, so no result below depends on them.
\end{itemize}

\subsection{Invariant polynomial model}

The model is a polynomial in the invariants of the 96-element group. The invariants are orbit sums of monomials; their counts up to degree 2, 4, 6, 8, 10 and 12 are 2, 9, 27, 68, 152 and 315.

Fit settings:
\begin{itemize}
  \item Data: energies and all six gradient components below 1~eV/f.u., 7 values per point, each orbit counted once (not weighted by its size).
  \item Gradients are scaled by a length of 0.1~\AA.
  \item Validation model: main\_V1 only (3{,}906 orbits). The refine\_V1 set (274 points off the grid lattice, 87 on grid nodes) and the later extra\_V1 set (Supplementary Section~\ref{sec:si_bfo_cov}) were never used in fitting it and serve as blind tests.
  \item Final model: main\_V1, refine\_V1 and extra\_V1 together (4{,}482 points below 1~eV/f.u.). All results below and in the main text use the final model; it changes every energy by at most 0.5~meV/f.u.\ and every amplitude by at most 0.001~\AA\ relative to the validation model.
\end{itemize}

\begin{table}[!htbp]
\centering
\caption{\textbf{Invariant polynomial fits to the BiFeO$_3$ grid (V1).} RMS errors in meV/f.u.; maximum errors in parentheses. Rows 1--3 are fitted to main\_V1 only; their blind tests are the 361-point refine set (274 points off the grid lattice, 87 on grid nodes) and the 253 extra\_V1 points. The hold-out removes 20\% of the orbits, with all their images, from the fit. The last row is the final model, fitted to all V1 data, with the hold-out drawn from the combined set. Changing the fit window to 600 or 1500~meV/f.u.\ changes the blind RMS to 0.88 and 1.51, respectively.}
\label{tab:bfo_fits}
\begin{tabular}{lccccc}
\toprule
degree (invariants) & train & blind refine & refine, $E<-500$ & blind extra\_V1 & 20\% hold-out \\
\midrule
8 (68)   & 5.2  & 9.4 (22)   & 8.0  & -- & -- \\
10 (152) & 1.6  & 3.4 (10.4) & 2.7  & -- & -- \\
12 (315) & 0.43 & 1.15 (5.0) & 0.76 & 0.57 (2.7) & 0.48 (5.1) \\
12 (315), final & 0.49 & -- & -- & -- & 0.51 (4.4) \\
\bottomrule
\end{tabular}
\end{table}

\begin{table}[!htbp]
\centering
\caption{\textbf{Stationary points of the final degree-12 model (V1).} Energies (meV/f.u.) are relative to cubic $Pm\bar3m$ ($E$) and to $R3c$ ($E-E_{R3c}$). The index is the number of negative eigenvalues of the full $6\times6$ Hessian. All 60 random starts converged to $R3c$. Energies are rounded to 0.1~meV/f.u.; the differences are taken from the unrounded values. The $R3c$ Hessian eigenvalues are 3.00 ($\times2$), 3.22, 6.01 ($\times2$) and 9.19~eV\,\AA$^{-2}$ per f.u.}
\label{tab:bfo_stat}
\begin{tabular}{lrrccl}
\toprule
phase & $E$ & $E-E_{R3c}$ & $|\mathbf P|,\ |\mathbf R|$ (\AA) & index & unstable direction \\
\midrule
$R3c$     & $-950.0$ & 0      & 0.519, 0.698 & 0 & -- \\
$Ima2$    & $-848.8$ & $+101.3$ & 0.488, 0.622 & 1 & joint $\mathbf P{+}\mathbf R$ rotation from $[110]$ towards $[11\pm1]$ \\
$R3m$     & $-738.9$ & $+211.2$ & 0.599, 0     & 1 & $\mathbf R\parallel[111]$ \\
$R\bar3c$ & $-675.9$ & $+274.2$ & 0, 0.791     & 1 & $\mathbf P\parallel[111]$ \\
$Imma$    & $-638.3$ & $+311.7$ & 0, 0.778     & 2 & $\mathbf P\parallel[110]$; $R_z$ \\
$Amm2$    & $-619.2$ & $+330.9$ & 0.541, 0     & 2 & $P_z$; $\mathbf R\parallel[110]$ \\
$I4cm$    & $-615.9$ & $+334.1$ & 0.336, 0.577 & 2 & joint $\mathbf P{+}\mathbf R$ towards $x$, $y$ \\
$P4mm$    & $-553.2$ & $+396.9$ & 0.547, 0     & 5 & \\
$I4/mcm$  & $-534.4$ & $+415.7$ & 0, 0.734     & 5 & \\
\bottomrule
\end{tabular}
\end{table}

The second- and fourth-order coefficients of the degree-12 model, including the three biquadratic $\mathbf P$--$\mathbf R$ couplings, are given in Supplementary Table~\ref{tab:bfo_coef}.

\begin{table}[!htbp]
\centering\small
\caption{\textbf{Landau coefficients of BiFeO$_3$.} Second- and fourth-order Taylor coefficients of the final degree-12 invariant model (all V1 data) at the cubic point, per formula unit, amplitudes $A_p$:
$E=\alpha_P P^2+\alpha_R R^2+\beta_1 (P^2)^2+\beta_2\sum_i P_i^4+\gamma_1 (R^2)^2+\gamma_2\sum_i R_i^4+\lambda_1 P^2R^2+\lambda_2 (\mathbf P\cdot\mathbf R)^2+\lambda_3\sum_i P_i^2R_i^2+O(6)$.
No trilinear coupling exists (Supplementary Section~\ref{sec:si_bfo_planes}), so the three $\lambda$ terms (the three independent components of the polar--tilt tensor of effective Hamiltonians~\cite{Kornev2006}) are the lowest-order couplings. Low- and high-order coefficients of a degree-12 fit are correlated; the directional quantities of Supplementary Section~\ref{sec:si_bfo_coupling} are the more robust measure of the coupling.}
\label{tab:bfo_coef}
\begin{tabular}{@{}llr@{}}
\toprule
Term & Meaning & Coefficient\\
\midrule
$\alpha_P P^2$ & soft polar mode ($a_P=2\alpha_P=-8.47$) & $-4.234$~eV\,Å$^{-2}$\\
$\alpha_R R^2$ & soft tilt ($a_R=-4.75$) & $-2.376$~eV\,Å$^{-2}$\\
$\beta_1 (P^2)^2$, $\beta_2\sum P_i^4$ & polar quartic & $+5.21$, $+5.06$~eV\,Å$^{-4}$\\
$\gamma_1 (R^2)^2$, $\gamma_2\sum R_i^4$ & tilt quartic & $+2.00$, $+1.69$~eV\,Å$^{-4}$\\
$\lambda_1 P^2R^2$ & isotropic competition & $+12.21$~eV\,Å$^{-4}$\\
$\lambda_2 (\mathbf P\cdot\mathbf R)^2$ & axis locking & $-5.78$~eV\,Å$^{-4}$\\
$\lambda_3 \sum P_i^2R_i^2$ & cubic anisotropy of the coupling & $-1.15$~eV\,Å$^{-4}$\\
trilinear ($\mathbf P\mathbf R\mathbf R$, $\mathbf P\mathbf P\mathbf R$) & forbidden by symmetry & 0\\
\midrule
\multicolumn{2}{@{}l}{effective biquadratic, $\mathbf P\parallel\mathbf R\parallel[111]$: $\lambda_1+\lambda_2+\lambda_3/3$} & $+6.05$~eV\,Å$^{-4}$\\
\multicolumn{2}{@{}l}{$\mathbf P\parallel[111]\perp\mathbf R\parallel[1\bar10]$: $\lambda_1+\lambda_3/3$} & $+11.8$~eV\,Å$^{-4}$\\
\multicolumn{2}{@{}l}{coupled Hessian at $R3c$, longitudinal / transverse $P$--$R$ element} & $+1.70$ / $-1.30$~eV\,Å$^{-2}$\\
\bottomrule
\end{tabular}
\end{table}

\paragraph{Fourth order is not enough.} A fourth-order expansion reproduces the ordering of the single-mode phases but not their energies: with the Taylor coefficients of Supplementary Table~\ref{tab:bfo_coef}, the fourth-order minima are 6--39\% too shallow and their amplitudes too small (for example $R\bar3c$: $-551$~meV/f.u.\ at 0.681~\AA, against $-676$ at 0.791~\AA\ in the degree-12 model; $R3m$: $-650$ at 0.554~\AA\ against $-739$ at 0.599~\AA). Over the whole six-dimensional window the training RMS falls from 87--143~meV/f.u.\ at degree 4 to 22--44 at degree 6, 5 at degree 8, 1.6 at degree 10 and 0.43 at degree 12. In the tilt-only subspace, the fourth-order rule that only one type of minimum exists still holds: $R\bar3c$ is the only tilt-only minimum, and $Imma$ and $I4/mcm$ are saddles (Supplementary Table~\ref{tab:bfo_stat}).

\subsection{Independent interpolant}

To check that no conclusion depends on the polynomial form, a second surface was built from the same data. It is a degree-8 invariant trend fitted to energies only (the 68 invariants and a constant) plus a thin-plate radial-basis interpolant~\cite{Duchon1977} of the residual (SciPy~\cite{Virtanen2020} \texttt{RBFInterpolator}, 100 neighbours). The radial-basis part acts on the data unfolded with all 96 operations, about 561{,}000 points. The surface passes through every DFT point to $10^{-12}$~meV and is exactly symmetric.

\begin{table}[!htbp]
\centering
\caption{\textbf{RBF interpolant: blind tests and comparison with the degree-12 model.} Errors are DFT minus prediction, in meV/f.u. (a) Blind tests of the RBF fitted to main\_V1 and refine\_V1 only; the 253 extra\_V1 points were computed after this surface had been built. (b) Final surfaces, both fitted to all V1 data (main\_V1, refine\_V1 and extra\_V1).}
\label{tab:bfo_rbf}
\begin{tabular}{lrr}
\toprule
\multicolumn{3}{l}{\textit{(a) blind tests}} \\
test & RMS & max \\
\midrule
held-out grid orbits (5\%, all 96 images removed) & 3.84 & 23.8 \\
\quad of which $E<-500$ & 0.45 & 1.7 \\
\quad of which $-500<E<0$ & 1.0--1.4 & 3.6 \\
refine set (361; 274 off-lattice) & 2.41 & 12.0 \\
\quad of which $E<0$ (261) & 1.69 & 6.6 \\
extra\_V1, plane checks (27) & 0.83 & 2.56 \\
extra\_V1, switching-path images incl.\ saddles (25) & 0.09 & 0.29 \\
extra\_V1, space-filling points (200) & 1.42 & 5.06 \\
extra\_V1, all (253) & 1.29 & 5.06 \\
\midrule
\multicolumn{3}{l}{\textit{(b) landscape, relative to $R3c$}} \\
point & degree-12 & RBF \\
\midrule
$R3c$ ($E$ vs cubic) & $-950.0$ & $-950.5$ \\
\quad $|\mathbf P|$, $|\mathbf R|$ (\AA) & 0.519, 0.698 & 0.519, 0.699 \\
$Ima2$ (joint 71$^\circ$ saddle) & 101.3 & 101.3 \\
$R3m$ ($\mathbf R=0$) & 211.2 & 210.5 \\
$R\bar3c$ ($\mathbf P=0$) & 274.2 & 274.7 \\
\bottomrule
\end{tabular}
\end{table}

Other interpolants performed worse on the refine set: a pure RBF in the raw coordinates gave an RMS of 22--39~meV/f.u., a quintic RBF in invariant coordinates 25, and a degree-6 trend plus RBF 12.6. Polynomial tails of degree 2 or higher inside \texttt{RBFInterpolator} are singular on the lattice grid. The final RBF, passing through all 7{,}319 V1 points (581{,}921 unfolded), changes the landscape values of part (b) by at most 0.1~meV/f.u.\ and the plane minima of Supplementary Table~\ref{tab:bfo_planes} by at most 0.3~meV/f.u.

\subsection{All high-symmetry $\mathbf P$--$\mathbf R$ planes}
\label{sec:si_bfo_planes}

\begin{figure}[!htbp]
  \centering
  \includegraphics[width=\textwidth]{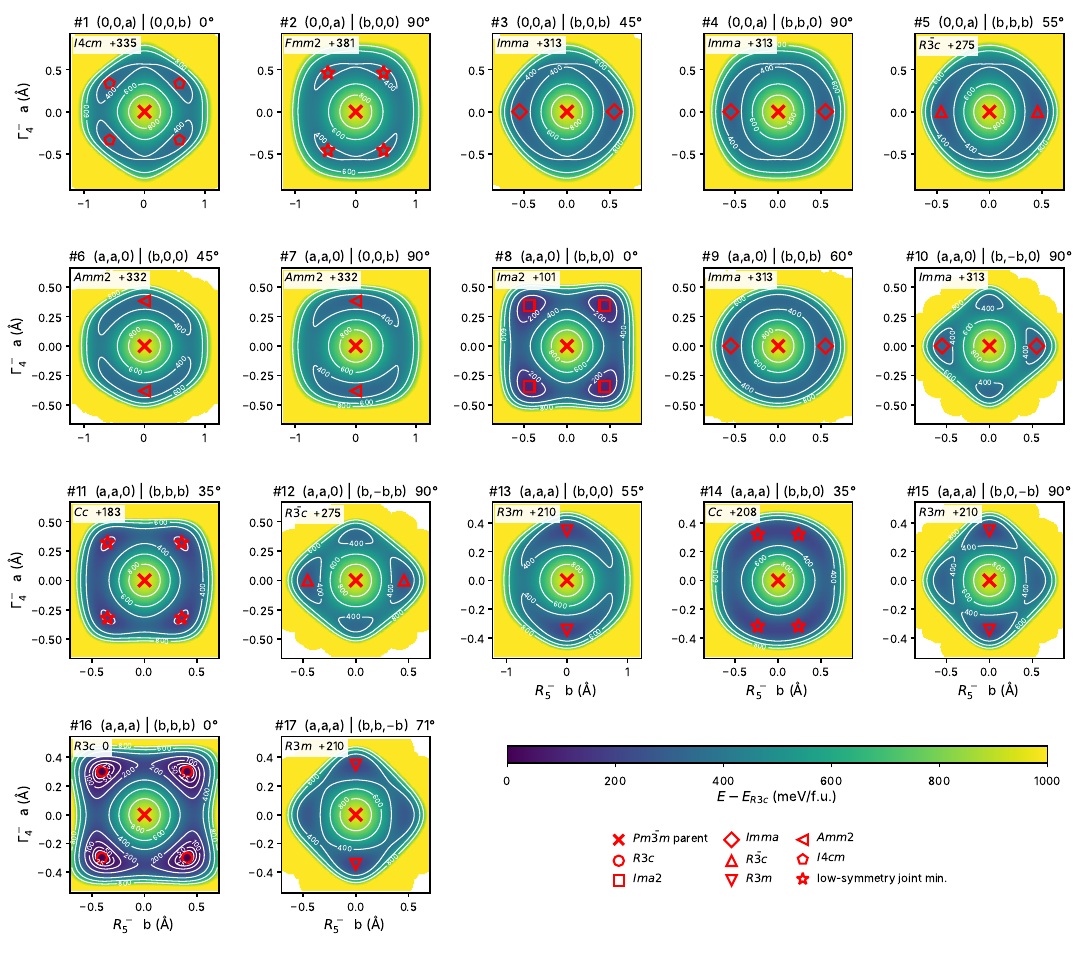}
  \caption{\textbf{Energy on all 17 inequivalent high-symmetry $\Gamma_4^-$--$R_5^-$ planes of BiFeO$_3$.}
  Each panel is the plane $\mathbf P=a\,\mathbf n_1$, $\mathbf R=b\,\mathbf n_2$. Each panel title gives $\mathbf n_1$ $|$ $\mathbf n_2$ and the angle between them. The axes give the per-component amplitudes $b$ ($R_5^-$, horizontal) and $a$ ($\Gamma_4^-$, vertical), so $|\mathbf P|=a|\mathbf n_1|$ and $|\mathbf R|=b|\mathbf n_2|$.
  The colour scale is $E-E_{R3c}$ in meV/f.u., common to all panels and clipped at 1000. White contours are at 25, 50, 100, 200, 400, 600 and 800~meV/f.u.
  The surface is the final RBF interpolant through all main\_V1, refine\_V1 and extra\_V1 DFT points, at the fixed cubic lattice ($a_c=3.9407$~\AA) with G-type antiferromagnetic order. White areas are more than 1.25 grid steps (one step is 0.152~\AA\ in $\mathbf P$ and 0.200~\AA\ in $\mathbf R$) from every DFT point and its 96 images, or outside the grid reach ($|\mathbf P|\le0.911$, $|\mathbf R|\le1.202$~\AA); they are not extrapolated. Grid nodes beyond the 1.6~\AA\ contact cut were not computed.
  The red $\times$ marks the cubic parent. Red open symbols mark the minimum of each plane, and the symbol shape gives its space group. The label in each panel gives that space group and its energy above $R3c$ (all values in Supplementary Table~\ref{tab:bfo_planes}).
  Joint minima, with both $a$ and $b$ nonzero, occur only on planes \#1, \#2, \#8, \#11, \#14 and \#16. These are the three planes with parallel axes, the two with axes $35^\circ$ apart and the perpendicular $\langle001\rangle$ plane \#2.}
  \label{fig:s_bfo_17planes}
\end{figure}

A plane through the six-dimensional space is defined by $\mathbf P=a\,\mathbf n_1$ and $\mathbf R=b\,\mathbf n_2$, with $\mathbf n_1$ and $\mathbf n_2$ chosen from $\langle001\rangle$ (3 axes), $\langle110\rangle$ (6) and $\langle111\rangle$ (4). With each axis counted once up to sign, this gives $13\times13=169$ axis pairs. Under the 96 operations they fall into 17 inequivalent classes (Supplementary Table~\ref{tab:bfo_planes}; the full list of 169 pairs is in \texttt{bfo\_PR\_combinations\_169.csv}). Main-text Fig.~\Mref{fig:bfo}b--e shows planes \#4, \#5, \#8 and \#16. Supplementary Fig.~\ref{fig:s_bfo_17planes} shows all 17.

\begin{table}[!htbp]
\centering
\small
\caption{\textbf{The 17 inequivalent $\Gamma_4^-$--$R_5^-$ planes.} Columns: $\theta$, the angle between $\mathbf n_1$ and $\mathbf n_2$ (modulo sign); pairs, the number of the 169 axis pairs in the class; $G_{\rm plane}$, the space group of a generic point on the plane; $G_{\min}$, the space group of the plane minimum; $a_{\min}$, $b_{\min}$, the per-component amplitudes at the minimum (\AA); $\Delta E_{\min}$, the minimum energy relative to $R3c$; $E_{\rm c}$, the coupling energy $E(a_P\mathbf n_1,b_R\mathbf n_2)-E(a_P\mathbf n_1,0)-E(0,b_R\mathbf n_2)$, where $a_P$ and $b_R$ are the single-mode minima on the two axes of the plane. Energies in meV/f.u., from the final RBF surface. Minima with $a$ or $b$ below 0.05~\AA\ are reported as lying on the axis.}
\label{tab:bfo_planes}
\resizebox{\linewidth}{!}{%
\begin{tabular}{rllrrllccrr}
\toprule
\# & $\mathbf n_1$ ($\Gamma_4^-$) & $\mathbf n_2$ ($R_5^-$) & $\theta$ ($^\circ$) & pairs & $G_{\rm plane}$ & $G_{\min}$ & $a_{\min}$ & $b_{\min}$ & $\Delta E_{\min}$ & $E_{\rm c}$ \\
\midrule
1  & $(0,0,1)$ & $(0,0,1)$   & 0    & 3  & $I4cm$ & $I4cm$      & 0.335 & 0.578 & $+335$ & $+1286$ \\
2  & $(0,0,1)$ & $(1,0,0)$   & 90   & 6  & $Fmm2$ & $Fmm2$      & 0.460 & 0.460 & $+381$ & $+733$ \\
3  & $(0,0,1)$ & $(1,0,1)$   & 45   & 12 & $Cm$   & $Imma$      & 0 & 0.551 & $+313$ & $+1331$ \\
4  & $(0,0,1)$ & $(1,1,0)$   & 90   & 6  & $Ima2$ & $Imma$      & 0 & 0.551 & $+313$ & $+894$ \\
5  & $(0,0,1)$ & $(1,1,1)$   & 54.7 & 12 & $Cc$   & $R\bar3c$   & 0 & 0.457 & $+275$ & $+1259$ \\
6  & $(1,1,0)$ & $(1,0,0)$   & 45   & 12 & $Cm$   & $Amm2$      & 0.383 & 0 & $+332$ & $+1164$ \\
7  & $(1,1,0)$ & $(0,0,1)$   & 90   & 6  & $Ima2$ & $Amm2$      & 0.382 & 0 & $+332$ & $+815$ \\
8  & $(1,1,0)$ & $(1,1,0)$   & 0    & 6  & $Ima2$ & $Ima2$      & 0.345 & 0.441 & $+101$ & $+510$ \\
9  & $(1,1,0)$ & $(1,0,1)$   & 60   & 24 & $P1$   & $Imma$      & 0 & 0.551 & $+313$ & $+1253$ \\
10 & $(1,1,0)$ & $(1,\bar1,0)$ & 90 & 6  & $Imm2$ & $Imma$      & 0 & 0.551 & $+313$ & $+3090$ \\
11 & $(1,1,0)$ & $(1,1,1)$   & 35.3 & 12 & $Cc$   & $Cc$        & 0.320 & 0.351 & $+183$ & $+717$ \\
12 & $(1,1,0)$ & $(1,\bar1,1)$ & 90 & 12 & $C2$   & $R\bar3c$   & 0 & 0.457 & $+275$ & $+2422$ \\
13 & $(1,1,1)$ & $(1,0,0)$   & 54.7 & 12 & $Cc$   & $R3m$       & 0.346 & 0 & $+210$ & $+1329$ \\
14 & $(1,1,1)$ & $(1,1,0)$   & 35.3 & 12 & $Cc$   & $Cc$        & 0.319 & 0.237 & $+208$ & $+1092$ \\
15 & $(1,1,1)$ & $(1,0,\bar1)$ & 90 & 12 & $Cm$   & $R3m$       & 0.346 & 0 & $+210$ & $+2615$ \\
16 & $(1,1,1)$ & $(1,1,1)$   & 0    & 4  & $R3c$  & $R3c$       & 0.300 & 0.404 & 0      & $+525$ \\
17 & $(1,1,1)$ & $(1,1,\bar1)$ & 70.5 & 12 & $Cc$ & $R3m$       & 0.346 & 0 & $+210$ & $+2526$ \\
\midrule
   &           &             &      & 169 &       &             &       &       &        &        \\
\bottomrule
\end{tabular}}
\end{table}

The table supports four observations.
\begin{enumerate}
  \item \textbf{Every plane is competitive.} $E_{\rm c}$ is positive on all 17 planes, so no pair of $\mathbf P$ and $\mathbf R$ axes is uncoupled. The coupling is weakest for parallel axes (\#8, $+510$; \#16, $+525$) and strongest for strongly misaligned axes (\#10, \#12, \#15 and \#17, $70.5$--$90^\circ$; $+2400$ to $+3100$~meV/f.u.).
  \item \textbf{Where the coupling is too strong, one order parameter collapses.} On 11 planes the minimum lies on an axis. The surviving single-mode phase is $R\bar3c$, $R3m$, $Imma$ or $Amm2$, whichever is lowest on that plane. This is the plane-by-plane form of the main-text statement that $\mathbf P$ and $\mathbf R$ cannot coexist along strongly misaligned axes.
  \item \textbf{Low-symmetry joint minima.} Planes \#11 and \#14 (both at $35^\circ$) have joint $Cc$ minima at $+183$ and $+208$~meV/f.u. On \#14 the joint minimum lies only 2~meV/f.u.\ below $R3m$. The $Cc$ points are minima within their planes, not stationary points of the full space. Both lie above the $Ima2$ saddle (Supplementary Table~\ref{tab:bfo_stat}), and neither lies on a switching path (Supplementary Table~\ref{tab:bfo_paths}).
  \item \textbf{Coupling at small amplitude.} At $|\mathbf P|=0.15$ and $|\mathbf R|=0.20$~\AA, $E_{\rm c}$ ranges from $+4.7$ to $+12.4$~meV/f.u.\ across the planes. Per component, the quartic coefficients are $+5.3$~eV\,\AA$^{-4}$ for $P_iR_i$ (same axis), $+12.2$ for $P_iR_j$ ($i\neq j$) and $2\lambda_2=-11.6$ for $P_iR_iP_jR_j$.
\end{enumerate}
No term bilinear or trilinear in $\mathbf P$ and $\mathbf R$ is allowed. A bilinear term is excluded because $\Gamma_4^-$ and $R_5^-$ belong to different points of the Brillouin zone. A trilinear term $\mathbf P\mathbf R\mathbf R$ would conserve crystal momentum ($2\mathbf k_R\equiv\Gamma$) but is odd under inversion, which reverses $\mathbf P$ and leaves $\mathbf R$ unchanged; $\mathbf P\mathbf P\mathbf R$ and $R^3$ are odd under the odd translation, which reverses $\mathbf R$ alone. Every invariant is therefore even in $\mathbf P$ and even in $\mathbf R$, and the lowest coupling is biquadratic (Supplementary Table~\ref{tab:bfo_coef}). Consequently, the two axes of every plane are lines of stationary single-mode solutions.

\subsection{Competition and axis locking}
\label{sec:si_bfo_coupling}

\paragraph{Magnitudes (along $[111]$).}
Each order parameter alone gives $R3m$ ($-738.9$~meV/f.u.\ at $|\mathbf P|=0.599$~\AA) or $R\bar3c$ ($-675.9$ at $|\mathbf R|=0.791$~\AA). Together they give $R3c$ at $-950.0$. If the two did not interact, the combined energy would be $-1414.8$. The difference is a coupling penalty of $+465$~meV/f.u.\ at the minimum, where $|\mathbf P|$ is reduced by 13.4\% and $|\mathbf R|$ by 11.8\%.

The optimal polar amplitude decreases steadily as the tilt grows:

\begin{tabular}{lccccc}
$|\mathbf R|$ (\AA) & 0 & 0.2 & 0.4 & 0.6 & 0.8 \\
$P^*(R)$ (\AA) & 0.599 & 0.592 & 0.572 & 0.538 & 0.500 \\
\end{tabular}

The optimal tilt likewise decreases as the polarization grows:

\begin{tabular}{lccccc}
$|\mathbf P|$ (\AA) & 0 & 0.2 & 0.4 & 0.5 & 0.6 \\
$R^*(P)$ (\AA) & 0.791 & 0.782 & 0.743 & 0.706 & 0.659 \\
\end{tabular}

The interaction energy $E(p,r)-E(p,0)-E(0,r)$ is positive everywhere and grows approximately as $p^2r^2$. At $p=0.52$ it is $+48$, $+173$, $+325$ and $+392$~meV/f.u.\ for $r=0.2$, 0.4, 0.6 and 0.7~\AA.

\paragraph{Directions.}
Rotating $\mathbf R$ rigidly away from $\mathbf P$, starting from $R3c$, costs the following energies; the cost is nearly isotropic about $[111]$:

\begin{tabular}{lccccccc}
misalignment & $5^\circ$ & $10^\circ$ & $20^\circ$ & $35^\circ$ & $55^\circ$ & $71^\circ$ & $90^\circ$ \\
energy (meV/f.u.) & $+7$ & $+28$ & $+113$ & $+339$ & $+715$ & $+975$ & $\approx+1070$ \\
\end{tabular}

The angular stiffness is $d^2E/d\theta^2=1830$~meV/f.u.\ per rad$^2$. If the magnitudes are allowed to relax at fixed misalignment, the costs are $+27$ at $10^\circ$ and $+208$ at $35^\circ$, where $|\mathbf R|$ falls to 0.35~\AA. At $55^\circ$ the tilt vanishes ($+211$, i.e.\ $R3m$), and at $90^\circ$ the polarization vanishes ($+310$).

\begin{table}[!htbp]
\centering
\caption{\textbf{Coupled Hessian at $R3c$} (eV\,\AA$^{-2}$ per f.u.), in the basis of components parallel ($\parallel$, along $[111]$) and perpendicular ($\perp$, along $[11\bar2]$) to the common axis. Longitudinal and transverse blocks decouple. The normalized off-diagonal elements are $+0.30$ (longitudinal) and $-0.29$ (transverse). The softest longitudinal mode, 3.22, has eigenvector $(P,R)=(0.30,-0.95)$: the tilt shrinks as the polarization grows (competition). The softest transverse mode, 3.00, has eigenvector $(-0.50,-0.87)$: $\mathbf P$ and $\mathbf R$ rotate together (locking). It is softer than rotating $\mathbf P$ alone (5.25) or $\mathbf R$ alone (3.75), and it is the start of the 71$^\circ$ path to the $Ima2$ saddle.}
\label{tab:bfo_hess}
\begin{tabular}{lrrrr}
\toprule
 & $P_\parallel$ & $R_\parallel$ & $P_\perp$ & $R_\perp$ \\
\midrule
$P_\parallel$ & 8.66 & $+1.70$ & 0 & 0 \\
$R_\parallel$ & $+1.70$ & 3.75 & 0 & 0 \\
$P_\perp$ & 0 & 0 & 5.25 & $-1.30$ \\
$R_\perp$ & 0 & 0 & $-1.30$ & 3.75 \\
\bottomrule
\end{tabular}
\end{table}

\subsection{Sampling completeness and validation set}
\label{sec:si_bfo_cov}

\paragraph{Coverage audit.}
Each of the 17 planes was checked against the six-dimensional data with all 96 images unfolded:
\begin{itemize}
  \item Every point with $E<0$ (relative to cubic) lies within 1.13 grid steps of a DFT point.
  \item The landscape rises steeply at the grid boundary: $E\ge+298$~meV/f.u.\ at the edge in $|\mathbf P|$ and $E\ge-237$ at the edge in $|\mathbf R|$, both far above $R3c$.
  \item Grid nodes next to the contact cut have $E\ge+765$~meV/f.u.
\end{itemize}
No low-energy region is therefore missing from the grid.

\paragraph{extra\_V1.}
To test the surface between grid nodes, 253 further static calculations were performed with the same settings (\texttt{gen\_extra.py}). They come in four groups:
\begin{itemize}
  \item 1 cubic reference.
  \item 27 plane checks on the 14 planes without a dense refine set. On each plane these are the RBF minimum, the point halfway to the origin and an interior point at $0.65\,a_P$, $0.65\,b_R$; duplicates were removed.
  \item 25 images along the four switching paths of Supplementary Table~\ref{tab:bfo_paths}, including the saddles.
  \item 200 points chosen by greedy farthest-point sampling in the region where the predicted energy is $E\le-300$~meV/f.u., with distances measured against all existing data and their images. These are the centres of the largest gaps in the grid.
\end{itemize}

\begin{table}[!htbp]
\centering
\caption{\textbf{extra\_V1 validation campaign.} QC and blind tests of the two surfaces built before the campaign: the RBF fitted to main\_V1 and refine\_V1, and the degree-12 validation model fitted to main\_V1 alone. Errors are DFT minus prediction, RMS (max) in meV/f.u.}
\label{tab:bfo_extra}
\begin{tabular}{lrcc}
\toprule
group & $n$ & RBF & degree 12 \\
\midrule
plane checks & 27 & 0.83 (2.56) & 1.08 (2.68) \\
switching-path images & 25 & 0.09 (0.29) & 0.70 (1.43) \\
space-filling (largest gaps) & 200 & 1.42 (5.06) & 0.43 (1.70) \\
all & 253 & 1.29 (5.06) & 0.57 (2.68) \\
\midrule
\multicolumn{4}{l}{converged 253/253; SCF steps 23--25; $|m_{\rm Fe}|=4.081$--4.133~$\mu_{\rm B}$; 0 QC flags} \\
\bottomrule
\end{tabular}
\end{table}

The largest RBF error, 5~meV/f.u., occurs at the centres of grid cells, where the degree-12 model does better (1.7~meV/f.u.). Along the switching paths, including the saddles, the RBF agrees with DFT to 0.3~meV/f.u.\ and the degree-12 model to 1.4~meV/f.u. No further densification of the grid is needed below $-300$~meV/f.u. Both final surfaces include these points.

\subsection{Switching paths}

Minimum-energy paths were computed with the string method~\cite{E2002String} (41 images) on the degree-12 model; the paths shown in Supplementary Fig.~\ref{fig:s_bfo_paths} were traced by steepest descent from each saddle (below). Domains are $(\mathbf P,\mathbf R)=(p\,\mathbf n,\,s\,r\,\mathbf n)$ with $\mathbf n\in\langle111\rangle$ and $s=\pm1$. The two signs of $s$ are degenerate and related by the odd translation.

\begin{table}[!htbp]
\centering
\caption{\textbf{Homogeneous switching barriers} (meV/f.u., final degree-12 model, V1; each barrier is the energy of the saddle in Supplementary Table~\ref{tab:bfo_stat}; the RBF values agree to 0.7~meV/f.u.). All routes are at zero field, in the fixed cubic cell, without domain walls. The last column gives the effect on the weak magnetization $\mathbf M\propto\mathbf D\times\mathbf L$ with $\mathbf D\propto\mathbf R$~\cite{Ederer2005}, inferred from symmetry and not computed.}
\label{tab:bfo_paths}
\begin{tabular}{lllr l}
\toprule
switch & route & saddle & barrier & $\mathbf M$ \\
\midrule
71$^\circ$, $\mathbf P$ and $\mathbf R$ together ($[111]\to[11\bar1]$) & one step & $Ima2$ & 101.3 & rotates with $\mathbf R$ \\
109$^\circ$, together ($[111]\to[\bar1\bar11]$) & two 71$^\circ$ steps & $Ima2$-type & 101.3 & rotates \\
180$^\circ$, $\mathbf P$ and $\mathbf R$ both reversed & three 71$^\circ$ steps & $Ima2$-type & 101.3 & reversed \\
180$^\circ$, $\mathbf P$ reversed, $\mathbf R$ fixed (direct)\textsuperscript{a} & through $\mathbf P=0$ & $R\bar3c$ & 274.2 & unchanged \\
180$^\circ$, $\mathbf R$ reversed, $\mathbf P$ fixed & through $\mathbf R=0$ & $R3m$ & 211.2 & reversed \\
71$^\circ$ of $\mathbf P$ with the relative sign $s$ flipped & $R$ reversal + 71$^\circ$ & $R3m$ & 211.2 & -- \\
\bottomrule
\multicolumn{5}{l}{\footnotesize \textsuperscript{a}The homogeneous path found by Xu \textit{et al.}~\cite{Xu2017} in effective-Hamiltonian MD with $\mathbf E$ antiparallel to $\mathbf P$.}\\
\end{tabular}
\end{table}

The paths avoid the 109$^\circ$ midpoint ($I4cm$) and the direct 180$^\circ$ midpoint ($Pm\bar3m$). An alternative way to reverse $\mathbf P$ while restoring the original tilt sense combines the steps above: reverse $\mathbf R$ through $R3m$ (211), then make three 71$^\circ$ steps (101). Its highest barrier is 211, which is lower than that of the direct route through $R\bar3c$ (274). This route was inferred from the computed saddles rather than obtained from a single string calculation.

Supplementary Fig.~\ref{fig:s_bfo_paths} shows the energy and the order-parameter magnitudes along the three elementary paths at both volumes, and the three ways of reversing $\mathbf P$ by 180$^\circ$. Each path was traced on the degree-12 model by steepest descent from its saddle, displaced by $2\times10^{-3}$~\AA\ along the single unstable eigenvector of the Hessian, to the two neighbouring $R3c$ domains; this gives the minimum-energy path exactly and reproduces the barriers of Supplementary Tables~\ref{tab:bfo_paths} and~\ref{tab:bfo_vol}.

\begin{figure}[!htbp]
\centering
\includegraphics[width=\linewidth]{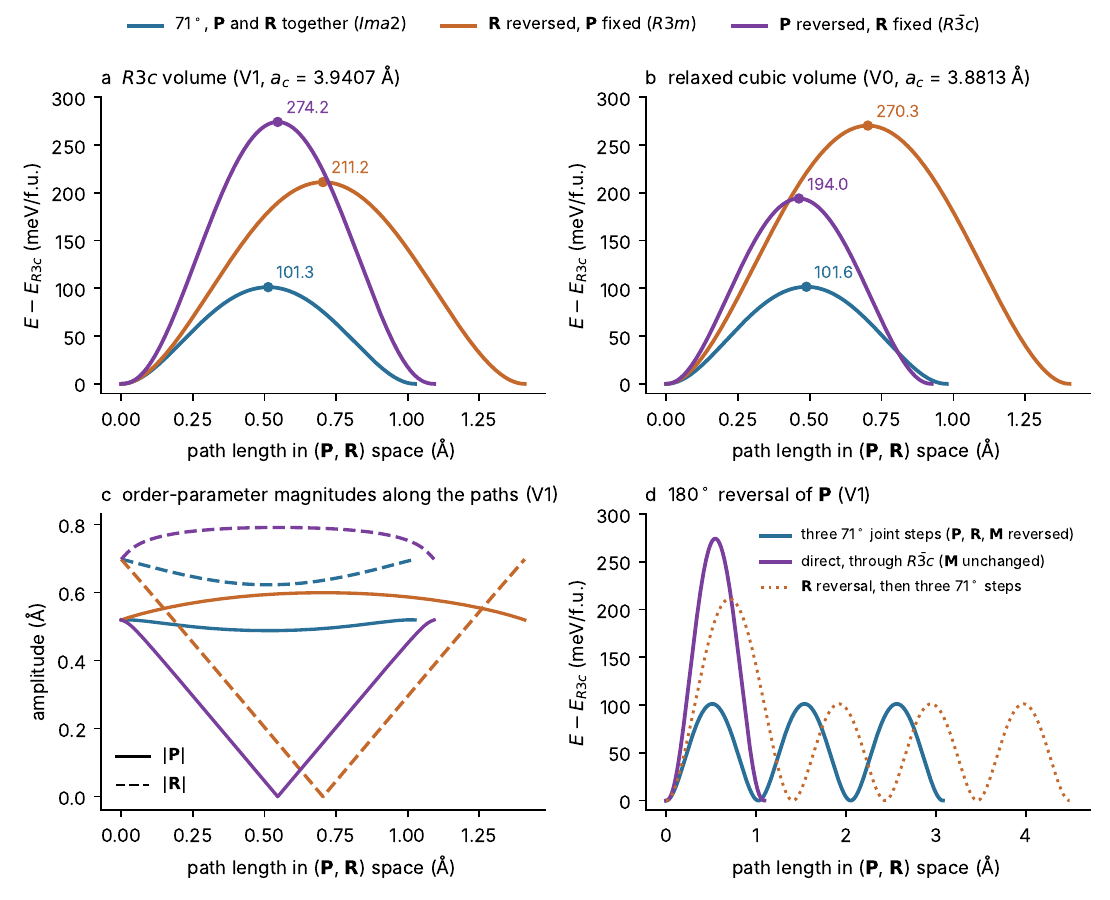}
\caption{\textbf{Homogeneous switching paths of BiFeO$_3$.} Minimum-energy paths on the degree-12 model, traced by steepest descent from each saddle along its unstable eigenvector to the neighbouring $R3c$ domains (fixed cubic cell, zero field, no domain walls; G-type order). Horizontal axis: arc length of the path in the six-dimensional ($\mathbf P$, $\mathbf R$) space, in \AA\ of translation-free $A_p$. \textbf{a},~Energy relative to $R3c$ at the $R3c$ volume (V1, $a_c=3.9407$~\AA) along the 71$^\circ$ joint rotation of $\mathbf P$ and $\mathbf R$ (blue; saddle $Ima2$, 101.3~meV/f.u.), the reversal of $\mathbf R$ at fixed $\mathbf P$ (orange; saddle $R3m$, 211.2) and the reversal of $\mathbf P$ at fixed $\mathbf R$ (purple; saddle $R\bar3c$, 274.2); dots mark the saddles. \textbf{b},~As \textbf{a} at the relaxed cubic volume (V0, $a_c=3.8813$~\AA): 101.6, 270.3 and 194.0~meV/f.u. \textbf{c},~$|\mathbf P|$ (solid) and $|\mathbf R|$ (dashed) along the three V1 paths, in the colours of \textbf{a}: the joint rotation keeps both order parameters near their $R3c$ values, whereas each reversal drives one of them through zero. \textbf{d},~Three ways of reversing $\mathbf P$ by 180$^\circ$ at V1: three successive 71$^\circ$ joint steps, which also reverse $\mathbf R$ and therefore the weak magnetization $\mathbf M$ (blue); the direct reversal through $R\bar3c$, which leaves $\mathbf R$ and $\mathbf M$ unchanged (purple); and the reversal of $\mathbf R$ through $R3m$ followed by three 71$^\circ$ steps, which reverses $\mathbf P$ and restores the original sense of $\mathbf R$ (orange dotted; highest barrier 211.2). The effect on $\mathbf M$ is inferred from symmetry~\cite{Ederer2005}, not computed. Barriers are homogeneous 0~K upper bounds.}
\label{fig:s_bfo_paths}
\end{figure}

The 25 DFT points along the paths (extra\_V1), including the saddles, agree with the RBF surface to 0.09~meV/f.u.\ RMS (maximum 0.29) and with the degree-12 model to 0.70 (maximum 1.43; Supplementary Table~\ref{tab:bfo_extra}); both final surfaces include them.

\subsection{Volume dependence}

\begin{table}[!htbp]
\centering
\caption{\textbf{Effect of volume.} Degree-12 fits at the $R3c$ volume (V1, $a_c=3.9407$~\AA) and at the relaxed cubic volume (V0, $a_c=3.8813$~\AA; fit RMS 0.28, hold-out 0.32~meV/f.u.). Energies in meV/f.u.\ relative to cubic at the same volume; barriers relative to $R3c$. In absolute terms the V0 parent lies 60.1~meV/f.u.\ below the V1 parent, and $R3c$ at V1 lies 56.3~meV/f.u.\ below $R3c$ at V0 (from the stored reference energies, $-32.5141$ and $-32.5742$~eV/f.u., and the $R3c$ energies of this table).}
\label{tab:bfo_vol}
\begin{tabular}{lll}
\toprule
 & V1 ($R3c$ volume) & V0 (cubic volume) \\
\midrule
$R3c$ ($|\mathbf P|$, $|\mathbf R|$) & $-950.0$ (0.519, 0.698) & $-833.6$ (0.451, 0.694) \\
$Ima2$ barrier (71$^\circ$) & $+101.3$ & $+101.6$ \\
$R3m$ & $-738.9$ & $-563.3$ \\
$R\bar3c$ & $-675.9$ & $-639.6$ \\
$\mathbf P$-reversal barrier ($R\bar3c$) & 274 & 194 \\
$\mathbf R$-reversal barrier ($R3m$) & 211 & 270 \\
\bottomrule
\end{tabular}
\end{table}

Compression weakens the polar mode much more than the tilt:
\begin{itemize}
  \item The $\mathbf P$-only well is 24\% shallower, and its optimum $P^*$ falls from 0.599 to 0.540~\AA.
  \item $|\mathbf P|$ in $R3c$ falls by 13\%, while $|\mathbf R|$ is unchanged.
  \item The $\mathbf R$-only well is only 5\% shallower.
\end{itemize}
As a result, the order of the single-mode phases reverses between the two volumes: $R3m<R\bar3c$ at V1 but $R\bar3c<R3m$ at V0. Whether reversing $\mathbf P$ or reversing the tilt is cheaper therefore depends on volume, whereas the joint 71$^\circ$ barrier is unchanged at 101~meV/f.u.

Compression also changes the Hessian index of two saddles ($Amm2$ from 2 to 3, $I4/mcm$ from 5 to 3), while $R3c$ stays the global minimum (60 of 60 starts) and $Ima2$ the lowest saddle; the gated rerun reproduces the V0 fit and all its stationary points (84 of 84 numbers of the BiFeO$_3$ rerun re-derived). Volume is not a grid coordinate: the two datasets are fixed-volume slices, and strain relaxation is not included in any barrier.

\subsection{Berry-phase polarization}
\label{sec:si_berry}

The polarization was computed with the Berry-phase method~\cite{KingSmith1993} along the untilted polar line, $\mathbf P=p\,(1,1,1)/\sqrt3$ with $\mathbf R=0$, from $p=0$ to 0.80~\AA\ in steps of 0.05~\AA\ (17 static runs).

Settings: the same as for the grid, plus \texttt{LCALCPOL}, \texttt{DIPOL} at the cell centre and \texttt{ISYM}~=~0; $\Omega=122.391$~\AA$^3$ (2~f.u.).

Branch: the centrosymmetric cubic point lies on the half-quantum branch, $P(0)=\tfrac12 e(\mathbf R_1+\mathbf R_2+\mathbf R_3)\cdot\hat n/\Omega=89.35$~$\mu$C\,cm$^{-2}$. This value is subtracted, as in Ref.~\cite{Neaton2005}, whose quantum is 184.2~$\mu$C\,cm$^{-2}$ against 178.7 at our volume.

Checks:
\begin{itemize}
  \item The ionic dipole computed directly from the structure reproduces the VASP value.
  \item The component of $\mathbf P$ perpendicular to $[111]$ is zero at every point.
  \item $E(p)$ agrees with the six-dimensional model to within 2.5~meV/f.u.\ along the whole line.
\end{itemize}

\begin{table}[!htbp]
\centering
\caption{\textbf{Berry-phase polarization along $\Gamma_4^-(a,a,a)$, $\mathbf R=0$} (V1, DFT only). $Z^*_{\rm mode}=(\Omega_{\rm fu}/e)\,dP/dp$ is the mode effective charge per f.u.\ per \AA\ of $A_p$. Its nominal-charge value (Bi$^{3+}$, Fe$^{3+}$, O$^{2-}$) for the same pattern is 4.57. The coercive field is the homogeneous value $E_c=\max(-dU/dp)/(V_{\rm fu}\,dP/dp)$.}
\label{tab:bfo_berry}
\begin{tabular}{rrrr}
\toprule
$p$ (\AA) & $P_{[111]}$ ($\mu$C\,cm$^{-2}$) & $E$ (meV/f.u.) & $Z^*_{\rm mode}$ ($e$) \\
\midrule
0.00 & 0     & 0        & 8.66 \\
0.10 & 22.5  & $-42.1$  & \\
0.20 & 43.8  & $-159.8$ & \\
0.30 & 63.7  & $-329.6$ & 7.28 \\
0.40 & 82.0  & $-516.1$ & \\
0.50 & 99.1  & $-672.9$ & \\
0.60 & 115.3 & $-740.0$ & 6.01 \\
0.70 & 130.6 & $-638.5$ & \\
0.80 & 145.1 & $-261.5$ & 5.38 \\
\midrule
\multicolumn{4}{l}{$P$ at $p=0.526$~\AA\ ($R3c$ polar amplitude): 103~$\mu$C\,cm$^{-2}$} \\
\multicolumn{4}{l}{$P$ at the $R3m$ minimum ($p=0.599$~\AA): 115~$\mu$C\,cm$^{-2}$} \\
\multicolumn{4}{l}{$E_c$, untilted line (DFT): $\approx27$~MV\,cm$^{-1}$ (at $p\approx0.37$~\AA)} \\
\multicolumn{4}{l}{$E_c$, tilt kept at $R3c$ (model, not DFT): $\approx11$~MV\,cm$^{-1}$} \\
\bottomrule
\end{tabular}
\end{table}

\paragraph{Comparison with the literature.}
\begin{itemize}
  \item \textbf{Polarization.} Neaton \textit{et al.}~\cite{Neaton2005} (LSDA$+U$) report 95~$\mu$C\,cm$^{-2}$ for $R3c$ along $[111]$. Our value of 103 is 8\% higher, as expected for a path without tilts, since tilting lowers the Born charges.
  \item \textbf{Born charges.} Their cubic Born charges (Bi 6.32, Fe 4.55, O $-3.06$), projected onto our $\Gamma_4^-$ pattern with O treated isotropically, give $Z^*_{\rm mode}=8.36$, against our 8.66 at $p\to0$. Their $R3c$ charges (Bi 4.37, Fe 3.49, O $-2.61$) give 6.27, against our 6.0--6.4 at $p=0.5$--0.6. The 30\% fall of $Z^*_{\rm mode}$ along $p$ therefore matches the drop of the Born charges from cubic to $R3c$, and a constant $Z^*$ should not be used.
  \item \textbf{Coercive field.} Xu \textit{et al.}~\cite{Xu2017} find $E_c=8$~MV\,cm$^{-1}$ for homogeneous switching in effective-Hamiltonian MD that includes strain and temperature. Measured single-crystal values are about 12~kV\,cm$^{-1}$~\cite{Lebeugle2007}, $\sim10^3$ times lower, because switching proceeds by nucleation.
\end{itemize}
Our homogeneous fields are 0~K, fixed-cell upper bounds.
A Berry-phase calculation along the tilted path, with $\mathbf R$ fixed at $R3c$, would test the $\sim$8\% reduction of $P$ by the tilts directly, and adding $-\mathbf E\cdot\mathbf P(p)\,V$ to the model would give field-dependent barriers for the $R\bar3c$ and $Ima2$ routes; both are left for future work.

\subsection{Relaxed anchors and comparison with the literature}

\begin{table}[!htbp]
\centering
\caption{\textbf{Fully relaxed phases (ISIF~3), $\Delta E$ relative to $R3c$ in meV/f.u.} The literature values were computed with the same functional and $U$ (PBEsol$+U$, $U=4$~eV, Dudarev, VASP) but in 40-atom cells with ENCUT 500~eV (ours: 10-atom cell, 600~eV). The $R3c$ volume is 61.22~\AA$^3$/f.u., against 61.47 in Ref.~\cite{Dieguez2011} ($-0.4$\%).}
\label{tab:bfo_lit}
\begin{tabular}{lrrrr}
\toprule
phase & this work & Di\'eguez~\cite{Dieguez2011} & Fedorova~\cite{Fedorova2022} & Di\'eguez PBE$+U$ / LDA$+U$ \\
\midrule
$R3c$ & 0 & 0 & 0 & 0 / 0 \\
$R3m$ & $+169$ & $+169$ & $+168$ & $+136$ / $+191$ \\
$R\bar3c$ & $+225$ & $+230$ & $+230$ & $+272$ / $+209$ \\
$Pm\bar3m$ & $+893$ & $+906$ & $+909$ & $+981$ / $+870$ \\
\bottomrule
\end{tabular}
\end{table}

All three functionals give the same order, $R3c<R3m<R\bar3c\ll Pm\bar3m$. The 13--16~meV/f.u.\ offset for the cubic phase (1.5\%) is plausibly due to the different ENCUT and to Pulay stress.

The fixed-lattice values of Fedorova \textit{et al.}\ are not directly comparable to our grid planes. Their values for $\mathbf P\parallel[111]$, $\mathbf R\parallel[111]$ and both together are $-580$, $-651$ and $-853$~meV/f.u., at the relaxed cubic volume. Ours are $-739$, $-676$ and $-950$, at the $R3c$ volume, which is 4.7\% larger. The difference is mostly in the polar mode, consistent with the volume dependence in Supplementary Table~\ref{tab:bfo_vol}.

Phases outside the $\Gamma_4^-\oplus R_5^-$ space lie close in energy: $Pnma$ at $+27$, supertetragonal $Cc$ at $+96$ and $P4mm$ at $+140$~meV/f.u.\ (Ref.~\cite{Dieguez2011}). Once the cell is free to strain, $R3m$ is therefore not the lowest polar-only phase. The landscape here describes the $\Gamma_4^-\oplus R_5^-$ subspace at fixed cubic shape.

\subsection{Magnetism and domain states}
\label{sec:si_bfo_magnet}

All calculations assume collinear G-type order without spin--orbit coupling, so the weak magnetization $\mathbf M$ is zero by construction. Its behaviour under switching is taken from the symmetry analysis of Ederer and Spaldin~\cite{Ederer2005}:
\begin{itemize}
  \item $E_{\rm DM}=-\mathbf D\cdot(\mathbf L\times\mathbf M)$ with $\mathbf D\parallel[111]$.
  \item The sign of $\mathbf D$ is set by the sense of the octahedral rotation, so $\mathbf D\propto\mathbf R$.
  \item Weak ferromagnetism is allowed in $R\bar3c$ and $R3c$ but not in $R3m$, so $\mathbf M$ comes from the tilts and not from $\mathbf P$.
\end{itemize}

\begin{table}[!htbp]
\centering
\caption{\textbf{Ederer--Spaldin domain states reached by the homogeneous routes} (zero field, fixed $\mathbf L$). States are labelled as in Ref.~\cite{Ederer2005}, Table~I. Their states 1--4 and 5--8 differ only in the relative sign of $\mathbf P$ and $\mathbf R$; they are exactly degenerate here, because no $\mathbf P\cdot\mathbf R$ term is allowed.}
\label{tab:bfo_states}
\begin{tabular}{llr}
\toprule
target & route & barrier (meV/f.u.) \\
\midrule
state 4: $\mathbf P$, $\mathbf D$, $\mathbf M$ reversed & three joint 71$^\circ$ steps ($Ima2$) & 101 \\
state 8: $\mathbf P$ reversed; $\mathbf D$, $\mathbf M$ unchanged & direct, through $R\bar3c$ & 274 \\
state 8, alternative & $\mathbf R$ reversal ($R3m$) + three 71$^\circ$ steps & 211 \\
\bottomrule
\end{tabular}
\end{table}

Ederer and Spaldin argued that state 8 is the most likely result of reversing $\mathbf P$, because reversing the rotations is energetically costly. In the homogeneous limit at zero field, our barriers instead favour state 4. Reversing $\mathbf R$ at fixed $\mathbf P$ is indeed costly (211~meV/f.u.), but rotating $\mathbf P$ and $\mathbf R$ together is cheap (101~meV/f.u.).

Which state an applied field selects depends on the field-energy gain along each route. A field exactly antiparallel to $\mathbf P$ favours the direct route found by Xu \textit{et al.} Resolving this requires adding the field term to the model and is left open here.

Spin canting, the cycloid and the magnetic anisotropy need non-collinear calculations with spin--orbit coupling.
A subset with spin--orbit coupling, computing $\mathbf M$ at $R3c$ and along the single-mode lines to test whether $\mathbf M(\mathbf R)\propto\mathbf R$ and $\mathbf M(\mathbf P)\approx0$, is left for future work.

\subsection{Structures}
\label{sec:si_bfo_struct}

The structure files, written by \texttt{make\_structures.py}, and the structures of main-text Fig.~\Mref{fig:bfo}a were generated from the six-dimensional basis. Each structure is provided in three cells:
\begin{itemize}
  \item the 10-atom fcc$(2a_c)$ cell used for DFT;
  \item the 30-atom hexagonal cell, $a_h=5.573$ and $c_h=13.651$~\AA\ at $a_c=3.9407$~\AA;
  \item the 40-atom $2\times2\times2$ cell.
\end{itemize}
All are at the fixed cubic lattice.

\begin{table}[!htbp]
\centering
\caption{\textbf{Structure files} (POSCAR; per-atom displacement tables in \texttt{\_displacements.txt}). Amplitudes are per component in \AA\ of $A_p$. Single-mode structures are at the single-mode minima of the degree-12 model. The coupled structures are at the plane minima of Supplementary Table~\ref{tab:bfo_planes}.}
\label{tab:bfo_structs}
\begin{tabular}{lllc}
\toprule
file & $\Gamma_4^-$ & $R_5^-$ & space group \\
\midrule
\texttt{parent\_Pm-3m} & 0 & 0 & $Pm\bar3m$ \\
\texttt{GM4-\_a00\_P4mm} & $(0.548,0,0)$ & 0 & $P4mm$ \\
\texttt{GM4-\_aaa\_R3m} & $(a,a,a)$, $|\mathbf P|=0.599$ & 0 & $R3m$ \\
\texttt{R5-\_b00\_I4mcm} & 0 & $(0.736,0,0)$ & $I4/mcm$ \\
\texttt{R5-\_bb0\_Imma} & 0 & $(0.551,0.551,0)$ & $Imma$ \\
\texttt{R5-\_bbb\_R-3c} & 0 & $(b,b,b)$, $|\mathbf R|=0.791$ & $R\bar3c$ \\
\texttt{GM4-\_aa0+R5-\_bb0\_Ima2} & $(0.345,0.345,0)$ & $(0.441,0.441,0)$ & $Ima2$ \\
\texttt{GM4-\_aaa+R5-\_bbb\_R3c} & $(0.300,0.300,0.300)$ & $(0.404,0.404,0.404)$ & $R3c$ \\
\bottomrule
\end{tabular}
\end{table}

Symmetrized CIFs (spglib) are provided for $Pm\bar3m$ and $R3c$. In the hexagonal setting of Kubel and Schmid, with Bi at the origin and $\mathbf P$ along $+c$, the $R3c$ model has:
\begin{itemize}
  \item Bi at 6a $(0,0,0)$;
  \item Fe at 6a $(0,0,0.2243)$;
  \item O at 18b $(0.4361,0.0172,0.9582)$.
\end{itemize}
The experimental values are Fe $z=0.2212$ and O $(0.443,0.012,0.952)$~\cite{Kubel1990}. The residual differences reflect the fixed cubic metric, which is not rhombohedrally relaxed.

\section{HfO$_2$ paths}\label{sec:si_hfo2}

Supplementary Table~\ref{tab:hfo2} lists the barriers and end-point energies of the paths of main-text Section~\Mref{sec:res_hfo2} and Fig.~\Mref{fig:hafnia}b, compared with our earlier work; Supplementary Table~\ref{tab:hfo2_set} gives the setting used for each transition.

\subsection{Full account of the HfO$_2$ results}\label{sec:si_hfo2_ext}
This subsection gives the complete account of the HfO$_2$ paths summarized in main-text Section~\Mref{sec:res_hfo2}, including the comparison with the expert paths.

HfO$_2$ (main-text Fig.~\Mref{fig:hafnia}a) tests transitions among several polymorphs rather than a single parent--child pair, which brings in the basis search, more than one reference phase and a pair of phases that are not group--subgroup related. From the tetragonal $P4_2/nmc$ phase, IRIS decomposes the transitions to the polar orthorhombic $Pca2_1$ (T1), the antipolar orthorhombic $Pbcn$ (T2) and the monoclinic $P2_1/c$ (T3); from $Pbcn$ it decomposes those to $Pca2_1$ (T4) and $P2_1/c$ (T5) (main-text Fig.~\Mref{fig:hafnia}a). For T2, the decomposition offered no supercell basis because the transition carries 11.3\% strain along the tetragonal $c$ axis, so $f_1$ fell back on its basis search and found one with $\det=k$ (Supplementary Fig.~\ref{fig:s_basis}e); the subgroup and index checks of $G_1$ are worked through for T1 in Supplementary Fig.~\ref{fig:s_index}b. The decompositions (Extended Data Table~\Mref{tab:expert}; amplitudes in Supplementary Table~\ref{tab:dec_hfo2}) show how the phases are related. From the tetragonal parent, $Pca2_1$ and $P2_1/c$ share the zone-boundary irreps $M_1$ and $M_3$: on crossed arms they bring in the polar $\Gamma_5^-$ and give $Pca2_1$, and on aligned arms the shear $\Gamma_5^+$ and give $P2_1/c$; $Pbcn$ needs $M_1$ alone. From $Pbcn$, the same two phases each follow a single $\Gamma$-point irrep, the polar $\Gamma_2^-$ or the non-polar $\Gamma_3^+$. The fully symmetric $\Gamma_1^+$ accompanies every transition as a secondary. $Pca2_1$ is therefore an improper ferroelectric with respect to the tetragonal phase, with the antipolar $M_1$ as primary, but a proper one with respect to $Pbcn$, with $\Gamma_2^-$ as primary: the mechanism depends on the reference, and IRIS makes that choice explicit. $Pca2_1$ and $P2_1/c$ are not group--subgroup related, so IRIS connects them through $Pbcn$, a supergroup of both (T6).

\begin{figure}[!htbp]
\centering
\includegraphics[width=\textwidth]{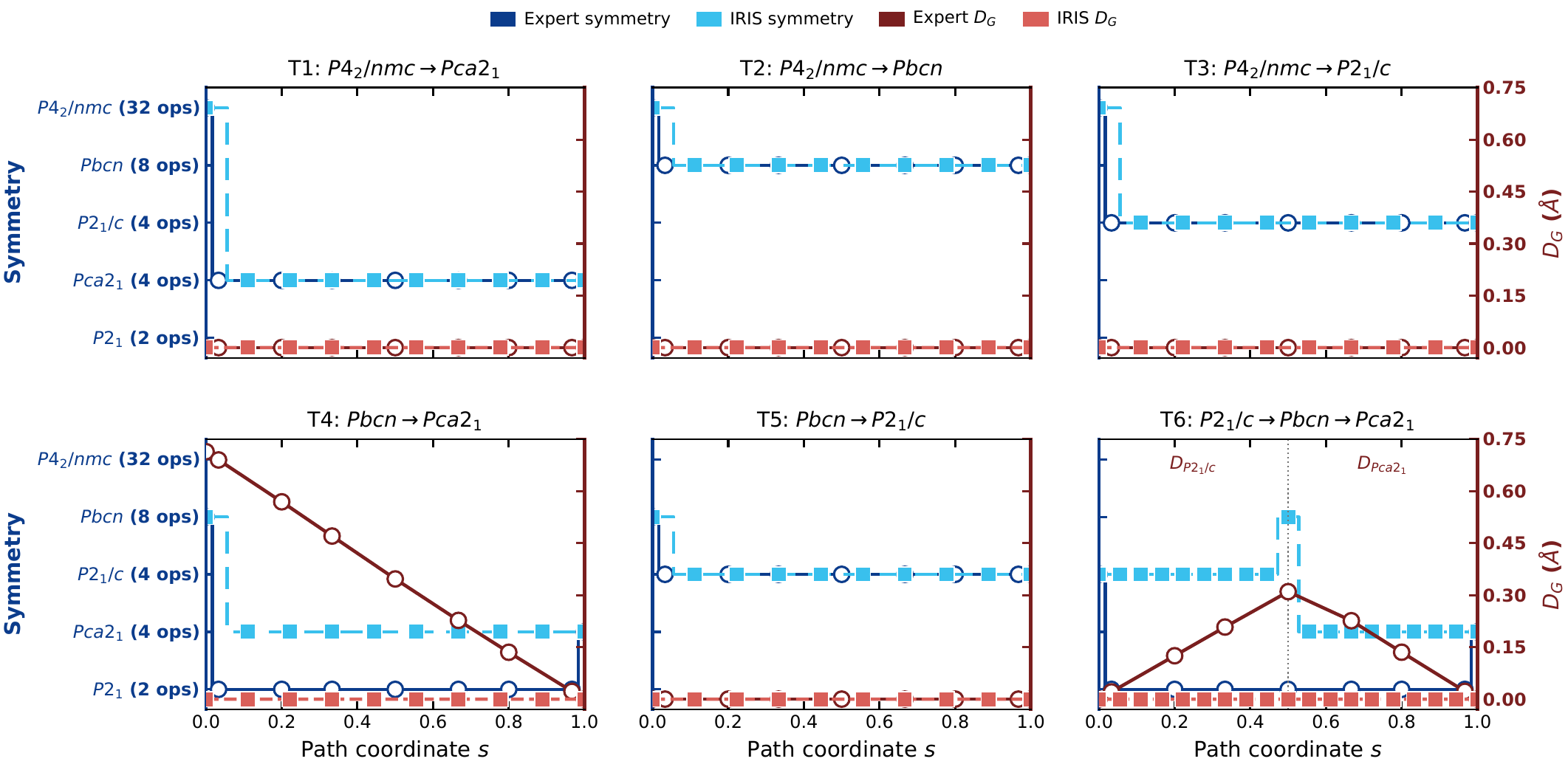}
\caption{\textbf{Symmetry and reference-symmetry distance $D_G$ along the six HfO$_2$ paths.} Blue: space group of each image identified with spglib (expert, dark blue circles; IRIS, light blue squares), with the number of symmetry operations in parentheses. Red: $D_G(X)$ in Å, the root-mean-square displacement of an image under the operations of its reference group $G$ (Supplementary eq.~\eqref{eq:si_DG}; right axis; expert, dark red circles; IRIS, light red squares). Lines join the images (expert, solid; IRIS, dashed). The path coordinate $s$ is the fraction along the path ($s=\lambda$ for IRIS). $D_G=0$ indicates that the reference symmetry is preserved, and larger values a greater departure from it. For T4, the expert path passes through $P2_1$, whereas IRIS remains on the $Pca2_1$ branch. For T6, $D_G$ is evaluated relative to $P2_1/c$ for $s\le0.5$ and to $Pca2_1$ for $s>0.5$ (dotted line: $Pbcn$, $s=0.5$).}
\label{fig:s_hfo2_sym}
\end{figure}

Along each path, every mode amplitude and the strain are scaled together from parent to child (main-text eq.~\Meqref{eq:path}; strain policy P3, Supplementary Fig.~\ref{fig:s_strain}; settings in Supplementary Table~\ref{tab:hfo2_set}). The energies (main-text Fig.~\Mref{fig:hafnia}b) give frozen-path barriers of 6.43 (T1), 24.68 (T2), 43.95 (T3), 80.50 (T4) and 52.37~meV/f.u.\ (T5; Supplementary Table~\ref{tab:hfo2}). The T1--T3 barriers are within 1.1~meV/f.u.\ of our earlier work~\cite{Manna2026hafnia}, and the end-point energies are consistent across paths: T2 followed by T4 reproduces T1, and T2 followed by T5 reproduces T3, to 0.06~meV/f.u. T6 rises to 123~meV/f.u., passes $Pbcn$ at $+71$~meV/f.u.\ and peaks at 151.5~meV/f.u.\ before reaching $Pca2_1$ at $+83$~meV/f.u., a third of the 445~meV/f.u.\ along the direct expert path. Extended Data Table~\Mref{tab:expert} compares these paths with paths built by an expert between the same phases (orange in main-text Fig.~\Mref{fig:hafnia}b), in which every structure is expressed in one common $P4_2/nmc$ basis rather than with respect to the reference phase of each transition, and the lattice is interpolated separately between the two end cells (Methods, main-text Section~\Mref{sec:paths}). The expert end structures were relaxed separately and only $Pbcn$ is shared, so the phase energies relative to $P4_2/nmc$ differ by 3--5~meV/f.u. For T1--T3 and T5 the two constructions stay on the same symmetry manifold, and for T1, T2 and T5 the barriers agree to a few meV/f.u. For T4 and T6 they do not. In the common basis, the $Pbcn$ and $Pca2_1$ components of $M_1$ ($M_1^{(60)}$ and $M_1^{(29)}$) change together, so the expert paths pass through $P2_1$, with two symmetry operations, whereas IRIS stays on the $Pca2_1$ branch for T4 and routes T6 through $Pbcn$; every IRIS image keeps the symmetry of its reference exactly ($D_G=0$; $D_G$ is defined in eq.~\eqref{eq:si_DG} below), whereas the expert T4 and T6 images depart from it by up to about 0.7~Å (Supplementary Fig.~\ref{fig:s_hfo2_sym}). These are the two paths on which the expert barriers are much higher: about 245 against 80.5~meV/f.u.\ for T4, and 445 against 151.5~meV/f.u.\ for T6. This comparison concerns the construction of the paths, not the language model, and the two constructions also differ in their end structures and lattice interpolation. T3 has the same intermediate symmetry in both constructions, yet the expert barrier is higher (about 72 against 44~meV/f.u.). The lag of the expert cell behind the atoms does not explain this alone, since T2 lags as much and costs only about 2~meV/f.u.\ more; a likely cause is that T3 alone contains a $\Gamma_5^+$ atomic mode, which couples bilinearly to the $\Gamma_5^+$ shear strain, so that a lagging shear leaves this coupling unrelieved. A single calculation of the expert structure at $\lambda=1/3$ in a cell interpolated with the atoms would test this. All barriers along straight paths are upper bounds; relaxed minimum-energy paths are left for future work. Across the four classes of transition, the same sequence of decomposition, certification, sizing, auditing and fitting produced every landscape (Extended Data Table~\Mref{tab:systems}).

\begin{table}[!htbp]
\centering\small
\caption{\textbf{HfO$_2$ straight irrep paths} (10 points each, $\lambda=k/9$). Energies are relative to the start of each path.}
\label{tab:hfo2}
\begin{tabularx}{\linewidth}{@{}XL{2.6cm}L{1.4cm}L{2.6cm}L{2.6cm}@{}}
\toprule
Path & Barrier (meV/f.u.) & at $\lambda$ & $\Delta E$ at end (meV/f.u.) & Earlier work~\cite{Manna2026hafnia}\\
\midrule
T1 $P4_2/nmc\to Pca2_1$ & 6.43 & 0.222 & $-87.38$ & $+6.1$\\
T2 $P4_2/nmc\to Pbcn$ & 24.68 & 0.333 & $-99.37$ & $+23.6$\\
T3 $P4_2/nmc\to P2_1/c$ & 43.95 & 0.333 & $-170.43$ & $+43.0$\\
T4 $Pbcn\to Pca2_1$ & 80.50 & 0.556 & $+12.05$ & --\\
T5 $Pbcn\to P2_1/c$ & 52.37 & 0.444 & $-71.03$ & --\\
T6 $P2_1/c\to Pbcn\to Pca2_1$ (T5 reversed, then T4) & 151.5 & T4 at 0.556 & $+83.08$ & --\\
\bottomrule
\end{tabularx}
\end{table}

Phase energies relative to $P4_2/nmc$ are $-170.43$ ($P2_1/c$), $-99.37$ ($Pbcn$) and $-87.38$~meV/f.u.\ ($Pca2_1$). The same crystal built in the different settings agrees to 0.000 ($P4_2/nmc$), 0.017 ($Pbcn$), 0.042 ($Pca2_1$) and 0.005~meV/f.u.\ ($P2_1/c$). Every nonzero difference is between a structure computed on the $5\times5\times5$ mesh (T1--T3) and the same structure on the $5\times6\times5$ mesh (T4, T5); on the same mesh the agreement is exact. Closure: $E(\mathrm{T2})+E(\mathrm{T4})=-87.32$ against $E(\mathrm{T1})=-87.38$, and $E(\mathrm{T2})+E(\mathrm{T5})=-170.40$ against $E(\mathrm{T3})=-170.43$~meV/f.u.; the gaps, $-0.060$ and $-0.023$~meV/f.u., are exactly the sums of the mesh differences ($0.0175+0.0425$ and $0.0175+0.0050$), so the closure measures $k$-point convergence rather than testing the energies independently. All 50 points used one static input apart from the mesh (PBE; PAW potentials for Hf (without semicore $5p$) and O; PREC Accurate, ENCUT 600~eV, $\Gamma$-centred $5\times5\times5$ mesh for T1--T3 and $5\times6\times5$ for T4 and T5, EDIFF $10^{-8}$, NELM 200, ALGO Normal, Gaussian smearing 0.05~eV, LREAL .FALSE., ISYM 0, NSW 0); all converged and are non-magnetic. Along each path, the strain is a single coordinate $s=\lambda$ snapped to the child metric; an earlier construction that scaled the strain with every irrep separately overstated it by a factor of 0.8 times the number of irreps on the path and is not used. The $Pbcn$ structure, the same as in our earlier work~\cite{Manna2026hafnia}, has a volume of 37.2~\AA$^3$ per f.u.

\begin{table}[!htbp]
\centering\small
\caption{\textbf{Settings (basis and origin relative to the parent) used for each HfO$_2$ transition.} $k$, cell multiplicity (basis determinant); $i=t\cdot k$, index of the subgroup (Supplementary Fig.~\ref{fig:s_index}).}
\label{tab:hfo2_set}
\begin{tabularx}{\linewidth}{@{}L{2.6cm}XL{3.8cm}@{}}
\toprule
Transition & Basis & Origin\\
\midrule
T1 137$\to$29 & $(0,0,1)(1,1,0)(-1,1,0)$, $k=2$, $i=8$ & $(1.75107,-0.25107,\tfrac54)$\\
T2 137$\to$60 & $(-1,1,0)(0,0,1)(1,1,0)$ (basis search), $k=2$, $i=4$ & $(\tfrac32,1,1)$\\
T3 137$\to$14 & $(1,-1,0)(1,1,0)(0,0,1)$, $k=2$, $i=8$ & $(1,1,\tfrac32)$\\
T4 60$\to$29 & $(0,1,0)(-1,0,0)(0,0,1)$, $k=1$, $i=2$ & $(\tfrac34,\tfrac34,-0.00107)$\\
T5 60$\to$14 & $(0,0,1)(1,0,0)(0,1,0)$, $k=1$, $i=2$ & $(0,0,0)$\\
\bottomrule
\end{tabularx}
\end{table}

\paragraph{Symmetry along the IRIS and expert paths.} Supplementary Fig.~\ref{fig:s_hfo2_sym} follows the space group of every image of the six paths, for the IRIS construction and for the expert construction of Extended Data Table~\Mref{tab:expert}, together with a continuous measure of how far each image is from the reference symmetry. For a reference group $G$ with $|G|$ operations, the reference-symmetry distance of a structure $X$ is
\begin{equation}\label{eq:si_DG}
D_G(X)=\sqrt{\frac{1}{|G|-1}\sum_{g\in G,\,g\neq e}\Delta_g(X)^2},
\end{equation}
where $\Delta_g(X)$ is the root-mean-square displacement between $X$ and its image under $g$, with atoms matched by species and displacements wrapped by the periodic boundary conditions. $D_G=0$ means that $X$ keeps every operation of $G$, and $D_G$ grows as $X$ departs from it. Along every IRIS path, $D_G=0$ at every image: each intermediate has the symmetry of the lower-symmetry end phase, and T6 passes through $Pbcn$ at $s=0.5$. The expert paths keep the reference symmetry for T1, T2, T3 and T5 (and agree with IRIS for T1, T2 and T5; T3 keeps the symmetry but has a higher barrier). For T4 and T6 the expert intermediates fall to $P2_1$ (two operations), and $D_G$ is nonzero along the path: for T4 it decreases from about 0.7~\AA\ at the $Pbcn$ end to zero at $Pca2_1$, and for T6 it rises to about 0.3~\AA\ at $s=0.5$. These are the two paths on which the expert barriers are much higher (Supplementary Section~\ref{sec:si_hfo2_ext}).

\subsection{A route for pairs without a group--subgroup relation}\label{sec:si_supergroup}

The HfO$_2$ paths show where the present workflow stops: T6 connects two phases that are not related by a group--subgroup relation, and the expert built it through a common supergroup chosen by hand, as T6 was composed here from T5 reversed and T4 (Supplementary Table~\ref{tab:hfo2}). Nothing in that construction is specific to hafnia, and it can be made a stage of the workflow rather than a stop. The blueprint, which we describe as a design and have not built, is: (i) given two phases $H_1$ and $H_2$ with no group--subgroup relation, enumerate their common supergroups $G$ by index with the minimal-supergroup tools of the Bilbao server (\textsc{supergroups}, \textsc{commonsuper}), each with the transformation matrices that bring $H_1$ and $H_2$ into its setting; (ii) for each candidate $G$, build the reference structure by averaging the two phases in that setting and score it by the strain and the largest atomic displacement needed to reach each phase, which is exactly the information $f_1$ already computes for one pair, so that candidates whose displacements exceed the basis-search cut-off are discarded as they are for a single transition; (iii) run $f_1$ and $f_2$ twice from the chosen $G$, once per child, as for the T1--T3 paths from $P4_2/nmc$, and compose the path through $G$; (iv) gate the result on three checks that the existing paths already pass: that the end-point energies agree with the direct calculations, that the reference phase is reachable from both children within the cut-off, and that the composed path closes with the other stored paths of the network ($G_5$ \texttt{CLOSURE}, \texttt{CROSS\_SETTING}, \texttt{NETWORK}; Section~\ref{sec:si_live}). The hand-built T6 is the test case: a correct implementation must recover $P4_2/nmc$ as the lowest-strain common supergroup of $Pbcn$ and $P2_1/c$ and reproduce the 151.5~meV/f.u.\ composite barrier. Where no common supergroup lies within the cut-off, the honest output is a refusal with the nearest candidates listed, which is the same stop the workflow makes today, but with its reason stated.

\section{Polar-origin convention}\label{sec:origin}

The vector representation of the parent point group ($\Gamma_4^-$ in $Pm\bar3m$, $\Gamma_2^-$ in $P6_3/mmc$, $\Gamma_5^-$ in $I4/mmm$ and $P4_2/nmc$) contains the acoustic translation. In a polar subgroup, the origin along the polar axis is free and the decomposition fixes it by convention. The mode vector of the polar irrep can therefore carry an arbitrary rigid translation of the whole crystal, which costs no energy. For BiFeO$_3$, an independent decomposition (doubling the cubic cell, matching atoms and splitting the displacement field into its $k=0$ and staggered parts) reproduces the $R_5^-$ amplitude to four digits and gives a $\Gamma_4^-$ amplitude of 1.0725~\AA\ with Bi fixed on its parent site but only 0.526~\AA\ with the mean displacement removed (main-text eq.~\Meqref{eq:trans}): a rigid shift along $[111]$ of 0.418~\AA\ per atom makes up 76\% of $\|\mathbf e\|^2$ ($f_T=0.76$), or 51\% of the amplitude. Before this was corrected, the amplitude calibration projected a translation-free child onto a translation-containing vector and gave a calibration ratio near 0.24 instead of 1, so unit amplitude did not reach the child. Because a translation changes no symmetry, the space-group check nevertheless passed. IRIS therefore removes the mean displacement from every mode vector (main-text eq.~\Meqref{eq:trans}), reports $A^{\mathrm{tf}}$ (eq.~\Meqref{eq:tf}) and $f_T$ (eq.~\Meqref{eq:fT}), records both values, and states the convention whenever amplitudes are compared with the literature. Zone-boundary irreps ($M_3^+$, $R_4^+$, $R_5^-$, $K_3$, $X_5^+$, $X_3^-$, $X_2^+$) have zero mean displacement and are unaffected.

\section{Evaluation of the agentic layer}\label{sec:si_eval}
This section gives the details behind main-text Section~\Mref{sec:autonomy} and Fig.~\Mref{fig:agentic}; the protocol is in main-text Methods, Section~\Mref{sec:eval}. Every experiment is a replay on the stored DFT energies: $f_3$ matches each frame to a stored structure, so no calculation is repeated and every outcome can be compared with the fault-free replay, whose numbers are those reported in the main text.

\paragraph{What each layer contributes: an ablation by layer.}\label{sec:si_layers} IRIS has four layers that could each be credited with its results: the deterministic nodes $f_1$--$f_6$ (the scripts that decompose, build, compute and fit), the gates $G_0$--$G_6$ (deterministic checks between them), the graph (the fixed order of nodes and gates, the checkpoints, the recovery order rule $\to$ sub-agent $\to$ human, and the audit log), and the language model (the intake and the sub-agents). The experiments of this section ablate them separately, and the attribution is as follows. (i) \emph{Nodes without gates.} The physics is in the nodes: the fault-free replays reproduce the published numbers with no gate failing and no model call (Extended Data Table~\Mref{tab:reruns}), and with every gate switched off the 24 harmless faults still complete with identical numbers; but on the 53 faulted configurations the same nodes return 14 silently wrong results, crash 15 times and are harmless 24 times (Supplementary Table~\ref{tab:si_faults}). The scientific validity on faulted inputs is therefore not a property of the nodes. (ii) \emph{Gates, one at a time.} Supplementary Table~\ref{tab:si_ablation} removes, for each fault, the gate that catches it, and also removes each gate that does not catch it (54 further runs, no change in outcome): no gate is redundant with the graph or with another gate, and removing the catching gate moves detection to a later gate at the cost of the work in between (up to 27{,}065 counterfactual DFT submissions for $G_4$) or, for 3 of 30 errors, lets the error through. Each gate's contribution is thus the fault class it alone stops at the cheapest point: $G_1$ the decomposition (6 faults), $G_2$ the frames (8), $G_3$ the job bookkeeping (2), $G_4$ the energies (24), $G_5$ the fit (13); $G_0$ and $G_6$ catch none of the 53 injected faults, because the injections pass schema validation by construction and $G_6$ re-derives rather than detects (its contribution is the audit, Supplementary Table~\ref{tab:si_audit}). (iii) \emph{Graph structure.} The graph adds nothing to the numbers: the fault-free replays give the same results with the sub-agent on or off, and every completed faulted run gives numbers identical to the fault-free one in both conditions (Supplementary Table~\ref{tab:si_interventions}). What it adds is measured separately: runs killed after $f_2$, $f_3$ or $f_5$ resume from their checkpoint with identical gate outcomes and no job submitted twice (4 of 4, \texttt{E7\_kill\_resume.csv}); the recovery order turns a gate failure into a stored rule (34 rule fixes in the 53 faulted runs), a sub-agent choice or a human question, whose counts are in Supplementary Table~\ref{tab:si_interventions}; and every reported number is re-derived from the audit log (Supplementary Table~\ref{tab:si_audit}). (iv) \emph{Language model.} With the model off (rules and human only) the 53 faulted runs complete as often (27) with 34 hand-offs; with the model on, 27 hand-offs (Supplementary Table~\ref{tab:si_interventions}); on the decision benchmark the stored rules alone are correct on 96\% of cases against 72--91\% for the models (Supplementary Tables~\ref{tab:si_models} and~\ref{tab:si_A}). The model's contribution is the plain-language intake (Supplementary Table~\ref{tab:si_E}) and seven fewer hand-offs to the human, not the validity of any result. In short, the physics comes from the deterministic nodes, the containment of errors from the gates, resumability and provenance from the graph, and the interface from the model; the paper's claim is the second of these, and the first and the fourth are ablated in the same tables.

\paragraph{Faults and outcomes.} The 53 faults (F001--F053) are injected only through the run configuration and recorded by $f_0$; data and code are never altered. They cover all six systems (Supplementary Table~\ref{tab:si_faults}). The 30 errors are failed or SCF-limited job parts (walltime or SCF step limit, some repeated after the rule has been applied), a rigid translation left in a mode vector, a single-mode check frame built on the wrong lattice, an inconsistent DFT setting, a wrong formula-unit count, a missing stored frame and cancelled jobs. The 23 limit faults tighten a fit, closure, coverage, symmetry or lattice limit, or narrow the atom-matching search, on data that are correct; they test that the gate enforces its limit. Outcomes (completed, stopped, silent, crash, harmless) are defined in Supplementary Table~\ref{tab:si_defs}. With all gates on, every fault was caught by the gate expected for it, and no run ended silent. With no gates, the 14 silent corruptions are the 13 job faults whose failed or SCF-limited parts were dropped from the fit and the HfO$_2$ formula-unit count; the crashes are faults that leave a later step without the input it needs. Supplementary Table~\ref{tab:si_ablation} gives the runs with one gate removed. Of the 27 error faults stopped by a later check, 23 were stopped by checks that also run in production ($G_5$, 21, of which 15 by \texttt{MISSING\_FRAMES} and one using the energies of the other HfO$_2$ paths; $G_4$, 1; $G_5$ after $G_3$ was removed, 1), 3 by the frame matching that exists only in replay ($G_3$, with $G_2$ removed), and 1 by a crash that the ablation itself causes; in a first replay, before \texttt{MISSING\_FRAMES}, the split had been 14, 12 and 1 (Section~\ref{sec:si_replay_found}).

Supplementary Table~\ref{tab:si_defs} collects the definitions used throughout this section, and Supplementary Table~\ref{tab:si_faultlist} lists every injected fault with its first gate failure, its handling and its outcome with and without the gates.

\begin{table}[!htbp]
\centering\footnotesize
\caption{\textbf{Definitions used in the evaluation of the agentic layer} (main-text Section~\Mref{sec:autonomy} and Methods, Section~\Mref{sec:eval}).}
\label{tab:si_defs}
\begin{tabularx}{\linewidth}{@{}>{\raggedright}p{3.3cm}>{\raggedright\arraybackslash}X@{}}
\toprule
Term & Definition\\
\midrule
\multicolumn{2}{@{}l}{\textit{Injected faults (Supplementary Table~\ref{tab:si_faultlist})}}\\
Error & A fault that corrupts a result if it passes: failed or SCF-limited job parts, a rigid translation left in a mode vector, a single-mode frame built on the wrong lattice, an inconsistent DFT setting or formula-unit count, a missing stored frame, cancelled jobs (30).\\
Limit fault & A fault that tightens a fit, closure, coverage, symmetry or lattice limit, or narrows the atom-matching search, on correct data; stopping is the correct outcome (23).\\
\multicolumn{2}{@{}l}{\textit{Outcome of a run}}\\
Completed & The run reaches $f_6$ and every re-derived number equals that of the fault-free replay.\\
Stopped & The run ends at a hand-off to the human.\\
Silent & The run completes with numbers that differ from the fault-free replay and nothing is flagged.\\
Crash & A node raises an error.\\
Harmless & The run completes with identical numbers although no gate acted.\\
Hand-off & Any stop that passes the run to the human, whether chosen by a rule, a sub-agent or the gate, counted whether or not a question is asked.\\
\multicolumn{2}{@{}l}{\textit{Recovery benchmark (53 cases)}}\\
Case & The exact prompt that a sub-agent receives at a real gate failure, with the menu of fixes of that failure.\\
Case classes & Rules off (29): a stored rule would fix it, with the rule table switched off; rule used up (6): the rule was applied and the failure returned; no rule (5): a code with no stored rule, where the human must decide (stop is the reference answer); decoy (11): the menu offers options that look related but do not address the failure (stop is the reference answer); classes were assigned when the cases were written and frozen with the labels; judgment (2): the borderline InMnO$_3$ selection (R5) and the CsSnI$_3$ hard secondary (D4). Held out: 16 cases added after the menus and prompts were final.\\
Correct & The reference option: for a rule-fixable failure the fix of that rule; for a decoy, stop; for a judgment call, the published decision.\\
Safe & Stop or another hand-off to the human when stop is not the reference: no damage, but a human intervention.\\
Harmful & Any other fix: it would change the result wrongly or waste calculations.\\
Invalid & An answer that is not a menu identifier.\\
Guided / bare prompt & With / without the one-line fix knowledge of the rule table for the failure code.\\
Consistency & Mean share of the ten repeats that give the modal answer.\\
\multicolumn{2}{@{}l}{\textit{Executed options (208 branches)}}\\
Branch & A failure resumed from its checkpoint with one menu option applied, all gates on unless stated.\\
Outcomes & Accepted and correct (within the frozen tolerance of the published numbers); equivalent (completed after a change of SCF mixing only, whose identical energies replay assumes); refused by the next gate, then repaired by a rule; rejected by a later gate; stopped by choice; accepted and wrong (none with the gates on).\\
\multicolumn{2}{@{}l}{\textit{Plain-language intake (50 requests)}}\\
Frozen request set & The 50 requests and their reference configurations, written and fixed with a hash before any model saw them.\\
Clear request & One of the 13 run configurations asked for in one of three styles: expert, terse or non-specialist (39).\\
Critical fields & Structures, transition, analysis type, strain policy, system profile and budget; a call is correct when all are right.\\
Valid but wrong & A configuration that passes $G_0$ (complete and internally consistent) but differs from the request in a critical field.\\
\bottomrule
\end{tabularx}
\end{table}

\begin{scriptsize}
\setlength{\tabcolsep}{3pt}
\begin{longtable}{@{}>{\raggedright}p{0.7cm}>{\raggedright}p{1.5cm}>{\raggedright}p{0.55cm}>{\raggedright}p{4.0cm}>{\raggedright}p{2.3cm}>{\raggedright}p{2.9cm}>{\raggedright}p{1.3cm}>{\raggedright\arraybackslash}p{1.2cm}@{}}
\caption{\textbf{The 53 injected faults.} Each fault is a single setting in the run configuration, recorded by $f_0$; data and code are never altered. Kind: E, error (corrupts a result if it passes); L, limit fault (tightens a threshold or narrows a search on correct data). Description: what the setting does. First gate: the first gate failure in the gated run, with its error code; for six InMnO$_3$ faults this is the published borderline decision of rule R5 (\texttt{R5\_\allowbreak{}BORDERLINE}), which the run meets before the injected fault, and the fault is then caught by the gate of its own handling. Handling: the sources of the fixes, in order (stored rule, or human; ``human (stop)'' ends the run with the human). Gated: completed (numbers identical to the fault-free replay) or stopped. Ungated: outcome with no gates, silent (numbers differ and nothing is flagged), crash or harmless. Summary by system in Supplementary Table~\ref{tab:si_faults}; definitions in Supplementary Table~\ref{tab:si_defs}.}\label{tab:si_faultlist}\\
\toprule
ID & System & Kind & Injected fault & First gate & Handling & Gated & Ungated\\
\midrule\endfirsthead
\toprule
ID & System & Kind & Injected fault & First gate & Handling & Gated & Ungated\\
\midrule\endhead
\bottomrule\endfoot
F001 & CsSnI$_3$ & E & job part 3 ends with WALLTIME & $G_4$ \texttt{WALLTIME} & rule \texttt{f4\_\allowbreak{}walltime\_\allowbreak{}resubmit} & completed & silent\\
F002 & CsSnI$_3$ & E & job part 7 ends with NELM (SCF step limit) & $G_4$ \texttt{NELM} & rule \texttt{f4\_\allowbreak{}nelm\_\allowbreak{}robust\_\allowbreak{}mixing} & completed & silent\\
F003 & CsSnI$_3$ & E & rigid translation of 0.05~\AA/atom left in the $M_3^+$ vector (removal switched off) & $G_2$ \texttt{TRANSLATION} & rule \texttt{f2\_\allowbreak{}translation} & completed & crash\\
F004 & CsSnI$_3$ & L & atom-matching basis search limited to dmax 0.3~\AA{} (CsSnI$_3$ needs 0.59~\AA) & $G_1$ \texttt{NO\_\allowbreak{}BASIS} & rule \texttt{f1\_\allowbreak{}no\_\allowbreak{}basis\_\allowbreak{}dmax\_\allowbreak{}ladder} & completed & harmless\\
F005 & CsSnI$_3$ & E & single-mode check on a frame built on the child's lattice (the No.~55 error of the CsSnI$_3$ campaign) & $G_2$ \texttt{SINGLE\_\allowbreak{}MODE\_\allowbreak{}SG} & rule \texttt{f2\_\allowbreak{}single\_\allowbreak{}mode\_\allowbreak{}sg\_\allowbreak{}fixed\_\allowbreak{}lattice} & completed & harmless\\
F006 & CsSnI$_3$ & L & G5 coverage demands 100 points per coefficient (800 $>$ 545 in the fit) & $G_5$ \texttt{THIN\_\allowbreak{}GRID} & rule \texttt{f5\_\allowbreak{}thin\_\allowbreak{}grid\_\allowbreak{}densify}; human (stop) & stopped & harmless\\
F007 & CsSnI$_3$ & E & stored frame 100 missing from the reference data & $G_3$ \texttt{UNMATCHED} & human (stop) & stopped & harmless\\
F008 & CsSnI$_3$ & L & dmax 0.3~\AA{} with a retry budget of 0 $\to$ straight to the human, who widens dmax & $G_1$ \texttt{NO\_\allowbreak{}BASIS} & human & completed & harmless\\
F009 & CsSnI$_3$ & E & all jobs cancelled while f3 waits & $G_3$ \texttt{JOBS\_\allowbreak{}CANCELLED} & human (stop) & stopped & crash\\
F010 & CsSnI$_3$ & E & WALLTIME again after the resubmit rule & $G_4$ \texttt{WALLTIME} & rule \texttt{f4\_\allowbreak{}walltime\_\allowbreak{}resubmit}; human (stop) & stopped & silent\\
F011 & CsSnI$_3$ & E & NELM again after the robust-mixing rule & $G_4$ \texttt{NELM} & rule \texttt{f4\_\allowbreak{}nelm\_\allowbreak{}robust\_\allowbreak{}mixing}; human (stop) & stopped & silent\\
F012 & Ca$_3$Ti$_2$O$_7$ & E & rigid translation (0.05~\AA/atom) injected into the Ca$_3$Ti$_2$O$_7$ $X_3^-$ vector, removal off (rules off) & $G_2$ \texttt{TRANSLATION} & rule \texttt{f2\_\allowbreak{}translation} & completed & crash\\
F013 & HfO$_2$ T1 & E & HfO$_2$ path job ends with WALLTIME (rules off) & $G_4$ \texttt{WALLTIME} & rule \texttt{f4\_\allowbreak{}walltime\_\allowbreak{}resubmit} & completed & crash\\
F014 & HfO$_2$ T1 & E & HfO$_2$ path job ends at NELM (rules off) & $G_4$ \texttt{NELM} & rule \texttt{f4\_\allowbreak{}nelm\_\allowbreak{}robust\_\allowbreak{}mixing} & completed & crash\\
F015 & HfO$_2$ T1 & L & HfO$_2$ closure T1 = T2 + T4 misses a tightened limit & $G_5$ \texttt{CLOSURE} & human (stop) & stopped & harmless\\
F016 & HfO$_2$ T1 & L & HfO$_2$ the same crystal differs between two settings beyond a tightened limit & $G_5$ \texttt{CROSS\_\allowbreak{}SETTING} & human (stop) & stopped & harmless\\
F017 & HfO$_2$ T1 & E & HfO$_2$ stored INCAR differs from the campaign template & $G_4$ \texttt{DFT\_\allowbreak{}INCONSISTENT} & human (stop) & stopped & harmless\\
F018 & HfO$_2$ T1 & L & HfO$_2$ points carry a magnetic moment above the (tightened) limit & $G_4$ \texttt{MAGNETIC} & human (stop) & stopped & harmless\\
F019 & HfO$_2$ T1 & E & HfO$_2$ configured formula units per cell disagree with the composition & $G_4$ \texttt{PER\_\allowbreak{}FU} & human (stop) & stopped & silent\\
F020 & BiFeO$_3$ & E & BiFeO$_3$ grid job part times out (rules off) --- the real campaign resubmitted 50 timed-out runs unchanged & $G_4$ \texttt{WALLTIME} & rule \texttt{f4\_\allowbreak{}walltime\_\allowbreak{}resubmit} & completed & silent\\
F021 & BiFeO$_3$ & E & BiFeO$_3$ grid job part hits the SCF step limit (rules off) --- the real campaign reran 15 runs with robust mixing & $G_4$ \texttt{NELM} & rule \texttt{f4\_\allowbreak{}nelm\_\allowbreak{}robust\_\allowbreak{}mixing} & completed & silent\\
F022 & BiFeO$_3$ & L & BiFeO$_3$ more SCF-limited runs than the declared limit (the real 20 near-metallic corners vs a limit of 10) & $G_4$ \texttt{SCF\_\allowbreak{}LIMITED} & human (stop) & stopped & harmless\\
F023 & ErMnO$_3$ & E & ErMnO$_3$ plane job part times out (rules off) & $G_4$ \texttt{WALLTIME} & rule \texttt{f4\_\allowbreak{}walltime\_\allowbreak{}resubmit} & completed & silent\\
F024 & ErMnO$_3$ & L & ErMnO$_3$ Meier fit RMS above a tightened limit (5.99 vs 5.0 meV/cell) & $G_5$ \texttt{RESIDUALS} & human (stop) & stopped & harmless\\
F025 & InMnO$_3$ (joint) & L & InMnO$_3$ joint fit max residual 12.06 $>$ 12 meV/cell (the real failure of the build; the limit was raised to 15) & $G_2$ \texttt{R5\_\allowbreak{}BORDERLINE} & human; human (stop) & stopped & crash\\
F026 & HfO$_2$ T5 & L & HfO$_2$ T5 child cell carries a $5.02^\circ$ rigid rotation; a rotation limit of $1^\circ$ is set & $G_2$ \texttt{PATH\_\allowbreak{}LATTICE} & human (stop) & stopped & harmless\\
F027 & HfO$_2$ T4 & E & HfO$_2$ T4 path job times out again after the resubmit rule (all 10 points in that part) & $G_4$ \texttt{WALLTIME} & rule \texttt{f4\_\allowbreak{}walltime\_\allowbreak{}resubmit}; human (stop) & stopped & crash\\
F028 & Ca$_3$Ti$_2$O$_7$ & L & no atom-matching basis within dmax 0.3~\AA, Ca$_3$Ti$_2$O$_7$ (rules off) & $G_1$ \texttt{NO\_\allowbreak{}BASIS} & rule \texttt{f1\_\allowbreak{}no\_\allowbreak{}basis\_\allowbreak{}dmax\_\allowbreak{}ladder} & completed & harmless\\
F029 & InMnO$_3$ & E & InMnO$_3$ plane job part hits the SCF step limit after R5 (rules off) & $G_2$ \texttt{R5\_\allowbreak{}BORDERLINE} & human; rule \texttt{f4\_\allowbreak{}nelm\_\allowbreak{}robust\_\allowbreak{}mixing} & completed & crash\\
F030 & ErMnO$_3$ & L & no atom-matching basis within dmax 0.3~\AA, ErMnO$_3$ (rules off) & $G_1$ \texttt{NO\_\allowbreak{}BASIS} & rule \texttt{f1\_\allowbreak{}no\_\allowbreak{}basis\_\allowbreak{}dmax\_\allowbreak{}ladder} & completed & harmless\\
F031 & HfO$_2$ T3 & L & no atom-matching basis within dmax 0.3~\AA, HfO$_2$ T3 (rules off) & $G_1$ \texttt{NO\_\allowbreak{}BASIS} & rule \texttt{f1\_\allowbreak{}no\_\allowbreak{}basis\_\allowbreak{}dmax\_\allowbreak{}ladder} & completed & harmless\\
F032 & BiFeO$_3$ & L & no atom-matching basis within dmax 0.3~\AA, BiFeO$_3$ (rules off) & $G_1$ \texttt{NO\_\allowbreak{}BASIS} & rule \texttt{f1\_\allowbreak{}no\_\allowbreak{}basis\_\allowbreak{}dmax\_\allowbreak{}ladder} & completed & harmless\\
F033 & Ca$_3$Ti$_2$O$_7$ & L & Ca$_3$Ti$_2$O$_7$ surface minimum closer to the grid edge than a widened margin (50 steps) & $G_5$ \texttt{COVERAGE} & human (stop) & stopped & harmless\\
F034 & ErMnO$_3$ & E & single-mode space group checked on a frame built on the child's lattice, ErMnO$_3$ (rules off) & $G_2$ \texttt{SINGLE\_\allowbreak{}MODE\_\allowbreak{}SG} & rule \texttt{f2\_\allowbreak{}single\_\allowbreak{}mode\_\allowbreak{}sg\_\allowbreak{}fixed\_\allowbreak{}lattice} & completed & harmless\\
F035 & ErMnO$_3$ & E & rigid translation injected into the ErMnO$_3$ $K_3$ vector, removal off (rules off) & $G_2$ \texttt{TRANSLATION} & rule \texttt{f2\_\allowbreak{}translation} & completed & crash\\
F036 & InMnO$_3$ & E & rigid translation injected into the InMnO$_3$ $K_3$ vector after R5 (rules off) & $G_2$ \texttt{R5\_\allowbreak{}BORDERLINE} & human; rule \texttt{f2\_\allowbreak{}translation} & completed & crash\\
F037 & Ca$_3$Ti$_2$O$_7$ & E & job part ends with WALLTIME, Ca$_3$Ti$_2$O$_7$ (rules off) & $G_4$ \texttt{WALLTIME} & rule \texttt{f4\_\allowbreak{}walltime\_\allowbreak{}resubmit} & completed & silent\\
F038 & InMnO$_3$ & E & job part ends with WALLTIME, InMnO$_3$ (rules off) & $G_2$ \texttt{R5\_\allowbreak{}BORDERLINE} & human; rule \texttt{f4\_\allowbreak{}walltime\_\allowbreak{}resubmit} & completed & crash\\
F039 & HfO$_2$ T3 & E & job part ends with WALLTIME, HfO$_2$ T3 (rules off) & $G_4$ \texttt{WALLTIME} & rule \texttt{f4\_\allowbreak{}walltime\_\allowbreak{}resubmit} & completed & crash\\
F040 & CsSnI$_3$ (D4) & E & job part ends with WALLTIME, CsSnI$_3$ (D4) (rules off) & $G_4$ \texttt{WALLTIME} & rule \texttt{f4\_\allowbreak{}walltime\_\allowbreak{}resubmit}; human & completed & silent\\
F041 & Ca$_3$Ti$_2$O$_7$ & E & job part ends at the SCF step limit, Ca$_3$Ti$_2$O$_7$ (rules off) & $G_4$ \texttt{NELM} & rule \texttt{f4\_\allowbreak{}nelm\_\allowbreak{}robust\_\allowbreak{}mixing} & completed & silent\\
F042 & ErMnO$_3$ & E & job part ends at the SCF step limit, ErMnO$_3$ (rules off) & $G_4$ \texttt{NELM} & rule \texttt{f4\_\allowbreak{}nelm\_\allowbreak{}robust\_\allowbreak{}mixing} & completed & silent\\
F043 & HfO$_2$ T5 & E & job part ends at the SCF step limit, HfO$_2$ T5 (rules off) & $G_4$ \texttt{NELM} & rule \texttt{f4\_\allowbreak{}nelm\_\allowbreak{}robust\_\allowbreak{}mixing} & completed & crash\\
F044 & Ca$_3$Ti$_2$O$_7$ & E & Ca$_3$Ti$_2$O$_7$ NELM again after the robust-mixing rule & $G_4$ \texttt{NELM} & rule \texttt{f4\_\allowbreak{}nelm\_\allowbreak{}robust\_\allowbreak{}mixing}; human (stop) & stopped & silent\\
F045 & ErMnO$_3$ & E & ErMnO$_3$ WALLTIME again after the resubmit rule & $G_4$ \texttt{WALLTIME} & rule \texttt{f4\_\allowbreak{}walltime\_\allowbreak{}resubmit}; human (stop) & stopped & silent\\
F046 & InMnO$_3$ & L & InMnO$_3$ grid too thin for the Meier fit after R5 (rules off) & $G_2$ \texttt{R5\_\allowbreak{}BORDERLINE} & human; rule \texttt{f5\_\allowbreak{}thin\_\allowbreak{}grid\_\allowbreak{}densify}; human (stop) & stopped & crash\\
F047 & ErMnO$_3$ & L & ErMnO$_3$ grid too thin for the Meier fit (rules off) & $G_5$ \texttt{THIN\_\allowbreak{}GRID} & rule \texttt{f5\_\allowbreak{}thin\_\allowbreak{}grid\_\allowbreak{}densify}; human (stop) & stopped & harmless\\
F048 & Ca$_3$Ti$_2$O$_7$ & L & Ca$_3$Ti$_2$O$_7$ surface mirror symmetry checked at a tightened tolerance ($10^{-6}$~meV/f.u.) & $G_5$ \texttt{MIRROR} & human (stop) & stopped & harmless\\
F049 & CsSnI$_3$ & L & CsSnI$_3$ plane fit RMS above a tightened limit & $G_5$ \texttt{RESIDUALS} & human (stop) & stopped & harmless\\
F050 & InMnO$_3$ & L & InMnO$_3$ plane fit CV error above a tightened limit (after R5) & $G_2$ \texttt{R5\_\allowbreak{}BORDERLINE} & human; human (stop) & stopped & crash\\
F051 & BiFeO$_3$ & L & BiFeO$_3$ degree-12 fit RMS above a tightened limit & $G_5$ \texttt{RESIDUALS} & human (stop) & stopped & harmless\\
F052 & HfO$_2$ T3 & L & HfO$_2$ closure T3 = T2 + T5 misses a tightened limit & $G_5$ \texttt{CLOSURE} & human (stop) & stopped & harmless\\
F053 & HfO$_2$ T3 & L & HfO$_2$ T3 child cell carries a $4.88^\circ$ rigid rotation; a rotation limit of $1^\circ$ is set & $G_2$ \texttt{PATH\_\allowbreak{}LATTICE} & human (stop) & stopped & harmless\\
\end{longtable}
\end{scriptsize}

\subsection{Recovery after a gate failure}\label{sec:si_recovery}
\begin{figure}[!htbp]
\centering
\includegraphics[width=\textwidth]{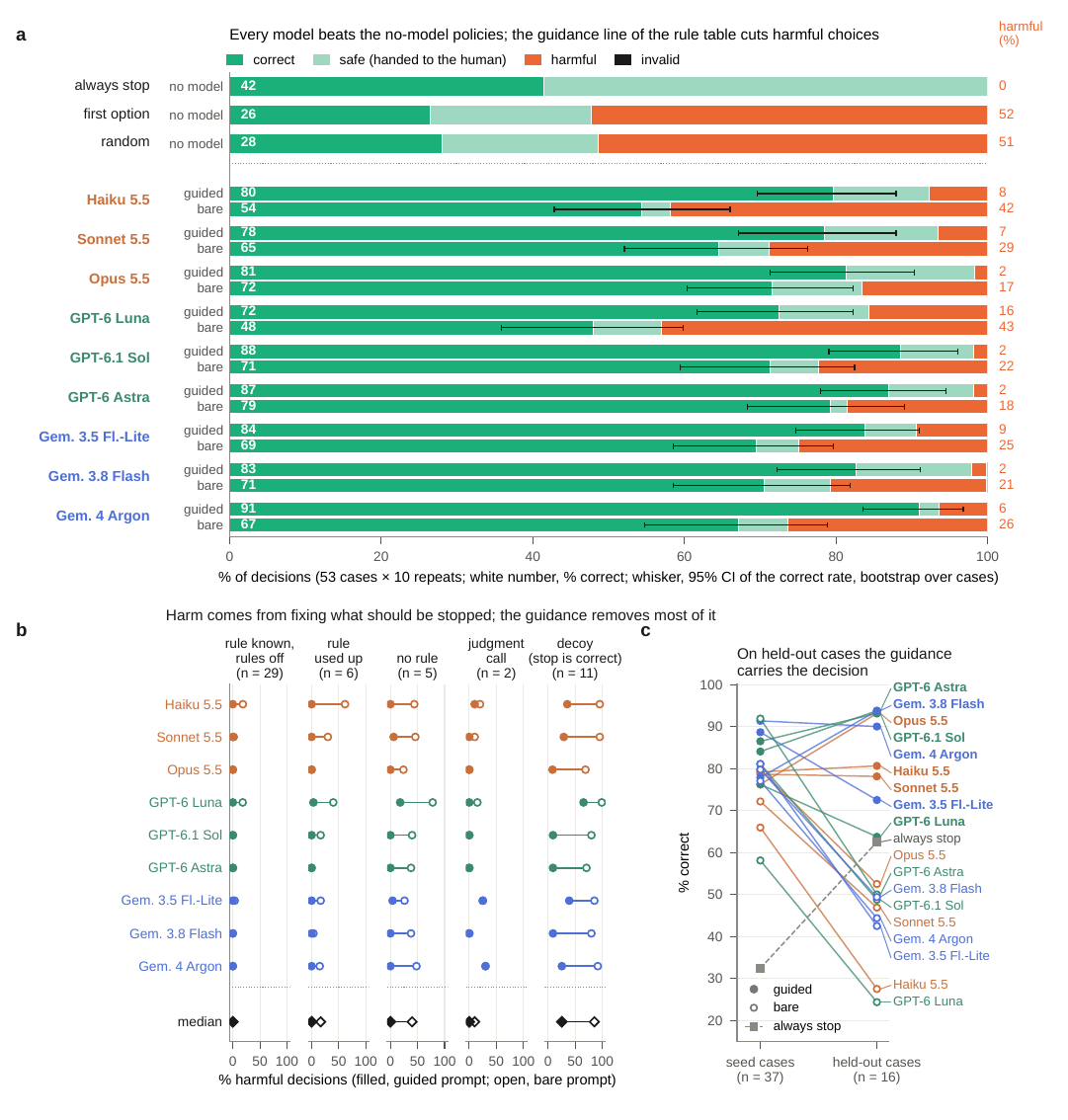}
\caption{\textbf{Language models on the recovery cases.} (a) Decisions on the 53 cases (ten repeats each) for three policies without a model (always stop, first option, random) and nine models, with the guided prompt (upper bar of each pair) and without the guidance line (bare, lower bar); dark green, correct; light green, safe (handed to the human); orange, harmful; black, invalid; white numbers, \% correct; orange numbers at the right, \% harmful; whiskers, 95\% confidence intervals of the correct rate from a bootstrap over cases. (b) Harmful rate by case class for each model (rule known but rules off, $n=29$; rule used up, 6; no rule, 5; judgment call, 2; decoy, where stop is correct, 11; colour: provider, orange Anthropic, green OpenAI, blue Google; filled: guided; open: bare; a line joins the bare and guided values of each model; bottom row, median over models, black diamonds, filled guided and open bare). (c) Correct rate on the 37 seed and 16 held-out cases, one line per model and prompt (filled, guided; open, bare; colours as in b), labelled by model at the right (bold, guided; regular, bare); grey squares, always stop.}
\label{fig:s_agentic_models}
\end{figure}

\begin{figure}[!htbp]
\centering
\includegraphics[width=\textwidth,height=0.82\textheight,keepaspectratio]{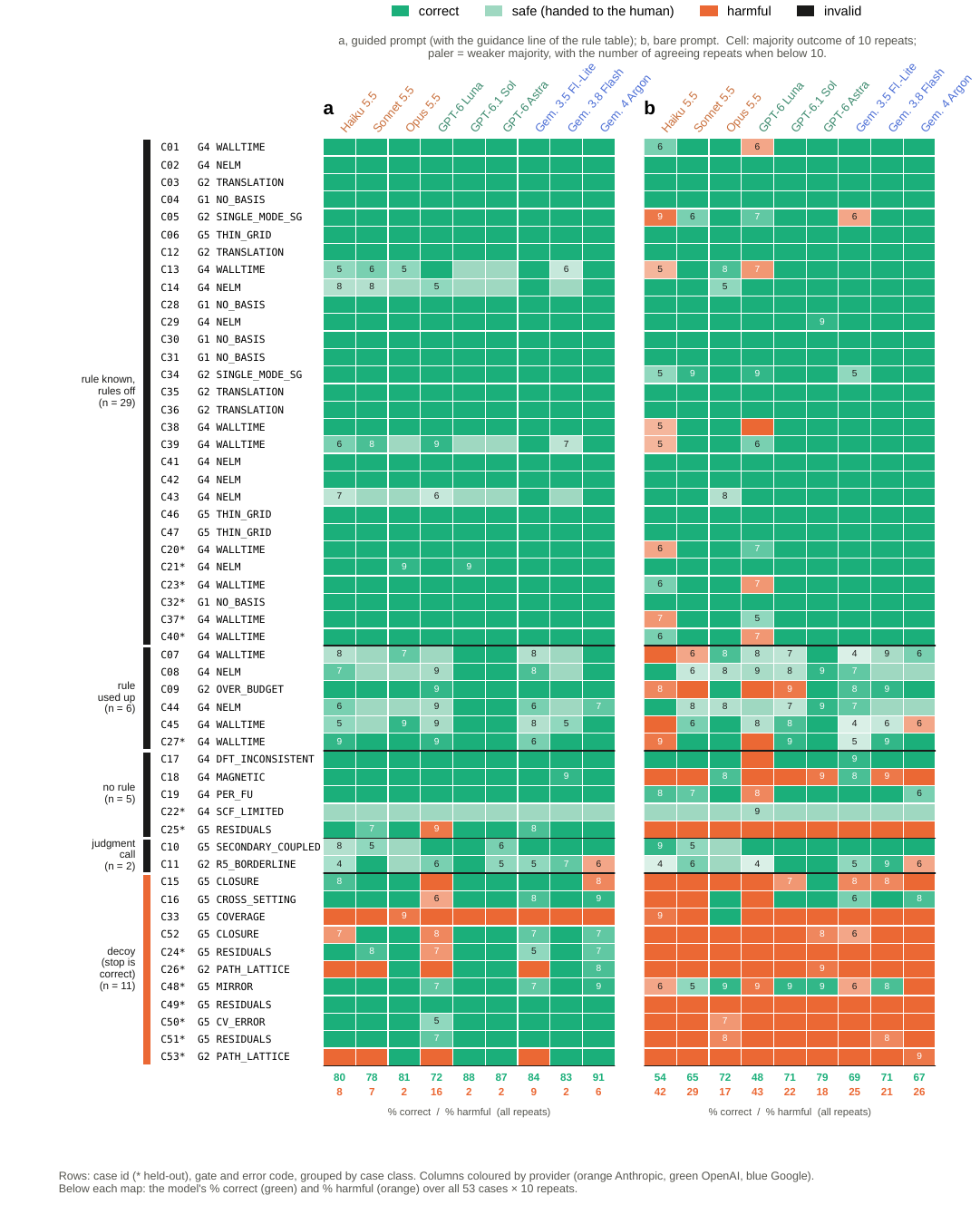}
\caption{\textbf{Sub-agent decisions per case.} Majority outcome of the ten repeats for every case (rows) and model (columns), with the guided (a) and bare (b) prompt. Each row is labelled with the case id (asterisk, held-out case), the gate and the error code, and rows are grouped by case class (rule known but rules off; rule used up; no rule; judgment call; decoy, where stop is correct). Dark green, correct; light green, safe (handed to the human); orange, harmful; black, invalid. A paler cell marks a weaker majority, with the number of agreeing repeats printed when fewer than ten. Below each map, the model's percentage of correct (green) and harmful (orange) decisions over all 53 cases and ten repeats. Model names are coloured by provider (orange Anthropic, green OpenAI, blue Google). Agreement and position bias are in Supplementary Fig.~\ref{fig:s_agentic_robust}.}
\label{fig:s_agentic}
\end{figure}

\begin{figure}[!htbp]
\centering
\includegraphics[width=\textwidth]{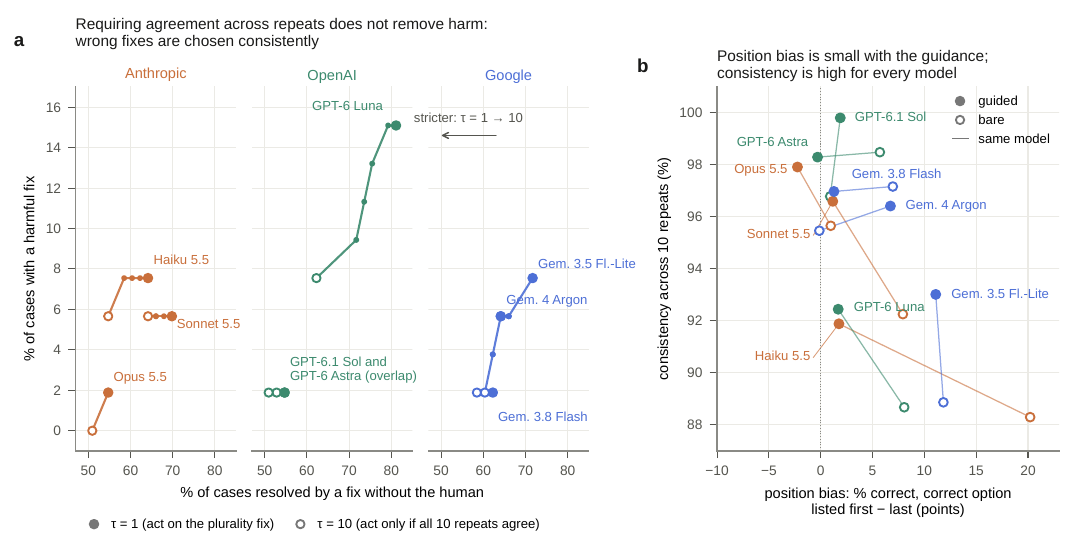}
\caption{\textbf{Agreement across repeats and position bias of the sub-agents.} (a) Escalation by agreement: for each model (guided prompt), a fix is applied only if at least $k$ of the ten repeats choose it, and the case is stopped otherwise; $k$ (written $\tau$ in the panel) runs from 1 (filled marker: act on the plurality fix) to 10 (open marker: act only when all ten repeats agree), one panel per provider; GPT-6.1 Sol and GPT-6 Astra coincide; horizontal axis, share of the 53 cases resolved by a fix without the human; vertical axis, share of cases with a harmful fix. Harm hardly falls as $k$ rises, because the wrong fixes are chosen consistently. (b) Position bias (correct rate with the correct option listed first minus listed last, in points) against consistency (mean share of the ten repeats giving the modal answer) for each model; colours by provider as in Supplementary Fig.~\ref{fig:s_agentic_models}b; filled, guided prompt; open, bare prompt; a thin line joins the guided and bare points of the same model (labels on the guided points).}
\label{fig:s_agentic_robust}
\end{figure}

When a gate fails, a fix must be chosen; this is the first place where a language model acts. On 53 benchmark cases built from real gate failures, nine models from three providers chose the correct fix (the reference option; harmful means any other fix, and stopping when stop is not the reference is safe; definitions in Supplementary Table~\ref{tab:si_defs}) in 72--91\% of decisions with rule-table guidance (1.7--16\% harmful) and in 48--79\% without it (17--43\% harmful), whereas the stored rules alone were correct in 96\% with none harmful, and a decision tree in 94\% (main-text Fig.~\Mref{fig:agentic}c, which also scores the same cases with the admissible menus of Section~\ref{sec:si_eligibility}; Supplementary Tables~\ref{tab:si_models} and~\ref{tab:si_A}; Supplementary Fig.~\ref{fig:s_agentic_models}). Model harm came almost entirely from failures that should be stopped and was repeated consistently, so majority voting over repeats~\cite{Wang2023SC} does not remove it (Supplementary Fig.~\ref{fig:s_agentic}; Supplementary Fig.~\ref{fig:s_agentic_robust}a). These results set the order of the recovery node: a stored rule is tried first; the model is consulted only where no rule applies and must choose from a listed menu that a gate re-checks; and the human remains the final step. In the faulted runs, the model completed the same 27 runs as the rules and human, and reduced hand-offs from 34 to 27 by answering the judgment calls as the human had (main-text Fig.~\Mref{fig:agentic}b; Supplementary Table~\ref{tab:si_interventions}). When all 208 menu options were executed in replay, no result outside tolerance was accepted with the gates on, against 20 silent wrong results among the 124 non-reference branches with the gates off (main-text Fig.~\Mref{fig:agentic}d; Supplementary Table~\ref{tab:si_C}). All 20 were stopped by checks that also run in production (Supplementary Table~\ref{tab:si_farr}); in a first replay, which had 22 wrong results, seven of them had been stopped only by the comparison with the stored campaign that exists in replay alone (Section~\ref{sec:si_replay_found}). Executing every recovery choice in replay shows how the three parts combine. With a stored rule first, the reference model for the rest and the gates on, 51.9\% of the 53 recovery cases of the benchmark (built from the faulted runs) completed correctly without the human, and no wrong result was accepted in replay. The same choices with the gates off completed 58.5\% but accepted a wrong result in 7.5\% (main-text Fig.~\Mref{fig:agentic}e; Supplementary Table~\ref{tab:si_stack}). The gates thus contain errors under the tested replay conditions, at the cost of a few correct completions. This does not show that no wrong result can be accepted in live production, nor that the gates reject a physically wrong but numerically well-converged landscape.

\paragraph{Benchmark cases and scoring.} The cases, their menus (Supplementary Table~\ref{tab:subagents}), classes and scoring are defined in main-text Methods, Section~\Mref{sec:eval}, and Supplementary Table~\ref{tab:si_defs}; for the two judgment calls the opposite of the published decision counts as harmful. The two invalid answers were both truncated by the provider. Position bias is the correct rate with the correct option listed first minus listed last.

\paragraph{Results by model and case.} Supplementary Table~\ref{tab:si_models} lists the nine models with their pinned identifiers, and Supplementary Fig.~\ref{fig:s_agentic} shows the outcome of every case for every model. With the guided prompt, a model's errors are concentrated on a few cases: the decoys C33 (all nine models densify the grid), C53 and C26 (four models rebuild the check frames), and the borderline R5 case C11, on which one model excludes the irreps that the published campaign kept. No guided model chose a harmful fix on any of the 29 cases that a stored rule fixes. Without the guidance, most models also fix decoys that they stop with it, and the failures without a rule are the most affected after the decoys; on the held-out cases the bare prompt falls below always stopping for every model. Requiring that $k$ of the ten repeats agree on a fix before acting, and stopping otherwise (a self-consistency vote~\cite{Wang2023SC} used for abstention, in the spirit of selective classification and generation~\cite{Geifman2017,Quach2024}), lowers the share of cases resolved without the human but hardly the share resolved harmfully (Supplementary Fig.~\ref{fig:s_agentic_robust}a): the wrong fixes are chosen consistently. Position bias, a known weakness of language models as multiple-choice selectors~\cite{Zheng2024MCQ}, is small for most models with the guided prompt ($-2$ to $+2$ points) but reaches 7 and 11 points for two Gemini models (Supplementary Fig.~\ref{fig:s_agentic_robust}b); the options were shuffled identically for all models, so it does not bias the comparison. Differences between models are judged from the case-bootstrap intervals (main-text Methods, Section~\Mref{sec:eval}). Against the reference model (guided), the differences in correct rate whose 95\% intervals exclude zero are Gemini~4 Argon, $+12.5$ points ($+2.8$ to $+23.2$), and GPT-6.1 Sol, $+10.0$ ($+0.6$ to $+19.8$); that of GPT-6 Astra, $+8.5$ ($-1.1$ to $+18.3$), includes zero. The harmful rate is lower than the reference's by 4.9 points ($0.6$ to $10.9$) for Claude Opus~5.5 and by 4.7 points ($0.4$ to $10.8$) for GPT-6.1 Sol, GPT-6 Astra and Gemini~3.8 Flash. Without the guidance, the reference model ranks seventh of the nine in correct rate (64.5\%), above Claude Haiku~5.5 and GPT-6 Luna. Costs per call were recorded only for one provider and are not compared.

\paragraph{Rules first, model as fallback.} In IRIS the sub-agent is never used alone: the recovery node tries a stored rule first and consults the model only where no rule applies. Scoring this order on the 53 cases, with the stored rule answering the 29 rule-fixable cases and the guided model answering the other 24 (decoys, used-up rules, failures without a rule and the two judgment calls), gives 74.7--96.2\% correct and 1.7--15.7\% harmful over the nine models (reference model 83.0\% and 6.6\%; GPT-6.1 Sol 96.2\% and 1.9\%), against 96.2\% and 0\% for the rules alone, which stop on every case without a rule. The combined order therefore matches the rules alone in the best case and adds harm on decoys in the others; what it adds is the two judgment calls, which the rules cannot answer and which the best models answer as the human did (both answers change the result: excluding the borderline irrep removes the $\Gamma_2^-$ axis of InMnO$_3$ and the run stops at $G_5$ with no $(Q,P)$ fit). On the recorded failures of this work a rule table is sufficient, and the model's place in recovery is the judgment calls and, by design, failures for which no rule exists yet, which the in-sample benchmark does not contain; the leave-one-system-out scoring below is a first measure of that case.

\paragraph{Transfer across systems.}\label{sec:si_loso} The rule table and the guidance lines of the sub-agent prompts were written from the six campaigns that the cases come from, so the 96.2\% of the rules alone is an in-sample number. Supplementary Table~\ref{tab:rules} records which campaign seeded each rule: of the six rules with a menu option, \texttt{NO\_BASIS} was seeded by HfO$_2$ alone, \texttt{SINGLE\_MODE\_SG} by CsSnI$_3$ alone, \texttt{WALLTIME} and \texttt{NELM} by BiFeO$_3$ alone, \texttt{TRANSLATION} by four campaigns and \texttt{THIN\_GRID} by none (a policy rule of Methods). Each case was therefore re-scored by a decider that had never seen its own system: for a case of system $S$, every rule seeded only by $S$ was removed from the rule table and the guidance sentence that transcribes it from the decision tree, and the retrieval memory excluded every case of $S$ (the variants of one campaign, such as HfO$_2$ T1--T5, count as one system; \texttt{eval\_extra/loso.py}, which wraps the frozen policies without changing them). A removed rule means no rule, hence \texttt{stop}, as in production. Supplementary Table~\ref{tab:si_loso} gives the result. The rules alone fall from 51 to 47 of 53 correct (88.7\%), and all four cases lost (the CsSnI$_3$ single-mode rebuild, the BiFeO$_3$ walltime and SCF-limit resubmissions, and the HfO$_2$ basis search) go to the human, none to a harmful fix; the decision tree loses the same four. Retrieval rises from 43 to 46, because leave-one-out had matched three first-failure cases to a used-up case of their own system with identical detail text. The table also makes visible what the in-sample number hid: 20 of the 53 cases already apply a rule to a system that did not seed it (the BiFeO$_3$ walltime and SCF rules on the other five campaigns, 13 cases; the HfO$_2$ basis rule on four others; the translation rule on CsSnI$_3$ and ErMnO$_3$; the CsSnI$_3$ single-mode rule on ErMnO$_3$), and every one of the 20 is answered correctly, because the rules are keyed by the error code of a deterministic gate, not by the material.

For the models, the same held-out sets remove the guidance sentence seeded by the case's own system from the system prompt (\texttt{eval\_extra/loso\_llm.py}; run on 29 October 2026, with the models of Supplementary Table~\ref{tab:si_models}); the user prompt, the menu and the option order are those of the benchmark. A sentence sits only in its gate's prompt, so seven cases are affected: the three $G_2$ cases of CsSnI$_3$ (C03, C05, C09, without the single-mode sentence), the three $G_4$ cases of BiFeO$_3$ (C20, C21, C22, without the walltime and SCF sentences) and the $G_1$ case of HfO$_2$ T3 (C31, without the basis sentence). Over the nine models and ten repeats (630 decisions), the models chose the correct option without being told in 79.0\% of decisions and a harmful one in 2.5\%, against 85.2\% and 0.2\% with the full guidance on the same seven cases and 74.4\% and 11.4\% with the bare prompt. What was lost is instructive. Every model still widened the basis search for the HfO$_2$ case (89 of 90 decisions) and resubmitted the BiFeO$_3$ walltime failure (82 of 90; the small Anthropic model switched to robust mixing in six repeats), and seven of nine rebuilt the CsSnI$_3$ single-mode frame on the fixed lattice in every repeat. The SCF-limit case divided the models: six kept robust mixing, while the two larger OpenAI models, no longer told that an SCF failure is resubmitted with robust mixing, stopped in 17 of 20 repeats, which is safe. The 16 harmful decisions come from the two small models (11 and 2) and the small OpenAI model (3), which, without the sentence, chose \texttt{exclude\_borderline} or \texttt{anchor\_only} for a single-mode mismatch or robust mixing for a walltime failure. The rule-less failure is thus answered correctly by most models most of the time, and the harm, where it occurs, is concentrated in the smallest models and is caught by the next gate in the option replay (Supplementary Table~\ref{tab:si_C}). This analysis separates the rules from the campaigns that seeded them; it does not separate the cases from their authors, and the fault taxonomy is still the one of this work (Supplementary Section~\ref{sec:si_limits}).

\begin{table}[!htbp]
\centering\scriptsize
\caption{Transfer across systems (leave-one-system-out). Top: the three deterministic deciders on the 53 cases, in sample and with every rule seeded only by the case's own campaign removed (decision tree: the guidance sentence that transcribes it; retrieval: the memory without the case's own system). Bottom: the nine models on the seven cases whose system prompt loses a guidance sentence (C03, C05, C09, C20, C21, C22, C31; ten repeats each, 70 decisions per model), with the full guided prompt, with the sentence seeded by the case's own system removed, and with the bare prompt. Correct / safe / harmful as in Supplementary Table~\ref{tab:si_defs}. Script \texttt{eval\_extra/loso.py}, \texttt{eval\_extra/loso\_llm.py}.}
\label{tab:si_loso}
\begin{tabular}{@{}lrrrrrr@{}}
\toprule
& \multicolumn{3}{c}{in sample} & \multicolumn{3}{c}{held-out system}\\
\cmidrule(lr){2-4}\cmidrule(lr){5-7}
Decider (53 cases) & correct & safe & harmful & correct & safe & harmful\\
\midrule
Rules alone & 51 (96.2\%) & 2 & 0 & 47 (88.7\%) & 6 & 0\\
Decision tree & 50 (94.3\%) & 2 & 1 & 46 (86.8\%) & 6 & 1\\
Retrieval (nearest case) & 43 (81.1\%) & 8 & 2 & 46 (86.8\%) & 5 & 2\\
\midrule
& \multicolumn{2}{c}{guided} & \multicolumn{2}{c}{own sentence removed} & \multicolumn{2}{c}{bare}\\
\cmidrule(lr){2-3}\cmidrule(lr){4-5}\cmidrule(lr){6-7}
Model (7 cases, 70 decisions) & correct & harmful & correct & harmful & correct & harmful\\
\midrule
Claude Haiku 5.5 & 60 & 0 & 45 & 11 & 37 & 23\\
Claude Sonnet 5.5 & 60 & 0 & 60 & 0 & 46 & 14\\
Claude Opus 5.5 & 59 & 0 & 59 & 0 & 60 & 0\\
GPT-6 Luna & 59 & 1 & 55 & 3 & 44 & 17\\
GPT-6.1 Sol & 59 & 0 & 51 & 0 & 51 & 9\\
GPT-6 Astra & 60 & 0 & 52 & 0 & 60 & 0\\
Gemini 3.5 Flash-Lite & 60 & 0 & 56 & 2 & 52 & 8\\
Gemini 3.8 Flash & 60 & 0 & 60 & 0 & 59 & 1\\
Gemini 4 Argon & 60 & 0 & 60 & 0 & 60 & 0\\
All nine (630) & 537 (85.2\%) & 1 (0.2\%) & 498 (79.0\%) & 16 (2.5\%) & 469 (74.4\%) & 72 (11.4\%)\\
\bottomrule
\end{tabular}
\end{table}

\paragraph{Why the models choose harmful fixes on decoys, and three remedies.}\label{sec:si_decoys} Almost every harmful decision in the benchmark is on one of the 11 decoys, failures whose correct answer is to stop because no listed fix addresses them. Reading the choices case by case (Supplementary Table~\ref{tab:si_decoys}) shows three patterns rather than random error. First, on the Ca$_3$Ti$_2$O$_7$ \texttt{COVERAGE} failure (C33: the grid minimum lies five steps from the grid edge, so the surface may continue downhill outside), 89 of 90 decisions across the nine models chose \texttt{densify}. The guidance line of the $G_5$ prompt is the cause: it says that a coverage failure means the grid is too thin, which is true of \texttt{THIN\_GRID} but not of a minimum near the edge, which needs a wider grid, an option the menu does not offer; the only correct answer is stop, and the budget check refuses the densified grid in the option replay. Second, on the two HfO$_2$ \texttt{PATH\_LATTICE} failures (C26, C53: the child cell carries a rigid rotation of about 5$^\circ$ relative to the parent, so the interpolated cells are not a pure strain path), 82 of 180 decisions chose \texttt{rebuild\_fixed\_lattice}: a lattice-sounding fix for a lattice-sounding failure that the $G_2$ guidance never mentions, since the sentence that names the fixed-lattice rebuild is about the single-mode space-group check. Third, on the tolerance failures of $G_5$ (\texttt{CLOSURE}, \texttt{CROSS\_SETTING}, \texttt{RESIDUALS}, \texttt{MIRROR}, \texttt{CV\_ERROR}; C15, C16, C24, C48, C49, C50, C51, C52), 73 of 630 decisions, almost all by the three small models and GPT-6 Luna, chose \texttt{accept\_secondary\_hold} when the detail showed a small overshoot (a closure gap of 0.06 against a tolerance of 0.01~meV per formula unit, an RMS of 0.098 against 0.01), reading ``accept'' as ``tolerate'' although that option exists for the \texttt{SECONDARY\_COUPLED} judgment call alone. The flagship models of OpenAI and Google and Claude Opus~5.5 made none of the second and third errors; all nine made the first, because the prompt told them to.

Each pattern has a one-sentence remedy in the guidance line, and none requires a new option or a change to the gates: for $G_2$, ``a child cell rotated relative to the parent (\texttt{PATH\_LATTICE}) is an input-structure problem that no listed fix repairs: stop''; for $G_5$, ``a minimum near the grid edge (\texttt{COVERAGE}) is not fixed by densifying: stop'' and ``\texttt{accept\_secondary\_hold} applies only to \texttt{SECONDARY\_COUPLED}; a tolerance exceeded by any margin is a failure, not a near miss: stop''. The general rule they instantiate is that every option on a menu must say which failure codes it is for, and that a guidance sentence must not describe two failures with one word. These sentences were written after the benchmark results were seen and tested on the same cases, so what follows is a post-hoc prompt-repair experiment, evidence that the failures were caused by the wording and not independent evidence that the revised prompts generalise (\texttt{eval\_extra/decoy\_remedy.py}, nine models, five repeats, 2{,}385 decisions, run on 29 October 2026). With the three sentences the harmful decisions on the decoys fall from 253 of 990 (25.6\%) to 3 of 495 (0.6\%): eight of the nine models make none, and the three that remain are Claude Haiku~5.5 densifying on a \texttt{CV\_ERROR} and a \texttt{CLOSURE} failure. On the non-decoys the models perform broadly comparably, 84.9\% against 83.8\% correct and 0.8\% against 0.3\% harmful; the remaining non-decoy harm is the R5 judgment call, where Gemini~4 Argon excludes the borderline irrep as it did before. The sub-agent's harmful rate, which the benchmark reported as 1.7--16\%, is therefore, on these cases, a property of three sentences in the prompts rather than of the models. The evaluated prompts remain the frozen ones; the three sentences are recorded with the test (\texttt{results/DECOY/remedy\_prompts.json}).

\paragraph{Eligibility conditions: the code restricts what is admissible, the model selects among it.}\label{sec:si_eligibility} A prompt sentence is the weaker fix, because it asks the model to read a rule it could ignore. The stronger one is to make the three problematic actions ineligible in code: each option of the recovery menus now names the error codes it is admissible for (\texttt{for:} in \texttt{recovery.yaml}, taken from the rule table and Methods: \texttt{densify} for \texttt{THIN\_GRID} only, \texttt{accept\_secondary\_hold} for \texttt{SECONDARY\_COUPLED} only, \texttt{rebuild\_fixed\_lattice} for \texttt{SINGLE\_MODE\_SG} only, and so on for every non-stop option), and the recovery node filters the menu by the failing code before any model or human sees it. A failure whose admissible menu is \texttt{stop} alone is decided in code and no model is asked. On the benchmark this is the case for 17 of the 53 cases, the eleven decoys, the five failures without a rule and the over-budget grid; the other 36 go to the model with a menu that contains only admissible options, and no case loses its correct option to the filter (checked before scoring). Rerun with the frozen prompts and the filtered menus (\texttt{eval\_extra/eligibility.py}, nine models, five repeats, 2{,}385 decisions of which 1{,}620 reach a model, run on 29 October 2026), the nine models are correct in 85.3--97.0\% of decisions (90.4\% pooled, against 72--91\% with the unrestricted menus) and harmful in 0.17\% (4 of 2{,}385; 0 for seven of the nine models), the four being the R5 judgment call, where two models exclude the borderline irrep, an admissible answer that the stored decision labels harmful. One correction was made after seeing the run and is reported as such: a first version of the conditions admitted \texttt{flag\_and\_continue} for \texttt{SCF\_LIMITED}, and 23 of the 27 harmful decisions of that run were that option on the one SCF-limited case; by the rule table an SCF-limited frame keeps its energy (\texttt{NELM\_AFTER\_ROBUST}), so there is nothing to continue without, and the option is now admissible for \texttt{WALLTIME} and \texttt{NELM} only, which makes that case code-decided (\texttt{results/ELIGIBILITY/}). The principle this implements is the one the gates already follow for numbers: the code decides what is scientifically admissible, and the language model chooses among admissible actions, so that a wrong choice can at most be a safe one. The option replay of Supplementary Table~\ref{tab:si_C}, which executed every option including the now-inadmissible ones, is what measured their consequences and justified the conditions; it is unaffected by them.

\begin{table}[!htbp]
\centering\scriptsize
\caption{The eleven decoys of the benchmark: what the nine models chose with the benchmark prompt (ten repeats each, 90 decisions per case; harmful choices and the option chosen), and with the three remedy sentences added to the guidance line (five repeats, 45 decisions per case; a post-hoc test on the same cases, Section~\ref{sec:si_decoys}). With the eligibility conditions of Section~\ref{sec:si_eligibility} every decoy's admissible menu is stop alone, so all eleven are decided in code with no model call (\texttt{results/DECOY/}, \texttt{results/ELIGIBILITY/}).}
\label{tab:si_decoys}
\begin{tabularx}{\linewidth}{@{}L{0.9cm}L{1.3cm}L{2.0cm}XL{2.3cm}L{1.7cm}@{}}
\toprule
Case & System & Gate, code & What the detail says & benchmark prompt: harmful (choice) & remedy sentences: harmful\\
\midrule
C33 & Ca$_3$Ti$_2$O$_7$ & $G_5$ COVERAGE & grid minimum 5 steps from the edge & 89/90 (densify 89) & 0/45\\
C26 & HfO$_2$ T5 & $G_2$ PATH\_LATTICE & child cell rotated 5.0$^\circ$ relative to the parent & 42/90 (rebuild\_\allowbreak{}fixed\_\allowbreak{}lattice 42) & 0/45\\
C53 & HfO$_2$ T3 & $G_2$ PATH\_LATTICE & child cell rotated 4.9$^\circ$ & 40/90 (rebuild\_\allowbreak{}fixed\_\allowbreak{}lattice 40) & 0/45\\
C52 & HfO$_2$ T3 & $G_5$ CLOSURE & end-point gap 0.023 against 0.01~meV/f.u. & 21/90 (accept 14, densify 7) & 0/45\\
C15 & HfO$_2$ T1 & $G_5$ CLOSURE & end-point gap 0.060 against 0.01 & 20/90 (accept 18, densify 2) & 1/45 (densify, Haiku)\\
C24 & ErMnO$_3$ & $G_5$ RESIDUALS & RMS 5.99 against 5.0~meV per cell & 17/90 (accept 17) & 0/45\\
C16 & HfO$_2$ T1 & $G_5$ CROSS\_\allowbreak{}SETTING & one structure differs between settings by 0.04~meV/f.u. & 9/90 (accept 9) & 0/45\\
C48 & Ca$_3$Ti$_2$O$_7$ & $G_5$ MIRROR & mirror asymmetry 0.42~meV/f.u. & 7/90 (accept 7) & 0/45\\
C50 & InMnO$_3$ & $G_5$ CV\_ERROR & CV RMS 4.26 against 0.01~meV per cell & 5/90 (accept 5) & 2/45 (densify, Haiku)\\
C51 & BiFeO$_3$ & $G_5$ RESIDUALS & RMS 0.49 against 0.1~meV/f.u. & 3/90 (accept 3) & 0/45\\
C49 & CsSnI$_3$ & $G_5$ RESIDUALS & RMS 0.098 against 0.01~meV/f.u. & 0/90 & 0/45\\
\midrule
All decoys & & & & 253/990 (25.6\%) & 3/495 (0.6\%)\\
Non-decoys (42 cases) & & & & correct 84.9\%, harmful 0.8\% & correct 83.8\%, harmful 0.3\%\\
\bottomrule
\end{tabularx}
\end{table}

\paragraph{The full recovery stack.} Main-text Fig.~\Mref{fig:agentic}e combines this order with the executed options of Supplementary Table~\ref{tab:si_C}. For each of the 53 recovery cases of the benchmark (built from the faulted runs; the judgment calls included), each recovery choice of the policy (the stored rule where one applies, otherwise the model with the guided prompt, in each of its ten repeats) was executed in replay, using the outcome of that menu option with all gates on and, for the 124 branches rerun with the gates switched off after the fix, without them; an option that was not rerun without gates is counted with its gated outcome (Supplementary Table~\ref{tab:si_stack}). With a stored rule first, the reference model for the rest and the gates on, 51.9\% of the cases completed correctly without the human, against 49.1\% for the stored rules alone, and none crashed or accepted a wrong result in replay; none was stopped only by a replay-only check (in a first replay 3.8\%, the \texttt{flag\_and\_continue} choices after a job failure, had been; they are now stopped by \texttt{MISSING\_FRAMES} at $G_4$). The same choices with the gates off completed 58.5\% but accepted a wrong result in 7.5\% and crashed in 1.9\%: switching the gates off moves 16.0 points out of the cases stopped by a gate or check (21.7\% to 5.7\%) into correct completions ($+6.6$), crashes ($+1.9$) and accepted wrong results ($+7.5$), while the stops for the human are unchanged (26.4\%). With GPT-6.1 Sol as the model, 52.8\% completed correctly with the gates on and 9.4\% were stopped by a gate or check, in line with its lower harmful rate in the benchmark (Supplementary Table~\ref{tab:si_models}). As in every option replay, a fix that would change the inputs of a calculation is credited with the stored result (Supplementary Section~\ref{sec:si_limits}).

\begin{table}[!htbp]
\centering\footnotesize\setlength{\tabcolsep}{3.5pt}
\caption{\textbf{The full recovery stack on the 53 recovery cases of the benchmark} (main-text Fig.~\Mref{fig:agentic}e). Each recovery choice of the policy (the stored rule where one applies, otherwise the model; guided prompt, ten repeats) is executed in replay with the outcome of that menu option with the gates on (Supplementary Table~\ref{tab:si_C}) or with the gates switched off after the fix (the 124 branches whose option is neither the reference nor stop); an option that was not rerun without gates is counted with its gated outcome. Percent of the 53 recovery cases (built from the faulted runs), averaged over the ten repeats where a model is used. Stopped for the human includes the limit faults, whose correct answer is to stop; stopped by a gate or check includes fixes refused or rejected after they were applied. $^{*}$Reference model.}
\label{tab:si_stack}
\begin{tabular}{@{}llrrrrrr@{}}
\toprule
Policy & Gates & \shortstack{completed\\correctly} & \shortstack{stopped for\\the human} & \shortstack{stopped by a\\production check} & \shortstack{stopped only by a\\replay-only check} & crash & \shortstack{wrong result\\accepted}\\
\midrule
Stored rules only & on & 49.1 & 45.3 & 5.7 & 0 & 0 & 0\\
Rules, then Claude Sonnet~5.5$^{*}$ & off & 58.5 & 26.4 & 5.7 & -- & 1.9 & 7.5\\
Rules, then Claude Sonnet~5.5$^{*}$ & on & 51.9 & 26.4 & 21.7 & 0 & 0 & 0\\
Rules, then GPT-6.1 Sol & on & 52.8 & 37.7 & 9.4 & 0 & 0 & 0\\
\bottomrule
\end{tabular}
\end{table}

\paragraph{False acceptances and false rejections.} The option replays can also be read as a test of the gates as a classifier, with the outcome of the same branch with the gates switched off after the fix as the ground truth (Supplementary Table~\ref{tab:si_farr}). A false acceptance is a branch accepted with the gates on whose result with the gates off is wrong; a false rejection is a branch stopped with the gates on whose result with the gates off is correct. On the 75 non-reference branches of errors, the gates accepted none of the 20 wrong results, and all 20 were stopped by checks that also run in production. In a first replay, which had 22 wrong results (the two InMnO$_3$ branches that excluded the borderline irrep then completed unchanged and now stop, Section~\ref{sec:si_replay_found}), 7 of the 22 had been stopped only by the comparison with the stored campaign that exists in replay alone, a false-acceptance rate of 7/22 with production checks alone; the \texttt{MISSING\_FRAMES} check brought it to 0/20 on this set. They stopped 11 of the 36 branches whose result was correct with the gates off; all 11 are faults that are harmless in replay only because their frames are linked to the stored energies (the CsSnI$_3$ and ErMnO$_3$ frames built on the strained cell, 8 branches, and the HfO$_2$ \texttt{INCAR} fault, 3), so in production they would have been correct rejections. On limit faults, the gates stopped all 33 branches whose result was correct, as the tightened limit requires; these are counted separately, because the limit, not the result, defines the correct outcome. Replay therefore bounds false acceptances from above only through the checks it shares with production, and it cannot measure false rejections in production; a production-relevant fault set is needed for both (Supplementary Section~\ref{sec:si_limits}).

\begin{table}[!htbp]
\centering\small
\caption{\textbf{The gates as a classifier on the 124 executed non-reference options} (Supplementary Table~\ref{tab:si_C}). Ground truth: the outcome of the same branch with the gates switched off after the fix (wrong: a wrong result, or a result from fewer frames, accepted; correct: numbers within tolerance or equivalent settings; no result: stopped by an error or without data). Gated: outcome with all gates on; replay-only: stopped only by the comparison with the stored campaign.}
\label{tab:si_farr}
\begin{tabular}{@{}llrrrr@{}}
\toprule
Fault kind & Without gates & Branches & \shortstack{accepted\\with gates} & \shortstack{stopped by a\\production check} & \shortstack{stopped only by a\\replay-only check}\\
\midrule
Error & wrong & 20 & 0 & 20 & 0\\
Error & correct & 36 & 25 & 11$^{a}$ & 0\\
Error & no result & 19 & 0 & 19 & 0\\
Limit & correct & 33 & 0 & 33$^{b}$ & 0\\
Limit & no result & 11 & 0 & 11 & 0\\
Judgment (R5, D4) & no result & 5 & 0 & 5$^{c}$ & 0\\
\midrule
All & & 124 & 25 & 99 & 0\\
\bottomrule
\end{tabular}
\begin{flushleft}\footnotesize $^{a}$Harmless only because replay links stored energies (8 frames built on the strained cell, 3 \texttt{INCAR} fault); correct rejections in production. $^{c}$Two are the InMnO$_3$ anchor-only answers to R5, stopped at $G_3$ because the anchor frame is not in the stored plane; in production that frame would be computed, so the stop is counted with the production checks (the frame request, not the matching, is the check). $^{b}$Required by the tightened limit.\end{flushleft}
\end{table}

\paragraph{What the replay found and changed.}\label{sec:si_replay_found}
The option replay was first run with the gates as they stood at the end of the campaigns. In that run seven of the 22 wrong results, all the \texttt{flag\_and\_continue} option after a failed job part on a large grid (CsSnI$_3$, four cases; BiFeO$_3$; ErMnO$_3$; CsSnI$_3$ with D4), were stopped only by the $G_6$ comparison with the stored campaign, which exists in replay alone: $f_4$ had marked the failed frames as accepted, $G_4$ had passed them, and the coverage test of $G_5$ (points per coefficient, position of the minimum) does not notice 15 missing nodes out of 545. In production the same branches would have completed with a fit on fewer frames than planned and nothing flagged. The gap has one cause, and one check closes it: a frame dropped from the planned grid now fails $G_4$ with \texttt{MISSING\_FRAMES} unless the run configuration carries a recorded human authorisation (\texttt{policy.fit.allow\_missing\_frames}), and the same test runs again at $G_5$ after the fit, counting every frame absent from it whatever removed it, so that it holds when $G_4$ is ablated; an authorised smaller grid is stated in every report. The first form of the check counted accepted flags only and was refined when the BiFeO$_3$ regression replay failed on it: the 20 SCF-limited corner frames of that campaign keep their energies and drop only their gradients under the stated policy, and are not missing frames. With the check in place, E1, E2, the 208 option branches, the 124 branches without gates and the gate ablation were re-run from scratch (the frozen evaluation inputs unchanged). Every number outside the affected branches is identical to the first run; the 24 \texttt{flag\_and\_continue} branches stop at $G_4$, Supplementary Table~\ref{tab:si_farr} has no replay-only column left among the errors, and in the ablation the nine faults stopped only at $G_6$ are stopped at $G_5$ (Supplementary Table~\ref{tab:si_ablation}). We report this as a case of the evaluation doing what it is for: the replay-only comparison is a reference answer the workflow must not depend on, and the ablation showed where it did.

The second change concerns the borderline selection R5. In the first replay the recorded answer (include, exclude or anchor only) was not read beyond letting the run continue, because the grid axes were taken from the configuration, which already contained $\Gamma_2^-$; the InMnO$_3$ branches that excluded it therefore completed with numbers identical to the reference, which the review of the evaluation rightly counted as no evidence. The answer is now applied inside the selection rules, where $f_2$ reads the active set, $G_2$ checks it (\texttt{R5\_NOT\_APPLIED} if the grid does not follow the recorded answer) and $f_6$ re-derives it: include confirms the rules' selection (the borderline active irreps stay axes, the sub-threshold ones stay out); exclude removes the borderline active irreps from the axes and holds them at zero, failing with \texttt{R5\_EXCLUDE\_BREAKS\_H} if the remaining set no longer generates the subgroup; anchor only does the same and adds one frame per removed irrep at the child amplitudes, whose energy relative to the nearest grid point is reported with the fit. For InMnO$_3$, $K_3$ alone generates $P6_3cm$, so excluding $\Gamma_2^-$ builds a one-axis grid on which the $(Q,P)$ model cannot be fitted, and the run stops at $G_5$ (\texttt{NO\_FIT}); the anchor-only option asks for the frame with both irreps at their child amplitudes, which the stored plane does not contain, so replay stops at $G_3$ and production would compute it. With this change the option replay was re-run once more: the four InMnO$_3$ R5 branches moved from completing unchanged to stopped, and no other branch changed; the clean replays, the faulted runs and the gate ablation are identical, since the published answer is include.

\paragraph{Faulted runs with the model on.} Supplementary Table~\ref{tab:si_interventions} gives the 53 faulted runs by system, with the model off and with the reference model. The completed and stopped runs are the same in both conditions, with identical numbers. With the model on, the sub-agent answered the four published decisions met in completed runs (R5 for InMnO$_3$, three times; D4 for CsSnI$_3$) as the human had, and in the stopped runs it answered R5 three more times and tried listed fixes before choosing stop (38 decisions in all, the three R5 answers included), which hands the run to the human without a further question being counted. The human was asked three times, for a missing frame, cancelled jobs and an exhausted retry budget, which go to the human by design ($G_3$ offers only stop, and an exhausted retry budget always goes to the human). Counting every hand-off to the human, whether or not a question is asked, gives 34 hand-offs with the model off and 27 with it on.

\begin{table}[!htbp]
\centering\scriptsize
\caption{The 53 injected faults by system and kind, with all gates on (gated) and with no gates (ungated). Errors corrupt a result if they pass; limit faults make a threshold stricter, or a search narrower, on correct data. Gated: completed (numbers identical to the fault-free replay) or stopped with the human; no gated run ended with a corrupted result. Ungated: silent (numbers differ from the fault-free replay and nothing is flagged), crash (a node raises an error) or harmless (numbers identical). The HfO$_2$ faults are spread over paths T1--T5 and the InMnO$_3$ faults include the joint fit.}
\label{tab:si_faults}
\begin{tabular}{@{}lrrrrrrrrrr@{}}
\toprule
& \multicolumn{6}{c}{errors} & \multicolumn{4}{c}{limit faults}\\
\cmidrule(lr){2-7}\cmidrule(lr){8-11}
& & \multicolumn{2}{c}{gated} & \multicolumn{3}{c}{ungated} & & \multicolumn{2}{c}{gated} & ungated\\
\cmidrule(lr){3-4}\cmidrule(lr){5-7}\cmidrule(lr){9-10}\cmidrule(lr){11-11}
System & $n$ & completed & stopped & silent & crash & harmless & $n$ & completed & stopped & crash / harmless\\
\midrule
CsSnI$_3$ & 9 & 5 & 4 & 5 & 2 & 2 & 4 & 2 & 2 & 0 / 4\\
Ca$_3$Ti$_2$O$_7$ & 4 & 3 & 1 & 3 & 1 & 0 & 3 & 1 & 2 & 0 / 3\\
InMnO$_3$ & 3 & 3 & 0 & 0 & 3 & 0 & 3 & 0 & 3 & 3 / 0\\
ErMnO$_3$ & 5 & 4 & 1 & 3 & 1 & 1 & 3 & 1 & 2 & 0 / 3\\
BiFeO$_3$ & 2 & 2 & 0 & 2 & 0 & 0 & 3 & 1 & 2 & 0 / 3\\
HfO$_2$ & 7 & 4 & 3 & 1 & 5 & 1 & 7 & 1 & 6 & 0 / 7\\
\midrule
All & 30 & 21 & 9 & 14 & 12 & 4 & 23 & 6 & 17 & 3 / 20\\
\bottomrule
\end{tabular}
\end{table}

\begin{table}[!htbp]
\centering\scriptsize
\caption{One gate removed at a time: for each fault, the gate that caught it in the gated run is removed and the run is repeated. Caught later: the run stopped at a later gate (the gate that then failed is listed). Extra runs: DFT runs on results that were then rejected, minus those of the gated run, from the submission records of $f_3$ (net of the resubmissions the gated run makes). For limit faults the data are correct, so removing the gate is harmless by construction, and the extra-runs count is negative because the gated run stops before using them. Replay-only: stopped by a check that a production run does not make, namely $G_3$ finding no stored structure for a wrongly built frame, or $G_6$ comparing with the stored campaign energies, which in replay it also uses for its second route; in production these faults would not be stopped there. $^{a}$The fourth, F036, crashed at $f_3$ because removing $G_2$ also removes the R5 decision of InMnO$_3$, so that no grid is built (an artefact of the ablation). $^{b}$Fifteen of the 21 $G_5$ catches are the \texttt{MISSING\_FRAMES} check (a fit on fewer frames than planned), five are the path analysis failing on a missing point, and one (F019) is the HfO$_2$ cross-setting check, which uses the energies of the other paths and so needs them computed in the same campaign. A further 54 runs remove, for each of the nine seed faults, a gate that did not catch it; none changed the outcome.}
\label{tab:si_ablation}
\begin{tabular}{@{}llrrrrll@{}}
\toprule
Kind & Gate removed & faults & caught later & of which replay-only & not caught (harmless) & caught by & extra runs\\
\midrule
error & $G_2$ & 6 & 4 & 3 & 2 & $G_3$ (3), crash at $f_3$ (1)$^{a}$ & 73\\
error & $G_3$ & 2 & 2 & 0 & 0 & $G_4$ (1), $G_5$ (1) & 0\\
error & $G_4$ & 22 & 21 & 0 & 1 & $G_5$ (21)$^{b}$ & 27{,}065\\
limit & $G_1$ & 6 & 0 & 0 & 6 & -- & 0\\
limit & $G_2$ & 2 & 0 & 0 & 2 & -- & 0\\
limit & $G_4$ & 2 & 0 & 0 & 2 & -- & $-10{,}317$\\
limit & $G_5$ & 13 & 0 & 0 & 13 & -- & $-16{,}327$\\
\bottomrule
\end{tabular}
\end{table}

\begin{table}[!htbp]
\centering\scriptsize
\caption{Sub-agent benchmark: nine models on the 53 cases, ten repeats per case and prompt variant (9{,}540 calls). Rates in \% of decisions; 95\% confidence intervals from a bootstrap over cases (repeats of a case resampled together). Guided: prompt with the one-line guidance of the rule table; bare: without it. Held-out: the 16 cases added after the prompts were final. Consistency: mean share of the ten repeats that give the modal answer. Position bias: correct rate with the correct option listed first minus listed last (points). Latency: median per call. $^{*}$Reference model. Policies without a model: always stop 41.5 correct, 0 harmful (held-out 62.5, 0); first option 26.4, 52.3; uniform random 28.0, 51.4.}
\label{tab:si_models}
\setlength{\tabcolsep}{3pt}
\resizebox{\linewidth}{!}{%
\begin{tabular}{@{}llrrrrrrrrrr@{}}
\toprule
& & \multicolumn{4}{c}{all 53 cases} & \multicolumn{2}{c}{held-out, correct} & \multicolumn{2}{c}{consistency} & pos.\ bias & latency\\
\cmidrule(lr){3-6}\cmidrule(lr){7-8}\cmidrule(lr){9-10}
Model & Identifier & \multicolumn{2}{c}{guided: correct (CI), harmful} & \multicolumn{2}{c}{bare: correct, harmful} & guided & bare & guided & bare & guided & guided (s)\\
\midrule
Claude Haiku 5.5 & \texttt{claude-haiku-5-5} & 79.6 (70--88) & 7.7 & 54.3 & 41.9 & 80.6 & 27.5 & 91.9 & 88.3 & +1.8 & 1.5\\
Claude Sonnet 5.5$^{*}$~\cite{ModelSonnet55} & \texttt{claude-sonnet-5-5} & 78.5 (67--88) & 6.6 & 64.5 & 28.9 & 78.1 & 46.9 & 96.6 & 92.3 & +1.2 & 2.0\\
Claude Opus 5.5~\cite{ModelOpus55} & \texttt{claude-opus-5-5} & 81.3 (71--90) & 1.7 & 71.5 & 16.6 & 93.1 & 52.5 & 97.9 & 95.7 & -2.2 & 3.0\\
GPT-6 Luna & \texttt{gpt-6-luna} & 72.5 (62--82) & 15.7 & 47.9 & 43.0 & 63.7 & 24.4 & 92.5 & 88.7 & +1.7 & 2.1\\
GPT-6.1 Sol~\cite{ModelGPT61Sol} & \texttt{gpt-6.1-sol} & 88.5 (79--96) & 1.9 & 71.3 & 22.3 & 93.1 & 48.8 & 99.8 & 96.8 & +1.9 & 2.8\\
GPT-6 Astra & \texttt{gpt-6-astra} & 87.0 (78--95) & 1.9 & 79.2 & 18.5 & 93.8 & 50.0 & 98.3 & 98.5 & -0.3 & 2.5\\
Gemini 3.5 Flash-Lite & \texttt{gemini-3.5-flash-lite} & 83.8 (75--91) & 9.4 & 69.4 & 24.9 & 72.5 & 42.5 & 93.0 & 88.9 & +11.1 & 0.8\\
Gemini 3.8 Flash & \texttt{gemini-3.8-flash} & 82.6 (72--91) & 1.9 & 70.6 & 20.6 & 93.8 & 49.4 & 97.0 & 97.2 & +1.3 & 3.5\\
Gemini 4 Argon~\cite{ModelGemini4Argon} & \texttt{gemini-4-argon} & 90.9 (84--97) & 6.4 & 67.2 & 26.4 & 90.0 & 44.4 & 96.4 & 95.5 & +6.7 & 3.7\\
\bottomrule
\end{tabular}}
\end{table}

\begin{table}[!htbp]
\centering\scriptsize
\caption{The 53 faulted runs by system, with the language model off (rules and human) and with the reference model as sub-agent. Human questions are split into the published decisions (rule R5 for InMnO$_3$, D4 for CsSnI$_3$) and those caused by the fault. With the model off, a stopped run counts one human question; with the model on, a run the sub-agent stops is handed to the human without a further question. Completed runs give numbers identical to the fault-free replay in both conditions.}
\label{tab:si_interventions}
\begin{tabular}{@{}lrrrrrrrrrr@{}}
\toprule
& & \multicolumn{5}{c}{model off} & \multicolumn{4}{c}{reference model on}\\
\cmidrule(lr){3-7}\cmidrule(lr){8-11}
System & runs & completed & rule fixes & \multicolumn{2}{c}{human: published, fault} & per run & completed & rule fixes & sub-agent & human\\
\midrule
CsSnI$_3$ & 13 & 7 & 9 & 1 & 7 & 0.61 & 7 & 9 & 9 & 3\\
Ca$_3$Ti$_2$O$_7$ & 7 & 4 & 5 & 0 & 3 & 0.43 & 4 & 6$^{a}$ & 5 & 0\\
InMnO$_3$ & 6 & 3 & 4 & 6 & 3 & 1.50 & 3 & 4 & 9 & 0\\
ErMnO$_3$ & 8 & 5 & 7 & 0 & 3 & 0.38 & 5 & 7 & 4 & 0\\
BiFeO$_3$ & 5 & 3 & 3 & 0 & 2 & 0.40 & 3 & 3 & 4 & 0\\
HfO$_2$ & 14 & 5 & 6 & 0 & 9 & 0.64 & 5 & 6 & 11 & 0\\
\midrule
All & 53 & 27 & 34 & 7 & 27 & 0.64 & 27 & 35 & 42 & 3\\
\bottomrule
\end{tabular}
\begin{flushleft}\footnotesize $^{a}$The additional rule fix with the model on is in F044, an SCF failure that returns after robust mixing: with the model on that run had 2 rule fixes and 2 sub-agent decisions before it stopped, against 1 rule fix with the model off.\end{flushleft}
\end{table}

\paragraph{Deterministic baselines, executed options and intake.} Supplementary Table~\ref{tab:si_A} compares the recovery policies without a language model. The rules alone and the decision tree miss only the two judgment calls, at which they stop; the tree's one harmful choice is the decoy C33, where it densifies the grid like every model, an option that the budget check then refuses (Supplementary Table~\ref{tab:si_C}). Retrieval misses the cases in which the same failure has returned after its fix, because it does not see the run history. Supplementary Table~\ref{tab:si_C} gives the consequence of every menu option. With all gates on, no branch completed with a wrong result; the wrong options were refused by the next gate or stopped by a later one. With the gates switched off after the fix, 20 of 124 branches completed with a wrong result and nothing flagged (22 in a first replay, before the R5 answer was applied; Section~\ref{sec:si_replay_found}). The replays also show that four kinds of label were too strict or wrong: a change of SCF mixing only (13 branches labelled harmful), an unchanged resubmission after an SCF failure that completes only because the scripted failures had run out (9), densifying a grid that the budget then refuses (3, labelled correct) and, in the first replay, excluding the borderline irrep in InMnO$_3$, which then had no effect because the answer to R5 was not applied (1; it now removes the axis and the run stops at $G_5$, consistent with the label); the frozen labels were not changed. Supplementary Table~\ref{tab:si_E} gives the intake results by model, run and request type for the four models, the rule-based parser (Section~\ref{sec:si_parser}) and the keyword lookup (every request, with its outcome, in Supplementary Table~\ref{tab:si_requests}; prompts in Supplementary Section~\ref{sec:si_prompts}), Supplementary Table~\ref{tab:si_loso} the transfer of the rules, the tree, retrieval and the models across systems (Section~\ref{sec:si_loso}), and Supplementary Fig.~\ref{fig:s_agentic_models} compares the nine models on the recovery cases with and without the guidance line.

\begin{table}[!htbp]
\centering\scriptsize
\caption{Deterministic recovery policies with the knowledge of the rule table (Methods, Section~\Mref{sec:eval}). Top: decisions on the 53 benchmark cases (one run each; rates in \%); the language models are in Supplementary Table~\ref{tab:si_models}. All four policies stop at the two judgment calls (C10, D4; C11, R5), which counts as safe. Bottom: the 53 faulted runs with each policy as sub-agent, beside the rules with a human and the reference model. Completed runs give numbers identical to the fault-free replay. Hand-offs count every question to the human and every run a sub-agent stops; extra runs are DFT runs beyond the fault-free replay, from the submission records of $f_3$ (replay count). Five retrieval runs that completed only because the scripted job failures had run out after a resubmission are counted as not completed (32 and 31 otherwise).}
\label{tab:si_A}
\begin{tabular}{@{}lrrrr@{}}
\toprule
Policy (53 cases) & correct & safe & harmful & held-out correct\\
\midrule
Rules only & 96.2 & 3.8 & 0.0 & 100.0\\
Decision tree & 94.3 & 3.8 & 1.9 & 100.0\\
Retrieval (leave-one-out) & 81.1 & 15.1 & 3.8 & 93.8\\
Retrieval (seed memory) & 83.0 & 13.2 & 3.8 & 93.8\\
\midrule
Policy (53 faulted runs) & completed & without hand-off & hand-offs & extra runs\\
\midrule
Rules + human & 27 & 22 & 34 & 770\\
Decision tree & 23 & 22 & 31 & 690\\
Retrieval (leave-one-out) & 27 & 26 & 22 & 940\\
Reference model (Claude Sonnet 5.5) & 27 & 26 & 27 & 770\\
\bottomrule
\end{tabular}
\end{table}

\begin{table}[!htbp]
\centering\scriptsize
\caption{Consequence of every menu option, executed from the checkpoint of the real failure (208 branches; Methods, Section~\Mref{sec:eval}), by the frozen label of the option. $^{a}$Identical energies assumed by replay: mixing settings do not change the converged energy except in magnetic or near-metallic systems (BiFeO$_3$ and the manganites are marked in the data). $^{b}$Twenty-four of the 76 are \texttt{flag\_and\_continue}, stopped at $G_4$ by \texttt{MISSING\_FRAMES} (in a first replay seven of them had been stopped only by the replay-only comparison of $G_6$); six depend on the stored energies of the other HfO$_2$ paths. Bottom: the 124 branches whose option is neither stop nor labelled correct, rerun with the gates switched off after the fork.}
\label{tab:si_C}
\begin{tabular}{@{}lrrrr@{}}
\toprule
Gates on (208 branches) & label correct & label safe & label harmful & all\\
\midrule
completed, correct & 21 & 3 & 9 & 33\\
completed, SCF mixing changed only$^{a}$ & 7 & 0 & 13 & 20\\
refused by the next gate, then repaired & 0 & 0 & 24 & 24\\
stopped by a later gate$^{b}$ & 3 & 5 & 68 & 76\\
stopped at $G_3$: frame not in the stored campaign$^{d}$ & 0 & 1 & 1 & 2\\
stopped by choice & 22 & 31 & 0 & 53\\
completed, wrong & 0 & 0 & 0 & 0\\
All & 53 & 40 & 115 & 208\\
\midrule
Gates off after the fork (124 branches) & & & & \\
\midrule
completed, correct or equivalent & & & & 69\\
completed, wrong (nothing flagged)$^{c}$ & & & & 20\\
stopped & & & & 35\\
\bottomrule
\multicolumn{5}{@{}p{0.92\linewidth}@{}}{$^{c}$15 \texttt{flag\_and\_continue} fits without the frames of the failed job (e.g.\ 922 of 962 energies for CsSnI$_3$: the 961 plane frames and the $P4/mbm$ anchor), 3 with the wrong number of formula units per cell kept (HfO$_2$), 2 with the injected rigid translation kept (InMnO$_3$, 91 of 225 energies). $^{d}$The anchor-only answer to R5 (InMnO$_3$) asks for a frame with both irreps at their child amplitudes, which the stored plane does not contain; production would compute it.}\\
\end{tabular}
\end{table}

\begin{table}[!htbp]
\centering\scriptsize
\caption{Plain-language intake (Methods, Section~\Mref{sec:eval}): 50 frozen requests, five repeats per model and one run each of the rule-based parser (parser v1, frozen before scoring; parser v2, one declared fix after it, Section~\ref{sec:si_parser}) and the keyword baseline. The model rows v1 and v2 are a different pair: v1 is the primary run and v2 the declared rerun with the system-profile descriptions rewritten (Section~\ref{sec:si_eval_ext}). Clear requests: calls with every critical field (structures, transition, analysis, strain policy, system profile, budget) right, overall (39 requests) and by style (13 each). Underspecified: a question, or the only stored analysis for U01. Out of scope: refused or asked. New material: valid configuration flagged as needing new calculations, with no profile of another material. Wrong field passed $G_0$: calls with a wrong critical field that $G_0$ accepted. v1, primary run; v2, declared rerun in which only the descriptions of the system profiles in the prompt were changed. $^{*}$Reference model.}
\label{tab:si_E}
\resizebox{\linewidth}{!}{%
\begin{tabular}{@{}llrrrrrrrrr@{}}
\toprule
Model & run & clear & expert & terse & non-specialist & underspecified & out of scope & new material & wrong field passed $G_0$ & unnecessary questions\\
\midrule
Claude Sonnet 5.5$^{*}$ & v1 & 186/195 & 56/65 & 65/65 & 65/65 & 23/25 & 15/15 & 15/15 & 9 & 0\\
 & v2 & 183/195 & 53/65 & 65/65 & 65/65 & 23/25 & 15/15 & 15/15 & 3 & 9\\
Claude Opus 5.5 & v1 & 185/195 & 55/65 & 65/65 & 65/65 & 25/25 & 15/15 & 15/15 & 9 & 1\\
 & v2 & 185/195 & 55/65 & 65/65 & 65/65 & 25/25 & 15/15 & 15/15 & 0 & 10\\
GPT-6.1 Sol & v1 & 180/195 & 50/65 & 65/65 & 65/65 & 25/25 & 15/15 & 15/15 & 15 & 0\\
 & v2 & 181/195 & 51/65 & 65/65 & 65/65 & 25/25 & 15/15 & 15/15 & 5 & 9\\
Gemini 4 Argon & v1 & 181/195 & 55/65 & 61/65 & 65/65 & 25/25 & 15/15 & 15/15 & 10 & 4\\
 & v2 & 194/195 & 65/65 & 64/65 & 65/65 & 25/25 & 15/15 & 15/15 & 0 & 1\\
Rule-based parser & v1 & 25/39 & 10/13 & 8/13 & 7/13 & 3/5 & 3/3 & 3/3 & 7 & 6\\
 & v2 & 29/39 & 10/13 & 10/13 & 9/13 & 3/5 & 3/3 & 3/3 & 1 & 8\\
Keyword baseline & v1 & 20/39 & 10/13 & 5/13 & 5/13 & 4/5 & 1/3 & 0/3 & 9 & 10\\
\bottomrule
\end{tabular}}
\end{table}

\subsection{The rule-based parser for the intake}\label{sec:si_parser}

The keyword lookup of Supplementary Table~\ref{tab:si_E} is built from the tables of system names, formulas, space groups and transition ids alone and leaves every other field to the policy default, so it is a floor, not a competitor. To test whether a deterministic interface could do the intake, we wrote a rule-based parser (\texttt{intake/parser.py}, about 290 lines) with a vocabulary for every field of the draft-intent schema, from the same information the models receive: the tables of \texttt{intake/library.py}, the schema and the system prompt of \texttt{intake/schema.py} (reproduced in Supplementary Section~\ref{sec:si_prompts}), and general crystallographic vocabulary. It was written without consulting the 50 requests (one, E01, had been seen in passing by its author while reading the intake code) and frozen with a hash (\texttt{intake/PARSER.sha256}) before any scoring. It answers each request once, deterministically, and its draft goes through the same build as a model's (resolution of defaults, configuration, $f_0$ and $G_0$) and the same scorer.

\paragraph{What it reads.} \emph{System}: the formula or system id; the names in the system table (hafnia); and the element-name reading of the formula, every element of the formula named in words with oxygen implied by the -ide, -ite or -ate name (cesium tin iodide, indium manganite, bismuth ferrite), so that ``hexagonal manganite'' alone is ambiguous and asks. A chemical formula not in the table (two or more element symbols with counts) is a new material with that formula. Two systems named, or none, ask. \emph{Structures}: space-group numbers after \# or No., the Hermann--Mauguin symbols of the transition table (with or without underscores), and, when no number or symbol is present, crystal-system words resolved on the chosen system's own transitions by the International Tables ranges (cubic, tetragonal, orthorhombic, monoclinic, hexagonal), with polar or non-polar deciding between two groups of the same crystal system (HfO$_2$: $Pca2_1$ against $Pbcn$); parent and child in order of mention. A transition id names everything. One structure only is resolved when the system has exactly one transition starting or ending there. A pair not in the table is out of scope. \emph{Analysis}: phrases for the four analyses tried from the most specific (six-component, 6D, invariant polynomial $\to$ grid6d; joint, hat, brim, wedge $\to$ joint fit; path, barrier, straight line $\to$ path; plane, 2D, energy surface, landscape, grid $\to$ plane); unstated, the policy default of the system. \emph{Strain policy}: fixed lattice or cell, strain at zero, clamped $\to$ P1; stated, parent or child cell $\to$ P2; strain-interpolated $\to$ P3; relaxed anchors $\to$ P4; bias statement $\to$ P5. \emph{System profile}: a profile id written out, else the stated functional (LDA, PBEsol, PBE), $+U$ scheme (Dudarev, Liechtenstein) and magnetic order (A-type, G-type, frustrated, ferromagnetic, non-spin-polarized) matched against the profile notes of the chosen system; exactly one match selects it, several ask, none leaves the default. \emph{Budget}: a number after at most, up to, budget, $\le$, or before runs, calculations, structures or points. \emph{Outputs}: D4, secondary, held at zero $\to$ secondary check; hat, brim; barrier; Landau, coefficients; stationary, minima, saddle; cross-setting, closure, network; energy surface, landscape. \emph{Out of scope}: molecular dynamics, phonons, other codes (Quantum ESPRESSO, LAMMPS, GPAW, CP2K, ABINIT, CASTEP, SIESTA), machine-learned potentials, Monte Carlo, finite temperature, NEB, experiment.

\paragraph{Result (v1, frozen).} On the 39 clear requests the parser produced a configuration with every critical field right in 25 (expert 10 of 13, terse 8 of 13, non-specialist 7 of 13), against 20 for the keyword lookup and 180--186 of 195 calls for the models; it asked on 3 of the 5 underspecified requests, refused all 3 out-of-scope requests and flagged all 3 new materials as needing new calculations without borrowing a profile (keyword lookup: 4 of 5, 1 of 3, 0 of 3). Seven of its clear-request configurations passed $G_0$ with a wrong field, six of them the same field: every InMnO$_3$ request (E10--E15, both configurations, all three styles) was given the profile \texttt{inmno3} (PBEsol$+U$, frustrated order) instead of the stored \texttt{inmno3\_aafm\_v2} (LDA$+U$, A-type). The cause is a bug, not vocabulary: the profile ids \texttt{cssni3}, \texttt{ca3ti2o7}, \texttt{inmno3}, \texttt{ermno3}, \texttt{bifeo3} and \texttt{hfo2} are also the system ids, so the rule ``a profile id written out selects that profile'' fired on the material's name; for the other five systems the profile of that name happens to be the stored default, so the bug was invisible there. The seventh wrong field is E06, a non-specialist request for the CsSnI$_3$ plane with the secondary check, whose phrase ``kept at zero'' is not in the D4 vocabulary (``held at zero'' is), so the plain plane configuration was built. One expert request was refused: in E07 (Ca$_3$Ti$_2$O$_7$) ``$\Gamma_5^-$ held at zero'' describes the trilinear mode, but the parser's D4 vocabulary read it as a request for the secondary check, for which this transition has no template. Six requests asked a question that the models did not need: the terse and non-specialist forms of the three HfO$_2$ paths that name one phase by its symbol and the other by its crystal system (``tetragonal $\to$ $Pbcn$'', ``$Pbcn$ $\to$ polar orthorhombic'', ``$Pbcn$ $\to$ monoclinic''); the parser resolves crystal-system words only when no symbol is present, so it found one structure, which starts or ends three stored transitions, and asked which. Of the underspecified requests, ``map the BiFeO$_3$ landscape'' (U01) was refused rather than asked, because ``landscape'' is in the plane vocabulary and BiFeO$_3$ has no plane template, and ``ErMnO$_3$, please'' (U05) was built as the default plane instead of asking whether the plane or the joint fit was wanted, since the parser cannot tell that a default is a choice. Supplementary Table~\ref{tab:si_parser} lists every miss.

\paragraph{Result (v2, one declared fix).} The profile bug is a one-line error with a clear specification (a profile id that is also a system id is the material's name, and the build then applies the system's default), so we fixed it and rescored, as the models were rescored in v2 after the profile descriptions were rewritten; no vocabulary was changed, both hashes are recorded and \texttt{PARSER\_V1=1} reproduces v1. With the fix the parser produced a correct configuration for 29 of the 39 clear requests (expert 10, terse 10, non-specialist 9 of 13): the four InMnO$_3$ requests that state no profile (E11, E12, E14, E15) now take the default, which is right, and the two that state ``LDA$+U$ A-AFM'' (E10, E13) now ask which profile, because ``A-AFM'' is not in the magnetic-order vocabulary and three InMnO$_3$ profiles use LDA$+U$. One wrong field passed $G_0$ (E06), one clear request was refused (E07) and eight were asked unnecessarily (the six HfO$_2$ paths and E10, E13).

\paragraph{What it shows.} The parser is within the range of the models on expert requests (77\% against 77--86\%), where every field is named in the vocabulary of the tables, and it never borrows a profile, never guesses a transition and refuses every out-of-scope request, because those behaviours are rules. It falls behind on terse and non-specialist requests, 77\% and 69\% against 94--100\% for the models even after the fix, and every one of its remaining misses is a phrasing its vocabulary does not contain: a transition named half by symbol and half by crystal system, a landscape that is a six-component grid, ``kept'' for ``held'', ``A-AFM'' for A-type. Each could be added, but each addition would be made after seeing the request that needed it, which is the regress a fixed vocabulary cannot escape and the reason the models' 65 of 65 on terse and non-specialist requests is the result that matters at the interface. The comparison is between a parser we wrote and models we did not train; a parser tuned on these requests, or written independently of us, was not tested (Supplementary Section~\ref{sec:si_limits}).

\paragraph{Guidelines for the intake.}\label{sec:si_intake_guidelines} The intake results, the v1$\to$v2 rerun and the parser study together give four rules for anyone who adapts the intake to new systems or new users. (i) Describe every option by its physics, never by its name: the only systematic intake error of the models, the wrong InMnO$_3$ profile in 9--10 calls per model (of the 9--15 calls per model that passed $G_0$ with a wrong field; the rest were a wrong analysis field on one ErMnO$_3$ request), came from profiles described in identical words, and fell to 0--5 when each was described by its functional, $+U$ scheme and magnetic order; the parser made the same error for a different reason (a profile named like the material). (ii) A valid but wrong choice passes $G_0$, which checks only that a configuration is complete and consistent; the option descriptions are therefore the guardrail, and a new option must be added with a description that distinguishes it from every existing one. (iii) Defaults are applied by code, not chosen by the model: a field the request does not state is left null, and the build fills it from the policy of the chosen system; this is why the terse requests, which state almost nothing, are handled in 65 of 65 calls. (iv) When two options match a request, the answer is a question, not a guess: the schema has a \texttt{clarify\_field}, the underspecified requests measure whether it is used, and the out-of-scope and new-material requests measure that the model refuses or flags rather than borrows a stored setting. A fixed vocabulary can implement (ii)--(iv) but not the paraphrase that (i) relies on, which is where the model earns its place.

\begin{table}[!htbp]
\centering\scriptsize
\caption{Every request the rule-based parser did not handle as the reference expects, in v1 and after the one declared fix (v2). Correct outcome: a configuration with every critical field right (clear requests), a question (underspecified), a refusal (out of scope) or a flagged valid configuration (new material). The models' outcome on the same request is from v1 (Supplementary Table~\ref{tab:si_E}).}
\label{tab:si_parser}
\begin{tabularx}{\linewidth}{@{}L{1.1cm}L{1.8cm}L{1.8cm}XL{2.0cm}L{1.9cm}@{}}
\toprule
Request & Style & Configuration & What the parser did, and why & v1 & v2\\
\midrule
E06 & non-specialist & cssni3\_d4 & ``kept at zero'' is not in the D4 vocabulary (``held at zero'' is): built the plain plane & wrong configuration, passed $G_0$ & same\\
E07 & expert & ca3ti2o7 & ``$\Gamma_5^-$ held at zero'' read as a request for the D4 secondary check, for which this transition has no template & refused & same\\
E10, E13 & expert & inmno3, inmno3\_joint & the material's name read as the profile \texttt{inmno3} (bug); after the fix, ``LDA$+U$ A-AFM'' matches three profiles, since ``A-AFM'' is not in the vocabulary & wrong profile, passed $G_0$ & asks which profile\\
E11, E12, E14, E15 & terse, non-specialist & inmno3, inmno3\_joint & the material's name read as the profile \texttt{inmno3} (bug); no profile stated, so the default is right after the fix & wrong profile, passed $G_0$ & correct\\
E29, E30 & terse, non-specialist & hfo2\_T2 & ``tetragonal $\to$ $Pbcn$'': the crystal-system word is resolved only when no symbol is present; one structure found, three transitions start or end there & asks which transition & same\\
E35, E36 & terse, non-specialist & hfo2\_T4 & ``$Pbcn$ $\to$ polar orthorhombic'': as above & asks which transition & same\\
E38, E39 & terse, non-specialist & hfo2\_T5 & ``$Pbcn$ $\to$ monoclinic'': as above & asks which transition & same\\
U01 & underspecified & (ask) & ``landscape'' is plane vocabulary; BiFeO$_3$ has no plane template & refused instead of asking & same\\
U05 & underspecified & (ask) & no analysis stated; the build's default (plane) applied, where the reference expects a question (plane or joint fit) & built the default & same\\
\bottomrule
\end{tabularx}
\end{table}

\subsection{Cost of the language-model layer}\label{sec:si_cost}

Every model call in this work is metered: the benchmark and intake runners record input, output and (where the provider reports it) reasoning tokens, the wall-clock latency of each call, and, for the providers whose list prices were recorded in \texttt{eval/models.yaml} on the run day, the dollar cost; the live chat recorded the model but did not meter its four intake calls. Supplementary Table~\ref{tab:si_cost} gives the totals per experiment (\texttt{eval\_extra/cost\_summary.py}; \texttt{results/COST/}). Three things stand out. The whole sub-agent benchmark, 9{,}540 decisions by nine models, cost \$12.74 for the 3{,}180 decisions of the three OpenAI models whose prices were recorded (\$0.17, \$2.01 and \$10.57 for the small, mid and flagship tiers, with reasoning tokens billed as output tokens), that is, between 0.02 and 1.0 cents per decision; the intake is dearer per call because the prompt carries the tables (2{,}000--3{,}700 input tokens; GPT-6.1 Sol \$1.47 for its 250 v1 calls, 0.6 cents each); the Anthropic and Google models were run on accounts without recorded list prices, so their dollar cost is not reported rather than estimated, but their token counts are of the same order (600--1{,}100 input and 50--500 output tokens per decision, the upper end being the reasoning tokens of Gemini~3.8 Flash). A decision takes 0.8--3.7~s at the median and up to 16~s at the 95th percentile (an intake call 2--7~s, up to 39~s for the slowest model), against job queue times of minutes and DFT campaigns of $10^2$--$10^4$ calculations; in the live BaTiO$_3$ campaign the four intake calls are invisible in the 34~min of wall time, which is dominated by the queue and the 2--7~min jobs. And in production the model is called far less often than in the benchmark: once per request at the intake, and once per gate failure without a stored rule, so a campaign that passes every gate, as BaTiO$_3$ did, makes one call. The rule table costs nothing and answers 96\% of the recorded failures; the model layer costs cents and seconds and answers the judgment calls and the failures the rules have not seen (Section~\ref{sec:si_loso}).

\begin{table}[!htbp]
\centering\scriptsize
\caption{Tokens, latency and cost of the language-model calls. Benchmark: the 53 cases, ten repeats, guided and bare prompts (1{,}060 decisions per model); tokens per decision averaged over both prompts, output including reasoning tokens where reported; latency with the guided prompt. Cost from the list prices recorded on the run day (OpenAI only; the Anthropic and Google accounts had no recorded price, so no cost is given). All rows from \texttt{results/COST/cost\_summary.csv} (\texttt{eval\_extra/cost\_summary.py}); reasoning tokens are billed as output tokens.}
\label{tab:si_cost}
\begin{tabular}{@{}lrrrrr@{}}
\toprule
Model & decisions & tokens in / decision & tokens out / decision & latency p50 / p95 (s) & cost (USD)\\
\midrule
\multicolumn{6}{@{}l}{\emph{Sub-agent benchmark (Section~\ref{sec:si_recovery})}}\\
Claude Haiku 5.5 & 1{,}060 & 1{,}090 & 184 & 1.5 / 3.3 & --\\
Claude Sonnet 5.5 & 1{,}060 & 1{,}090 & 135 & 2.0 / 5.1 & --\\
Claude Opus 5.5 & 1{,}060 & 1{,}090 & 197 & 3.0 / 6.5 & --\\
GPT-6 Luna & 1{,}060 & 599 & 200 & 2.1 / 4.3 & 0.17\\
GPT-6.1 Sol & 1{,}060 & 599 & 70 & 2.8 / 4.9 & 2.01\\
GPT-6 Astra & 1{,}060 & 599 & 80 & 2.5 / 4.3 & 10.57\\
Gemini 3.5 Flash-Lite & 1{,}060 & 604 & 49 & 0.8 / 1.0 & --\\
Gemini 3.8 Flash & 1{,}060 & 604 & 478 & 3.5 / 11.5 & --\\
Gemini 4 Argon & 1{,}060 & 828 & 81 & 3.7 / 15.2 & --\\
All nine & 9{,}540 & & & & 12.74$^{\ast}$\\
\midrule
\multicolumn{6}{@{}l}{\emph{Intake (50 requests, five repeats, v1; four models; the v2 rerun is the same size)}}\\
Claude Sonnet 5.5 (v1) & 250 & 3{,}742 & 209 & 2.0 / 5.0 & --\\
Claude Opus 5.5 (v1) & 250 & 3{,}742 & 263 & 3.3 / 6.5 & --\\
GPT-6.1 Sol (v1) & 250 & 2{,}071 & 175 & 4.2 / 6.8 & 1.47\\
Gemini 4 Argon (v1) & 250 & 2{,}535 & 245 & 6.9 / 39.1 & --\\
\midrule
\multicolumn{6}{@{}l}{\emph{Transfer, guidance sentence removed (7 cases, ten repeats; nine models)}}\\
nine models & 630 & 687--1{,}240 & 48--267 & 0.8--3.7 / 1.0--10.5 & 0.91 (three priced models)\\
\bottomrule
\end{tabular}
\begin{flushleft}\footnotesize $^{\ast}$Three priced models, 3{,}180 decisions; the components are rounded.\end{flushleft}
\end{table}

\subsection{Failure modes and their recovery paths}\label{sec:si_taxonomy}

Supplementary Tables~\ref{tab:si_taxonomy} and~\ref{tab:si_taxonomy_c} regroup every failure mode met in this work, in the campaigns, the faulted runs and the benchmark, into three kinds by where it arises: at the intake, where a request is ambiguous or a valid choice is wrong; at the crystallographic gates $G_1$--$G_3$, where a decomposition, a frame or a replay link is rejected; and at the compute and analysis gates $G_4$--$G_6$, where a calculation has not converged, a frame is missing, or a fit or a re-derivation fails its tolerance. For each the table gives the code, who answers it in the recovery order (a stored rule, the listed-option sub-agent, or the human), the fix, and where it was seen. The division is also a division of trust: the first kind can pass $G_0$ undetected and is prevented by the option descriptions (Section~\ref{sec:si_intake_guidelines}); the second is caught by deterministic checks that compare two independent routes to the same structure; the third is caught by thresholds, and is where the judgment calls and the decoys live.

\begin{table}[!htbp]
\centering\scriptsize
\caption{Failure modes by kind, with the recovery path of each: (a) intake and (b) crystallographic gates; (c) in Supplementary Table~\ref{tab:si_taxonomy_c}. Who: the tier of the recovery order that answers it (rule, the stored rule table; sub-agent, a listed-option choice by the model; human). Seen: C, the six campaigns (Extended Data Table~\Mref{tab:errors}); F, the 53 injected faults (Supplementary Table~\ref{tab:si_faults}); B, the 53 benchmark cases (Supplementary Table~\ref{tab:si_A}); E, the 50 intake requests (Supplementary Table~\ref{tab:si_E}); L, the live BaTiO$_3$ campaign.}
\label{tab:si_taxonomy}
\begin{tabularx}{\linewidth}{@{}L{2.2cm}L{2.7cm}L{1.2cm}XL{1.4cm}@{}}
\toprule
Failure & Code (gate) & Who & Fix & Seen\\
\midrule
\multicolumn{5}{@{}l}{\emph{(a) Intake: request ambiguous, or a valid but wrong choice}}\\
Request names no system, or several & \texttt{AMBIGUOUS} ($G_0$) & code & a question to the user (\texttt{clarify\_\allowbreak{}field}); never a guess & E (5 requests, 23--25/25 asked)\\
Request outside the stored tables (other engine, dynamics, no stored route) & \texttt{OUT\_\allowbreak{}OF\_\allowbreak{}SCOPE} ($G_0$) & code & refusal with reason & E (3/3)\\
Material without a stored campaign & \texttt{NEEDS\_\allowbreak{}NEW\_\allowbreak{}DFT} ($G_0$) & code & valid configuration flagged for new calculations; no borrowed profile & E (3/3); L (4 runs)\\
Valid but wrong field (profile, analysis) & passes $G_0$ & prevention only & describe every option by its physics (v2: 9--15 $\to$ 0--5 errors per model) & E\\
Model output not in the schema & \texttt{SCHEMA} ($G_0$) & rule & re-ask with the validator's errors & E (0 in 1{,}000 calls)\\
\midrule
\multicolumn{5}{@{}l}{\emph{(b) Crystallographic gates: decomposition, frames, replay links}}\\
No supercell basis within the cut-off & \texttt{NO\_\allowbreak{}BASIS} ($G_1$) & rule & widen $d_{\max}$ (1.0, 1.5, 2.0~Å), then the basis search & C (HfO$_2$); F, B\\
Space-group assignment unstable or engines disagree & \texttt{SG\_\allowbreak{}LADDER}, \texttt{SG\_\allowbreak{}DISAGREE} ($G_1$) & rule / human & adopt the tightest tolerance at which the engines agree; else human & C\\
Mode norms inconsistent with the cell count & \texttt{DECOMP\_\allowbreak{}NORM} ($G_1$) & human & recount primitive cells from the parent CIF & C (Ca$_3$Ti$_2$O$_7$)\\
Atoms permuted between parent and child & \texttt{ATOM\_\allowbreak{}MATCH} ($G_2$) & rule & species-constrained re-assignment & C (Ca$_3$Ti$_2$O$_7$)\\
Rigid translation in a polar vector & \texttt{TRANSLATION} ($G_2$) & rule & subtract the mean displacement, rescale & C (4 systems); F, B\\
Exported vectors off scale & \texttt{CALIBRATION} ($G_2$) & rule & calibrate with $s_i$ & C (3 systems)\\
Single-mode frame has the wrong space group & \texttt{SINGLE\_\allowbreak{}MODE\_\allowbreak{}SG} ($G_2$) & rule & rebuild on the fixed parent lattice & C (CsSnI$_3$); F, B\\
Active set does not generate $H$; polar child with no polar irrep & \texttt{R3\_\allowbreak{}FAIL}, \texttt{R4\_\allowbreak{}NO\_\allowbreak{}POLAR} ($G_2$) & rule / human & add the largest polar irrep (R4); else human & --\\
Borderline selection & \texttt{R5\_\allowbreak{}BORDERLINE} ($G_2$) & sub-agent / human & include, exclude or anchor only (applied in the rules) & C (InMnO$_3$); F, B\\
Grid over budget & \texttt{OVER\_\allowbreak{}BUDGET} ($G_2$) & human & reduce the grid (D3, not implemented) or raise the budget & F, B\\
Child cell rotated relative to the parent & \texttt{PATH\_\allowbreak{}LATTICE} ($G_2$) & human & input-structure problem; no listed fix & B (decoy)\\
Frame not in the stored campaign (replay) & \texttt{UNMATCHED} ($G_3$) & human & compute it in production; stop in replay & F, B\\
Stored frame carries a rigid shift & \texttt{RIGID\_\allowbreak{}SHIFT} ($G_3$) & code & match modulo translation up to 0.05~Å & C (InMnO$_3$)\\
\midrule
\bottomrule
\end{tabularx}
\end{table}

\begin{table}[!htbp]
\centering\scriptsize
\caption{Failure modes by kind (continued): (c) compute and analysis gates. Columns as in Supplementary Table~\ref{tab:si_taxonomy}.}
\label{tab:si_taxonomy_c}
\begin{tabularx}{\linewidth}{@{}L{2.2cm}L{2.7cm}L{1.2cm}XL{1.4cm}@{}}
\toprule
Failure & Code (gate) & Who & Fix & Seen\\
\midrule
\multicolumn{5}{@{}l}{\emph{(c) Compute and analysis gates: convergence, missing data, fit and re-derivation tolerances}}\\
Job killed at the time limit & \texttt{WALLTIME} ($G_4$) & rule & resubmit the failed parts unchanged & C (BiFeO$_3$, 50 runs); F, B\\
SCF not converged within NELM & \texttt{NELM} ($G_4$) & rule & resubmit with robust mixing & C (BiFeO$_3$, 15); F, B\\
Still not converged after resubmission & \texttt{NELM\_\allowbreak{}AFTER\_\allowbreak{}ROBUST} ($G_4$) & rule & keep $E_0$ as SCF-limited, drop from gradient fits & C (BiFeO$_3$, 20)\\
Frames of a failed part missing from the fit & \texttt{MISSING\_\allowbreak{}FRAMES} ($G_4$, $G_5$) & human & resubmit, or authorise the smaller grid (every report then says so) & F (option replay, 24 branches)\\
Inconsistent inputs, magnetism, formula units & \texttt{DFT\_\allowbreak{}INCONSISTENT}, \texttt{MAGNETIC}, \texttt{PER\_\allowbreak{}FU} ($G_4$) & human & configuration error; no listed fix & F, B\\
Too few points per coefficient & \texttt{THIN\_\allowbreak{}GRID} ($G_5$) & rule & return to $f_2$, halve the step (budget permitting) & F, B\\
Minimum near the grid edge & \texttt{COVERAGE} ($G_5$) & human & extend $Q_{\max}$; densifying does not help & B (decoy)\\
Fit tolerance exceeded (RMS, CV, mirror, closure, cross-setting, network) & \texttt{RESIDUALS}, \texttt{CV\_\allowbreak{}ERROR}, \texttt{MIRROR}, \texttt{CLOSURE}, \texttt{CROSS\_\allowbreak{}SETTING}, \texttt{NETWORK} ($G_5$) & human & model or data problem; no listed fix & B (decoys); L (network passed)\\
Secondary held at zero would lower the energy & \texttt{SECONDARY\_\allowbreak{}COUPLED} ($G_5$) & sub-agent / human & hold with a caveat (D4) or add the axis & C (CsSnI$_3$); F, B\\
A reported number does not re-derive & \texttt{MISMATCH}, \texttt{NOT\_\allowbreak{}REDERIVED} ($G_6$) & rule / human & re-run $f_5$ once; never edited by hand & F (replay only)\\
\bottomrule
\end{tabularx}
\end{table}

\subsection{Prompt gallery}\label{sec:si_prompts}
The language model receives two kinds of prompt: the intake prompt of $f_0$, which turns a request into a draft configuration, and the prompt of a recovery sub-agent, which chooses one fix from the menu of a gate failure. Every prompt is a fixed text file (\texttt{prompts/*.md}), frozen with its SHA-256 digest before any evaluation call; the same text is sent to every model, and the gallery below is generated from these files by the code (\texttt{python -m iris\_agent.prompts --gallery}), not edited by hand. Boxes are coloured by who writes the text: \textcolor{iSys}{\textbf{blue}}, system prompt (fixed by IRIS); \textcolor{iUser!80!black}{\textbf{grey}}, user message (a template filled by the code, or the request as typed); \textcolor{iGuide}{\textbf{orange}}, gate-specific lines; \textcolor{iOut}{\textbf{green}}, what the model returns; \textcolor{iGate}{\textbf{violet}}, the verdict of the gate or the message shown to the human. In every prompt the model returns one choice from a fixed list, or a configuration that $G_0$ then checks; it never computes, decides whether a check passes, routes the workflow or writes a number into the results.

\paragraph{Intake.} The system prompt lists the stored systems, transitions, analyses, strain policies, system profiles and outputs, and asks for structured output whose fields can take only these values; the only number the model may give is the budget, and it may instead ask one question or decline (Methods, main-text Section~\Mref{sec:eval}). The request is sent as the user message, unchanged. The primary run (v1) used the prompt of Box~S1, whose SHA-256 (\texttt{ed9e4930c59d1109\ldots}) matches the v1 freeze; the model's answer must follow a JSON schema frozen with it. The tables in the prompt are generated from the stored system, transition and profile files, so the prompt changes only when these files do; in the declared rerun (v2) only the twelve lines describing the system profiles changed, from a short chemical description (for example ``InMnO$_3$, Mn$^{3+}$ d4 magnetic'' for all six InMnO$_3$ profiles) to the functional, magnetic order and $+U$ scheme read from each profile file (for example ``\texttt{inmno3\_aafm\_v2}: InMnO$_3$, LDA+U (Dudarev, $U_\mathrm{eff}$), A-type AFM (layers)'').

\begin{iPrompt}{iSys}{Box S1 \textbullet\ Intake system prompt of Experiment E, v1 (SHA-256 \texttt{ed9e4930\ldots}); the request follows as the user message}
You are the intake step (f0) of IRIS, a workflow that maps symmetry-mode energy landscapes from stored DFT campaigns. Turn the user's request into a draft intent by choosing ids from the tables below.

Rules:
- Fill a field only when the request states or clearly implies it; otherwise return null (code then uses the policy default of the chosen system and analysis).
- You never set DFT parameters (cutoffs, k-points, U, magnetic moments, functionals) or any number other than the run budget. Choose only existing system profiles and policy options.
- If a required choice (system, transition, or what to compute) cannot be decided from the request, set clarify_field to that field and write ONE short question; do not guess.
- If the request cannot be served by these tables (no listed transition, an engine other than the stored VASP DFT campaigns, molecular dynamics, ...), set out_of_scope = true and give the reason.
- parent_sg / child_sg: the space-group numbers of the parent and child, when stated or implied.
- A material without a stored campaign: system = "new", its formula, parent_sg / child_sg and the analysis. Choose a system profile only if one exists for that same material; never use the profile of another material. Such a request needs new DFT; it is not out of scope.

Systems:
  cssni3: CsSnI3, ABX3 halide perovskite
  ca3ti2o7: Ca3Ti2O7, n = 2 Ruddlesden-Popper layered perovskite
  inmno3: InMnO3, hexagonal manganite (YMnO3-type)
  ermno3: ErMnO3, hexagonal manganite (YMnO3-type)
  bifeo3: BiFeO3, ABO3 perovskite multiferroic
  hfo2: HfO2 (hafnia), fluorite-derived oxide
  new: any other material (give its formula)

Stored transitions (id: system, parent -> child):
  cssni3_221_62: cssni3, #221 -> #62, Pm-3m -> Pnma, tilts M3+ (+) R4+
  ca3ti2o7_139_36: ca3ti2o7, #139 -> #36, I4/mmm -> A2_1am, rotations X3- (+) X2+ (GM5- induced)
  inmno3_194_185: inmno3, #194 -> #185, P6_3/mmc -> P6_3cm, trimerization K3 (+) polar GM2-
  ermno3_194_185: ermno3, #194 -> #185, P6_3/mmc -> P6_3cm, trimerization K3 (+) polar GM2-
  bifeo3_221_161: bifeo3, #221 -> #161, Pm-3m -> R3c, polar GM4- (+) rotation R5- (six components)
  hfo2_137_29: hfo2, #137 -> #29, P4_2/nmc (tetragonal) -> Pca2_1 (polar orthorhombic)
  hfo2_137_60: hfo2, #137 -> #60, P4_2/nmc (tetragonal) -> Pbcn (non-polar orthorhombic)
  hfo2_137_14: hfo2, #137 -> #14, P4_2/nmc (tetragonal) -> P2_1/c (monoclinic)
  hfo2_60_29: hfo2, #60 -> #29, Pbcn (non-polar orthorhombic) -> Pca2_1 (polar orthorhombic)
  hfo2_60_14: hfo2, #60 -> #14, Pbcn (non-polar orthorhombic) -> P2_1/c (monoclinic)

Analyses:
  plane: 2D energy surface of two order parameters on a grid (+ Landau fit where the system has one)
  joint_fit: one Landau model fitted jointly to the plane and the polar (hat) wedge
  grid6d: six-component grid (polar + rotation, 3 components each), invariant-polynomial fit
  path: straight-line irrep path between the two structures (10 points)

Strain policies:
  P1: fixed lattice
  P2: stated cell [the stored campaigns state it: parent cell for CsSnI3, Ca3Ti2O7, BiFeO3; child cell for the manganites]
  P3: strain-interpolated
  P4: relaxed anchors [not implemented in this build]
  P5: fixed-lattice bias statement [always applied with a fixed lattice, not a separate construction]

System profiles (id: material, note):
  bifeo3: BiFeO3, Fe3+ d5 high-spin, G-type AFM
  ca3ti2o7: Ca3Ti2O7, n=2 Ruddlesden-Popper layered perovskite (Ca3Ti2O7-type): double TiO6 perovskite
  cssni3: CsSnI3, ABX3 halide perovskite (corner-sharing SnI6 octahedra)
  ermno3_aafm: ErMnO3, Mn3+ d4 magnetic
  hfo2: HfO2, fluorite-derived oxide (NOT a perovskite)
  inmno3: InMnO3, Mn3+ d4 magnetic
  inmno3_aafm: InMnO3, Mn3+ d4 magnetic
  inmno3_aafm_v2: InMnO3, Mn3+ d4 magnetic
  inmno3_aafm_v2_lt1: InMnO3, Mn3+ d4 magnetic
  inmno3_fm: InMnO3, Mn3+ d4 magnetic
  inmno3_frust_v2: InMnO3, Mn3+ d4 magnetic
  yino3: YInO3, hexagonal RXO3 (YMnO3-type): layers of corner-sharing InO5 trigonal bipyramids s

Outputs:
  energy_surface: the energies on the grid / path
  landau_coefficients: fitted Landau / invariant coefficients
  stationary_points: minima and saddles of the fitted model
  barrier: path barrier and end energy
  d4_secondary_check: D4: 1D scans of a secondary held at zero, to check it would not lower the energy
  hat_and_brim: Mexican-hat height and brim barrier (joint fit)
  path_network_checks: cross-setting and closure checks against the other stored hafnia paths

Answer only with the JSON object of the schema.
\end{iPrompt}

\paragraph{Intake calls.} Four calls of the reference model (Claude Sonnet~5.5, first repeat) show the three outcomes; for each field, $f_0$ records whether the value came from the model or from the policy defaults (Supplementary Table~\ref{tab:si_requests} lists all 50 requests).

\begin{iCall}{E02 \textbullet\ terse request \textbullet\ configuration}
\iReq{map the CsSnI3 octahedral-tilt landscape}
\iAns{structures CsSnI$_3$ $221\to62$, transition \texttt{cssni3\_221\_62}, system profile \texttt{cssni3} (model); analysis plane, strain policy P1, budget 1{,}100 (defaults)}
\iVer{\texttt{OK}: equals the reference configuration; its replay reproduced all 43 numbers of the fault-free replay.}
\end{iCall}
\begin{iCall}{U01 \textbullet\ underspecified request \textbullet\ configuration}
\iReq{map the BiFeO3 landscape}
\iAns{structures BiFeO$_3$ $221\to161$ (model); six-dimensional grid, P1, budget 10{,}400 (defaults); no question asked}
\iVer{\texttt{OK}: scored correct, since the six-dimensional grid is the only stored analysis for BiFeO$_3$ (Supplementary Table~\ref{tab:si_E}).}
\end{iCall}
\begin{iCall}{O02 \textbullet\ out of scope \textbullet\ declined}
\iReq{Map the CsSnI3 tilt plane with Quantum ESPRESSO and a hybrid functional.}
\iAns{declines: ``IRIS only serves stored VASP DFT campaigns; a Quantum ESPRESSO engine with a hybrid functional is not available, and DFT settings cannot be set by the intake step.''}
\iVer{\texttt{OUT\_OF\_SCOPE}: nothing is run.}
\end{iCall}
\begin{iCall}{N01 \textbullet\ new material \textbullet\ flagged}
\iReq{PbTiO3 Pm-3m (\#221) -\textgreater{} P4mm (\#99): polar GM4- landscape, energy surface of two polarization components (P\_z vs P\_x) on a fixed cubic cell.}
\iAns{formula PbTiO$_3$, $221\to99$, plane, P1, no stored profile borrowed: ``no stored data for PbTiO3: the campaign needs new DFT''}
\iVer{\texttt{NEEDS\_NEW\_DFT}: a valid request for a material with no stored calculations; the human decides.}
\end{iCall}

\paragraph{Recovery sub-agents.} A sub-agent is called only when a gate fails and no stored rule matches, or the rule is used up. There is one sub-agent per gate; all seven share the system prompt, user template and output schema below and differ only in the two orange lines, the checks of their gate and the guidance (Box~S4). The bare prompt of the benchmark is the guided prompt with the \texttt{Guidance:} line removed (Supplementary Table~\ref{tab:si_defs}). Any answer that is not one of the listed option identifiers is ignored and the human is asked; the menus are in Supplementary Table~\ref{tab:subagents}. Boxes S2--S5 are verbatim.

\begin{iPrompt}{iSys}{Box S2 \textbullet\ System prompt of the $G_4$ sub-agent (guided)}
You are the G4 recovery sub-agent of IRIS, a workflow that maps symmetry-mode energy landscapes.
Step f4 (Reviewer) produced an output that failed gate G4. The gate is plain Python; its verdict
is final and you cannot change it. You choose which listed fix to try next.

Gate G4 checks: energies converged (OSZICAR |dE| < EDIFF in fewer than NELM steps) or flagged.
Guidance: Walltime: resubmit unchanged. SCF step limit: resubmit with robust mixing. If a fix was already tried for the same failure, choose between flagging the frames and stopping. Inconsistent settings, magnetism or configuration values are for the human.

Rules:
- Answer with JSON {"id": ..., "reason": ...} and nothing else.
- "id" must be exactly one of the option ids you are given. Choose "stop" if none fits or if the
  same fix has already failed (see the history).
- The reason is one sentence. It may quote numbers from the input; it must not introduce new ones.
- Do not propose new calculations, new options, or edits to any number.
\end{iPrompt}

\begin{iPrompt}{iUser}{Box S3 \textbullet\ User template, shared by all sub-agents (filled by the code)}
System: {system}
Failed: {node} {node_name}, gate {gate}, code {code}
Detail: {detail}
Checks run by the gate:
{tests}

Retries used on this step: {attempts} of {max_retry}
What has been tried so far in this run:
{history}

Options (choose one id):
{options}
\end{iPrompt}

\begin{iBox}{iGuide}{Box S4 \textbullet\ Gate-specific lines of the seven sub-agent prompts}
\scriptsize\setlength{\tabcolsep}{3pt}\renewcommand{\arraystretch}{1.15}
\begin{tabularx}{\linewidth}{@{}l>{\raggedright}p{1.6cm}>{\raggedright}X>{\raggedright}p{2.5cm}>{\raggedright\arraybackslash}X@{}}
Gate & Step & \texttt{Gate }$G_i$\texttt{ checks:} & Codes it can raise & \texttt{Guidance:} (dropped in the bare prompt)\\\midrule
$G_0$ & f0 Intake & config valid: schema, CIF files exist and have the expected space groups and checksums, budget, write location. & \texttt{SCHEMA, \allowbreak{}CIF\_MISSING, \allowbreak{}CIF\_UNREADABLE, \allowbreak{}CIF\_SG, \allowbreak{}CIF\_CHANGED, \allowbreak{}BUDGET, \allowbreak{}PATHS} & A config problem is for the human to correct (re-asking the intake model arrives with build step 6).\\
$G_1$ & f1 Crystallographer & decomposition valid: H is a subgroup of G, basis determinant equals the index k, ISODISTORT space group equals spglib. & \texttt{NO\_BASIS, \allowbreak{}NOT\_SUBGROUP, \allowbreak{}BASIS\_DET, \allowbreak{}SG\_DISAGREE} & If no basis was found, widening the atom-matching search is the usual fix. A pair that is not group-subgroup, or space groups that disagree, cannot be fixed here.\\
$G_2$ & f2 Path Planner & grid valid: c1 the active set generates H; c2 atom matching, translation-free vectors, calibration, strict single-mode space group; c3 budget and full descent-tree coverage; every frame's space group and minimum distance. & \texttt{R3\_FAIL, \allowbreak{}R5\_BORDERLINE, \allowbreak{}TRANSLATION, \allowbreak{}SINGLE\_MODE\_SG, \allowbreak{}CALIBRATION, \allowbreak{}ATOM\_MATCH, \allowbreak{}OVER\_BUDGET, \allowbreak{}TREE\_COVERAGE, \allowbreak{}FRAME\_SG, \allowbreak{}D\_MIN} & A strict single-mode space-group mismatch is usually the frame being built on the strained movie cell: rebuild on the fixed parent lattice. A rigid translation left in a polar vector is removed by subtracting the mean displacement. R5\_BORDERLINE (a small margin between the smallest active and the largest inactive irrep) is a modelling choice: include, exclude or anchor the borderline irreps. Over budget cannot be fixed here.\\
$G_3$ & f3 Runner & jobs finished: none queued or running; in replay, every frame matched to a stored computed structure. & \texttt{JOBS\_PENDING, \allowbreak{}UNMATCHED} & Replay never computes a missing frame; unmatched frames are flagged and a human decides.\\
$G_4$ & f4 Reviewer & energies converged (OSZICAR \textbar{}dE\textbar{} \textless{} EDIFF in fewer than NELM steps) or flagged. & \texttt{WALLTIME, \allowbreak{}NELM, \allowbreak{}ENERGY\_RANGE} & Walltime: resubmit unchanged. SCF step limit: resubmit with robust mixing. If a fix was already tried for the same failure, choose between flagging the frames and stopping. Inconsistent settings, magnetism or configuration values are for the human.\\
$G_5$ & f5 Analyst & fit acceptable: cross-validated error, residuals, coverage of the fit region. & \texttt{CV\_ERROR, \allowbreak{}RESIDUALS, \allowbreak{}THIN\_GRID, \allowbreak{}COVERAGE} & A coverage failure means the grid is too thin where the fit needs it: go back to f2 and densify. Failures that concern the model or the stored data are not fixed by densifying.\\
$G_6$ & f6 Skeptic & every number in the report re-derived by re-running the tools. & \texttt{NOT\_REDERIVED, \allowbreak{}MISMATCH} & A number that does not re-derive is never edited by hand: re-run f5 or stop.\\
\end{tabularx}
\end{iBox}
\noindent\footnotesize The boxes reproduce the frozen prompts verbatim. Two points for the reader: the parenthesis in the $G_0$ guidance refers to a build step of the implementation and was part of the prompt as evaluated; and the ``codes it can raise'' lists name the main codes of each gate as written when the prompts were frozen, whereas the benchmark also poses further codes of the same gates (\texttt{PATH\_LATTICE} at $G_2$; \texttt{DFT\_INCONSISTENT}, \texttt{MAGNETIC}, \texttt{PER\_FU}, \texttt{SCF\_LIMITED} at $G_4$; \texttt{CLOSURE}, \texttt{CROSS\_SETTING}, \texttt{MIRROR}, \texttt{SECONDARY\_COUPLED} at $G_5$), which the model sees in the failure detail of the user prompt; these are the no-rule, judgment and decoy cases of Supplementary Table~\ref{tab:si_defs} and Section~\ref{sec:si_decoys}.\normalsize

\begin{iPrompt}{iOut}{Box S5 \textbullet\ Output schema of every sub-agent (\texttt{id} must be one of the listed options)}
{
 "type": "object",
 "required": [
  "id",
  "reason"
 ],
 "properties": {
  "id": {
   "type": "string",
   "description": "one id from the options list"
  },
  "reason": {
   "type": "string",
   "maxLength": 300
  }
 }
}
\end{iPrompt}

\begin{iPrompt}{iGate}{Box S6 \textbullet\ Message shown to the human when the recovery node escalates (not sent to a model)}
IRIS paused at {node} {node_name}: gate {gate} failed with {code}.
Detail: {detail}
Why this came to you: {why_human}
Retries used on {node}: {attempts}

Options:
{options}

Answer with one of:
  python -m iris_agent.run --config {config_path} --thread {thread_id} --resume '{{"option": "<id>"}}'
  python -m iris_agent.run --config {config_path} --thread {thread_id} --resume '{{"fix": {{...}}}}'
Choosing "stop" ends the run; the audit log keeps the full trail.
\end{iPrompt}

\paragraph{Two recovery cases.} C01 (rules off): job part 3 of the CsSnI$_3$ grid ends at the walltime (fault F001), $G_4$ \texttt{WALLTIME}. Menu: resubmit the failed parts unchanged; resubmit them with robust mixing; flag the frames and continue without them; stop and ask the human. Reference: resubmit, the fix of the stored rule \texttt{f4\_walltime\_resubmit}, which is also what the guidance line of Box~S4 says. All nine models chose it in all ten repeats with the guided prompt. C33 (decoy): the Ca$_3$Ti$_2$O$_7$ surface minimum lies closer to the grid edge than a widened margin of 50 steps (fault F033), $G_5$ \texttt{COVERAGE}. Menu: go back to $f_2$ and densify the grid; D4, keep the secondary at zero and report the energy it would release; stop and ask the human. Reference: stop; the fault is a limit fault that tightens the coverage margin on correct data, so the run should go to the human (Supplementary Table~\ref{tab:si_faultlist}). With the guided prompt every model chose to densify (89 of 90 calls; one Opus~5.5 repeat stopped), as did the decision tree, following the $G_5$ guidance line literally; in replay the densified run was stopped by a later check (Supplementary Table~\ref{tab:si_C}). Boxes S7--S10 give the two prompts exactly as sent and three stored answers each. Every call stores the first 16 hexadecimal digits of the SHA-256 of its system and user prompt; these reproduce from the text of the boxes (\texttt{69e8dbb11ead1aec} for C01 and \texttt{1dd64ee5ef614e21} for C33), and the prompt was the same for all nine models. The reasons given for C33 quote the gate detail and the guidance line of Box~S4: the decoy works because the guidance prescribes densifying for every coverage failure, while here the failure comes from an injected margin of 50 steps that no grid of the stored extent meets, so the run should go to the human.

\begin{iPrompt}{iUser}{Box S7 \textbullet\ C01, user message as sent (guided prompt, repeat 0; system prompt: Box~S2)}
System: cssni3
Failed: f4 Reviewer, gate G4, code WALLTIME
Detail: 40 frames from failed jobs (WALLTIME), first: frame 120
Checks run by the gate:
[
 {
  "test": "converged_or_flagged",
  "status": "fail",
  "code": "WALLTIME",
  "detail": "40 frames from failed jobs (WALLTIME), first: frame 120"
 }
]

Retries used on this step: 0 of 3
What has been tried so far in this run:
nothing yet

Options (choose one id):
[
 {
  "id": "robust_mixing",
  "text": "Resubmit the failed parts with robust mixing"
 },
 {
  "id": "flag_and_continue",
  "text": "Flag the frames and continue without them"
 },
 {
  "id": "resubmit",
  "text": "Resubmit the failed parts unchanged"
 },
 {
  "id": "stop",
  "text": "Stop and ask the human"
 }
]
\end{iPrompt}
\begin{iBox}{iOut}{Box S8 \textbullet\ C01, stored answers of repeat 0 (all correct)}
\footnotesize
\noindent\textbf{Sonnet~5.5} \,(\texttt{claude-sonnet-5-5}; frozen label: correct)\\
\texttt{\{"id": "resubmit", "reason": "The failure code is WALLTIME and nothing has been tried yet (0 of 3 retries used), so guidance says to resubmit the 40 failed frames unchanged."\}}\par\smallskip
\noindent\textbf{GPT-6.1 Sol} \,(\texttt{gpt-6.1-sol}; frozen label: correct)\\
\texttt{\{"id": "resubmit", "reason": "The jobs failed due to WALLTIME, so resubmit the failed parts unchanged as no fix has been tried yet."\}}\par\smallskip
\noindent\textbf{Gemini~4 Argon} \,(\texttt{gemini-4-argon}; frozen label: correct)\\
\texttt{\{"id": "resubmit", "reason": "Because 40 frames failed due to WALLTIME and nothing has been tried yet, the guidance is to resubmit the failed parts unchanged."\}}
\end{iBox}

\begin{iPrompt}{iUser}{Box S9 \textbullet\ C33, user message as sent (guided prompt, repeat 0; system prompt: Box~S2 with the $G_5$ lines of Box~S4)}
System: ca3ti2o7
Failed: f5 Analyst, gate G5, code COVERAGE
Detail: grid minimum [-0.9192388155, -0.2121320344] Å lies 5.0 steps from the grid edge (< 50): the surface may continue downhill outside
Checks run by the gate:
[
 {
  "test": "coverage_complete",
  "status": "pass"
 },
 {
  "test": "minimum_inside_grid",
  "status": "fail",
  "code": "COVERAGE",
  "detail": "grid minimum [-0.9192388155, -0.2121320344] \u00c5 lies 5.0 steps from the grid edge (< 50): the surface may continue downhill outside"
 },
 {
  "test": "mirror_symmetry",
  "status": "pass"
 },
 {
  "test": "script_agrees",
  "status": "pass"
 }
]

Retries used on this step: 0 of 3
What has been tried so far in this run:
nothing yet

Options (choose one id):
[
 {
  "id": "accept_secondary_hold",
  "text": "D4: keep the secondary at zero and report the energy it would release as a caveat"
 },
 {
  "id": "densify",
  "text": "Go back to f2 and densify the grid"
 },
 {
  "id": "stop",
  "text": "Stop and ask the human"
 }
]
\end{iPrompt}
\begin{iBox}{iOut}{Box S10 \textbullet\ C33, stored answers of repeat 0 (all harmful: the reference is stop)}
\footnotesize
\noindent\textbf{Sonnet~5.5} \,(\texttt{claude-sonnet-5-5}; frozen label: harmful)\\
\texttt{\{"id": "densify", "reason": "The COVERAGE failure says the grid minimum lies only 5.0 steps from the grid edge (\textless{} 50) so the surface may continue downhill outside, which is a grid-coverage problem that the guidance says to fix by going back to f2 and densifying, and nothing has been tried yet."\}}\par\smallskip
\noindent\textbf{GPT-6.1 Sol} \,(\texttt{gpt-6.1-sol}; frozen label: harmful)\\
\texttt{\{"id": "densify", "reason": "The gate reports a COVERAGE failure because the grid minimum is too close to the edge, and densifying has not yet been tried."\}}\par\smallskip
\noindent\textbf{Gemini~4 Argon} \,(\texttt{gemini-4-argon}; frozen label: harmful)\\
\texttt{\{"id": "densify", "reason": "The gate failed with code COVERAGE because the grid minimum [-0.9192388155, -0.2121320344] \AA{} lies 5.0 steps from the grid edge, so we must go back to f2 and densify."\}}
\end{iBox}

\begin{scriptsize}
\setlength{\tabcolsep}{3pt}
\begin{longtable}{@{}>{\raggedright}p{0.6cm}>{\raggedright}p{1.2cm}>{\raggedright}p{6.3cm}>{\raggedright}p{4.1cm}>{\raggedleft}p{0.95cm}>{\raggedright\arraybackslash}p{1.1cm}@{}}
\caption{\textbf{The 50 intake requests.} Every request exactly as given to the models (v1, the primary run; the requests are identical in v2), written and fixed with a hash before any model saw them (Methods, main-text Section~\Mref{sec:eval}). Requests E01--E39 ask for the 13 run configurations of Extended Data Table~\Mref{tab:reruns}, each in three styles (expert, terse, non-specialist); U01--U05 are underspecified, O01--O03 out of scope and N01--N03 name materials with no stored calculations. Two expert requests come with Materials Project identifiers attached. Expected: the reference configuration (system, parent$\to$child space-group numbers, analysis, strain policy, system profile when it differs from the system name, run budget; other accepted values are listed in the frozen request file), or the expected outcome. Models: calls with the expected outcome, of 20 (four models, five repeats; for clear requests, every critical field right). Keyword: whether the keyword lookup gave the expected outcome. Totals and rates by style in Supplementary Table~\ref{tab:si_E}; prompts in Supplementary Section~\ref{sec:si_prompts}.}\label{tab:si_requests}\\
\toprule
ID & Style & Request (verbatim) & Expected & Models & Keyword\\
\midrule\endfirsthead
\toprule
ID & Style & Request (verbatim) & Expected & Models & Keyword\\
\midrule\endhead
\bottomrule\endfoot
E01 & expert & CsSnI3 Pm-3m (\#221) -\textgreater{} Pnma (\#62): M3+ (+) R4+ tilt plane, fixed cubic cell with strain at zero, cssni3 profile, Landau fit, at most 1100 runs. \textit{[attached: mp-614013; mp-568570]} & CsSnI$_3$ 221$\to$62; plane; P1; 1{,}100 & 20/20 & yes\\
E02 & terse & map the CsSnI3 octahedral-tilt landscape & CsSnI$_3$ 221$\to$62; plane; P1; 1{,}100 & 20/20 & yes\\
E03 & non-spec. & Why does cubic CsSnI3 distort into its orthorhombic low-temperature phase? Compute the energy surface of the tilting. & CsSnI$_3$ 221$\to$62; plane; P1; 1{,}100 & 20/20 & yes\\
E04 & expert & CsSnI3 \#221 -\textgreater{} \#62, M3+/R4+ plane at fixed cubic cell, cssni3 profile, \textless{}= 1100 runs, plus the D4 hard-secondary check: X5+ scans at the origin, on the ring and near the child (0.8 of the child primaries). & CsSnI$_3$ 221$\to$62; plane; P1; 1{,}100; + X$_5^+$ check & 20/20 & no\\
E05 & terse & CsSnI3 tilt landscape + check the X5+ held at zero & CsSnI$_3$ 221$\to$62; plane; P1; 1{,}100; + X$_5^+$ check & 20/20 & no\\
E06 & non-spec. & Compute the tilting energy map of CsSnI3, and check whether a distortion we leave out (kept at zero) would lower the energy if we let it in. & CsSnI$_3$ 221$\to$62; plane; P1; 1{,}100; + X$_5^+$ check & 20/20 & no\\
E07 & expert & Ca3Ti2O7 I4/mmm (\#139) -\textgreater{} A2\_1am (\#36): X3- rotation x X2+ tilt energy surface, fixed cell, GM5- held at zero, ca3ti2o7 profile, \textless{}= 1400 runs. \textit{[attached: mp-4163]} & Ca$_3$Ti$_2$O$_7$ 139$\to$36; plane; P1; 1{,}400 & 20/20 & yes\\
E08 & terse & map the Ca3Ti2O7 rotation-tilt energy surface & Ca$_3$Ti$_2$O$_7$ 139$\to$36; plane; P1; 1{,}400 & 20/20 & yes\\
E09 & non-spec. & Ca3Ti2O7 is said to be a hybrid improper ferroelectric. Compute the energy map of the two octahedral rotations that make it polar. & Ca$_3$Ti$_2$O$_7$ 139$\to$36; plane; P1; 1{,}400 & 20/20 & yes\\
E10 & expert & P6\_3/mmc -\textgreater{} P6\_3cm InMnO3, K3-GM2- plane at Phi = 0, fixed (child) lattice, LDA+U A-AFM profile, Meier-form fit, \textless{}= 230 runs. & InMnO$_3$ 194$\to$185; plane; P1; \texttt{inmno3\_aafm\_v2}; 230 & 0/20 & yes\\
E11 & terse & map the InMnO3 trimerization landscape & InMnO$_3$ 194$\to$185; plane; P1; \texttt{inmno3\_aafm\_v2}; 230 & 16/20 & yes\\
E12 & non-spec. & How does InMnO3 become ferroelectric? Compute the energy surface. & InMnO$_3$ 194$\to$185; plane; P1; \texttt{inmno3\_aafm\_v2}; 230 & 20/20 & yes\\
E13 & expert & InMnO3 \#194 -\textgreater{} \#185: joint fit of the K3-GM2- plane (Phi = 0) and the hat wedge (Phi 0-30 deg), fixed child cell, LDA+U A-AFM profile, \textless{}= 420 runs; report the hat height and brim barrier. & InMnO$_3$ 194$\to$185; joint fit; P1; \texttt{inmno3\_aafm\_v2}; 420 & 1/20 & no\\
E14 & terse & InMnO3 joint plane + hat fit & InMnO$_3$ 194$\to$185; joint fit; P1; \texttt{inmno3\_aafm\_v2}; 420 & 20/20 & no\\
E15 & non-spec. & For InMnO3, fit one energy model to both the flat trimerization-polarization map and the energies at different trimerization angles, and tell me how high the Mexican hat is. & InMnO$_3$ 194$\to$185; joint fit; P1; \texttt{inmno3\_aafm\_v2}; 420 & 20/20 & no\\
E16 & expert & P6\_3/mmc -\textgreater{} P6\_3cm ErMnO3, K3-GM2- plane at Phi = 0 plus the GM2- axis, fixed child lattice, LDA+U A-AFM profile (ermno3\_aafm), Meier-form fit, \textless{}= 240 runs. & ErMnO$_3$ 194$\to$185; plane; P1; \texttt{ermno3\_aafm}; 240 & 15/20 & yes\\
E17 & terse & map the ErMnO3 trimerization landscape & ErMnO$_3$ 194$\to$185; plane; P1; \texttt{ermno3\_aafm}; 240 & 20/20 & yes\\
E18 & non-spec. & How does ErMnO3 become ferroelectric? Compute the energy surface. & ErMnO$_3$ 194$\to$185; plane; P1; \texttt{ermno3\_aafm}; 240 & 20/20 & yes\\
E19 & expert & ErMnO3 \#194 -\textgreater{} \#185: joint fit of the K3-GM2- plane (Phi = 0, with the GM2- axis) and the hat wedge, fixed child cell, ermno3\_aafm profile, \textless{}= 420 runs; hat height and brim barrier. & ErMnO$_3$ 194$\to$185; joint fit; P1; \texttt{ermno3\_aafm}; 420 & 20/20 & no\\
E20 & terse & ErMnO3 joint plane + hat fit & ErMnO$_3$ 194$\to$185; joint fit; P1; \texttt{ermno3\_aafm}; 420 & 20/20 & no\\
E21 & non-spec. & For ErMnO3, fit one energy model to both the flat trimerization-polarization map and the energies at different trimerization angles, and tell me how high the Mexican hat is. & ErMnO$_3$ 194$\to$185; joint fit; P1; \texttt{ermno3\_aafm}; 420 & 20/20 & no\\
E22 & expert & BiFeO3 Pm-3m (\#221) -\textgreater{} R3c (\#161): six-component GM4- (+) R5- grid on the 10-atom fcc(2a\_c) cell, fixed cubic lattice at the child volume, PBEsol+U G-AFM profile, invariant-polynomial fit, \textless{}= 10400 runs. & BiFeO$_3$ 221$\to$161; 6D grid; P1; 10{,}400 & 20/20 & yes\\
E23 & terse & full 6D polarization-rotation energy grid for BiFeO3 & BiFeO$_3$ 221$\to$161; 6D grid; P1; 10{,}400 & 20/20 & yes\\
E24 & non-spec. & In BiFeO3 the polarization and the rotation of the oxygen octahedra can each point in three directions. Compute the energy as a function of all six and find the stable phases. & BiFeO$_3$ 221$\to$161; 6D grid; P1; 10{,}400 & 20/20 & yes\\
E25 & expert & HfO2 P4\_2/nmc (\#137) -\textgreater{} Pca2\_1 (\#29): straight-line irrep path, amplitudes and strain scaled together, 10 points, hfo2 (PBE) profile, \textless{}= 20 runs. & HfO$_2$ 137$\to$29; path; P3; 20 & 20/20 & yes\\
E26 & terse & tetragonal -\textgreater{} polar orthorhombic HfO2 path & HfO$_2$ 137$\to$29; path; P3; 20 & 20/20 & no\\
E27 & non-spec. & How much energy does it take for tetragonal hafnia to turn into the ferroelectric orthorhombic phase? Compute the energy along the way. & HfO$_2$ 137$\to$29; path; P3; 20 & 20/20 & no\\
E28 & expert & HfO2 P4\_2/nmc (\#137) -\textgreater{} Pbcn (\#60): straight-line irrep path, amplitudes and strain scaled together, 10 points, hfo2 (PBE) profile, \textless{}= 20 runs. & HfO$_2$ 137$\to$60; path; P3; 20 & 20/20 & yes\\
E29 & terse & tetragonal -\textgreater{} Pbcn HfO2 path & HfO$_2$ 137$\to$60; path; P3; 20 & 20/20 & no\\
E30 & non-spec. & How much energy does it take for tetragonal hafnia to turn into its non-polar orthorhombic (Pbcn) phase? Compute the energy along the way. & HfO$_2$ 137$\to$60; path; P3; 20 & 20/20 & no\\
E31 & expert & HfO2 P4\_2/nmc (\#137) -\textgreater{} P2\_1/c (\#14): straight-line irrep path, amplitudes and strain scaled together, 10 points, hfo2 (PBE) profile, \textless{}= 20 runs. & HfO$_2$ 137$\to$14; path; P3; 20 & 20/20 & yes\\
E32 & terse & tetragonal -\textgreater{} monoclinic HfO2 path & HfO$_2$ 137$\to$14; path; P3; 20 & 20/20 & no\\
E33 & non-spec. & How much energy does it take for tetragonal hafnia to turn into the monoclinic ground state? Compute the energy along the way. & HfO$_2$ 137$\to$14; path; P3; 20 & 20/20 & no\\
E34 & expert & HfO2 Pbcn (\#60) -\textgreater{} Pca2\_1 (\#29): straight-line irrep path, amplitudes and strain scaled together, 10 points, hfo2 (PBE) profile, \textless{}= 20 runs. & HfO$_2$ 60$\to$29; path; P3; 20 & 20/20 & yes\\
E35 & terse & Pbcn -\textgreater{} polar orthorhombic HfO2 path & HfO$_2$ 60$\to$29; path; P3; 20 & 20/20 & no\\
E36 & non-spec. & How much energy does it take for the non-polar orthorhombic (Pbcn) phase of hafnia to turn into the ferroelectric orthorhombic phase? Compute the energy along the way. & HfO$_2$ 60$\to$29; path; P3; 20 & 20/20 & no\\
E37 & expert & HfO2 Pbcn (\#60) -\textgreater{} P2\_1/c (\#14): straight-line irrep path, amplitudes and strain scaled together, 10 points, hfo2 (PBE) profile, \textless{}= 20 runs. & HfO$_2$ 60$\to$14; path; P3; 20 & 20/20 & yes\\
E38 & terse & Pbcn -\textgreater{} monoclinic HfO2 path & HfO$_2$ 60$\to$14; path; P3; 20 & 20/20 & no\\
E39 & non-spec. & How much energy does it take for the non-polar orthorhombic (Pbcn) phase of hafnia to turn into the monoclinic ground state? Compute the energy along the way. & HfO$_2$ 60$\to$14; path; P3; 20 & 20/20 & no\\
U01 & under-spec. & map the BiFeO3 landscape & ask one question, or the only stored analysis (six-dimensional grid) & 20/20 & yes\\
U02 & under-spec. & hafnia transition & ask one question & 20/20 & yes\\
U03 & under-spec. & compute the energy barrier for HfO2 going to the monoclinic phase & ask one question & 20/20 & yes\\
U04 & under-spec. & map the octahedral tilt landscape of the perovskite & ask one question & 20/20 & yes\\
U05 & under-spec. & ErMnO3, please & ask one question & 18/20 & no\\
O01 & out of scope & Compute the group-subgroup path from monoclinic HfO2 (P2\_1/c, \#14) up to tetragonal P4\_2/nmc (\#137), with \#14 as the parent. & decline or ask & 20/20 & yes\\
O02 & out of scope & Map the CsSnI3 tilt plane with Quantum ESPRESSO and a hybrid functional. & decline or ask & 20/20 & no\\
O03 & out of scope & Run 20 ps of NVT molecular dynamics on BiFeO3 at 600 K and give me the polarization autocorrelation. & decline or ask & 20/20 & no\\
N01 & new mat. & PbTiO3 Pm-3m (\#221) -\textgreater{} P4mm (\#99): polar GM4- landscape, energy surface of two polarization components (P\_z vs P\_x) on a fixed cubic cell. & valid; needs new DFT & 20/20 & no\\
N02 & new mat. & YInO3 P6\_3/mmc -\textgreater{} P6\_3cm: K3-GM2- plane at Phi = 0, fixed child cell. & valid; needs new DFT & 20/20 & no\\
N03 & new mat. & Ca3Mn2O7 I4/mmm (\#139) -\textgreater{} A2\_1am (\#36): X3- rotation x X2+ tilt energy surface, fixed cell. & valid; needs new DFT & 20/20 & no\\
\end{longtable}
\end{scriptsize}

\subsection{Full account of the gated replays and evaluation}\label{sec:si_eval_ext}
This subsection gives the complete account of the gated replays, fault injection, recovery benchmark, option replays and intake test summarized in main-text Section~\Mref{sec:autonomy}.

To test the workflow as a whole rather than step by step, IRIS was rerun end to end on the stored DFT energies. In this replay mode $f_3$ matches every frame it builds to a stored structure and links its energy, so that no calculation is repeated and an unmatched frame stops the run; every other step, from the decomposition to the re-derivation of the reported numbers by $f_6$, runs as described in Methods. All six systems have been rerun, in 13 run configurations (Extended Data Table~\Mref{tab:reruns}). Every frame was matched to a stored structure (fractional coordinates within $5\times10^{-5}$, after removing rigid shifts of at most 0.015~Å in the older InMnO$_3$ frames; for BiFeO$_3$, every atom within $4\times10^{-5}$~Å of the position built from the six vectors), every gate passed except where the two human decisions described below were required, and $f_6$ re-derived every reported number by an independent route (in replay, the energies of that route are read from the stored campaign files). The reruns reproduce the landscapes and Landau models of main-text Sections~\Mref{sec:res_perov}--\Mref{sec:res_bfo}. For BiFeO$_3$, IRIS rebuilt the six-component plan from its own vectors and its own 96 operations, with the same 9{,}998 and 3{,}306 orbits, the same 6{,}724 and 2{,}873 orbits kept by the contact cut and the same representatives as the stored campaign; it linked all 10{,}307 energies and 10{,}287 gradient vectors (the 20 SCF-limited runs keep their energies only), checked the 95 symmetry controls (equal to 0.0015~meV/f.u.), and refitted the degree-12 model with coefficients identical to the published ones to round-off, the same nine stationary points and the same hold-out error. The CsSnI$_3$ coefficients agree to $10^{-9}$, and the manganite coefficients differ only in the third decimal, because the stored grid coordinates were rounded to four decimals; refitting on the rounded coordinates reproduces the published values exactly. Two human decisions were needed. The first was the borderline InMnO$_3$ selection of rule R5, for which the human kept $\Gamma_2^-$ as an axis; the answer is applied inside the selection rules, and the alternatives have measurable consequences (Section~\ref{sec:si_replay_found}). The second came from the hard-secondary scans of CsSnI$_3$ (D4): $X_5^+$ is stiff at the parent and on the $M_3^+$ axis, but at 0.8 of the child primaries the trilinear $M_3^+R_4^+X_5^+$ term gives it a linear term, and relaxing it releases 1.9~meV/f.u., more than the 1.4~meV/f.u.\ ring corrugation; $G_5$ therefore flagged it, and the human kept $X_5^+$ at zero on the plane with this caveat (Supplementary Table~\ref{tab:perov}, Supplementary Fig.~\ref{fig:s_x5}). The reruns also exposed two errors that the original Ca$_3$Ti$_2$O$_7$ campaign had not caught, both now corrected: the plane axes had been normalized for the two-cell supercell of the decomposition instead of the four primitive cells of the frame, which the norm check of $G_1$ flagged (Extended Data Table~\Mref{tab:errors}), and a ranking of the irreps on raw amplitudes would have taken the polar $\Gamma_5^-$, 75\% of which is a rigid translation, as a primary instead of $X_2^+$ (Methods).

The gates were then tested by injecting faults through the run configuration, never by altering data or code. A seed set of nine faults on CsSnI$_3$ (F001--F009) covers the error classes of Extended Data Table~\Mref{tab:errors} and the compute failures of the campaigns. Five were fixed by a stored rule (a walltime kill and an SCF failure, both resubmitted through $f_3$; a rigid translation left in a mode vector; a decomposition without a basis; and the No.~55 single-mode frame), and in each case the final numbers were identical to those of the fault-free run; the other four went to the human, as designed (a grid refinement that exceeded the budget, a missing stored frame, cancelled jobs and an exhausted retry budget). The full set of 53 faults over the six systems (Methods, main-text Section~\Mref{sec:eval}) contains 30 errors, which corrupt a result if they pass (failed or SCF-limited job parts, a rigid translation left in a mode vector, a single-mode frame built on the wrong lattice, an inconsistent DFT setting or formula-unit count, a missing frame, cancelled jobs), and 23 limit faults, which make a threshold stricter, or a search narrower, on data that are correct, and so test that each limit is enforced. With all gates on, every fault was caught by the gate expected for it ($G_4$ 24, $G_5$ 13, $G_2$ 8, $G_1$ 6, $G_3$ 2; for the six InMnO$_3$ faults after the published R5 decision at $G_2$, which the first-gate column of Supplementary Table~\ref{tab:si_faultlist} lists) and none ended with a corrupted result (main-text Fig.~\Mref{fig:agentic}d; Supplementary Table~\ref{tab:si_faults}). Of the 30 errors, 21 were fixed by a stored rule (four of these runs also passed through the published R5 or D4 decision) and completed with numbers identical to those of the fault-free replay; the other 9 stopped with the human, five because the failure returned after its rule had been applied and four because no rule applies (a missing frame, cancelled jobs, and an inconsistent DFT setting and formula-unit count in HfO$_2$). Of the 23 limit faults, 6 completed with identical numbers after the basis search was widened (by the stored rule in five, by the human in the sixth, whose retry budget had been set to zero) and 17 stopped with the human, as a tightened limit should. With the gates off, the same 53 faults gave 14 silent corruptions (failed or SCF-limited parts of a job dropped from the fit without notice, which in the CsSnI$_3$ seed set left the fit on 530 and 515 of its 545 nodes and shifted the coefficients by about $10^{-3}$, and a wrong formula-unit count in HfO$_2$), 15 crashes and 24 harmless outcomes, most of them limit faults whose data were fine. Removing one gate at a time (107 runs) shows where each gate earns its place (main-text Fig.~\Mref{fig:agentic}d; Supplementary Table~\ref{tab:si_ablation}). On the nine seed faults, removing a gate that did not catch the fault never changed a run (54 of 54). With the catching gate of an error fault removed, 27 of the 30 were still stopped by a later check, but only after the work had been done; 23 of them by a check that also runs in production: 21 at $G_5$ (15 by \texttt{MISSING\_FRAMES}, the fit refusing a grid with frames missing; 5 by the path analysis failing on a missing point; 1 by the cross-setting check, which uses the energies of the other HfO$_2$ paths and so needs them computed in the same campaign), 1 at $G_4$ and 1 at $G_5$ after $G_3$ was removed. Three were stopped by a check that exists only in replay, at $G_3$, where a wrongly built frame has no stored structure to match. The last one crashed for a reason the ablation itself creates: removing $G_2$ also removes the R5 decision of InMnO$_3$, so no grid is built. In a first replay, before the \texttt{MISSING\_FRAMES} check, 9 more had been stopped only at $G_6$, which in replay compares with the stored campaign: 8 job faults that, with no gates at all, corrupted the result silently, and the missing frame of F007 (Section~\ref{sec:si_replay_found}). The remaining 3 were not caught: an inconsistent INCAR setting in HfO$_2$, and the No.~55 frame and its ErMnO$_3$ counterpart. They were harmless in replay only because the stored energies are correct or the frame goes unmatched; in production the No.~55 frame was computed, and cost 961 runs before it was found in the first CsSnI$_3$ pass. Before these faults were stopped, the runs exceeded those of the gated run by 27{,}138 DFT runs, most of them on the BiFeO$_3$ grid; this is a replay-based count of counterfactual submissions, taken from the submission records of $f_3$ and net of the resubmissions that the gated run makes, not a measured saving of production compute. For the three faults stopped only by a replay check, a production run would have spent these runs and accepted the result. Replay therefore understates the value of $G_2$. Limit faults are harmless when their gate is removed, by construction. Runs killed after $f_2$, $f_3$ or $f_5$ (CsSnI$_3$) or after $f_3$ (HfO$_2$ T1) and resumed in a new process gave reports, gate outcomes and job submissions identical to the uninterrupted runs (4 of 4).

The recovery sub-agents were evaluated on 53 benchmark cases, each the prompt of a real gate failure with its menu of fixes (Methods, main-text Section~\Mref{sec:eval}), answered ten times by each of nine models from three providers, and against deterministic policies given the same knowledge (main-text Fig.~\Mref{fig:agentic}c; Supplementary Table~\ref{tab:si_models} and~\ref{tab:si_A}, Supplementary Fig.~\ref{fig:s_agentic_models}). With the guided prompt, which carries the one-line fix knowledge of the rule table, the models chose the correct fix in 72--91\% of decisions and a harmful one in 1.7--16\%, against 42\% correct for always stopping and 51--52\% harmful for the first listed or a random option; without the guidance every model did worse (48--79\% correct, 17--43\% harmful), most of all on the 16 held-out cases, where it fell below always stopping. The rules alone and the decision tree did better than every model. The stored rules alone, stopping whenever no rule applies, were correct in 96\% of decisions with none harmful, and a decision tree written from the same guidance lines in 94\%, with one harmful choice; both missed only the two judgment calls, where they stop. Retrieving the fix of the most similar previous case was correct in 81\%. These are in-sample numbers, since the rules were seeded by the campaigns the cases come from; scored with every rule seeded by a case's own campaign removed, the rules are correct in 89\%, with every lost case going to the human and none to a wrong fix, and the models, with the same sentence removed from their prompt, chose the right fix without being told in 79\% of decisions on the seven cases affected, with 2.5\% harmful (Section~\ref{sec:si_loso}). Harmful choices by the models came almost entirely from decoys, failures whose correct answer is to stop: with guidance no model chose a harmful fix on any of the 29 failures that a stored rule fixes, whereas on decoys the median harmful rate was 25\%. Wrong fixes were chosen consistently, so that requiring most of the ten repeats to agree before acting hardly lowered harm (Supplementary Fig.~\ref{fig:s_agentic_robust}a). The benchmark therefore measures recovery from recorded failures, for which a rule table is the better tool; it does not test failures of kinds not seen when the rules and prompts were written.

In the 53 faulted runs (main-text Fig.~\Mref{fig:agentic}b; Supplementary Table~\ref{tab:si_A} and~\ref{tab:si_interventions}), the language model, retrieval and the rules with a human completed the same 27 runs, all with numbers identical to the fault-free replay; the decision tree completed 23, because it stops at the two judgment calls, which the model answered as the human had. Counted as every hand-off to the human, whether or not a question is asked, the human was involved 34 times with rules alone, 31 with the tree, 27 with the model and 22 with retrieval, which in exchange submitted the most runs (940 against 770). The model's measured contribution to recovery is therefore the judgment calls, both of which change the result: the D4 answer keeps $X_5^+$ at zero with a caveat, and the R5 answer is applied inside the selection rules, so that excluding $\Gamma_2^-$ removes the axis and the run stops at $G_5$ without a $(Q,P)$ fit, and the anchor-only option asks for frames the stored plane does not contain (Section~\ref{sec:si_replay_found}; see the option replays below).

Every option on every menu was then executed: each of the 53 failures was resumed from its checkpoint once per option, 208 branches in all, with every gate on (main-text Fig.~\Mref{fig:agentic}d; Supplementary Table~\ref{tab:si_C}). No branch completed with a result outside the frozen tolerance of the published numbers. Of the 208, 33 completed correctly, 20 completed after a change of SCF mixing only, whose identical energies are assumed by replay, 24 chose a wrong option that the next gate refused before a rule repaired the run, 76 were stopped by a later gate, 2 asked for a frame that replay does not have (the R5 anchor frames of InMnO$_3$) and 53 stopped by choice. Twenty-four of the 76 drop the frames of a failed job and continue; all are stopped at $G_4$ by \texttt{MISSING\_FRAMES}, seven of which had been stopped only by the replay-only comparison of $G_6$ in a first replay (Section~\ref{sec:si_replay_found}). With the gates switched off after the fix, the 124 branches whose option is neither the reference nor stop gave 20 wrong results with nothing flagged: fits without the frames of a failed job, a wrong number of formula units, and a rigid translation left in the frames. The replays also corrected the reference labels in four ways: changing only the SCF mixing does not change converged energies (13 branches labelled harmful), an unchanged resubmission after an SCF failure completed only because the scripted failures had run out (9), densifying the grid was refused by the budget in three cases where it was labelled correct, and, in a first replay, excluding the borderline irrep in InMnO$_3$ changed nothing because the R5 answer was not read beyond letting the run continue; that gap is closed (Section~\ref{sec:si_replay_found}): the answer is now applied, excluding $\Gamma_2^-$ builds a one-axis grid on which the $(Q,P)$ model cannot be fitted and the run stops at $G_5$, and the anchor-only option asks for a frame the stored plane does not contain. Executing the recovery policy itself on the 53 recovery cases of the benchmark, with a stored rule first and the reference model for the rest, completed 51.9\% of the cases correctly without the human and accepted no wrong result with the gates on, against 58.5\% completed and a wrong result accepted in 7.5\% with the gates off (main-text Fig.~\Mref{fig:agentic}e; Supplementary Table~\ref{tab:si_stack}).

Finally, the intake step was built and tested on 50 frozen requests (Methods, main-text Section~\Mref{sec:eval}): each of the 13 run configurations of Extended Data Table~\Mref{tab:reruns} asked for in three styles (expert, terse and non-specialist), 5 underspecified requests, 3 out of scope and 3 for materials with no stored calculations, each answered five times by four models (1{,}000 calls) and once each by a rule-based parser and a keyword baseline (main-text Fig.~\Mref{fig:agentic}a; Supplementary Table~\ref{tab:si_E}). On clear requests the models produced a configuration with every critical field right in 180--186 of 195 calls (92--95\%), against 25 of 39 for a rule-based parser and 20 of 39 for the keyword lookup; on terse and non-specialist requests in 65 of 65 calls (61 of 65 for one model on terse requests), against 8 and 7 of 13 for the parser and 5 of 13 for the lookup. The parser was written from the schema, the tables and the system prompt without consulting the requests and frozen before scoring (Section~\ref{sec:si_parser}); six of its seven wrong fields came from one bug, the material's name read as a choice of the profile of the same name, and a second version with that one fix reaches 29 of 39 (10, 10 and 9 of 13 by style). On expert requests it is within the range of the models (10 of 13 against 50--56 of 65 calls); its remaining misses are requests that name a transition by one symbol and one crystal-system word, the BiFeO$_3$ landscape asked for as a plane, and a D4 request phrased in words it has no rule for, which is what a fixed vocabulary cannot anticipate. The models asked a question, or built the only stored analysis, in 23--25 of the 25 calls per model on underspecified requests, refused or asked on every out-of-scope request and flagged every new material as needing new calculations without borrowing another material's settings, where the parser managed 3 of 5, 3 of 3 and 3 of 3 and the keyword lookup 4 of 5, 1 of 3 and 0 of 3. Every configuration with all critical fields right was run end to end in replay: 752 calls (732 on the 50 frozen requests and 20 on the underspecified request U01, once its field was supplied) gave the 13 configurations of Extended Data Table~\Mref{tab:reruns} and reproduced the published numbers exactly, and 60 new-material calls stopped as needing new calculations. The errors that matter are valid but wrong choices, which $G_0$ cannot catch: 9--15 calls per model passed $G_0$ with a wrong field. Of these, 9 or 10 per model came from two InMnO$_3$ requests (E10, E13) for which the prompt described the six InMnO$_3$ profiles identically, so that the models chose a PBEsol$+U$ profile when LDA$+U$ was asked for; the other 5, all by GPT-6.1 Sol, chose a wrong analysis for E16 (Supplementary Table~\ref{tab:si_requests}). A declared rerun in which only the profile descriptions were changed, to name the functional and the magnetic order, removed most of these errors (0--5 per model); the models then asked which of the two LDA$+U$ variants to use, which also raised the unnecessary questions on clear requests from 0--4 to 1--10 per model (Supplementary Table~\ref{tab:si_E}), a trade of a few wrong profiles for a few questions. Taken together, the language model earns its place at the interface, turning plain-language requests into checked configurations, which a keyword lookup cannot do and a rule-based parser does only for requests phrased as it expects, and in the judgment calls; on recorded failures, stored rules do better, and the gates keep out the wrong results of either.

\subsection{Extended discussion of the agentic layer and limitations}\label{sec:si_limits}
This subsection gives in full the discussion of the agentic layer and the limitations summarized in the main-text Discussion.

Because the language model chooses only among listed fixes and answers from a human become new rules, the system improves by accumulating rules rather than by retraining. Its contribution can therefore be measured directly, against deterministic policies given the same knowledge and against a run with the model switched off. The answer is specific (main-text Section~\Mref{sec:autonomy}). At the interface, the model turns plain-language requests, terse or non-specialist, into checked configurations that reproduce the published numbers, which a keyword lookup cannot do and a rule-based parser does for expert phrasing only, and it recognizes requests it should question, decline or flag as needing new calculations. In recovery from recorded failures, stored rules and a decision tree written from the same guidance choose better than any of nine models, whose harmful choices were concentrated on failures that should be stopped and were made consistently, so that agreement between repeated answers is no safeguard; there, the model's measured role is to answer the judgment questions as the human did. In both places the safeguard is the gate that checks every choice again: executing every recovery option showed no wrong result accepted with the gates on, against 20 of 124 with them off, while a valid but wrong choice at the intake, such as the wrong electronic-structure profile, passes $G_0$ and must be prevented by describing each option by its physics. This places IRIS at the opposite end from tool-calling agents that plan and route a workflow themselves~\cite{Yao2023,Zou2025,Pham2025}: restricting the model to listed options makes each of its decisions checkable and its contribution measurable, which is the kind of rigorous assessment that benchmarks of scientific agents call for~\cite{Chen2025}, and it keeps the safeguards in code rather than in the prompt~\cite{Tang2025}.

\paragraph{Limitations.} Landscapes are computed on a fixed lattice by default, so strain coupling enters only through explicit strain-interpolated or relaxed anchors; we report the resulting bias where it matters. The HfO$_2$ frozen paths give upper bounds on the barriers, and the BiFeO$_3$ minimum-energy paths are computed on the fitted model at fixed cell, at zero field, without strain relief or domain walls. Of the policies for grids too large to sample fully, D4 and D5 are implemented for the cases shown but have not yet been tested broadly, and D3 is not yet implemented. The threshold $A_{\mathrm{thr}}$ (0.05~Å here) is a policy choice, and selections near it are flagged rather than decided automatically (rule R5). Magnetic order is fixed in every calculation, so spin--lattice couplings that require the spin--orbit interaction are not part of the landscapes. The intake was tested on 50 requests written by us with the stored configurations in view, not by independent users, and the chat front end was not evaluated; a valid but wrong choice at the intake passes $G_0$. The end-to-end evaluation replays stored energies, so a fault that would cause new calculations in production, such as a wrongly built frame, is seen in replay only as a missing frame (main-text Section~\Mref{sec:autonomy}); the benchmark cases come from the six systems of this work, and the held-out cases were chosen to cover them rather than drawn at random; the leave-one-system-out scoring of Section~\ref{sec:si_loso} separates the rules from the campaigns that seeded them, but the cases and the fault taxonomy were written by the authors, so transfer to independently authored cases and to unseen failure mechanisms is not tested. The deterministic policies were written after the aggregate model results had been seen, although from the prompts and frozen cases only. The consequences of recovery options were measured in replay, where resubmitted runs take their stored energies, so that a fix that changes the inputs of a calculation is credited with the stored result. The trilinear $\Gamma_5^-$ direction of Ca$_3$Ti$_2$O$_7$ is not yet mapped. Some checks exist only in replay ($G_3$ matching every frame to a stored structure, and the $G_6$ comparison with the stored campaign energies), so faults stopped only by them would not be stopped there in production; after the \texttt{MISSING\_FRAMES} check these are the three wrongly built frames caught at $G_3$ when $G_2$ is removed (Supplementary Table~\ref{tab:si_ablation}). The intake was compared with a keyword lookup and with a rule-based parser written by us from the schema; a parser tuned on the requests, or an independently written one, was not tested. The prospective runs are the four BaTiO$_3$ campaigns of Section~\ref{sec:si_live}.

\section{Prospective BaTiO$_3$ campaign}\label{sec:si_live}

\subsection{Runs}
\begin{figure}[!tp]
\centering
\includegraphics[width=\textwidth]{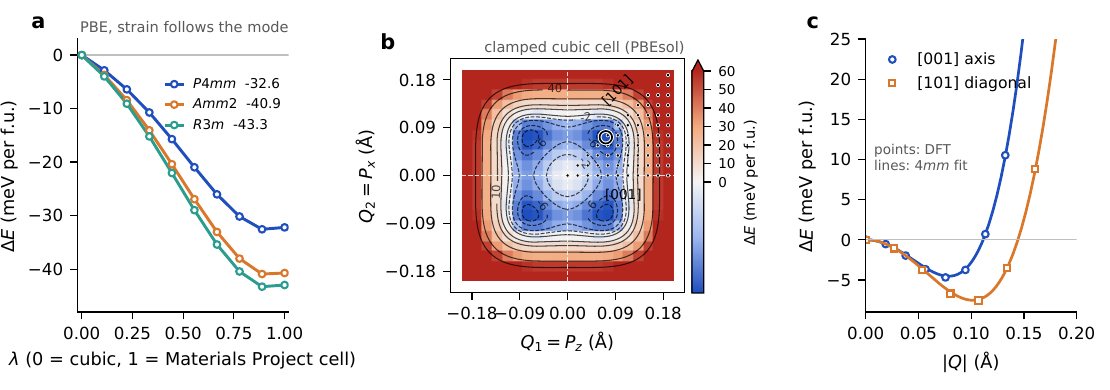}
\caption{\textbf{The prospective BaTiO$_3$ campaign.} Every energy was computed by IRIS in live mode from typed requests (main-text Section~\Mref{sec:live}).
\textbf{a},~The three polar paths from cubic $Pm\bar{3}m$ to $P4mm$ (T1), $Amm2$ (T2) and $R3m$ (T3), ten frames each, PBE at the Materials Project (PBE) cells, with the strain interpolated alongside the $\Gamma_4^-$ mode; $\lambda=1$ is the Materials Project cell. Minima at $\lambda=0.89$: $-32.6$, $-40.9$ and $-43.3$~meV per formula unit, the known order $R<O<T$ (Supplementary Table~\ref{tab:si_batio3_paths}).
\textbf{b},~The $(P_z,P_x)$ plane of $\Gamma_4^-$ at the fixed cubic cell, a separate run with PBEsol at the PBE cubic cell (Section~\ref{sec:si_live_settings}): a $21\times21$ grid rebuilt by $4mm$ symmetry from the 66 computed orbit representatives (black dots, $0\le Q_2\le Q_1$; Supplementary Table~\ref{tab:si_batio3_plane_points}); contours are the fitted $4mm$ Landau expansion (Supplementary Table~\ref{tab:si_batio3_plane}); the circle marks its minimum, $-7.5$~meV per formula unit on the $[101]$ diagonal. The colour scale is clipped at 60~meV (the corner reaches 221~meV).
\textbf{c},~Cuts of \textbf{b} along the $[001]$ axis and the $[101]$ diagonal: computed points and the fit (PBEsol, fixed PBE cell). The anisotropy favours the diagonal, the fixed-cell picture of refs~\cite{cohen1992origin,kingsmith1994first}. Because \textbf{a} and \textbf{b} were run with different functionals, their well depths are reported separately and are not subtracted to quantify the strain contribution.}
\label{fig:s_live}
\end{figure}

Four live runs make up the campaign reported in main-text Section~\Mref{sec:live}: three straight-line paths from cubic $Pm\bar{3}m$ to $P4mm$, $Amm2$ and $R3m$, with the strain interpolated alongside the mode (policy P3), and one $(P_z,P_x)$ plane of $\Gamma_4^-$ at the clamped cubic cell. Each run started from a typed request in the chat interface (main-text Methods, Section~\Mref{sec:livemode}); the intake model was Claude Sonnet~5.5 in every run, and the only human input was the confirmation to submit. Supplementary Table~\ref{tab:si_batio3_runs} lists the runs and Supplementary Fig.~\ref{fig:s_live} shows the results. On disk the $Amm2$ and $R3m$ runs carry the tag T3 (directories \texttt{T3\_221-38} and \texttt{T3\_221-160}, the tag given at launch); in the text they are called T2 and T3 by phase.


The deposited campaign directory also holds two rehearsal runs and one repeat, kept because they are part of the record. The rehearsals (threads \texttt{PL} and \texttt{PLM}) are \emph{mock} runs of the plane: the intake, the Materials Project fetch, FINDSYM and ISODISTORT, the frame construction, every gate and the re-derivation ran exactly as in the live runs, but the PBS scheduler and VASP were replaced by a fake scheduler and a synthetic energy function (\texttt{tests/live\_fakes}), and the page and the audit log flag the run as \emph{MOCK}. Mock mode serves two purposes. It exercises a campaign end to end, including the job-submission and collection code, without spending cluster time: the plane was rehearsed twice at no cost before its 66 calculations were submitted, and the rehearsals exposed a defect in the construction of the second component (the generating rotation had to be taken about an atomic site, Section~\ref{sec:si_live_found}) that would otherwise have cost a wasted batch. It also separates the two sources of failure a prospective run can have: a mock run that passes shows that the crystallography, the grid, the gates and the executor's bookkeeping are right, so a failure in the subsequent live run can only come from the calculations or the cluster. The repeat (\texttt{PL2}) is an independent live run of the plane, started from a fresh request after PL1 had finished; it gave the same 66 energies and the same fitted coefficients as PL1 to every printed digit, the reproducibility of the whole chain from request to Landau fit, not of VASP alone.

\begin{table}[!htbp]\centering
\caption{\textbf{The four live runs.} Materials Project identifiers as fetched (the API returned alphabetic identifiers, decoded to the classic form); cells as fetched; wall time from the first audit entry to the last, which includes the ISODISTORT calls (3--5~min) and the queue; jobs are PBS parts of five static calculations each. Parent in all runs: mp-2998, $Pm\bar{3}m$, $a=4.0077$~\AA, 7.5~meV/atom above the hull. Every run passed all seven gates on the first attempt and put no decision to the human other than the confirmation to submit.}\label{tab:si_batio3_runs}
\small\begin{tabular}{lllrrrrl}
run & child & MP id / cell & f.u. & frames & jobs & wall (min) & $f_6$\\\hline
T1 & $P4mm$ (99) & mp-5986, 5 atoms & 1 & 10 & 2 & 6.6 & 63 of 63\\
T2 & $Amm2$ (38) & mp-5777, 10 atoms & 2 & 10 & 2 & 11.1 & 64 of 64\\
T3 & $R3m$ (160) & mp-5020, 15 atoms (hex.) & 3 & 10 & 2 & 6.1 & 65 of 65\\
PL1 & $P4mm$ (99) & mp-5986, 5 atoms & 1 & 66 & 14 & 10.0 & 351 of 351\\\hline
\multicolumn{8}{l}{totals: 96 calculations, 20 jobs, 34~min, $f_6$ 543 of 543}\\
\end{tabular}\end{table}

\subsection{Settings and a note on the functionals}\label{sec:si_live_settings}
All calculations used VASP with PAW potentials Ba\_sv, Ti and O, a plane-wave cut-off of 600~eV, electronic convergence to $10^{-7}$~eV, Gaussian smearing of 0.05~eV, no spin polarization and no ionic relaxation; the $\Gamma$-centred $k$-mesh is set on the parent cell by $k_i=\lceil|\mathbf{b}_i|/0.20\,\text{\AA}^{-1}\rceil$ ($8\times8\times8$ for the 5-atom cell) and kept for every frame of a run. The three paths used PBE, the functional with which the Materials Project cells were relaxed. The plane was launched from a separate configuration template that still carried PBEsol from an earlier rehearsal, so the plane is PBEsol at the PBE cubic cell: its cubic reference energy is $-39.4035$~eV against $-39.6052$~eV for PBE (Supplementary Tables~\ref{tab:si_batio3_paths} and \ref{tab:si_batio3_plane_points}). Each run is valid on its own: the paths give the ordering of the three wells with the strain interpolated, the plane the anisotropy of the polar mode at a fixed cell (the sign of $c-2b$ is the one both functionals give for BaTiO$_3$). What the two runs cannot do together is quantify the contribution of strain to the well depths, since the depth of the fixed-cell well ($-7.5$~meV per formula unit) is a PBEsol number at a PBE volume; the main text therefore reports the two sets of depths separately and attributes the role of strain to the literature, not to a subtraction of these runs. We report both runs as made. No gate flagged the mismatch: $G_4$ checks that INCAR, KPOINTS and POTCAR are identical \emph{within} a run, and the network check of $G_5$ compared the cubic reference energy per formula unit across the \emph{paths} of a campaign (agreement to 0.04--0.11~meV per formula unit here, Supplementary Table~\ref{tab:si_batio3_paths}) but did not include the plane. This is a boundary of the verification as it ran: the gates validated each campaign but did not enforce consistency between independently configured runs that are then read together. Two checks were added afterwards, and neither was active during the prospective campaign. The energy check: the plane's origin is now compared with $\lambda=0$ of every finished path of the campaign (code \texttt{NETWORK}; here the gap is 202~meV per formula unit against a tolerance of 1). The metadata check (\texttt{SETTINGS\_MISMATCH}, 31 October 2026; \texttt{nodes/f5\_plane2.py}, \texttt{gates.py}, \texttt{tests/test\_live\_mode.py}): before any energy of the plane is compared with the paths, the INCAR tags that fix the functional and the numerical settings (\texttt{GGA}, \texttt{ENCUT}, \texttt{EDIFF}, \texttt{ISMEAR}, \texttt{SIGMA}, \texttt{ISPIN}, the $+U$ tags, \texttt{PREC}, \texttt{LREAL}) and the POTCAR data sets of the plane's origin and of $\lambda=0$ of every other path are read from the written inputs and must be identical; a mismatch fails $G_5$ with the tags listed, and the run is handed to the human with the individual results intact: the code refuses the joint interpretation, not either result. On the recorded inputs of this campaign the check reports the three paths and the plane as individually valid and the cross-run comparison as incompatible (\texttt{GGA}: \texttt{PS} against \texttt{PE}), which is the diagnostic that would have stopped the comparison before it was written.

\subsection{Paths}
\begin{table}[!htbp]\centering\caption{\textbf{Live BaTiO$_3$ paths (PBE).} $\Delta E$ in meV per formula unit relative to the cubic point of each path; $E_0$ is the VASP total energy of the computed cell (1, 2 and 3 formula units) at $\lambda=0$ and at the minimum. All 30 calculations converged.}\label{tab:si_batio3_paths}
\small\begin{tabular}{c|rrr}
$\lambda$ & $P4mm$ (T1, 5 atoms) & $Amm2$ (T2, 10 atoms) & $R3m$ (T3, 15 atoms)\\\hline
0.000 & 0.000 & 0.000 & 0.000\\
0.111 & -2.850 & -3.722 & -3.987\\
0.222 & -6.410 & -8.424 & -9.083\\
0.333 & -10.726 & -14.061 & -15.203\\
0.444 & -15.707 & -20.390 & -22.033\\
0.556 & -20.965 & -26.925 & -28.977\\
0.667 & -26.023 & -33.048 & -35.393\\
0.778 & -30.158 & -38.016 & -40.433\\
0.889 & -32.552 & -40.907 & -43.270\\
1.000 & -32.195 & -40.715 & -42.957\\
\hline $E_0(\lambda=0)$ (eV/cell) & -39.60517 & -79.21041 & -118.81582\\
$E_0(\lambda=0)$ per f.u. (eV) & -39.605167 & -39.605203 & -39.605273\\
\end{tabular}\end{table}

The minima are at $\lambda=0.889$ for all three paths, $-32.552$, $-40.907$ and $-43.270$~meV per formula unit; the end points ($\lambda=1$, the Materials Project cells) lie 0.2--0.4~meV above the minima. No path has a barrier (monotonic soft-mode descent). The three profiles share the cubic reference: the energy per formula unit of the 5-, 10- and 15-atom cubic cells is $-39.605167$, $-39.605203$ and $-39.605273$~eV, so the three settings agree pairwise to 0.04--0.11~meV per formula unit (tolerance 1~meV); the $G_5$ network check of T3 against T1 and T2 saw 0.11 and 0.07~meV per formula unit.

\subsection{Plane}
The second component $\mathbf{e}_2$ is the image of the child's $\Gamma_4^-$ field $\mathbf{e}_1$ under the parent's four-fold rotation about $y$, taken about the first atomic site for which the rotation maps the ISODISTORT frame onto itself (recorded in the grid file); $G_2$ verified $|\mathbf{e}_2|/|\mathbf{e}_1|=1$ and every atom onto an equal atom. The grid is $21\times21$ over $[-q_\mathrm{max},q_\mathrm{max}]^2$ with $q_\mathrm{max}=1.2$ times the translation-free amplitude of the $P4mm$ child, 0.158~\AA\ (the result panel of Supplementary Fig.~\ref{fig:s_live_page_plane} prints the raw amplitude, 0.184~\AA), so $q_\mathrm{max}=0.189$~\AA, steps of 0.0189~\AA; the 441 points fall into 66 orbits under $4mm$ (Burnside: $(441+21+21+1+21+1+1+21)/8=66$; brute force: 66), of which one representative with $0\le Q_2\le Q_1$ was computed (Supplementary Table~\ref{tab:si_batio3_plane_points}). The weighted least-squares fit of the $4mm$-invariant expansion and the minimum are in Supplementary Table~\ref{tab:si_batio3_plane}.

\begin{table}[!htbp]\centering
\caption{\textbf{Landau fit of the clamped-cell plane (PBEsol).} $E=a(Q_1^2+Q_2^2)+b(Q_1^4+Q_2^4)+cQ_1^2Q_2^2+d(Q_1^6+Q_2^6)+e(Q_1^4Q_2^2+Q_1^2Q_2^4)$, meV per formula unit, $Q$ in \AA, $E(0,0)=0$, weights = orbit multiplicities. The minimum is searched numerically in the wedge $0\le Q_2\le Q_1$. Because $d,e<0$ the fitted surface turns over beyond the sampled range; it is used inside the grid only, which $G_5$ enforces (minimum inside the grid). The coefficients are the clamped (fixed-strain) ones and are not to be compared with stress-free phenomenological $\alpha_{ij}$ tables without the electrostrictive correction.}\label{tab:si_batio3_plane}
\small\begin{tabular}{lr}
$a$ (meV/\AA$^2$) & $-1480.2$\\
$b$ (meV/\AA$^4$) & $121\,577.3$\\
$c$ (meV/\AA$^4$) & $53\,545.7$\\
$d$ (meV/\AA$^6$) & $-177\,414.9$\\
$e$ (meV/\AA$^6$) & $-415\,973.0$\\
$c-2b$ (meV/\AA$^4$) & $-189\,608.9$ (diagonal favoured)\\
rms residual (meV) & 0.120\\
fitted minimum & $-7.539$~meV at $(Q_1,Q_2)=(0.072,0.072)$~\AA, on the $[101]$ diagonal\\
lowest computed point, axis / diagonal & $-4.696$ (P10, $(4,0)$) / $-7.569$ (P14, $(4,4)$)\\
$f_6$ & 351 of 351 numbers re-derived and equal\\
\end{tabular}\end{table}

\begin{longtable}{c|cc|c|r|r}
\caption{\textbf{Live BaTiO$_3$ $(P_z,P_x)$ plane at the clamped cubic cell (PBEsol; thread PL1).} The 66 orbit representatives of the $21\times21$ grid under $4mm$, $0\le Q_2\le Q_1$, grid index $(a,b)$, $Q_i=0.0189\,a_i$~\AA, orbit multiplicity $m$, VASP total energy $E_0$ of the 5-atom cell and $\Delta E$ relative to the cubic point. Sum of multiplicities $=441$. Every calculation converged; the independent repeat PL2 gave the same 66 energies to all printed digits.}\label{tab:si_batio3_plane_points}\\
point & $a$ & $b$ & $m$ & $E_0$ (eV) & $\Delta E$ (meV/f.u.)\\\hline\endfirsthead
point & $a$ & $b$ & $m$ & $E_0$ (eV) & $\Delta E$ (meV/f.u.)\\\hline\endhead
P00 & 0 & 0 & 1 & -39.40351 & 0.000\\
P01 & 1 & 0 & 4 & -39.40406 & -0.545\\
P02 & 1 & 1 & 4 & -39.40459 & -1.078\\
P03 & 2 & 0 & 4 & -39.40548 & -1.970\\
P04 & 2 & 1 & 8 & -39.40599 & -2.477\\
P05 & 2 & 2 & 4 & -39.40731 & -3.799\\
P06 & 3 & 0 & 4 & -39.40719 & -3.673\\
P07 & 3 & 1 & 8 & -39.40765 & -4.141\\
P08 & 3 & 2 & 8 & -39.40886 & -5.347\\
P09 & 3 & 3 & 4 & -39.41022 & -6.710\\
P10 & 4 & 0 & 4 & -39.40821 & -4.696\\
P11 & 4 & 1 & 8 & -39.40863 & -5.112\\
P12 & 4 & 2 & 8 & -39.40968 & -6.167\\
P13 & 4 & 3 & 8 & -39.41081 & -7.293\\
P14 & 4 & 4 & 4 & -39.41108 & -7.569\\
P15 & 5 & 0 & 4 & -39.40727 & -3.761\\
P16 & 5 & 1 & 8 & -39.40763 & -4.117\\
P17 & 5 & 2 & 8 & -39.40851 & -4.996\\
P18 & 5 & 3 & 8 & -39.40935 & -5.842\\
P19 & 5 & 4 & 8 & -39.40926 & -5.752\\
P20 & 5 & 5 & 4 & -39.40702 & -3.506\\
P21 & 6 & 0 & 4 & -39.40282 & 0.697\\
P22 & 6 & 1 & 8 & -39.40311 & 0.407\\
P23 & 6 & 2 & 8 & -39.40379 & -0.278\\
P24 & 6 & 3 & 8 & -39.40433 & -0.815\\
P25 & 6 & 4 & 8 & -39.40384 & -0.322\\
P26 & 6 & 5 & 8 & -39.40111 & 2.401\\
P27 & 6 & 6 & 4 & -39.39468 & 8.836\\
P28 & 7 & 0 & 4 & -39.39299 & 10.523\\
P29 & 7 & 1 & 8 & -39.39321 & 10.303\\
P30 & 7 & 2 & 8 & -39.39369 & 9.821\\
P31 & 7 & 3 & 8 & -39.39391 & 9.607\\
P32 & 7 & 4 & 8 & -39.39299 & 10.525\\
P33 & 7 & 5 & 8 & -39.38976 & 13.749\\
P34 & 7 & 6 & 8 & -39.38278 & 20.734\\
P35 & 7 & 7 & 4 & -39.37031 & 33.208\\
P36 & 8 & 0 & 4 & -39.37568 & 27.833\\
P37 & 8 & 1 & 8 & -39.37583 & 27.683\\
P38 & 8 & 2 & 8 & -39.37611 & 27.406\\
P39 & 8 & 3 & 8 & -39.37599 & 27.520\\
P40 & 8 & 4 & 8 & -39.37465 & 28.867\\
P41 & 8 & 5 & 8 & -39.37092 & 32.592\\
P42 & 8 & 6 & 8 & -39.36339 & 40.126\\
P43 & 8 & 7 & 8 & -39.35035 & 53.164\\
P44 & 8 & 8 & 4 & -39.32984 & 73.674\\
P45 & 9 & 0 & 4 & -39.34848 & 55.032\\
P46 & 9 & 1 & 8 & -39.34856 & 54.952\\
P47 & 9 & 2 & 8 & -39.34864 & 54.875\\
P48 & 9 & 3 & 8 & -39.34821 & 55.305\\
P49 & 9 & 4 & 8 & -39.34645 & 57.063\\
P50 & 9 & 5 & 8 & -39.34224 & 61.272\\
P51 & 9 & 6 & 8 & -39.33418 & 69.328\\
P52 & 9 & 7 & 8 & -39.32061 & 82.899\\
P53 & 9 & 8 & 8 & -39.29959 & 103.918\\
P54 & 9 & 9 & 4 & -39.26890 & 134.610\\
P55 & 10 & 0 & 4 & -39.30867 & 94.844\\
P56 & 10 & 1 & 8 & -39.30869 & 94.825\\
P57 & 10 & 2 & 8 & -39.30858 & 94.932\\
P58 & 10 & 3 & 8 & -39.30786 & 95.656\\
P59 & 10 & 4 & 8 & -39.30572 & 97.796\\
P60 & 10 & 5 & 8 & -39.30107 & 102.447\\
P61 & 10 & 6 & 8 & -39.29254 & 110.976\\
P62 & 10 & 7 & 8 & -39.27850 & 125.013\\
P63 & 10 & 8 & 8 & -39.25705 & 146.464\\
P64 & 10 & 9 & 8 & -39.22600 & 177.514\\
P65 & 10 & 10 & 4 & -39.18285 & 220.661\\
\end{longtable}

\subsection{Gate and job log}
Supplementary Table~\ref{tab:si_batio3_log} gives the timeline of the plane run from its audit log; the paths follow the same sequence with two job parts each. The wall time is dominated by the ISODISTORT calls in $f_1$ and by the queue: the 14 parts of the plane ended between 1.5 and 4.7~min after submission, and the 5-atom static calculations themselves take about 12~s each on one node of 32 cores.

\begin{table}[!htbp]\centering
\caption{\textbf{Timeline of the plane run PL1} (local time, 2026-10-28). Every gate passed on its first attempt; the recovery node was never entered.}\label{tab:si_batio3_log}
\small\begin{tabular}{llp{9cm}}
time & node / gate & event\\\hline
13:00:44 & $f_0$ & intake from the typed request (Claude Sonnet~5.5): BaTiO$_3$, 221$\to$99, plane\\
13:00:45 & $f_0$ & Materials Project: parent mp-2998, child mp-5986; run configuration written\\
13:00:48 & human & confirmation to submit (\texttt{y}); live executor started\\
13:00:50 & $G_0$ & PASS, 11/11 tests\\
13:00:50--13:06:01 & $f_1$ & FINDSYM + ISODISTORT~\cite{Stokes2005,Campbell2006} live; $G_1$ PASS, 5/5 tests\\
13:06:01 & $f_2$, $G_2$ & 66 orbit representatives built and certified; $G_2$ PASS, 3/3 tests\\
13:06:02 & $f_3$ & 14 PBS parts submitted; graph interrupted, waiting on jobs\\
13:07:38--13:10:44 & $f_3$ & job-end events for parts 13; 0; 1; 2, 4, 5, 7, 9, 11, 12; 3, 6, 8, 10 (no failures)\\
13:10:44 & $G_3$--$G_6$ & jobs PASS 1/1; energies PASS 4/4; profile PASS 3/3; re-derivation PASS 1/1\\
13:10:45 & end & run completed; Landau fit, minimum and $f_6$ report (351 of 351) printed\\
\end{tabular}\end{table}

\subsection{Chat interface and transcript}
\begin{figure}[!tp]
\centering
\includegraphics[width=\textwidth,height=0.8\textheight,keepaspectratio]{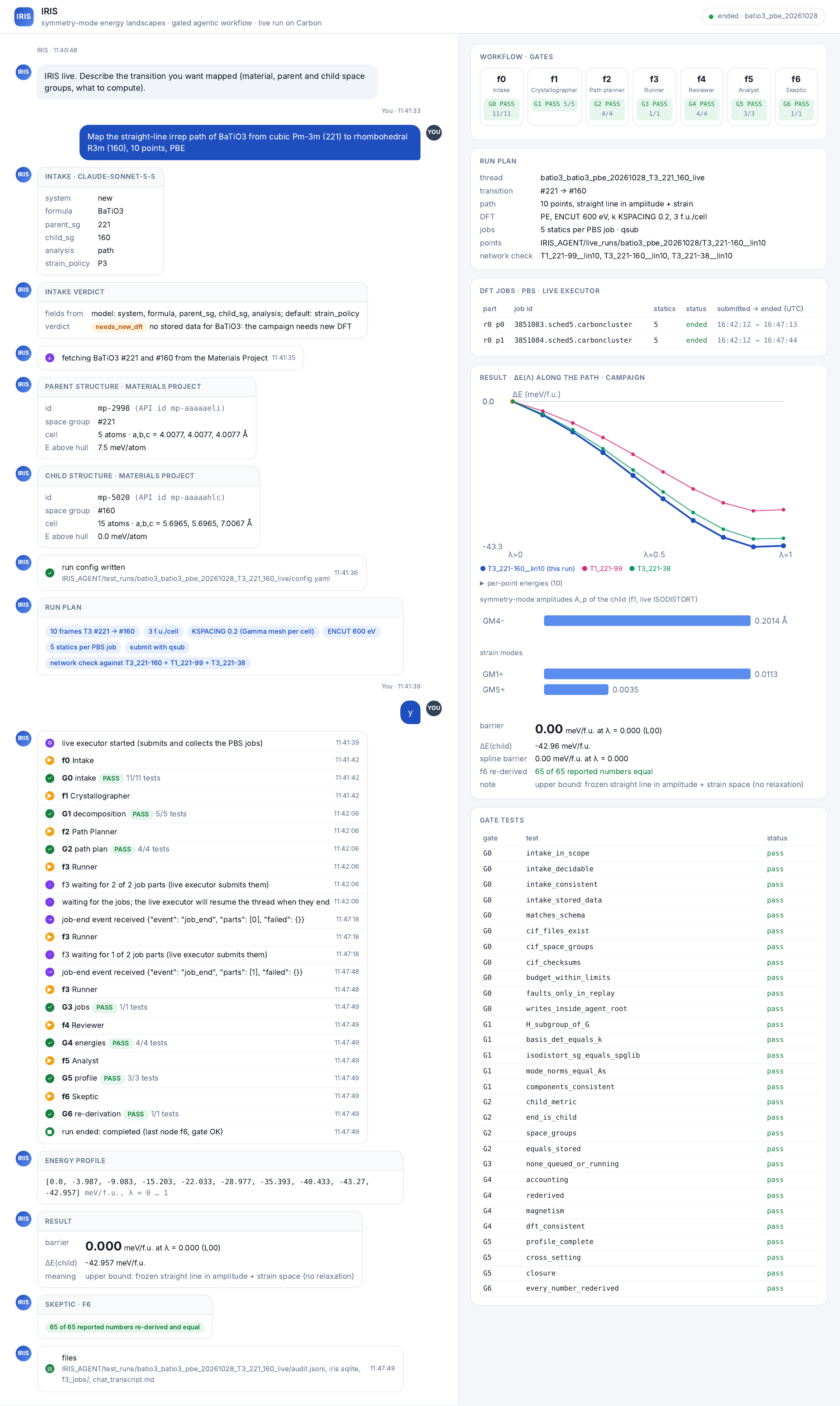}
\caption{\textbf{The IRIS chat page at the end of the live $R3m$ path run (T3)}, saved from the browser as the run ended and reproduced without editing (the page is a single HTML document served by the run; zoom in to read). Left column, the conversation: the typed request, the intake card with the model's fields, the intake verdict, the two Materials Project structure cards, the run plan, the confirmation to submit, then one line per node start, gate result and job event with its time stamp, and the result cards. Right column, the workflow board with the seven nodes and their gate results, the run plan with the network check against the finished T1 and T2 paths, the PBS job table from the live executor, the result panel, which overlays this profile on the two other paths of the campaign and shows the symmetry-mode amplitudes of the child from the live ISODISTORT call ($\Gamma_4^-$ 0.2014~\AA; strains $\Gamma_1^+$, $\Gamma_5^+$), the barrier and end-point energies and the $f_6$ count (65 of 65), and the full list of gate tests.}
\label{fig:s_live_page_path}
\end{figure}

\begin{figure}[!tp]
\centering
\includegraphics[width=\textwidth,height=0.84\textheight,keepaspectratio]{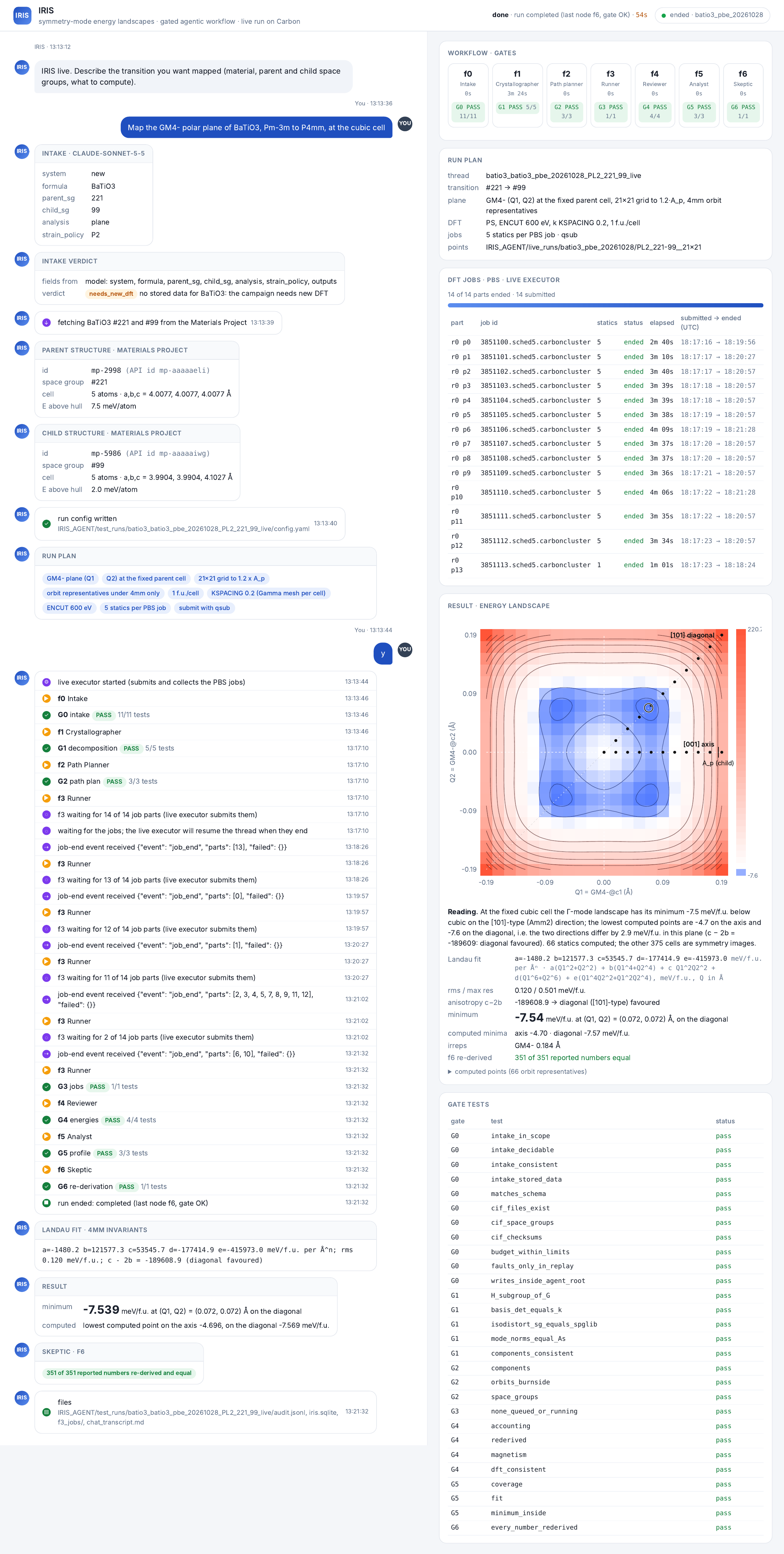}
\caption{\textbf{The IRIS chat page at the end of the live plane run} (the repeat PL2, identical to PL1 in every number), saved and reproduced as in Supplementary Fig.~\ref{fig:s_live_page_path}. The later version of the page adds a status line in the header (last node, outcome, elapsed time), the wall time of each node on the board ($f_1$, the ISODISTORT call, 3~min 24~s), a progress bar and per-part elapsed times in the job panel (14 PBS parts of five static calculations), and, in the result panel, the heat map rebuilt by symmetry from the 66 computed points (dots) with the contours of the fitted $4mm$ Landau expansion, the written reading, the coefficients, the minimum and the $f_6$ count (351 of 351).}
\label{fig:s_live_page_plane}
\end{figure}

Supplementary Figs.~\ref{fig:s_live_page_path} and~\ref{fig:s_live_page_plane} show the chat page itself at the end of two of the runs, the $R3m$ path and the plane, saved from the browser as each run ended. The transcript of the $P4mm$ path run, as written by the chat interface (\texttt{chat\_transcript.md}; paths in the deposited directory shortened), is reproduced in Supplementary Fig.~\ref{fig:s_live_transcript}. The transcripts of the other runs have the same structure and are deposited with the data.

\begin{figure}[!tp]
\centering
\providecolor{chatIris}{HTML}{1F4FBF}\providecolor{chatHuman}{HTML}{C2410C}\providecolor{chatGate}{HTML}{15803D}\providecolor{chatMute}{HTML}{6B7280}
\newcommand{\chatI}[2]{\par\noindent\hangindent=5.6em\hangafter=1\textcolor{chatMute}{[#1]}\hspace{0.5em}\textcolor{chatIris}{\textbf{IRIS}}\hspace{0.5em}#2}
\newcommand{\chatN}[2]{\par\noindent\hangindent=5.6em\hangafter=1\textcolor{chatMute}{[#1]}\hspace{0.5em}\textcolor{chatIris}{\textbf{IRIS}}\hspace{0.5em}\textcolor{chatMute}{#2}}
\newcommand{\chatG}[2]{\par\noindent\hangindent=5.6em\hangafter=1\textcolor{chatMute}{[#1]}\hspace{0.5em}\textcolor{chatIris}{\textbf{IRIS}}\hspace{0.5em}\textcolor{chatGate}{#2}}
\newcommand{\chatH}[2]{\par\noindent\hangindent=5.6em\hangafter=1\textcolor{chatMute}{[#1]}\hspace{0.5em}\textcolor{chatHuman}{\textbf{you}}\hspace{0.5em}\textcolor{chatHuman}{#2}}
\begin{tcolorbox}[enhanced,width=\textwidth,colback=white,colframe=chatIris!60,colbacktitle=chatIris!10,coltitle=black,
  fonttitle=\sffamily\bfseries\footnotesize,title={IRIS chat \,\textcolor{chatMute}{\mdseries live run T1 \,\textperiodcentered\, BaTiO$_3$ $Pm\bar{3}m\to P4mm$ \,\textperiodcentered\, Carbon, 2026-10-28}},
  boxrule=0.5pt,arc=1.5mm,left=2mm,right=2mm,top=1.5mm,bottom=1.5mm,fontupper=\scriptsize\ttfamily]
\chatI{10:38:50}{IRIS live. Describe the transition you want mapped (material, parent and child space groups, what to compute).}
\chatI{10:40:18}{intake (claude-sonnet-5-5): system=new, formula=BaTiO3, parent\_sg=221, child\_sg=99, analysis=path, strain\_policy=P3}
\chatI{10:40:18}{sources: model: system, formula, parent\_sg, child\_sg, analysis; default: strain\_policy | verdict: needs\_new\_dft --- no stored data for BaTiO3: the campaign needs new DFT}
\chatI{10:40:18}{fetching BaTiO3 \#221 and \#99 from the Materials Project -> live\_inputs/batio3\_pbe\_20261028/T1}
\chatI{10:40:20}{\quad parent: mp-2998 \#221, 5 atoms, a,b,c = [4.0077, 4.0077, 4.0077] \AA, E above hull 7.5 meV/atom}
\chatI{10:40:20}{\quad child: mp-5986 \#99, 5 atoms, a,b,c = [3.9904, 3.9904, 4.1027] \AA, E above hull 2.0 meV/atom}
\chatI{10:40:20}{config written: test\_runs/\ldots\_T1\_221\_99\_live/config.yaml}
\chatI{10:40:20}{plan: 10 frames T1 \#221 -> \#99, 1 f.u./cell, KSPACING 0.2 (Gamma mesh per cell), ENCUT 600 eV, 5 statics per PBS job, submit with qsub}
\chatH{10:40:27}{y}
\chatI{10:40:27}{live executor started (pid 2502509)}
\chatN{10:40:29}{f0 Intake starts}
\chatG{10:40:29}{G0 (intake): PASS, 11/11 tests}
\chatN{10:40:29}{f1 Crystallographer starts}
\chatG{10:43:59}{G1 (decomposition): PASS, 5/5 tests}
\chatN{10:43:59}{f2 Path Planner starts}
\chatG{10:43:59}{G2 (path plan): PASS, 4/4 tests}
\chatN{10:43:59}{f3 Runner starts}
\chatI{10:43:59}{f3 waiting for 2 of 2 job parts (live executor submits them)}
\chatI{10:46:35}{resumed with \{"event": "job\_end", "parts": [0], "failed": \{\}\}}
\chatI{10:47:05}{resumed with \{"event": "job\_end", "parts": [1], "failed": \{\}\}}
\chatG{10:47:05}{G3 (jobs): PASS, 1/1 tests}
\chatN{10:47:05}{f4 Reviewer starts}
\chatG{10:47:05}{G4 (energies): PASS, 4/4 tests}
\chatN{10:47:05}{f5 Analyst starts}
\chatG{10:47:05}{G5 (profile): PASS, 3/3 tests}
\chatN{10:47:05}{f6 Skeptic starts}
\chatG{10:47:05}{G6 (re-derivation): PASS, 1/1 tests}
\chatI{10:47:05}{run ended: completed (last node f6, gate OK)}
\chatI{10:47:05}{profile dE(lambda) meV/f.u.: [0.0, -2.85, -6.41, -10.726, -15.707, -20.965, -26.023, -30.158, -32.552, -32.195]}
\chatI{10:47:05}{barrier 0.000 meV/f.u. at lambda = 0.000 (L00); dE(end) = -32.195 meV/f.u.; upper bound: frozen straight line in amplitude + strain space (no relaxation)}
\chatI{10:47:05}{f6: 63 of 63 reported numbers re-derived and equal}
\end{tcolorbox}
\caption{\textbf{Chat transcript of the live $P4mm$ path run (T1)}, as written by the chat interface (\texttt{chat\_transcript.md}; directory paths shortened). Grey, timestamps and node starts; blue, IRIS; orange, the human's only input; green, gate results. Identifiers are shown in their classic form; in this first run the page printed the API's alphabetic identifiers (mp-aaaaaeli, mp-aaaaaiwg), which were decoded in the later runs.}\label{fig:s_live_transcript}
\end{figure}

\subsection{What the live runs found}\label{sec:si_live_found}
Three defects surfaced in the live rehearsals before the reported runs and were fixed in code, not by stored rules: the basis-determinant test of $G_1$ rejected the centred conventional cells of $Amm2$ and hexagonal $R3m$ (it now accepts $\det=k\,Z_\mathrm{child}/Z_\mathrm{parent}$ alongside $\det=k$), and the symmetry-mode forms of ISODISTORT's hexagonal $R3m$ output (\texttt{Dx,-Dx,Dz} on the 9b site) had been applied as independent components, which broke the mirror of the rebuilt frame and was caught by $G_1$ as a species-count mismatch; and the rotation that generates the second component of the plane was first taken about the origin of the frame, which is not a rotation axis of the perovskite when ISODISTORT shifts the origin, so the mock run of the plane stopped at $f_2$ until the rotation was taken about an atomic site. A first path run with PBEsol on the PBE-relaxed cells put the $P4mm$ end 3.7~meV per formula unit \emph{above} cubic, a smooth and well-converged profile that is physically consistent, since PBEsol prefers a smaller volume than the PBE cells provide, and wrong for the question; the functional was changed to PBE for the paths. As discussed in the main text, none of these is a gate failure in the sense of Extended Data Table~\Mref{tab:errors}: the gates check crystallography, convergence and fits, not whether the request was answered with the intended physics.


\subsection{Supplementary Video 1}
\textbf{Supplementary Video 1 $|$ A prospective IRIS run on BaTiO$_3$.} Screen recording of the live campaign on the Carbon cluster, captured as one frame every 2~s and played at 20 times real time. Left, the IRIS chat; right, the workflow board with each node's state, gate result and wall time, the run plan, the PBS job table and the result panel (energy profile for a path, heat map with the fitted Landau contours for the plane). The request is typed at the start of each run; $f_0$ fetches the structures from the Materials Project, $f_1$ calls ISODISTORT, $f_2$ builds and certifies the frames, $f_3$ submits the PBS jobs and collects them, $f_4$--$f_6$ check, fit and re-derive. No frame was removed or reordered; the only human input is the confirmation to submit.

\section{Code and data repository}\label{sec:si_repo}

The code, the frozen evaluation set and the results behind every evaluation table and figure of this work are in one repository, \url{https://github.com/sukritimanna/IRIS} (MIT licence; the version used for this work will be archived on Zenodo under a DOI on publication). Supplementary Table~\ref{tab:si_repo} maps its directories to the sections of the paper, and Supplementary Table~\ref{tab:si_repo_results} maps each evaluation table and figure to the file that produced it. The stored DFT campaigns (structures, frames, energies, audit logs) and the run directories of the live BaTiO$_3$ campaign are deposited separately (Data availability); every replay in the repository reads them from the path \texttt{paths.repo\_root} of its configuration.

\begin{table}[!htbp]
\centering\scriptsize
\caption{Layout of the IRIS repository.}
\label{tab:si_repo}
\begin{tabularx}{\linewidth}{@{}L{3.6cm}X@{}}
\toprule
Directory & Contents (paper section)\\
\midrule
\texttt{iris\_agent/} & the workflow: nodes $f_0$--$f_6$, gates $G_0$--$G_6$ (\texttt{gates.py}), recovery, the replay and live engines, the chat and web interface (Methods; Supplementary Sections~\ref{sec:si_gates}, \ref{sec:si_live})\\
\texttt{rules/}, \texttt{recovery.yaml}, \texttt{recovery\_live.yaml} & the mode-selection rules R1--R5 (\texttt{select\_modes.py}) and the stored recovery rule table with the sub-agent menus (Supplementary Tables~\ref{tab:rules}, \ref{tab:subagents})\\
\texttt{prompts/} & the intake and gate sub-agent prompts, verbatim as evaluated (Supplementary Section~\ref{sec:si_prompts})\\
\texttt{configs/} & the 13 run configurations of the replays (Extended Data Table~\Mref{tab:reruns}) and the BaTiO$_3$ live configurations\\
\texttt{eval/} & the frozen evaluation set: 53 sub-agent cases, 53 faults, the model list, \texttt{FROZEN.sha256}\\
\texttt{policies/} & rules only, decision tree and retrieval, frozen before scoring (\texttt{FROZEN.sha256})\\
\texttt{consequences/} & the option replay, \texttt{fork.py} (Supplementary Table~\ref{tab:si_C})\\
\texttt{intake/} & the plain-language intake, the 50 frozen requests, the keyword lookup, the rule-based parser (\texttt{parser.py}, \texttt{PARSER.sha256}; Section~\ref{sec:si_parser})\\
\texttt{eval\_extra/} & post-freeze analyses that wrap the frozen evaluation without changing it: leave-one-system-out (\texttt{loso.py}, \texttt{loso\_llm.py}; Section~\ref{sec:si_loso}), the decoy remedy test (\texttt{decoy\_remedy.py}), the eligibility conditions (\texttt{eligibility.py}; Section~\ref{sec:si_eligibility}) and the cost summary (\texttt{cost\_summary.py}; Section~\ref{sec:si_cost})\\
\texttt{tests/} & the pytest suite: replay regression, gates, recovery, live mode (148 tests without the stored campaigns, 158 with them)\\
\texttt{results/} & the CSVs behind the evaluation (Supplementary Table~\ref{tab:si_repo_results})\\
\texttt{scripts/figures/} & the script that draws main-text Fig.~\Mref{fig:agentic} from \texttt{results/}\\
\texttt{live\_demo/} & the setup, validation, recording and archiving scripts and the runbook of the prospective campaign (Supplementary Section~\ref{sec:si_live}); cluster identifiers are placeholders\\
\texttt{docs/} & the changelog of every change made after the evaluation freeze of 19 October 2026, the manual checks and the API-key setup\\
\bottomrule
\end{tabularx}
\end{table}

\begin{table}[!htbp]
\centering\scriptsize
\caption{Results files behind the evaluation tables and figures.}
\label{tab:si_repo_results}
\begin{tabularx}{\linewidth}{@{}L{5.2cm}X@{}}
\toprule
File (under \texttt{results/}) & Table or figure\\
\midrule
\texttt{frozen\_2026-10-19/A/A\_cases.csv}, \texttt{A\_runs.csv} & deterministic deciders on the cases and in the faulted runs (Supplementary Table~\ref{tab:si_A}; Fig.~\Mref{fig:agentic}b,c)\\
\texttt{summary/S5\_3\_*.csv} & the nine models on the 53 cases (Supplementary Table~\ref{tab:si_models}; Supplementary Figs.~\ref{fig:s_agentic_models}, \ref{fig:s_agentic}, \ref{fig:s_agentic_robust})\\
\texttt{summary/E1\_runs.csv}, \texttt{E2\_faults\_gated.csv}, \texttt{E3\_*.csv}, \texttt{E4\_faults\_ungated.csv}, \texttt{E7\_*.csv} & clean replays, faulted runs with and without gates, interventions, audit completeness and kill--resume (Extended Data Table~\Mref{tab:reruns}; Supplementary Tables~\ref{tab:si_faults}, \ref{tab:si_interventions})\\
\texttt{final\_2026-10-29/C/C\_options.csv}, \texttt{C2\_options.csv} & every recovery option executed, gates on and gates off after the fork (Supplementary Tables~\ref{tab:si_C}, \ref{tab:si_farr}, \ref{tab:si_stack}; Fig.~\Mref{fig:agentic}d,e); the first replay is \texttt{frozen\_2026-10-19/C/}\\
\texttt{summary/E9\_gate\_ablation.csv} (per-fault verdicts and the extra runs on an unusable result); \texttt{final\_2026-10-29/\allowbreak{}E9\_gate\_ablation.csv} & gate ablation (Supplementary Table~\ref{tab:si_ablation})\\
\texttt{LOSO/loso\_summary.csv}, \texttt{loso\_cases.csv}, \texttt{loso\_llm\_summary.csv}, \texttt{loso\_llm\_cases.csv}, \texttt{loso\_prompts.json} & transfer across systems (Supplementary Table~\ref{tab:si_loso})\\
\texttt{DECOY/decoy\_remedy\_summary.csv}, \texttt{decoy\_remedy\_cases.csv}, \texttt{remedy\_prompts.json} & the post-hoc prompt repair on the decoys (Supplementary Table~\ref{tab:si_decoys})\\
\texttt{ELIGIBILITY/eligibility\_\allowbreak{}cases\_run1.csv}, \texttt{eligibility\_\allowbreak{}summary\_run1.csv} & the eligibility run as made; the corrected conditions are applied at scoring by \texttt{eval\_extra/eligibility.py --summarize} and by the figure script (Section~\ref{sec:si_eligibility}; main-text Fig.~\Mref{fig:agentic}c)\\
\texttt{COST/cost\_summary.csv} & tokens, latency and cost per experiment and model (Supplementary Table~\ref{tab:si_cost})\\
\texttt{frozen\_2026-10-19/E/E\_calls.csv}, \texttt{E\_summary.csv}; \texttt{E/parser\_v2/}; \texttt{summary/E\_summary\_v5.csv} & intake: the four models and the keyword lookup; the rule-based parser (v2; v1 with \texttt{PARSER\_V1=1}); the summary with the parser row read by the figure script (Supplementary Tables~\ref{tab:si_E}, \ref{tab:si_parser}; Fig.~\Mref{fig:agentic}a)\\
\bottomrule
\end{tabularx}
\end{table}

\clearpage
\IRISsibibliography

\end{document}